\documentclass[10pt,twoside]{epnt1p}
\usepackage{color}
\usepackage{epsfig, floatflt}
\usepackage{graphics}
\usepackage[normalem]{ulem}

\usepackage{amssymb}
\usepackage{amsmath}
\usepackage{array}

\usepackage{xspace}   %%FRK needed for macro \DS

\usepackage{textcomp}

\usepackage{url}

\newcommand{\lsim}{\raisebox{-0.13cm}{~\shortstack{$<$ \\[-0.07cm] $\sim$}}~}
\newcommand{\gsim}{\raisebox{-0.13cm}{~\shortstack{$>$ \\[-0.07cm] $\sim$}}~}

\def\nm{\nu_\mu}

\def\nmnt{\nu_\mu \leftrightarrow \nu_\tau}

\def\s2t{\sin^2 2 \theta}
\def\Dm2{\Delta m^2}

\def\ll{\langle}

\newcommand{\Dmq}{\Delta m^2}
\newcommand{\Dlt}{\Delta\delta}

\newcommand{\gwig}{\mbox{\,\raisebox{.3ex}
    {$>$}$\!\!\!\!\!$\raisebox{-.9ex}{$\sim$}}\,}

\newcommand{\egzk}{E_{\rm GZK}}
\newcommand{\rarr}{\rightarrow}
\newcommand{\sig}{\sigma^{\rm CC}_{\nu N}}

\newcommand{\Esh}{E^{\rm sh}_{\rm th}}
\newcommand{\Eth}{E^\tau_{\rm th}}

\newcommand{\lmin}{l_{\rm min}}

\newcommand{\dmin}{d_{\rm min}}
\newcommand{\dmax}{d_{\rm max}}

\newcommand{\zthin}{z_{\rm thin}}

\newcommand{\zc}{z_{\rm cloud}}

\newcommand{\numu}{\nu_\mu}
\newcommand{\nutau}{\nu_\tau}
\newcommand{\nue}{\nu_e}

\newcommand{\bwide}{\begin{widetext}}
\newcommand{\ewide}{\end{widetext}}
\newcommand{\barr}{\begin{array}}
\newcommand{\earr}{\end{array}}
\newcommand{\ba}[1]{\begin{eqnarray} \label{(#1)}}
\newcommand{\ea}{\end{eqnarray}}

\newcommand{\vf}{\varphi}

\renewcommand*{\eqref}[1]{Eq.~(\ref{eq:#1})}

\newcommand*{\figref}[1]{Fig.~\ref{fig:#1}}
\newcommand*{\figlab}[1]{\label{fig:#1}}

\newcommand*{\seclab}[1]{\label{sec:#1}}
\newcommand{\Omit}[1]{}
\def\beq{\begin{equation}}
\def\eeq{\end{equation}}
\def\bea{\begin{eqnarray}}
\def\eea{\end{eqnarray}}

\def\VYP#1#2#3{{\bf #1}, #3, #2}  % Volume, page (Year)

\def\PLB#1#2#3{Phys.~Lett.~B~\VYP{#1}{#2}{#3}}

\def\PRD#1#2#3{Phys.~Rev.~D~\VYP{#1}{#2}{#3}}
\def\PRL#1#2#3{Phys.~Rev.~Lett.~\VYP{#1}{#2}{#3}}

\newcommand{\E}{\rm EHEC\nu}
\newcommand{\C}{\rm {C}\nu{\rm B}}

\newcommand{\DS}[1]{$\mathsf{#1}$\xspace}
\newcommand{\fracnew}[2]
           {\protect\frac{{#1}_{\protect\vphantom{!_a}}}{{#2}^{\protect\vphantom{a}}}}

\newcommand{\bdi}{\begin{displaymath}}
\newcommand{\edi}{\end{displaymath}}
\newcommand{\bfi}{\begin{figure}}
\newcommand{\efi}{\end{figure}}

\newcommand{\beqa}{\begin{eqnarray}}
\newcommand{\eeqa}{\end{eqnarray}}

\newcommand {\BN}     {Bogoliubov--Nambu}

\newcommand {\SM}     {Standard Model}
\renewcommand{\tabcolsep}{0.75pc}  % enlarge column spacing
\renewcommand{\arraystretch}{1.1}  % enlarge line spacing
\newcommand{\ud}{\mathrm{d}}

\begin{document}
\pagenumbering{roman}

%%%%%%%%%%%%%%%%%%%%%%%%%%%%%%%%%%%%%%%%%%%%%%%%%%%%%% Front Page 1%%%%%%%%%%%%%%
%\vspace*{-3cm}
%\hspace*{-4.3cm}
%\includegraphics{cover.pdf}

\newpage
%%%%%%%%%%%%%%%%%%%%%%%%%%%%%%%%%%%%%%%%%%%%%%%%%%%%%% Front Page 1%%%%%%%%%%%%%%
\thispagestyle{empty}

\vspace*{3cm}
\begin{center}
{\Large  Proceedings of the first\\

Workshop on Exotic Physics with Neutrino Telescopes\\
\vspace*{1cm}

 EPNT06}

\vspace*{1cm}
  Uppsala, Sweden, September 20-22, 2006

%\vspace*{6cm}
\vfill
 editor\\
Carlos P. de los Heros\\
Uppsala University\\

%\vspace*{5cm}
%\vfill
%\includegraphics[height=0.2\linewidth,width=\linewidth]{Topimage.eps}
\end{center}

\newpage
\thispagestyle{empty}
\enlargethispage{1cm}
\vspace*{18cm}
\noindent Proceedings of the first Workshop on Exotic Physics with Neutrino
Telescopes\\
\noindent Carlos P. de los Heros, editor\\
\noindent Uppsala University, 2007.\\
\noindent ISBN 978-91-506-1913-3\\

\noindent cover photo: ``Untitled'', oil and mixed media on
canvas. \copyright\, C. de los Heros 2003.

%%%%%%%%%%%%%%%%%%%%%%%%%%%%%%%%%%%%%%%%%%%%%%%%%%%%%% Foreword %%%%%%%%%%%%%%
\newpage
\section*{Foreword}

The first {\it Workshop on Exotic Physics with Neutrino Telescopes} 
was held in Uppsala, Sweden, between September 20-22, 2006. 
 The main idea behind the organization of the workshop was to discuss
 the possibilities of using large--scale neutrino telescopes to study physics
topics beyond the search for astrophysical point sources of neutrinos, which 
constitutes the main rationale for the construction of such detectors.
 The label 'exotics' is therefore to be understood in this spirit. 
Indeed many of the topics discussed in the workshop 
were not 'exotic' at all in a generic sense.\\

 The workshop was based on several review talks giving an overview 
of relevant subjects, and contributed talks from experiments
and theorists touching upon more detailed developments. \\

 The smaller neutrino detectors, MACRO, Baksan and Super-K and the
 large--scale neutrino telescopes already in operation, Baikal and
 AMANDA, reviewed the status of their analyses and results on 
 topics ranging from particle searches to tests of fundamental laws, 
 proving that their detection techniques can produce competitive limits  
in comparison with present accelerator results. The next--generation 
 experiments, IceCube at the South Pole and ANTARES and NESTOR in the Mediterranean
 (with their km$^3$ successor, KM3NET, already under R\&D), will be able to improve the current 
limits by a few orders of magnitude, although we all expect that such 
detectors will lead to a discovery rather than limit improvements.\\

 The theory talks provided  guidelines for experimental searches,
 and covered the signatures expected in neutrino telescopes from 
different dark matter candidates, micro black holes, models with extra dimensions, 
non--standard neutrino oscillation scenarios and 
new neutrino interactions. A wealth of 
possibilities for neutrino detectors to search for.\\

 The view of the 'exotic' universe is a comprehensive one, with
 results from satellite, accelerator and laboratory experiments
 complementing each  other. The program included reviews from
 accelerator,  underground labs and space--based projects whose research topics
 overlap with those of neutrino telescopes. \\

 The workshop was held in the Polhem auditorium of the {\AA}ngstr\"om 
Laboratory, the complex housing all physics departments of Uppsala
 University, and brought together about 50 scientists from 16
 countries.\\

 The support from the Swedish Research Council (Vetenskapsr{\aa}det)
 and the Faculty of Physics of Uppsala University are kindly
 acknowledged.\\

\noindent Carlos P. de los Heros\\
\noindent Editor\\

\noindent Uppsala, October 2006.

%%%%%%%%%%%%%%%%%%%%%%%%%%%%%%%%%%%%%% reset counters %%%%%%%%%%%%%%
%%
\setcounter{section}{0}
\setcounter{subsection}{0}
\setcounter{figure}{0}
\setcounter{table}{0}
\newpage

%%%%%%%%%%%%%%%%%%%%%%%%%%%%%%%%%%%%%%%%%% committees %%%%%%%%%%%%%%
\section*{Local Organizing Committee}
\begin{figure}[th]
\begin{minipage}[t]{0.5\linewidth}
\flushleft
Carlos P. de los Heros (chair)\\
\noindent Olga Botner
\end{minipage}\hfill
\begin{minipage}[t]{0.5\linewidth}
\flushleft
\noindent Allan Hallgren\\
\noindent Inger Ericson (secretary)
\end{minipage}
\end{figure}

\vspace*{1cm}

\section*{List of Participants}
\begin{figure}[th]
\begin{minipage}[t]{0.5\linewidth}
\flushleft
M. Ahlers,  Germany   \\                            
I. Albuquerque,   Brazil  \\                        
R. Aloisio,    Italy  \\                    
A. Belias,   Greece \\                  
O. Botner,     Sweden  \\             
A. Bouchta,    Sweden \\                
T. Burgess,   Sweden\\         
E. Castorina,   Italy  \\                           
B. Christy,    USA  \\                        
J. Conrad,    Sweden  \\                                
A. Davour,   Sweden     \\                 
C. P. de los Heros,  Sweden    \\ 
Z.~A.~Dzhilkibaev,    Russia  \\   
J. Edsj\"o,  Sweden \\                    
R. Ehrlich,    USA  \\   
E. Flyckt,   France   \\    
M. Giorgini,   Italy \\  
M.~C.~Gonz\'alez-Garc\'{i}a,  USA/Spain  \\ 
A.   Hallgren,  Sweden   \\            
K.   Han,   New Zealand  \\                   
D. Hardtke,  USA   \\                                  
D. Hooper,  USA    \\                     
D. Hubert,  Belgium   \\                             
P. O. Hulth,  Sweden   \\                   
K. Hultqvist,   Sweden 
\end{minipage}\hfill
\begin{minipage}[t]{0.5\linewidth}
\flushleft
J. I. Illana,   Spain    \\                            
G. Ingelman,  Sweden  \\                        
D. Jonsson,  Sweden               \\                    
J. Kameda,   Japan   \\                  
F.  Klinkhamer,   Germany    \\             
M. Kowalski,  Germany      \\                
A.  Kusenko,    USA \\         
A. Lipniacka,    Norway      \\                           
J. Madsen,   Denmark  \\                        
A.  Olivas,   USA     \\                                
E. Osipova,      Russia   \\                 
S. Palomares-Ruiz,    UK       \\ 
A. Pohl,      Sweden \\   
A. Rizzo,      Belgium \\                    
L. Schrempp,      Germany  \\   
O.  Scholten,    The Netherlands  \\    
U. Schwanke,   Germany  \\  
T. Sumner,   UK     \\ 
O. Suvorova, Russia  \\            
Y. Takenaga,  Japan    \\                   
U. Tengblad,     Sweden  \\                                  
C. Walck, Sweden   \\                     
G. Wikstrom,  Sweden  \\                             
H. Wissing, Germany  \\                   
\end{minipage}
\end{figure}

\newpage
%%%%%%%%%%%%%%%%%%%%%%%%%%%%%%%%%%%%%%%%%% programme %%%%%%%%%%%%%%

\section*{Workshop Programme}
\vspace*{0.5cm}
{\bf Wednesday sept 20}\\
   08:30-09:00 REGISTRATION\\
   09:00-09:15 Welcome and introduction (C. de los Heros)\\
   {\it chair: C. de los Heros}\\
   09:15-09:50 Review of the status and perspectives of accelerator searches (A. Lipniacka)\\
   09:55-10:30 Review of direct DM and other particle searches (T. Sumner)\\
   10:30-11:00 registration(cont.) + coffee/tea\\
   {\it chair: D. Hardtke}\\
   11:00-11:20 Summary of searches for exotic phenomena with Super-K (J. Kameda)\\
   11:25-11:45 Summary of MACRO results on exotic physics (M. Giorgini)\\
   11:50-12:10 Results from Baksan (O. Suvorova)\\
   12:15-12:35 Calculating upper limits and claiming discovery in searches for exotic physics (J. Conrad)\\
   12:40-13:45 LUNCH\\
   {\it chair: T. Sumner}\\
   13:45-14:20 Kaluza-Klein dark matter (D. Hooper)\\
   14:25-14:45 Nearly vertical muons from the lower hemisphere in the Baikal neutrino experiment (Z. Dzhilkibaev)\\
   14:50-15:10 WIMP searches with AMANDA (D. Hubert)\\
   15:15-15:45 coffee/tea\\
   {\it chair: A. Kusenko}\\
   15:45-16:05 DARKSUSY: a dark matter Monte Carlo (J. Edsj\"o)\\
   16:10-16:30 Direct detection of physics betond the Standard Model (I. Albuquerque)\\
   16:35-17:10 Review of DM searches with gamma ray telescopes and balloons (U. Schwanke)\\
   17:15-17:45 Strangelets and nuclearites--an overview (J. Madsen)\\

\vspace*{0.5cm}

{\bf Thursday sept 21}\\
   {\it chair: U. Schwanke}\\
   09:15-09:50 Q-ball properties and signatures (A. Kusenko)\\
   09:55-10:15 Q-ball searches with Super-K (Y. Takenaga)\\
   10:20-10:40 Searches for magnetic monopoles, Q-balls and nuclearites with AMANDA-II (H. Wissing)\\
   10:45-11:15 coffee/tea\\
   {\it chair: A. Hallgren}\\
   11:15-11:35 Search for relativistic magnetic monopoles with the Baikal neutrino telescope (E. Osipova)\\
   11:40-12:00 Exotic physics searches with ANTARES (E. Castorina)\\
   12:05-12:25 Perspectives of searches for monopoles, WIMPs and exotica with IceCube (D. Hardtke)\\
   12:30-12:50 NESTOR (A. Belias)\\
   12:55-14:10 LUNCH\\
   {\it chair: P. O. Hulth}\\
   14:10-14:45 UHE neutrino signatures in top-down scenarios (R. Aloisio)\\
   14:50-15:10 Radio detection of UHE neutrinos off the Moon (O. Scholten)\\
   15:15-15:35 Probing the variation of relic neutrino masses at neutrino observatories (L. Schrempp)\\
   15:40-16:10 coffee/tea\\
   {\it chair: K. Hultqvist}\\
   16:10-16:30 Neutrino and neutralino showers at horizons (D. Fargion)\\
   16:35-16:55 Tests of Lorentz invariance with AMANDA-II (A. Olivas)\\
   17:00-17:35 Micro black hole properties and signatures (M. Kowalski)\\
   19:30 workshop dinner\\
\vspace*{0.5cm}

{\bf Friday sept 22}\\
   {\it chair: I. Albuquerque}\\
   09:15-09:50 Non-standard neutrino oscillation scenarios (M. C. Gonz\'alez-Garc\'{i}a)\\
   09:55-10:15 Emergent relativity: neutrinos as probes of the underlying theory (F. Klinkhamer)\\
   10:20-10:40 TeV gravity at neutrino telescopes (J. I. Illana)\\
   10:45-11:15 coffee/tea\\
   {\it chair: C. de los Heros}\\
   11:15-11:35 Neutrino cross sections above 1019 eV (S. Palomares-Ruiz)\\
   11:40-12:00 Strongly interacting neutrinos at ultra-high energies (M. Ahlers)\\
   12:05    closing remarks (C. de los Heros)\\
   12:10 LUNCH\\
   14:00-15:30 Guided visit to the university museum (Gustavianum) and
   the cathedral\\

\newpage
%%%%%%%%%%%%%%%%%%%%%%%%%%%%%%%%%%%%%%%%%%%%%%%%%%%%%% TOC %%%%%%%%%%%%%%%%%%%%%%%%%%
\section*{Table of Contents}
\vspace*{0.5cm}
\noindent {\it Foreword} \hfill iii\par
\noindent {\it Local Organizing Committee} \hfill iv\par
\noindent {\it List of Participants} \hfill iv\par
\noindent {\it Workshop Programme} \hfill v\\

\noindent {\bf Review of the status and perspectives of accelerator
  searches\footnote{No contribution received}}\hfill 1\par
A. Lipniacka \\[0.2cm]
\noindent {\bf Direct dark matter searches} \hfill 2\par
T.~J. Sumner \\[0.2cm]
\noindent {\bf Summary of searches for exotic phenomena with Super-K}\hfill 9\par
J. Kameda \\[0.2cm]
\noindent {\bf Summary of MACRO results on exotic physics}\hfill 14\par
M. Giorgini \\[0.2cm]
\noindent {\bf Results with the Baksan neutrino telescope}\hfill 19\par
M.~M. Boliev, A.~V. Butkevich, S.~P. Mikheyev, O.~V. Suvorova\\[0.2cm]
\noindent {\bf Discovery and upper limits in search for } \par
\noindent {\bf exotic physics with neutrino telescopes}\hfill 24\par
J. Conrad \\[0.2cm]
\noindent {\bf Kaluza-Klein dark matter }\hfill 29\par
D. Hooper  \\[0.2cm]
\noindent {\bf Nearly vertical muons from the lower hemisphere} \par 
\noindent {\bf in the Baikal neutrino experiment}\hfill 34\par 
K. Antipin {\it et al} \\[0.2cm]
\noindent {\bf Neutralino searches with AMANDA and Icecube--}\par 
\noindent {\bf past, present and future}\hfill 39\par 
D. Hubert \\[0.2cm]
\noindent {\bf DARKSUSY: a dark matter Monte Carlo$^{1}$}\hfill 44\par 
J. Edsj\"o \\[0.2cm]
\noindent {\bf Direct detection of physics beyond the Standard Model}\hfill 45\par 
I.~F.~M. Albuquerque \\[0.2cm]
\noindent {\bf Dark matter searches with balloon experiments and}\par
\noindent {\bf gamma ray telescopes}\hfill 50\par
U. Schwanke \\[0.2cm]
\noindent {\bf Strangelets, nuclearites, Q-balls--a brief overview}\hfill 57\par 
J. Madsen \\[0.2cm]
\noindent {\bf Properties and signatures of supersymmetric Q-balls}\hfill 63\par 
A. Kusenko \\[0.2cm]
\newpage
\noindent {\bf Search for neutral Q-balls in Super-Kamiokande II }\hfill 70\par
Y. Takenaga  \\[0.2cm]
\noindent {\bf Status of searches for magnetic monopoles, Q-balls and} \par
\noindent {\bf nuclearites with the AMANDA-II detector}\hfill 75\par 
A. Pohl, H. Wissing \\[0.2cm]
\noindent {\bf Search for relativistic magnetic monopoles with the } \par
\noindent {\bf Baikal neutrino telescope}\hfill 80\par 
K. Antipin {\it et al} \\[0.2cm]
\noindent {\bf Dark matter searches with the ANTARES neutrino telescope}\hfill 85\par
E. Castorina \\[0.2cm]
\noindent {\bf Exotic physics with IceCube }\hfill 89 \par
D. Hardtke \\[0.2cm]
\noindent {\bf NESTOR}\hfill 94\par
A. Belias \\[0.2cm]
\noindent {\bf Ultra high energy neutrino signatures in top-down scenarios}\hfill 99\par 
R. Aloisio \\[0.2cm]
\noindent {\bf Radio detection of UHE neutrinos off the Moon}\hfill 106\par 
O. Scholten  {\it et al} \\[0.2cm]
\noindent {\bf Probing the variation of relic neutrino masses with} \par
\noindent {\bf extremely high-energy cosmic neutrinos}\hfill 111\par 
L. Schrempp \\[0.2cm]
\noindent {\bf Black holes at neutrino telescopes: bounds on TeV-gravity}\hfill 116\par 
M. Kowalski \\[0.2cm]
\noindent {\bf Non-standard neutrino oscillations at IceCube}\hfill 122\par
M.~C. Gonz\'alez-Garc\'{i}a \\[0.2cm]
\noindent {\bf Testing Lorentz invariance using atmospheric neutrinos } \par
\noindent {\bf and AMANDA-II}\hfill 131\par 
J.~L. Kelley \\[0.2cm]
\noindent {\bf Emergent relativity: neutrinos as probe of the underlying theory }\hfill 136\par
F.~R. Klinkhamer \\[0.2cm]
\noindent {\bf Probing TeV gravity at neutrino telescopes}\hfill 142\par 
J.~I. Illana, M. Masip, D. Meloni \\[0.2cm]
\noindent {\bf Inferring neutrino cross sections above 10$^{19}$ eV}\hfill 147\par 
S. Palomares-Ruiz \\[0.2cm]
\noindent {\bf Exotic neutrino interactions in cosmic rays}\hfill 152\par 
M. Ahlers

%%%%%%%%%%%%%%%%%%%%%%%%%%%%%%%% CONTRIBUTIONS  %%%%%%%%%%%%%%%%%%%%%%%%%%
% reset the number page and start with arabic
\newpage
\setcounter{page}{1}
\pagenumbering{arabic}

\begin{frontmatter}

\title{Review of the status and perspectives of accelerator searches}

\author[address1]{Anna Lipniacka}

\address[address1]{Department of Particle Physics, University of Bergen, Norway}

\end{frontmatter}
\vspace*{5cm}
\begin{center}
 No contribution received
\end{center}

\setcounter{section}{0}
\setcounter{subsection}{0}
\setcounter{figure}{0}
\setcounter{table}{0}
\setcounter{equation}{0}
\newpage

%%%%%%%%%%%%%%%%%%%%%%%%%%%%%%%%%%%%%%%%%%%%%%%%%%% Title, authors and addresses
\begin{frontmatter}


\title{Direct Dark Matter Searches}

% use optional labels between square brackets to link authors explicitly to addresses:
% If more than one author, keep a comma between the author tags

\author[address1]{T. J. Sumner}
%\author[address2]{C. O. Author}

\address[address1]{Blackett Laboratory, Imperial College London, \\
Prince Consort Road, London. SW7 2BW. UK \\ email:
t.sumner@imperial.ac.uk}
%\address[address2]{Department of Something Else, Univesity2}

\begin{abstract}
The main goal of this article is to give an update on the
experimental status of direct dark matter searches.  Before doing
that the evidence for dark matter will be very briefly reviewed
along with a few words about the candidate solutions to the dark
matter problem. The experimental requirements for detection of the
most favoured candidate, the neutralino, will be given, which then
leads into a review of the current status and future plans for
experiments across the world.
\end{abstract}

% \begin{keyword}
% dark matter, direct detection
%\PACS
% \end{keyword}

\end{frontmatter}

%%%%%%%%%%%%%%%%%%%%%%%%%%%%%%%%%%%%%%%%%%%%%%%%%%%%%% MAIN TEXT
\section{\label{sec:intro} The Evidence for [Cold] Dark Matter}

The evidence for the existence of dark matter is now extensive and
present on all scale-lengths from galaxies up to the Universe as a
whole.  The direct observational evidence has steadily accumulated
since the first observations by Oort and Zwicky in the 1930s and
includes galaxy rotation curves, galaxy cluster velocity dispersion,
cluster x-ray gas, gravitational lensing, and cluster streaming
motions.  In addition we now have a powerful suite of information
being used to produce models of the Universe as a whole, which start
with primordial density perturbations, as seen from microwave
anisotropies measured by COBE and WMAP, and then look to see how
these evolve under the action of gravity to form the structures we
see today.  Part of this process requires a knowledge of the
structure and evolution of space-time itself, which is informed by
high-redshift populations of standard tracers, such as supernova
type 1a.  The evolution process itself is then modeled using
sophisticated N-body simulations and the resulting predicted galaxy
distributions are then confronted with real data from large survey
projects.  Typical models of the Universe use a dozen or so
parameters, amongst which are the total matter density, $\Omega_m$,
the baryonic matter density, $\Omega_b$, and the hot dark matter
density, $\Omega_\nu$.  The quality of the fits between predictions
and data, such as the microwave background anisotropy angular power
spectrum and the matter density power spectrum (as seen from galaxy
distributions) is now extremely good and they provide strong
constraints on combinations of parameters used in the modeling.  For
example Tegmark et al.~\cite{teg06} use microwave anisotropy data
from WMAP and large scale structure data from SDSS to deduce that
the total mass density in the Universe is $\Omega_m = 0.24\pm 0.02$.
The baryonic mass density, which contains all the normal matter, can
only contribute $\Omega_b = 0.042\pm 0.002$.  Whilst this is nicely
consistent with big-bang nucleosynthesis models, it clearly leaves
the need for some other dominant mass component at the several sigma
level.  This `dark matter' is usually divided into two classes, hot
and cold, depending on whether it was relativistic in the earl
Universe.  Neutrinos are a natural candidate for hot dark matter and
the modeling gives $\Omega_\nu < 0.024$, which then leaves most of
the unseen dark matter in some non-baryonic form which is not
neutrinos or the like; i.e `cold dark matter'.

\section{\label{sec:instructions} The Galactic Candidates}

N-body simulations of structure growth in the Universe show that
galaxies themselves are also cold dark matter dominated which
produces a consistent solution for the ubiquitous flat rotation
curves.  In looking for candidate explanations for the cold dark
matter in galaxies, and hence in our Milky Way, it is natural to
first focus on possibilities which are already motivated for other
reasons and which can provide dark matter both for the Universe as a
whole and for galaxies.  Two prime candidates exist.  These are new
particles called the axion and the neutralino.  The first of these
was postulated in the 1970s to suppress strong CP violation, which
should otherwise have been much prevalent. Recently~\cite{zavat06}
there has been much excitement caused by a positive claim from a
laboratory search for axion-like interactions.   Unfortunately the
inferred particle properties are many orders of magnitude away from
what was expected for cosmologically relevant axions and additional
observations and theoretical development are needed to consolidate
the result and its relevance to dark matter.  The second well
motivated particle candidate is the so-called neutralino, which is
the lightest new particle coming out of supersymmetry.  The
properties of supersymmetric weakly interacting massive particles
(WIMPs) can be such that they were produced in the early Universe
and that the lightest of these remains with us today in
cosmologically important quantities. In addition it turns out that
their weak scattering cross-sections make them potentially
detectable in laboratory based direct searches.

Before discussing neutralino detection in detail below it is useful
to also mention the possibility of resolving the dark matter problem
not be adding new mass but by modifying the way in which gravity
and/or the inertial response to gravity works.  This was first
suggested as MOdified Newtonian Dynamics, MOND, by
Milgrom~\cite{mil83} and, as a phenomenological approach, works very
well~\cite{mil06}.  Bekenstein~\cite{bek04} has since produced a
more formal theoretical basis for MOND, including the relativistic
sector.  This formulation received a tough test recently through the
observation of two galaxy cluster which have passed through each
other causing the main baryonic matter component, hot cluster gas,
to separate from the cold dark matter component as seen through its
lensing effects~\cite{clowe06}. Paramos and Bertolami~\cite{para06}
summarise the current attempts to recover this behaviour from within
MOND.

\section{\label{sec:neut} Neutralino Detection}
The dominant interaction between neutralinos, $\chi$, and any target
medium will be its elastic scattering from a nucleus.  The energy
imparted to the recoiling nucleus is then the means of direct
detection of these particles. Indirect detection through measuring
the annihilation products (neutrinos, $\gamma$-rays and other
particles) from $\chi\bar{\chi}$ will be covered elsewhere.

%To save space use the minipage environement to place two figures in
%a row, as in the examples provided in this file.\par

\begin{figure*}[t]
%\begin{minipage}[t]{0.48\linewidth}
\centering\epsfig{file=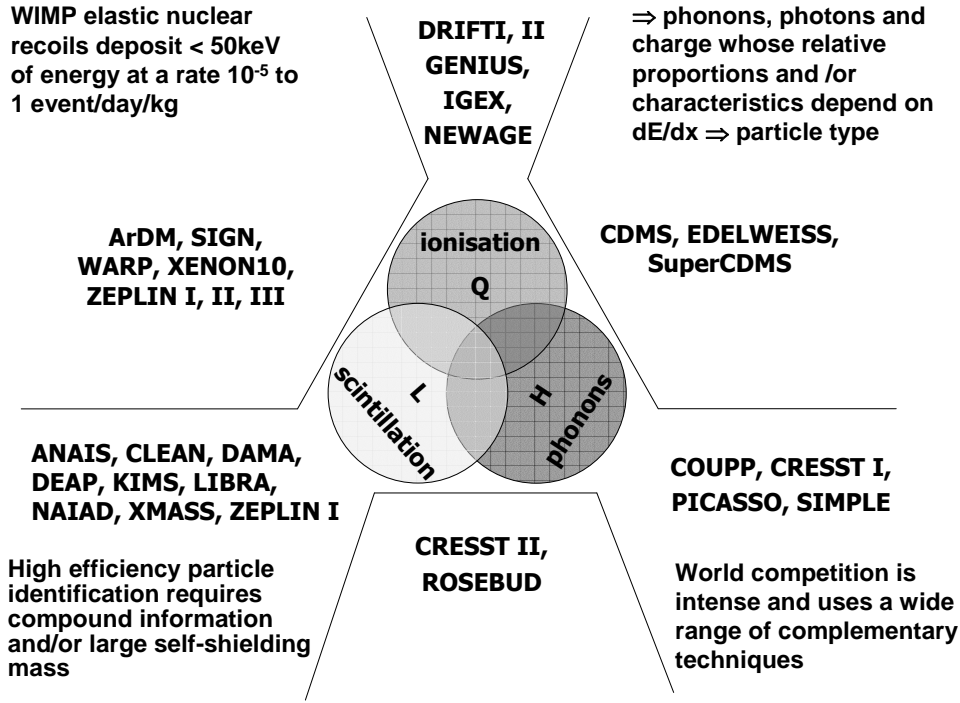,angle=270,width=\linewidth}
\caption{Illustrating the possible techniques and diversity of
direct searches} \label{sumner_fig1}
%\end{minipage}\hfill
%\begin{minipage}[t]{0.48\linewidth}
%\centering\epsfig{file=techs.eps,height=\linewidth,width=\linewidth}
%\caption{Caption to figure 2} \label{fig2}
%\end{minipage}
\end{figure*}

When a neutralino scatters elastically from a target nucleus it
recoils with a kinetic energy typically below $\sim50\,$keV.  As
shown in figure~\ref{sumner_fig1} this energy deposit can result in three
measurable effects; photon emission, charge production and/or phonon
generation.  Experiments can attempt to measure any of these and
many experiments try to measure combinations of them simultaneously
to obtain better diagnostic information of the cause of each event
as the relative efficiency of channeling energy into the three
signal channels depends on $dE/dx$ and hence particle species.  In
the following each technique is briefly reviewed in turn before
presenting a summary of current status.

\subsection{\label{sec:scint} Scintillation detectors}

In the class of scintillation detectors the experiments compare the
shape of the background energy spectrum to that expected from WIMP
nuclear recoils.  In addition they often make use of other
information contained within the scintillation signal, such as an
annual modulation in the rate/spectrum or pulse shape
discrimination.  The DAMA experiment used $\sim100\,$kg of NaI and
searched for, and found, an annual modulation of the expected form
at 6.3$\sigma$~\cite{bern00}. No supporting evidence has yet been
seen by any other experiment and the same team has a new experiment,
LIBRA, with a $300\,$kg array~\cite{bern06}. Similar experiments
with NaI include ANAIS and NAIAD. ANAIS~\cite{amare06} is in
development as a new $107\,$kg experiment at CanFranc.
NAIAD~\cite{ahmed03,alner05} is a completed experiment run at Boulby
which used pulse shape discrimination but not annual modulation.
ELEGANT V~\cite{yosh00} also used a large mass NaI searching for
annual modulation but did not achieve sufficient sensitivity to
check the DAMA result.  They also tried looking for a spectral line
from an inelastic scattering process.  Another crystal scintillator
being developed into a high mass instrument is CsI.  The KIMS
experiment has just completed a pilot run with a $6.6\,$kg detector
giving an upper limit which just reaches the top of the DAMA
positive detection area~\cite{lee06}.

Another class of scintillator is the liquid noble gases.  These
exhibit particle discrimination through the scintillation
time-constant.  The light emission comes from a radiative
dissociation of an excited dimer.  However the dimer can be in
either a singlet or triplet excited state and the decay lifetimes in
these two states differs.  The relative populations ending up in the
singlet and triplet states depends on $dE/dx$.  This time constant
discrimination in xenon has already been used by the ZEPLIN I
experiment~\cite{alner05b}.  Argon and neon can also be used in this
mode, and indeed show somewhat greater time constant ratios.
DEAP(argon)~\cite{boulay06} and CLEAN(neon)~\cite{nikkel06} are two
instruments in advanced development.  Liquid noble gases offer
excellent prospects for scaling to much larger mass detectors. XMASS
is intending to build an $800\,$kg instrument relying on
self-shielding to provide an inner background-free volume, thus no
longer needing particle discrimination.  Currently the project has
results from a $100\,$kg prototype~\cite{namba05}

\subsection{\label{sec:scion} Hybrid scintillation/ionisation detectors}

All the experiments shown as scintillation/ionisation hybrids in
figure~\ref{sumner_fig1} are based on the noble gases.  In fact most use
two-phase systems in which the main target mass is in liquid phase.
Prompt scintillation is created at the point of interaction in the
liquid.  However any ionisation released is drifted through the
liquid phase towards the surface by an applied electric field. At
the surface the field is strong enough for the electrons to be
extracted into the gas phase.  Here they either cause proportional
electroluminescence giving another burst of light or they can be
directly measured as charge. The argon instruments are
ArDM~\cite{rubbia06} and WARP~\cite{szelc06}.  Both of these build
on the heritage from ICARUS and aim to develop tonne-scale targets.
ArDM is building a $600\,$kg prototype using many innovative design
features and using an existing containment vessel.  WARP has built a
2.3\,litre prototype and has already some encouraging first results.
XENON10~\cite{aprile06}, ZEPLIN II~\cite{alner05c} and ZEPLIN
III~\cite{araujo06} are xenon based instruments. XENON10 has a
target mass of $\sim10\,$kg and is operating in Gran Sasso. ZEPLIN
III has a target mass of $\sim40$kg and is operating in Boulby.
XENON10 has a somewhat better light collection for the scintillation
signal as it has photomultiplers immersed in the liquid phase as
well as in the gas phase. Both use bulk PTFE as a structural
material with insulation and with good reflectivity at the UV xenon
wavelength.  Results are expected from both experiments imminently.
ZEPLIN III has completed its build and test phase~\cite{akimov06}
and is now ready for deployment in the Boulby laboratory.  It has
several novel engineering and physics design features which probably
give it a better grasp than either XENON10 or ZEPLIN II.

Finally a relative newcomer is SIGN~\cite{white05} which is
proposing to use high pressure neon gas in a conventional
cylindrical wire chamber configuration.  CsI coatings are to be used
to convert the scintillation signal into charge.

\subsection{\label{sec:cryo} Hybrid cryogenic detectors}
The hybrid cryogenic detectors combine a bolometric signal channel
with either ionisation or scintillation. The technology demands
small segmented detectors and this is currently limiting the total
target masses to modest values ($<10\,$kg).

CDMS~\cite{akerib06} and EDELWEISS~\cite{sang05} use ionisation
detectors with thermal sensing elements bonded to them. CDMS uses Ge
and Si, whereas EDELWEISS only uses Ge. CDMS is midway through a
second generation experiment, CDMS II, and has published results
from mid-term operations.  Plans are well developed for the next
phase SuperCDMS~\cite{akerib06b} to increase the target mass to
$25\,$kg. EDELWEISS has completed a first phase experiment and is
now readying its second generation for operation with a partial
complement of detectors.

CRESST II~\cite{pet06} and ROSEBUD~\cite{amare06b} use scintillators
with thermal sensing elements.  The scintillation light is proximity
coupled into photodetectors.  CRESST has operated a 300\,g CaWO$_4$
crystal in Gran Sasso and has established useful limits from it.
They now have a second generation instrument allowing up to $10\,$kg
of target crystal waiting for full deployment. ROSEBUD has
demonstrated a 46\,g prototype using a BGO crystal.

\subsection{\label{sec:therm}Bubble chambers}
Figure~\ref{sumner_fig1} shows a few experiments in the phonon only
segment.  Two of these are worthy of discussion here as they are
progressing rapidly.  These are COUPP~\cite{bolte06} and
PICASSO~\cite{barn05}, which are bubble chambers.  Contrary to the
bolometers they not specifically sensitive to the total energy
deposit but rather to the local $dE/dx$ in the track of the
recoiling nucleus.

COUPP is a heavy-liquid bubble chamber and which uses CF$_3$I
allowing a good sensitivity to both spin-dependent and
spin-independent scattering cross-sections.  A 2\,kg prototype is
currently operating underground at 300\,m water equivalent at
Fermilab.

PICASSO use superheated droplets of C$_4$F$_{10}$ and looks for
bubble nucleation caused by elastic WIMP scattering from the F
nuclei.  This makes the experiment particularly effective for
searching for spin-dependent scattering.  The experiment is
operating at 6000\,m water equivalent depth in SNOLAB.  An earlier
stage has been completed with some 19\,g of F.  The current phase is
scaling up to 2\,kg active mass of F.

\begin{figure*}[t]
\begin{minipage}[t]{0.48\linewidth}
\centering\epsfig{file=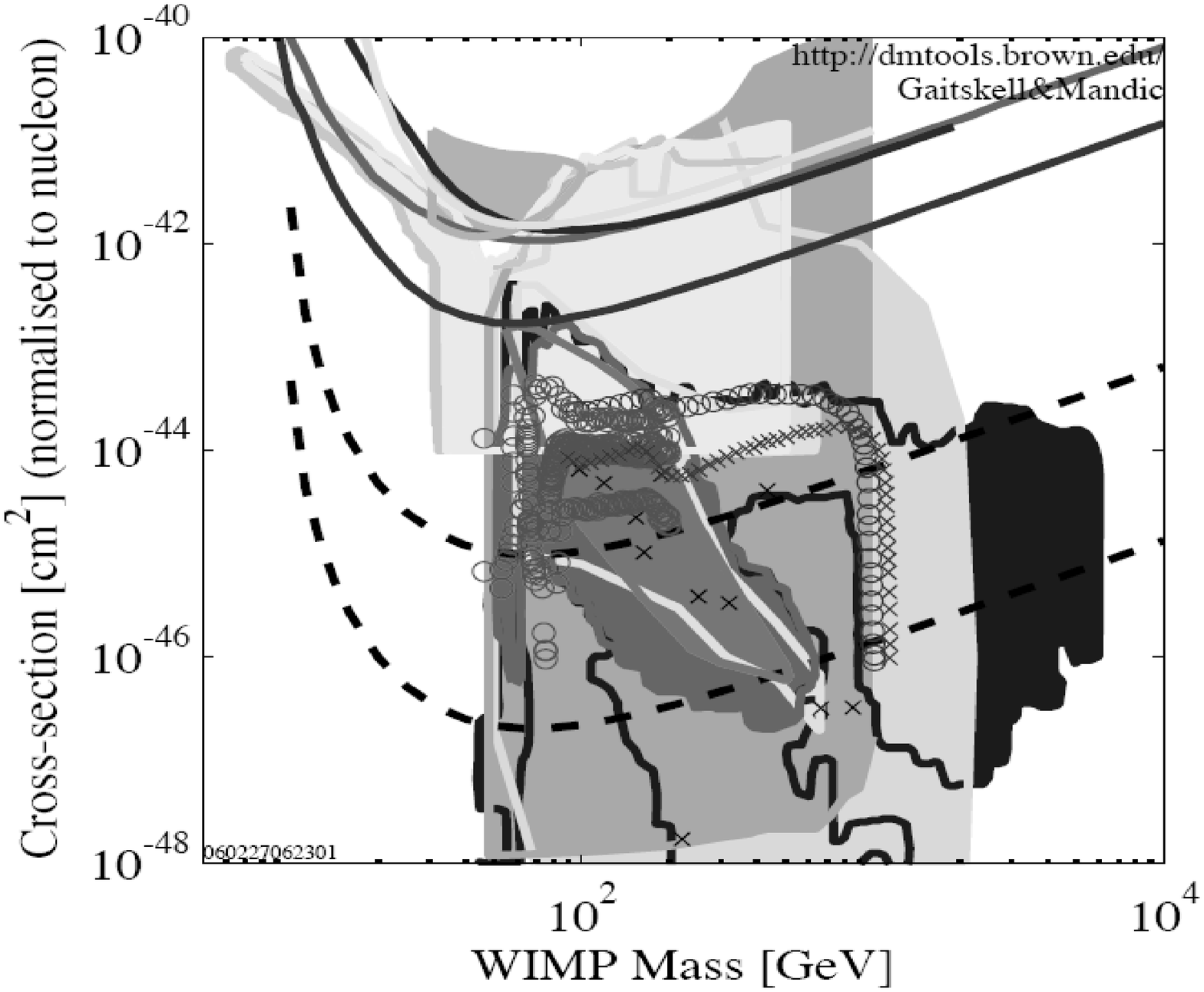,height=\linewidth,width=\linewidth,angle=270}
\label{sumner_fig2} \caption{The current status of direct dark matter
searches for spin-independent interactions.}
\end{minipage}\hfill
\begin{minipage}[t]{0.48\linewidth}
\centering\epsfig{file=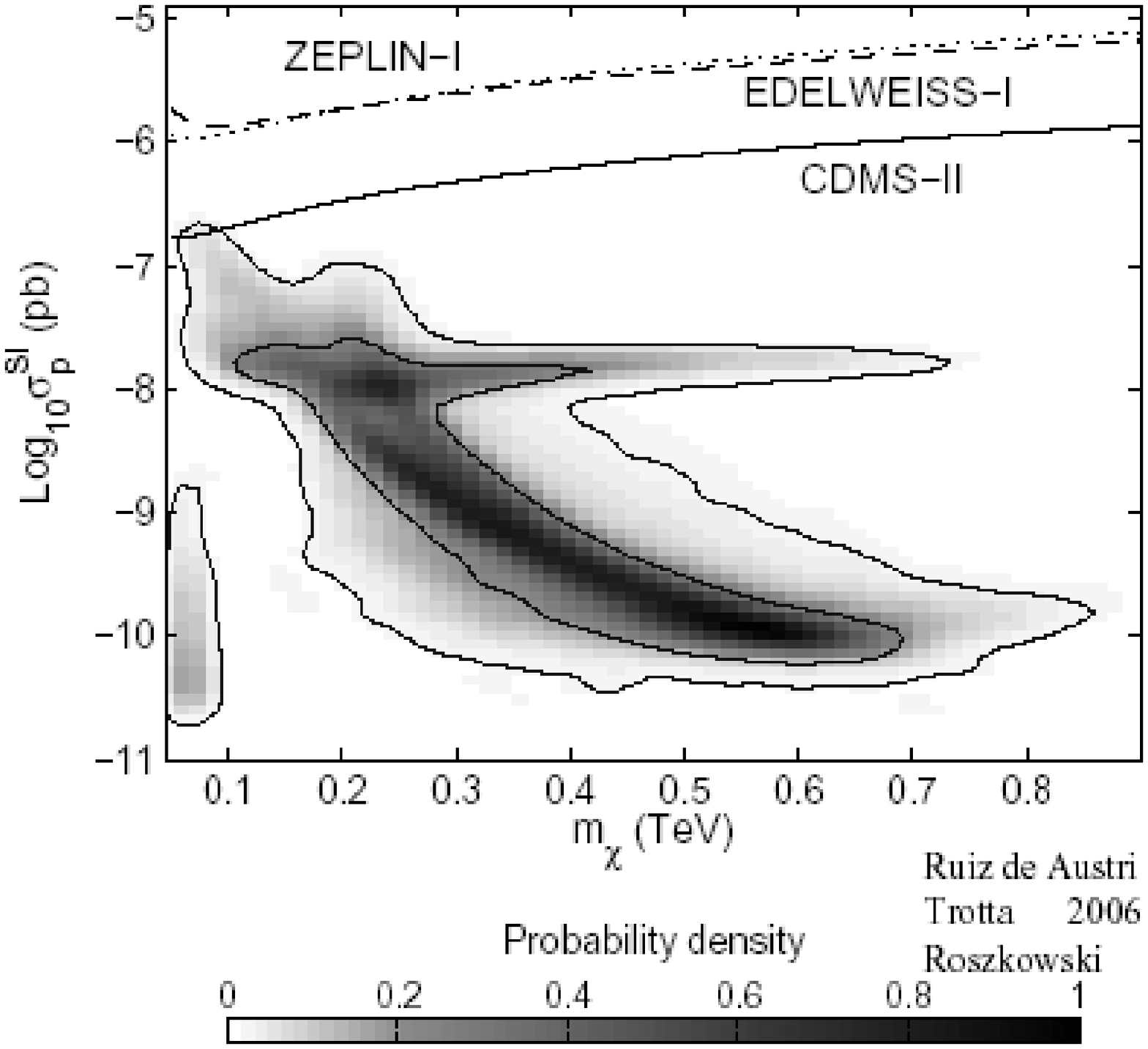,height=\linewidth,width=\linewidth,angle=270}
\label{sumner_fig3} \caption{Theoretical parameter space with associated
probabilities determined by Monte Carlo simulations~\cite{rui06}.}
\end{minipage}
%\caption{Or you can also make a common caption expanding the whole
%  text width. Note that in this case, the two plots are numbered as a
%single figure}
\end{figure*}

\subsection{\label{sec:dirc}Directional gas time projection
chambers} A critical diagnostic to bring to play once a detection
appears secure is to study the directional signature on the WIMP
velocity distribution imparted by the motion of the solar system as
it orbits around the Galactic centre.  This will be modulated by
both the orbit of the Earth around the Sun and, in detector
coordinates, but the Earth's rotation.  These will be very powerful
signatures to confirm the Galactic origin of the signals and,
ultimately to probe the local WIMP velocity distribution as an
astrophysical tool.  To be able to determine a WIMP direction
requires the nuclear recoil track to be extended allowing the track
direction and sense to be recovered.  This points to low-pressure
gas detectors with very good position sensitivity and fine energy
resolution to allow the track to be sampled along its length for
head-tail determination.  Two projects are pursuing this goal;
DRIFT~\cite{alner05d} and NEWAGE~\cite{seki06}.  DRIFT uses 1m$^3$
modules containing CS$_2$ gas at 40\,torr pressure.  It is now in
second generation prototype implementation operating in the Boulby
mine at 3000\,m water equivalent depth. The NEWAGE project is
developing a micro-TPC ($10\times10\times10\,$cm$^3$) using CF$_4$
gas at 150\,torr.  Current tests are aimed at characterizing the
response of this and Ar-C$_2$H$_6$ to nuclear recoils.

\section{\label{sec:status} Current Status}

Figure~\ref{sumner_fig2} summarizes the current status of direct dark
matter searches in the context of a standard model of the Galaxy,
which fixes the local dark matter density and velocity distribution
and allows each experiment to be assessed in terms of the
spin-independent WIMP scattering cross-section normalized to an
individual nucleon.  The solid curves correspond to the experimental
upper limits.  Drawing a vertical line at a mass of $100\,$GeV the
lowest line is that from CDMS II.  Going upwards the next curve
reached is from ZEPLIN I, then EDELWEISS and CRESST II.  These last
three curves all pass through the DAMA positive detection allowed
region, with ZEPLIN I just making a marginal contact.  CDMS II is
well below the DAMA region and hence excludes it in the context of
the model assumptions. The DAMA region is shown as a filled area,
just above a cross-section of $10^{-42}$\,cm$^2$ and between masses
of $\sim40$ to $\sim80$\,GeV.  Other filled areas are regions
populated by the many different supersymmetric models which are
possible.  These predictions span many orders of magnitude in
cross-section and the dashed lines then indicate the extra reach the
experiments will have as they progress through 100\,kg to 1000\,kg
target masses.

Figure~\ref{sumner_fig3} gives more hope to the experimentalists by
concentrating on a constrained minimal supersymmetry model (CMSSM)
parameter space~\cite{rui06}.  Using a Markov Chain Monte Carlo
simulation the authors have assigned a probability rating to each
region of parameter space.  Although this does not rule out the need
for tonne-scale targets to reach down to the $10^{-10}$\,pb level it
does indicate that future improvements in experiments will rapidly
start to probe new space with significant probabilities of a
positive detection and a number of such experiments are indeed
poised to make inroads into the space.

%%%%%%%%%%%%%%%%%%%%%%%%%%%%%%%%%%%%%% reset.txt counters %%%%%%%%%%%%%%
%%
%%%%%%% do not change below here  %%%%%%%%%%%%%%%%%%%%%%%%%%%%%
%%
\setcounter{section}{0}
\setcounter{subsection}{0}
\setcounter{figure}{0}
\setcounter{table}{0}
\setcounter{equation}{0}
\newpage

%%%%%%%%%%%%%%%%%%%%%%%%%%%%%%%%%%%%%%%%%%%%%%%%%%% Title, authors and addresses
\begin{frontmatter}


\title{Summary of search for exotic phenomena with Super-Kamiokande}

% use optional labels between square brackets to link authors explicitly to addresses:
% If more than one author, keep a comma between the author tags

\author[address1]{Jun Kameda},
%\author[address2]{for Super-Kamiokande collaboration}

\address[address1]{Institute of Cosmic Ray Research, University of
  Tokyo, Japan }
%\address[address2]{Department of Something Else, Univesity2}

\begin{abstract}
 In this paper, I present the results on exotic phenomena search in Super-Kamiokande.
\end{abstract}

% \begin{keyword}

%\PACS 
% \end{keyword}

\end{frontmatter}

%%%%%%%%%%%%%%%%%%%%%%%%%%%%%%%%%%%%%%%%%%%%%%%%%%%%%% MAIN TEXT
\section{\label{sec:intro} Introduction}
 
The Super-Kamiokande detector is a cylindrically-shaped
water Cherenkov detector with 50 kiloton of ultra-pure water.
It is located about 1000m underground in the Kamioka Observatory in the
Kamioka mine in Gifu Prefecture, Japan. 
Super-Kamiokande is a multi purpose experiment:
Neutrino physics using atmospheric neutrino and solar neutrinos,
search for astrophysical neutrinos, nucleon decay search,
exotic particle search (Q-ball, WIMPs, magnetic monopoles, etc.), and so on.
In this paper I report the results of indirect WIMPs search and new physics 
search using atmospheric neutrinos
(CPT violation, Lorentz invariant violation).

The observed atmospheric neutrino events in Super-Kamiokande can be categorized
into four types:(1) All visible particles are contained in the detector, we
call these Fully-Contained (FC) events, (2) at least one visible particle escaped
from the detector, called Partially-Contained (PC) events,
(3) The neutrino induced muons entering and passing through
the detector from below, we call these upward-through going muons (UT)
(4) The neutrino induced muons entering and stopping the detector from below,
 called upward stopping muons (US).
Details of the event selections and reconstructions are described in~\cite{super-k_full_paper}.
%The event (3) and (4) are induced by energetic muon neutrinos
%in the rock surrounding the detector.
%To reduce the huge background from the cosmic ray muons, we select only upward-going
%events in (3) and (4).

\section{\label{sec:instructions} Indirect WIMPs search using neutrino-induced muons}
Weak Interacting Massive Particles (WIMPs) are one of the plausible candidates
 of dark matter.
The lightest supersymmetric particle (LSP) of supersysmmetric theories is the most
theoretically well developed WIMP candidate.
Many supersymmetric theories predict LSPs with mass from several 10s GeV to TeV region.
The WIMPs orbiting in the galactic halo will be captured by celestial objects like
the Sun and the Earth, and will annihilate in pairs to the secondaries
($c\overline{c}$, $b\overline{b}$, $\tau\overline{\tau}$,..).
High energy neutrinos will be produced as a decay product of the secondary particles.
A good channel to observe such high energy neutrinos is a 
entering muons induced by the neutrino interactions in the rock surrounding the
detector, because of their large target volume. 
The backgrounds are the produced high energy muons induced by the atmospheric neutrinos.

\begin{figure*}[t]
\begin{minipage}[t]{0.48\linewidth}
\centering\epsfig{file=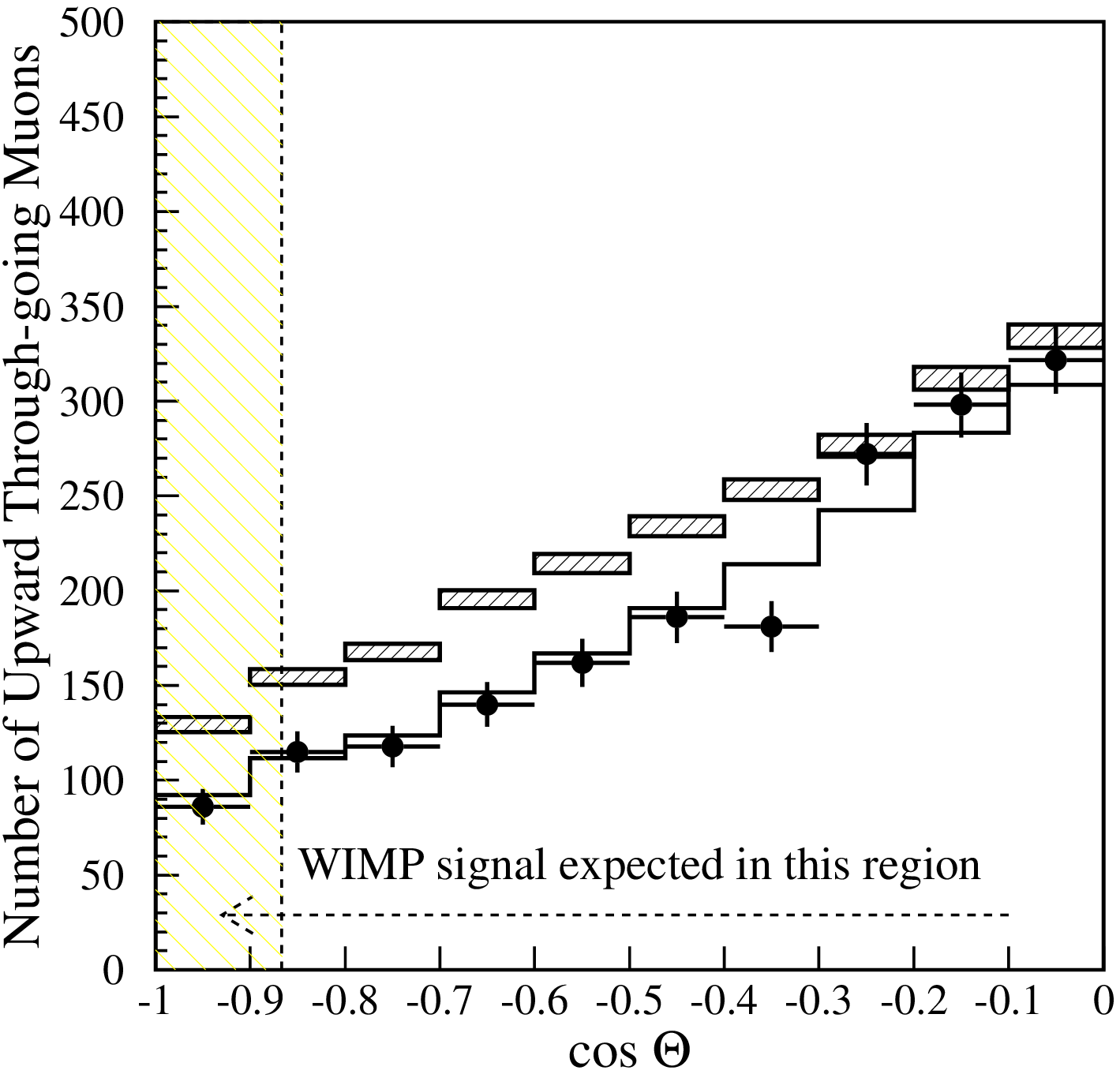,height=\linewidth,width=\linewidth}
\caption{Zenith angle distribution of upward through-going muons 
with respect to the center of the Earth. The circles show data, Boxes
and solid line show the MC prediction without and with neutrino oscillation,
respectively. ($\Delta{m}^2,\sin^22\theta$)=($2\times10^{-3},1.0$) was assumed.}

\label{fig:super-k_wimp_zenith_earth}
\end{minipage}\hfill
\begin{minipage}[t]{0.48\linewidth}
\centering\epsfig{file=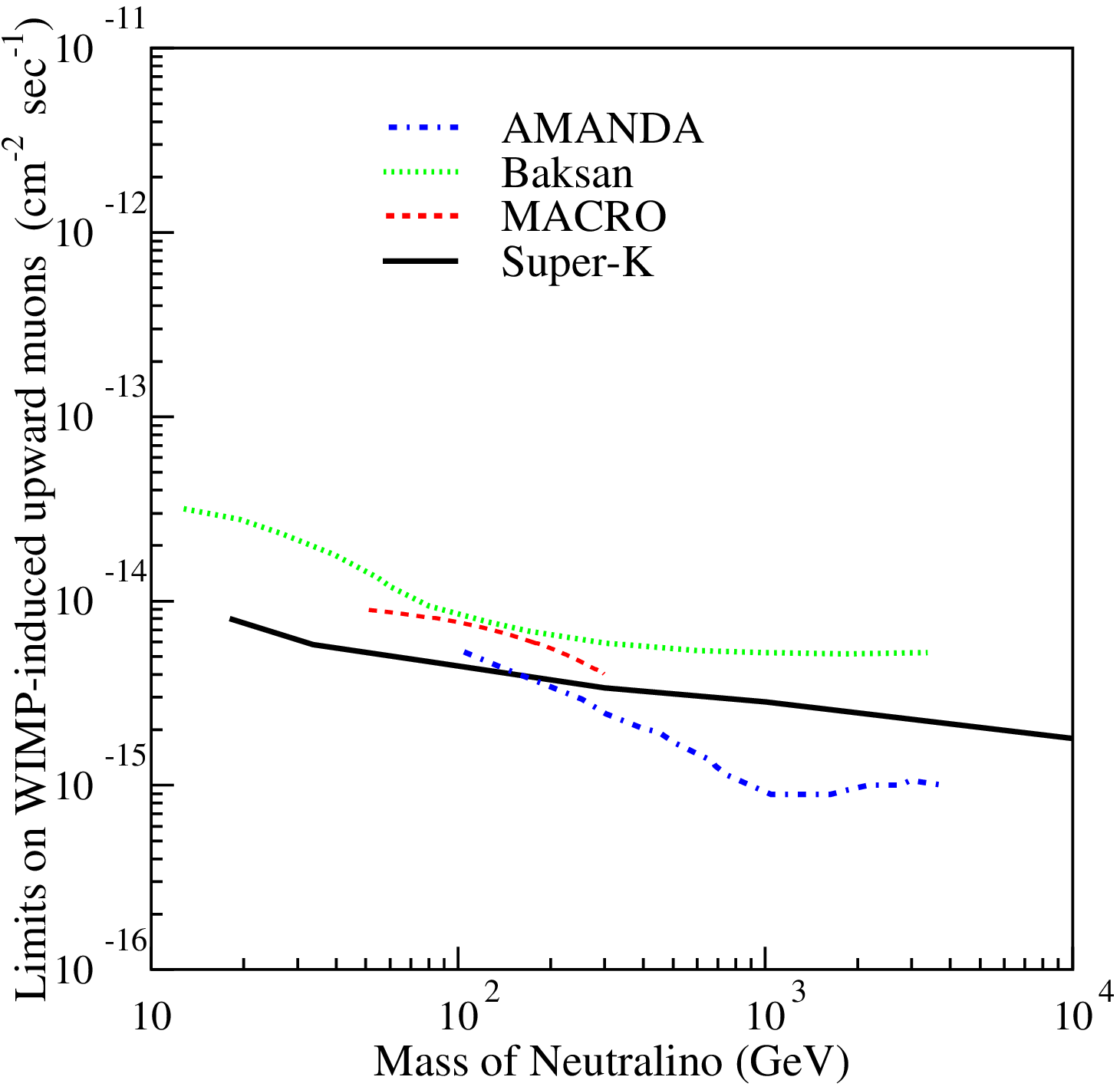,height=\linewidth,width=\linewidth}
\caption{Super-K WIMP-induced upward-muon flux limits from the Earth as a function
of WIMP mass along with those from the other experiments
\cite{super-k_wimp_flux_macro,super-k_wimp_flux_baksan,super-k_wimp_flux_amanda_earth}.}
\label{fig:super-k_wimp_limit_earth}
\end{minipage}
\end{figure*}

Figure~\ref{fig:super-k_wimp_zenith_earth} shows the zenith angle of the
1679.6 live days of the upward-going muons. No significant excess was found in the signal region.

\begin{figure*}[t]
\begin{minipage}[t]{0.48\linewidth}
\centering\epsfig{file=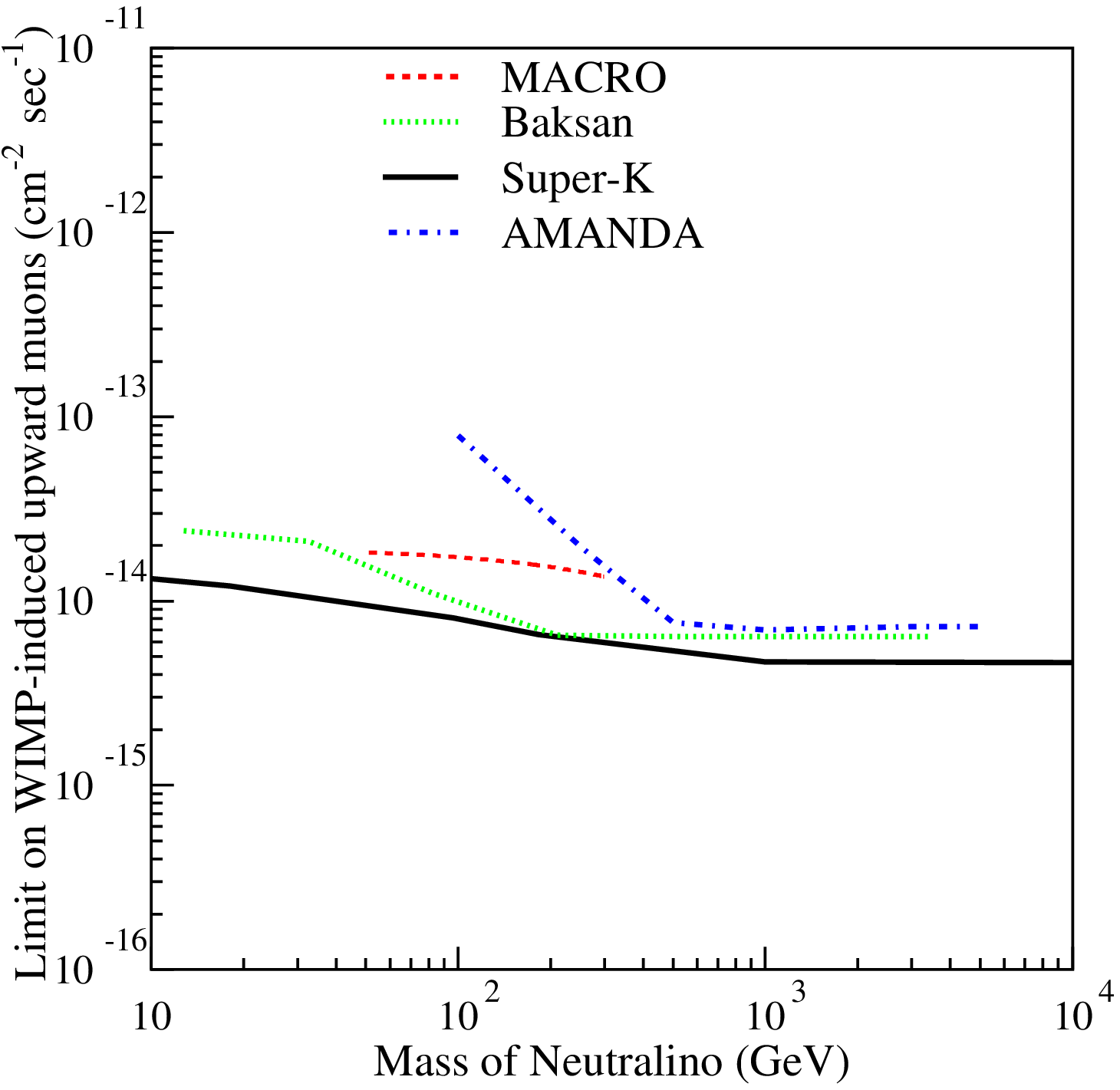,height=\linewidth,width=\linewidth}
\caption{Super-K WIMP-induced upward-muon flux limits from the Sun as a function
of WIMP mass along with those from the other experiments
\cite{super-k_wimp_flux_macro,super-k_wimp_flux_baksan,super-k_wimp_flux_amanda_sun}.}
\label{fig:super-k_wimp_limit_sun}
\end{minipage}\hfill
\begin{minipage}[t]{0.48\linewidth}
\centering\epsfig{file=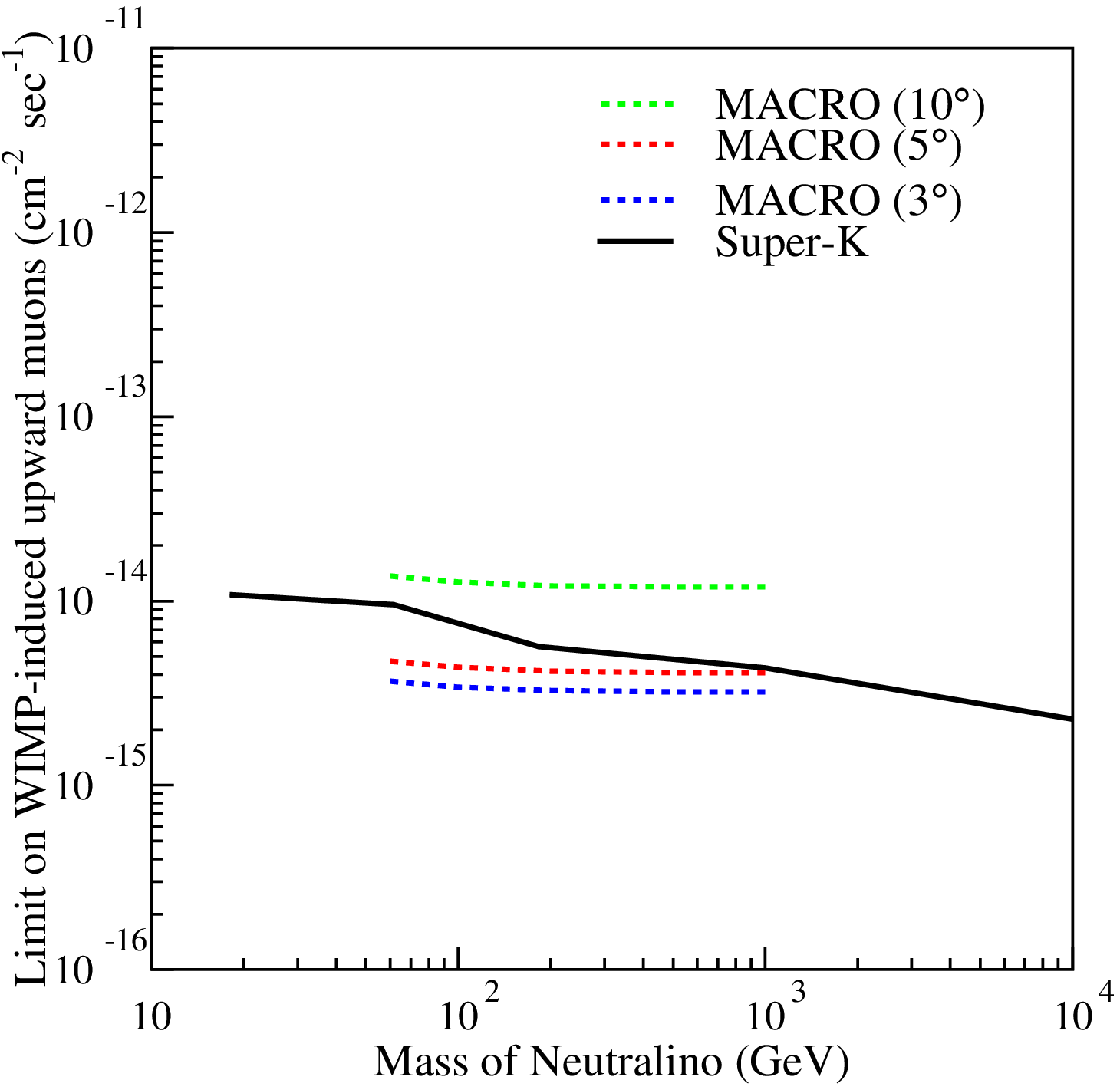,height=\linewidth,width=\linewidth}
\caption{Super-K WIMP-induced upward-muon flux limits from the Galactic center
as a function of WIMP mass along with those from the other experiments
\cite{super-k_wimp_flux_macro}}
\label{fig:super-k_wimp_limit_gal}
\end{minipage}
\end{figure*}

The signal window size which includes 90\% of the neutrino--induced muons is a function of 
neutrino energy, because of the finite scattering angle of neutrinos, muon multiple
scattering, and the finite size of the WIMP annihilation region in the Earth.
The neutrino energy from WIMPs annihilation depends on the WIMPS masses, 
so the flux limit depends on the WIMPs masses.
We employed the result of the Monte Carlo simulation done in Ref.~\cite{super-k_wimp_kam}
to calculate the angular windows which contain 90\% of the signal for various WIMPs
masses.
Figs~\ref{fig:super-k_wimp_limit_earth}-\ref{fig:super-k_wimp_limit_gal}
show the flux limits as a function of the WIMP mass from the Sun,
the Earth, and the Galactic center, respectively.

The limit of the indirect WIMPs search can be translated and compared to
the results from the direct search experiments using the calculated ratio of
the direct-to indirect detection rates~\cite{super-k_wimp_calc_ratio}.
The calculated ratio has uncertainties in the model dependence of the WIMP annihilation,
and we used the maximum value of the ratio to get conservative limits.

\begin{figure*}[t]
\begin{minipage}[t]{0.48\linewidth}
\centering\epsfig{file=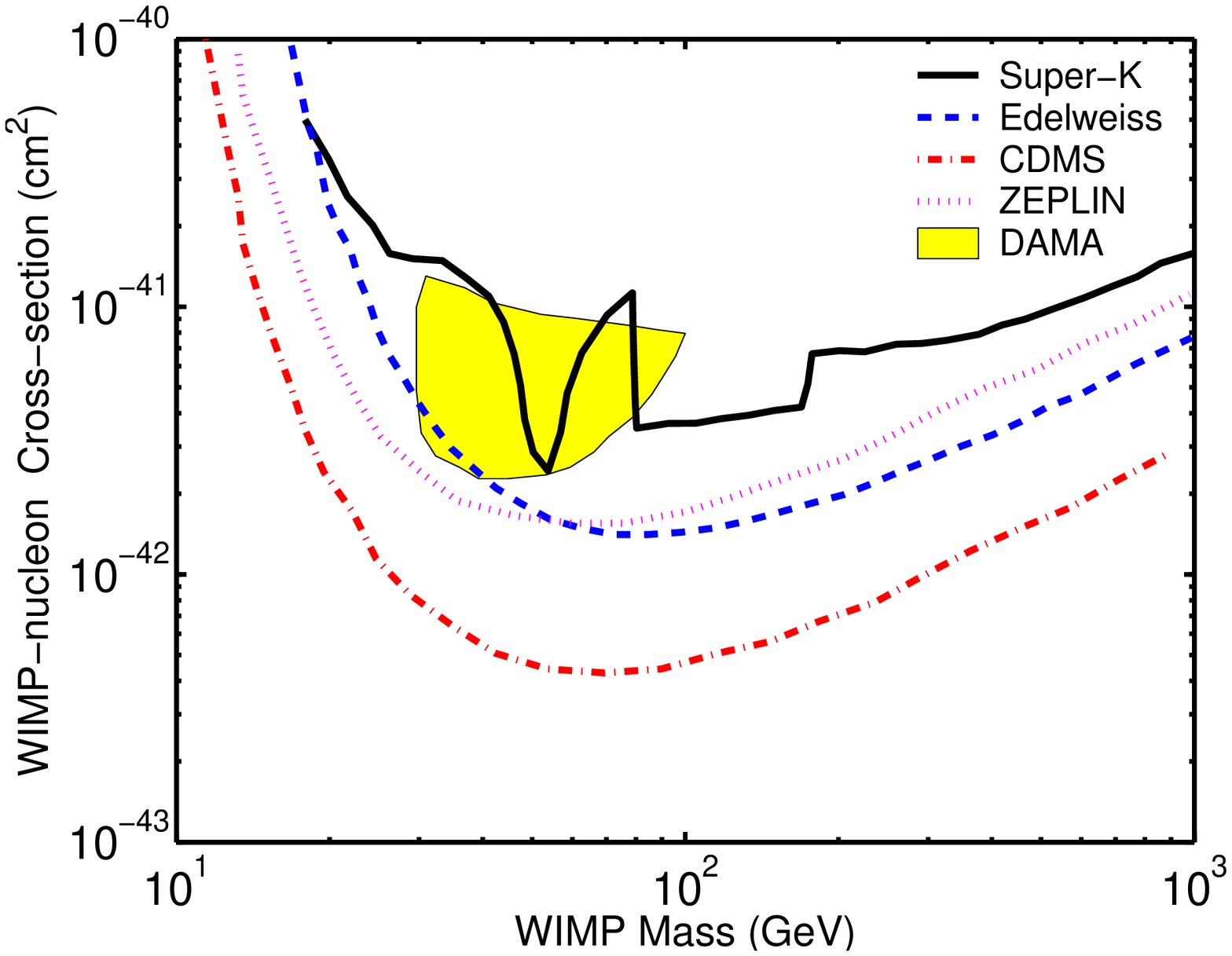,height=\linewidth,width=\linewidth}
\caption{Super-K 90\% C.L. excluded region in WIMP parameter space
(solid line) for WIMP with scalar coupling. DA-MA 3$\sigma$ allowed region(filled)
\cite{super-k_wimp_direct_DAMA},90\% C.L. excluded region from CDMS(dot-dashed)
\cite{super-k_wimp_direct_CDMS}, EDELWEISS(dashed)\cite{super-k_wimp_direct_EDELWEISS},
and ZEPLIN(dotted)\cite{super-k_wimp_direct_ZEPLIN} are also shown.}
\label{fig:super-k_wimp_limit_scalar}
\end{minipage}\hfill
\begin{minipage}[t]{0.48\linewidth}
\centering\epsfig{file=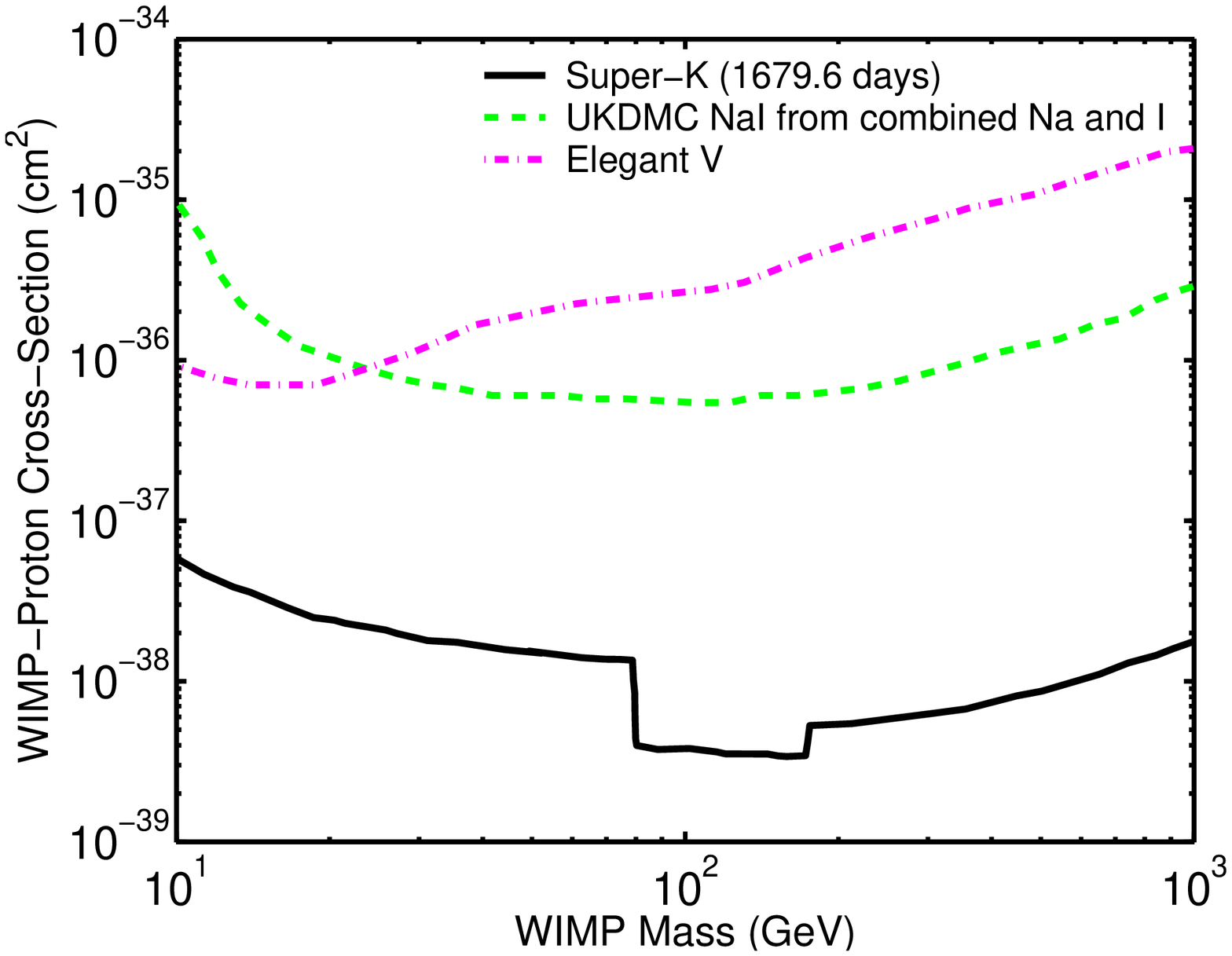,height=\linewidth,width=\linewidth}
\caption{Super-K 90\% C.L. excluded region in WIMP parameter space
(solid line) for WIMP with axial vector coupling. The
90\% C.L. excluded region from UKDMC(dashed)\cite{super-k_wimp_direct_UKDMC},
ELEGANT(dot-dashed)\cite{super-k_wimp_direct_ELEGANT} are also shown.}
\label{fig:super-k_wimp_limit_axial}
\end{minipage}
\end{figure*}

Figs.~\ref{fig:super-k_wimp_limit_scalar} and \ref{fig:super-k_wimp_limit_axial}
show the 90\% excluded regions in WIMP parameter space
from Super-Kamiokande and the results of the other experiments for scalar
coupling and axial vector coupling, respectively. 
The sum of the flux from the Sun and the Earth has been used
for scalar coupling, and only the flux from the Sun was used for axial vector coupling.
We had excluded a significant portion of the favored region from the DAMA experiment
in scalar coupling, and largely pushed the limit on axial vector coupling.

\section{\label{sec:plots} Search for new physics on neutrino using atmospheric neutrinos}
The observed deficit pattern of the atmospheric neutrino is perfectly explained by the
$\nu_\mu \rightarrow \nu_\tau$ neutrino oscillation due to neutrino masses~\cite{super-k_full_paper}.
But the finite masses of the neutrino is not the only possible source of neutrino
oscillation.
In general, survival probability of the neutrino oscillation can be written as:
\begin{equation}
P(\nu_{\mu} \rightarrow \nu_{\mu}) =
 1-\sin^22\theta\sin^2\left(\beta\cdot\frac{L}{E^n}\right),
\end{equation}
where $\theta$ is a mixing angle of neutrinos, $\beta$ is the
frequency of the neutrino oscillation, and the power index $n$ is
a parameter which depends on each theory. 
Examples of these theories are:
violation of Lorentz invariance~\cite{super-k_loen:theory:nu_osc_VLI}
(n = -1), violation of the equivalence principle~\cite{super-k_loen:theory:nu_osc_VEP}(n = -1
),
CPT violation~\cite{super-k_loen:theory:nu_osc_cpt}(n = 0),
and coupling to space-time torsion fields~\cite{super-k_loen:theory:nu_osc_torsion}(n = 0).
In the case of `standard oscillation', the index $n$~is~1.

\begin{figure*}[t]
\begin{minipage}[t]{0.48\linewidth}
\centering\epsfig{file=sk-loen_chi2.epsi,height=\linewidth,width=\linewidth}
\caption{The $\chi^2$ distribution as a function of energy index $n$.}
\label{super-k_loen:fig:chi2}
\end{minipage}\hfill
\begin{minipage}[t]{0.48\linewidth}
\centering\epsfig{file=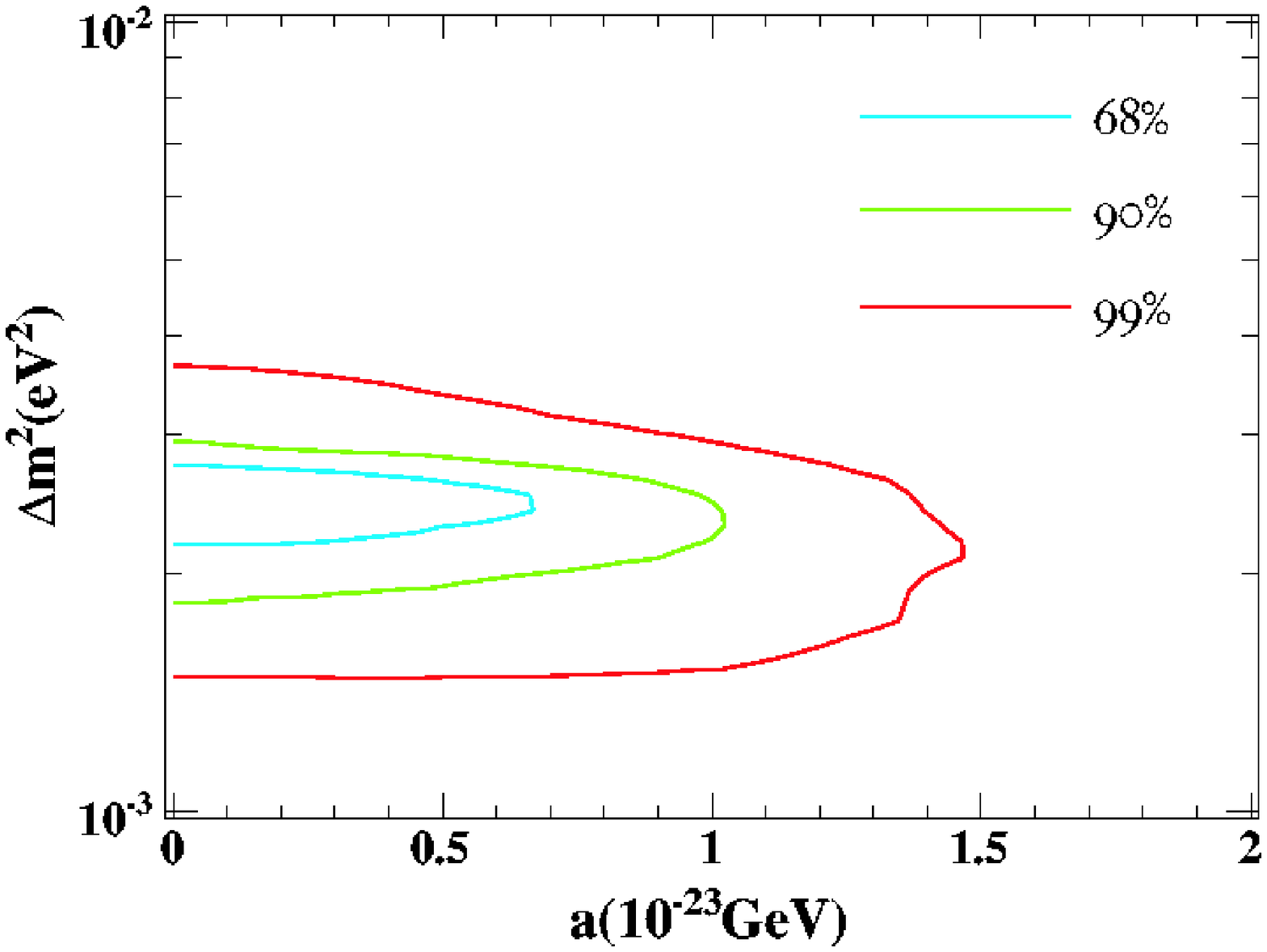,height=\linewidth,width=\linewidth}
\caption{The 68\%,90\%,99\% C.L. allowed region on ($\Delta{a}$,$\Delta{m}^2$) space.}
\label{super-k_cptv_limit}
\end{minipage}
\end{figure*}

We carried $\chi^2$ test in three dimensional parameter space
($\sin^2{2\theta}$,$\beta$,$n$) to find the best value of $n$. 
1144 days FC + PC, 1138 days of UT, and 1117 days of US were used for this test.
Fig.~\ref{super-k_loen:fig:chi2} shows the $\chi^2$ 
distribution as a function of $n$.
The $\chi^2$ is minimized by optimizing the $\sin^2{2\theta}$, $\beta$
and the systematic parameters.
We fitted a parabola function near the minimum $\chi^2$ region, and 
obtained $n = 1.14 \pm 0.11$, which is consistent with the neutrino
oscillation induced by finite neutrino masses ($n=1.0$).
%The other scenario highly excluded as a dominant source of the neutrino oscillation.

%\subsection{CPT violation and Lorentz violation of neutrinos}
The new physics cannot be a dominant source of the neutrino oscillation,
but there are possibility of the sub-dominant source of the neutrino deficit.
We tested two scenarios as a sub-dominant mechanism.
The survival probability of the neutrino with neutrino masses and CPT violating
term can be written as:
\begin{equation}
P(\nu_{\mu} \rightarrow \nu_{\mu}) =
 1-\sin^22\theta\sin^2\left(\left(\frac{\Delta{m^2}}{4E}+\Delta{a}\right)L\right),
\end{equation}
where $\Delta{a}$ is the CPT violation parameter.
Fig.~\ref{super-k_cptv_limit} shows the allowed region of the parameter space
using 1489.2 live days FC+PC, 1645.9 live days of upward-going muons.
The definition of the $\chi^2$ is shown in \cite{super-k_full_paper}.
The result is consistent with no CPT violating term, and 
we got the limit of the $\Delta{a} < 10^{-23}$ at 90\% C.L.

\begin{figure*}[t]
\begin{minipage}[t]{0.48\linewidth}
\centering\epsfig{file=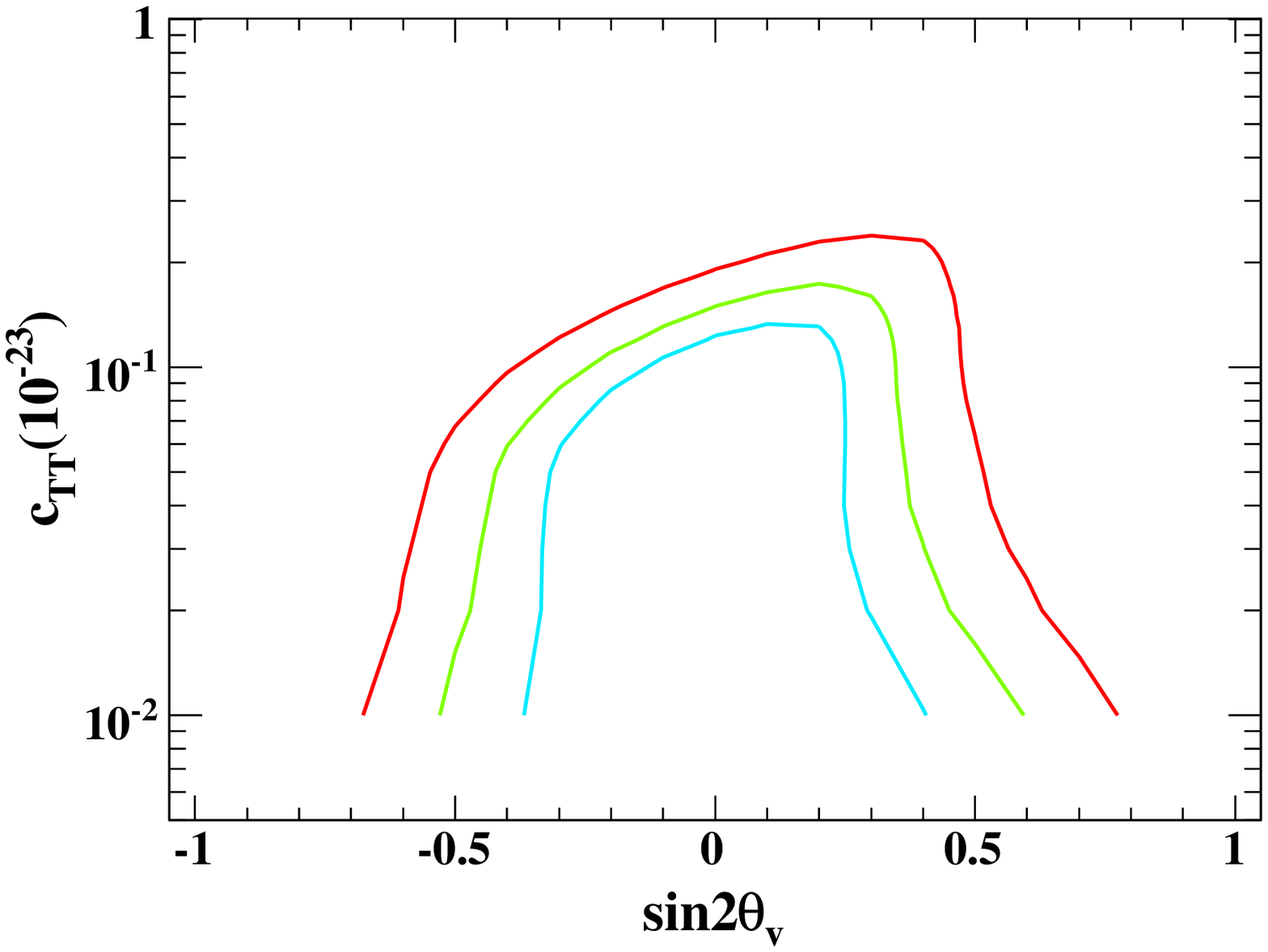,height=\linewidth,width=\linewidth}
\caption{68\%, 90\%, and 99\% allowed region of the parameter space for the
sign in $\Omega$ is plus case.}
\label{super-k_liv_limit_0}
\end{minipage}\hfill
\begin{minipage}[t]{0.48\linewidth}
\centering\epsfig{file=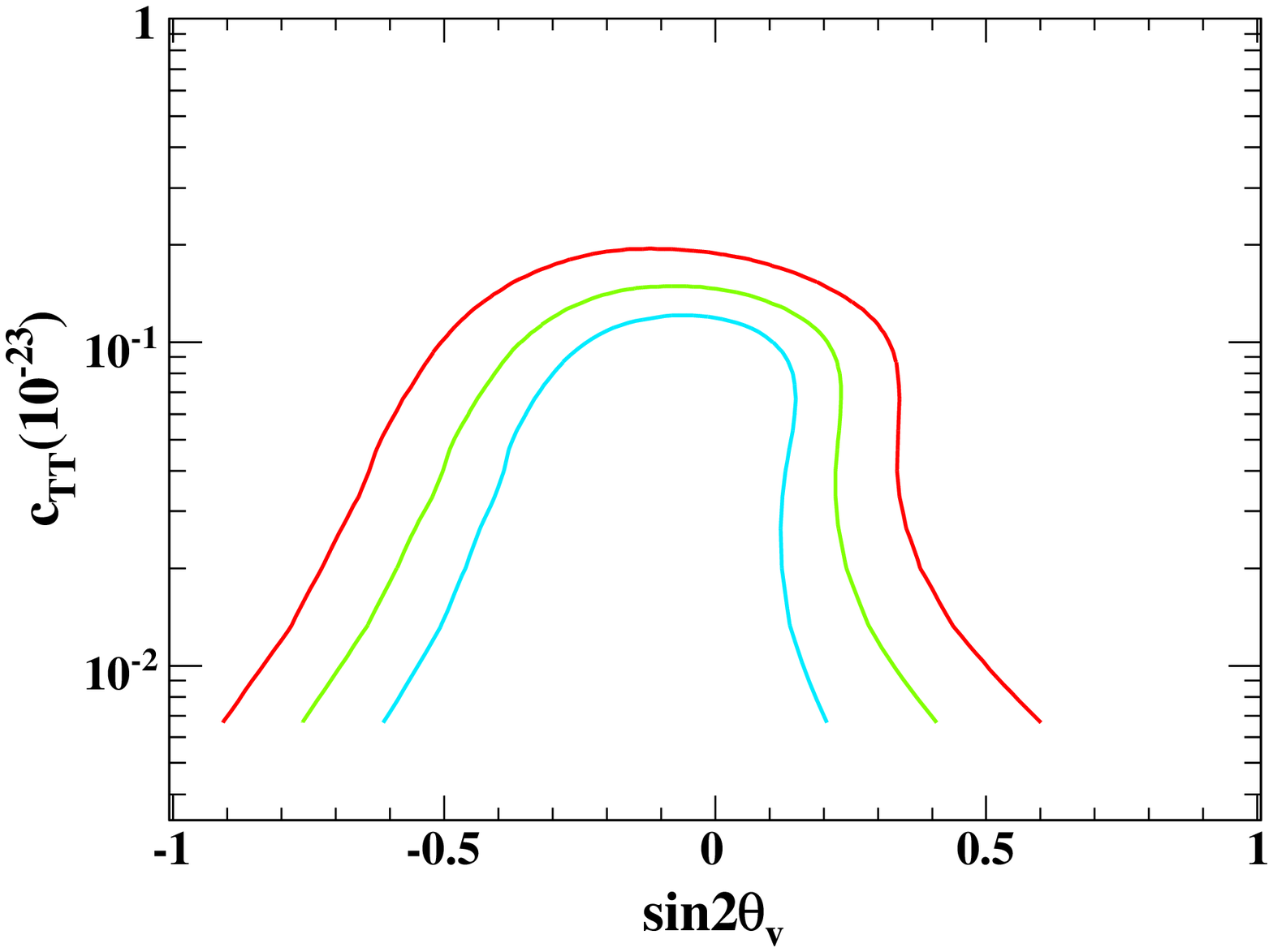,height=\linewidth,width=\linewidth}
\caption{68\%, 90\%, and 99\% allowed region of the parameter space for the
sign in $\Omega$ is minus case.}
\label{super-k_liv_limit_1}
\end{minipage}
\end{figure*}

A type of Lorentz violating term can also introduce neutrino
oscillations. 
The survival probability of the neutrino with neutrino masses and a CPT violating
term can be written:
\begin{equation}
P(\nu_{\mu} \rightarrow \nu_{\mu}) =
% 1-\sin^22\Theta\sin^2\left(1.27\sqrt{(\Delta{m}^2L/E)^2\pm4c^{TT}\sin^2\theta_{v}LE+4(c^{TT}LE)^2}\right),
 1-\sin^22\Theta\sin^2\left(\Omega\right)
\end{equation}
where $\tan^2{\Theta}=\left(1+(E/E_c)^2\sin2\theta_{v}\right)/\left((E/E_{c})^2
\cos2{\theta_v}\right)$, $E_{c}=\sqrt{\Delta{m}^2/2c^{TT}}$, 
$\Omega =1.27\sqrt{(\Delta{m}^2L/E)^2\pm4c^{TT}\sin2\theta_{v}LE+4(c^{TT}LE)^2}$, 
$c^{TT}$ is the difference of the attainable velocity of neutrinos,
$\theta_{v}$ is a mixing angle of the velocity eigenstate,
respectively. 
The plus/minus sign in $\Omega$ corresponds to the relative phase of the mixing of
velocity eigenstate and mass eigenstate, and we consider both signs.
We scanned the ($c^{TT}$,$\sin2\theta_{v}$) space assuming
$\Delta{m}^2=2.4\times10^{-3}$eV${^2}$, and full mixing of the mass eigenstate,
which is the best fit value from the neutrino oscillation due to neutrino masses.
Figs.~\ref{super-k_liv_limit_0} and \ref{super-k_liv_limit_1}
show the allowed regions of the parameter space for both sign.
The result is consistent with no Lorentz invariant violating term, and
we got the limit of  $c^{TT} <10^{-24}$ at 90\% C.L. for both cases.

\section*{Acknowledgments}
{\small 
The author would like to thank Super-Kamiokande collaboration.
}

%%%%%%%%%%%%%%%%%%%%%%%%%%%%%%%%%%%%%% reset.txt counters %%%%%%%%%%%%%%
%%
%%%%%%% do not change below here  %%%%%%%%%%%%%%%%%%%%%%%%%%%%%
%%
\setcounter{section}{0}
\setcounter{subsection}{0}
\setcounter{figure}{0}
\setcounter{table}{0}
\setcounter{equation}{0}
\newpage

\begin{frontmatter}

\title{Summary of MACRO results on exotic physics}

\author[address1]{M. Giorgini},

\address[address1]{Department of Physics, University of Bologna 
and INFN \\ Viale C. Berti Pichat 6/2, I-40127 Bologna, Italy}

\begin{abstract}
MACRO was a multi-purpose experiment that took data from 1989 to 2000, at 
the underground Laboratory of Gran Sasso (Italy). MACRO gave important 
results/limits concerning: (i) the oscillation of atmospheric 
neutrinos, also in the non-conventional scenario of violations of 
Lorentz invariance, (ii) the searches for exotic particles (supermassive 
GUT magnetic monopoles, nuclearites, WIMPs), (iii) muon physics and 
astrophysics. A summary of the MACRO results will be presented and 
discussed, focusing the attention on the exotica searches.
\end{abstract}

\end{frontmatter}

\section{\label{sec:intro} Introduction}
MACRO was a large area multipurpose underground detector \cite{r1} 
designed to search for rare events and rare phenomena in the cosmic radiation.
The detector was located in Hall B of the underground Gran Sasso Laboratory 
(Italy). It was optimised to search for the supermassive magnetic monopoles 
\cite{mono,pdecay} predicted by Grand Unified Theories (GUT).
The experiment obtained important results on atmospheric neutrino oscillations 
\cite{high,low,sterile,scatt,ultimo} and performed neutrino astronomy studies 
\cite{nuastro}, indirect searches for WIMPs \cite{wimps},
 search for low energy stellar gravitational collapse neutrinos \cite{grcol}, 
studies of the high energy underground muon flux (which is an indirect tool 
to study the primary cosmic ray composition and high energy hadronic 
interactions \cite{cr}), searches for fractionally charged particles (LIPs) 
\cite{lips} and other rare particles that may exist in the cosmic radiation. 

The detector started data taking in 1989 and it was running until December 
2000. The apparatus had global dimensions of $76.6 \times 12 \times 9.3$ 
m$^3$ and was composed of three sub-detectors: liquid scintillation counters, 
limited streamer tubes and nuclear track detectors. Each one of them could 
be used in ``stand-alone'' and in ``combined'' mode.
It may be worth to stress that all the physics and
astrophysics items listed in the 1984 proposal were covered and good results 
were obtained on each of them.
 
\section{\label{sec:nu-osc} Atmospheric neutrino oscillations}

MACRO detected $\nm$-induced muon events in 4 different topologies. 

The {\it upthroughgoing muons} come from $\nm$ interactions in the rock below 
the detector; the $\nm$'s have a median energy of $\sim 50$ GeV.

\begin{figure*}[t]
\begin{minipage}[t]{0.4\linewidth}
\hspace{-0.5cm}
\epsfig{file=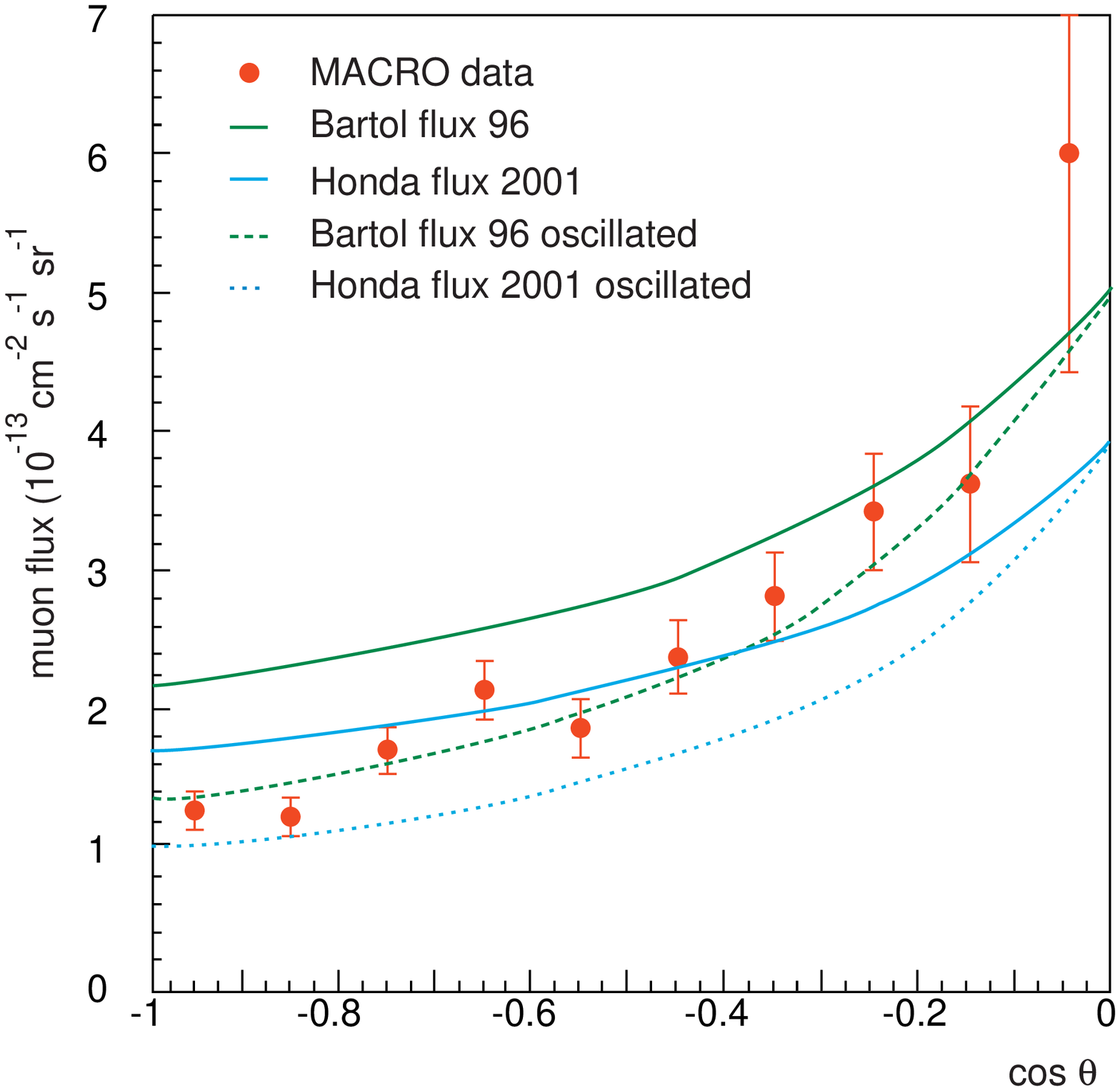,width=5.8cm,height=4.6cm}
\caption{Comparison of the MACRO upward-throughgoing muons (black circles) 
with the predictions of the Bartol96 and of the Honda2001 MC oscillated 
and non oscillated fluxes (oscillation parameters $\Dm2 = 2.3 \cdot 10^{-3}$ 
eV$^2$ and $\s2t_m =1$).}
\label{fig:zenith}
\end{minipage}\hfill
\begin{minipage}[t]{0.55\linewidth}
\centering\epsfig{file=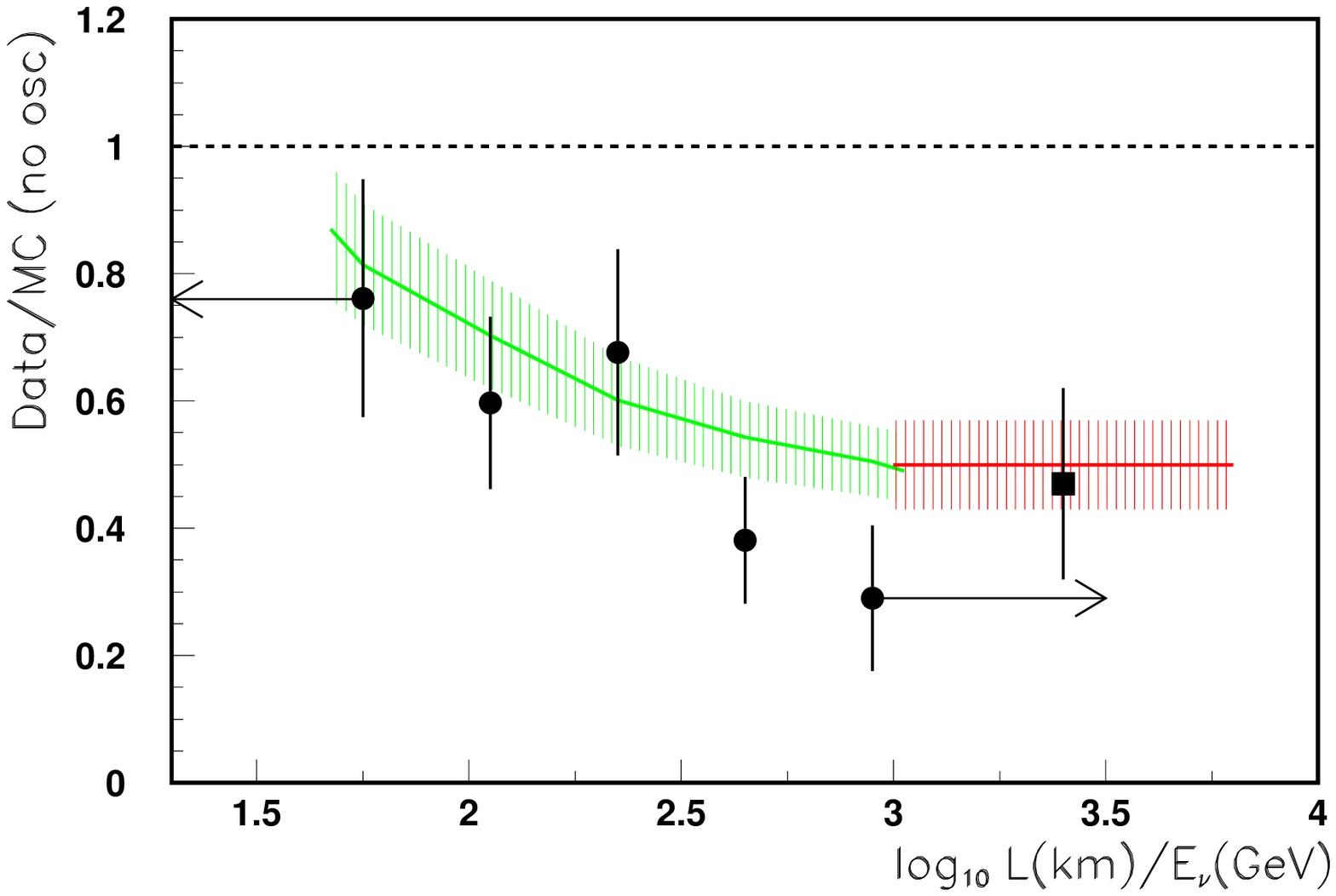,width=\linewidth}
\caption{Ratio Data/MC$_{\mbox{no~osc}}$ 
as a function of the estimated $L/ E_\nu$ for the upthroughgoing muon 
sample (black points). The solid line is the MC expectation assuming 
$\Dm2 = 2.3 \cdot 10^{-3}$ eV$^2$ and $\s2t_m = 1$. The last point 
(black square) is obtained from the IU sample.}
\label{fig:le}
\end{minipage}
\end{figure*}

Fig. \ref{fig:zenith} shows the zenith distribution of the measured 
902 upthroughgoing muons (black circles) compared with two 
MonteCarlo (MC) predictions: the Bartol96 \cite{bartol} flux with and 
without oscillations (the dashed and solid lines, respectively) and the 
Honda2001 flux \cite{honda}. The FLUKA MC predictions \cite{fluka} agree 
perfectly with the Honda2001.  

For a subsample of $\sim 300$ upthroughgoing events, we estimated the muon 
energy through Multiple Coulomb Scattering in the rock absorbers in the
lower apparatus \cite{scatt}. The evaluated resolution on $E_\nu$ is 
$\sim 100 \%$. The parent neutrino path length is $L \sim 2 R_E \cos \Theta$, 
where $R_E$ is the Earth radius. Fig. \ref{fig:le} shows the 
ratio data/MC as a function of the estimated $L/E_\nu$ for the 
upthroughgoing muon sample. 
The black circles are data/Bartol96 MC (assuming no oscillations); the
solid line is the oscillated MC prediction for $\Dm2 = 2.3 \cdot 10^{-3}$ 
eV$^2$ and $\s2t_m =1$. The shaded region represents the simulation 
uncertainties. The last point (black square) is obtained
from semicontained upward going muons.

The low energy events ($IU$, $ID+UGS$ \cite{low}) are produced 
by parent $\nm$ interacting inside the lower detector, or by upgoing muons 
stopping in the detector. The median energy of the parent neutrino is 
$\sim 3-4$ GeV for all topologies. In both cases, the zenith distributions are 
in agreement with the oscillation prediction with the optimised parameters
\cite{ultimo}. 

In order to reduce the effects of systematic uncertainties in the MC 
absolute fluxes we used the following three independent ratios 
\cite{ultimo}:
\begin{enumerate}
        \item [(i)] High Energy data: vertical/horizontal ratio,
        $R_{1} = N_{vert}/N_{hor}$
        \item [(ii)] High Energy data: low energy/high energy ratio,
        $R_{2} = N_{low}/N_{high}$
        \item [(iii)] Low Energy data: 
        $R_{3} = (Data/MC)_{IU}/(Data/MC)_{ID+UGS}$
\end{enumerate}
Combining the three independent results, the no oscillation hypothesis 
is ruled out at the $\sim 5 \sigma$ level.

To evaluate the hypothesis of oscillation for different values of $\Dm2$ and 
 $\s2t_m$, the Feldman-Cousins \cite{fel_cou} procedure was used and the 
corresponding $90\%$ C.L. region for the $\nmnt$ oscillation is 
given in ref \cite{ultimo}. The best fit is reached at $\Dm2 = 
2.3 \cdot 10^{-3}$ eV$^2$ and $\s2t_m = 1$.

\subsection{\label{sec:lorentz} Search for exotic contributions to atmospheric 
neutrino oscillations}
\vspace{-0.3cm}
MACRO searched for ``exotic'' contributions to standard mass-induced
atmospheric neutrino oscillations, arising from a possible violation of Lorentz
invariance (VLI), using two different and complementary analyses. The first 
approach uses the low energy ($E_\nu < 28$ GeV) and the high energy 
($E_\nu > 142$ GeV) samples. The mass neutrino
oscillation parameters have the values given in Sect. \ref{sec:nu-osc} and 
 we mapped the evolution of the $\chi^2$ estimator
in the plane of the VLI parameters $\Delta v$ and $\sin^2 2 \theta_v$. No
$\chi^2$ improvement was found, so we applied the Feldman-Cousins 
\cite{fel_cou} method to determine $90\%$ C.L. limits on the parameter:
$|\Delta v| < 3 \cdot 10^{-25}$ \cite{lorentz}.

The second approach exploits a data subsample
characterised by intermediate neutrino energies. It is based on the maximum
likelihood technique and 
considers the mass neutrino oscillation parameters varying in the $90\%$ C.L. 
border \cite{ultimo}. The obtained $90\%$ C.L. limit 
on the $\Delta v$ parameter is also around $10^{-25}$ \cite{praga} .

\section{\label{sec:nu-astrophysical} Neutrinos from astrophysical sources}
\subsection{\label{sec:HE} Search for astrophysical HE muon neutrinos}  
\vspace{-0.3cm}
High energy $\nm$'s are expected to come from several galactic 
and extra-galactic sources. An excess of events was searched for around 
the positions of known sources in $3^{\circ }$ (half width) angular bins. The 
$90\%$ C.L. upper limits on the muon fluxes from specific celestial sources
were in the range $10^{-15} \div 10^{-14}$ cm$^{-2}$ s$^{-1}$ 
\cite{mudiffu}. A search for 
time coincidences of the upgoing muons with $\gamma$-ray bursts was also 
made. No statistically significant time correlation was found \cite{nuastro}.

A different analysis was made for the search for a diffuse astrophysical 
neutrino flux, using a dedicated method to select higher energy upthroughgoing
muons. The flux upper limit was set at the level of $1.5 \cdot 10^{-14}$ 
cm$^{-2}$ s$^{-1}$ \cite{nudiffu}.

\subsection{\label{sec:WIMPs} Indirect searches for WIMPs}
\vspace{-0.3cm} 
Weakly Interacting Massive Particles (WIMPs) could be part of the
galactic dark matter; they could be intercepted by celestial bodies,
slowed down and trapped in their centres, where WIMPs and anti-WIMPs could
annihilate and yield neutrinos of GeV or TeV energy,
in small angular windows from their centres. One WIMP candidate 
is the lowest mass neutralino.

To look for a WIMP signal, we searched for upthroughgoing muons from the 
Earth centre, using $10^{\circ} \div 15^{\circ}$ cones around the Nadir; the 
$90\%$ C.L. muon flux limits are $0.8 \div 1.4 \cdot 10^{-14}$ 
cm$^{-2}$ s$^{-1}$ \cite{wimps}. These limits, when compared with 
the predictions of a supersymmetric model, eliminate a sizable range 
of parameters used in the model. 

A similar procedure was used to search for $\nm$ from the
Sun: the muon upper limits are at the level of about 
$1.5 \div 2 \cdot 10^{-14}$ cm$^{-2}$ s$^{-1}$ \cite{wimps}. 

\vspace{-0.3cm}
\subsection{\label{sec:GC} Neutrinos from stellar gravitational collapses} 
\vspace{-0.3cm}
A stellar gravitational collapse of the core of a massive star is expected 
to produce a large burst of all types of neutrinos and antineutrinos with 
energies of $ 5 \div 60$ MeV and with a duration of $\sim 10$ s. No stellar 
gravitational collapses in our Galaxy were observed from 1989 to 2000 
\cite{grcol}. 

\begin{figure*}[t]
\begin{minipage}[t]{0.47\linewidth}
\hspace{-0.5cm}
\epsfig{file=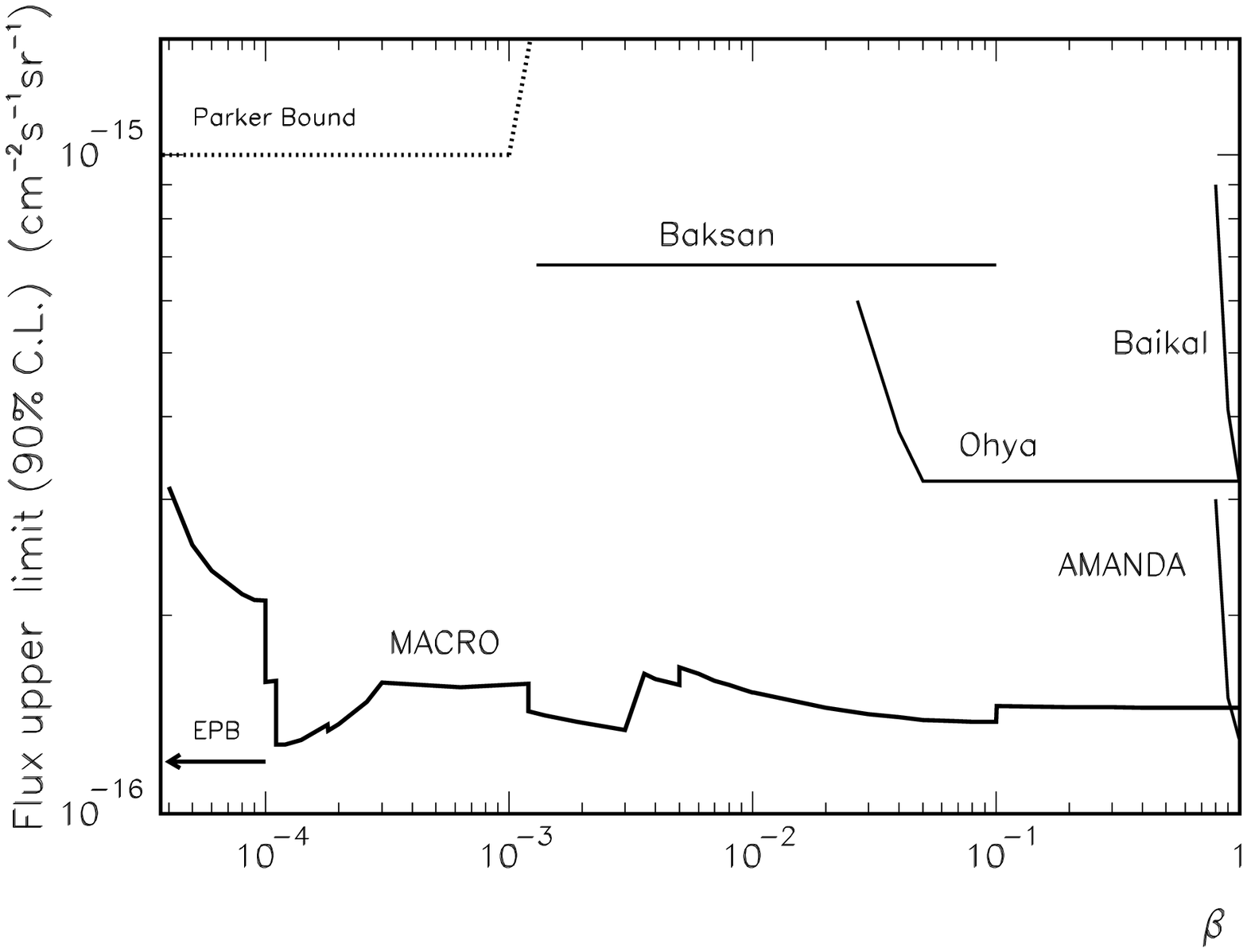,width=6.6cm,height=6.5cm}
\caption{$90\%$ C.L. upper limit obtained by MACRO for an isotropic flux 
of GUT MMs with $g=g_D$ compared with direct limits given by other 
experiments.}
\label{fig:mm}
\end{minipage}\hfill
\begin{minipage}[t]{0.49\linewidth}
\centering\epsfig{file=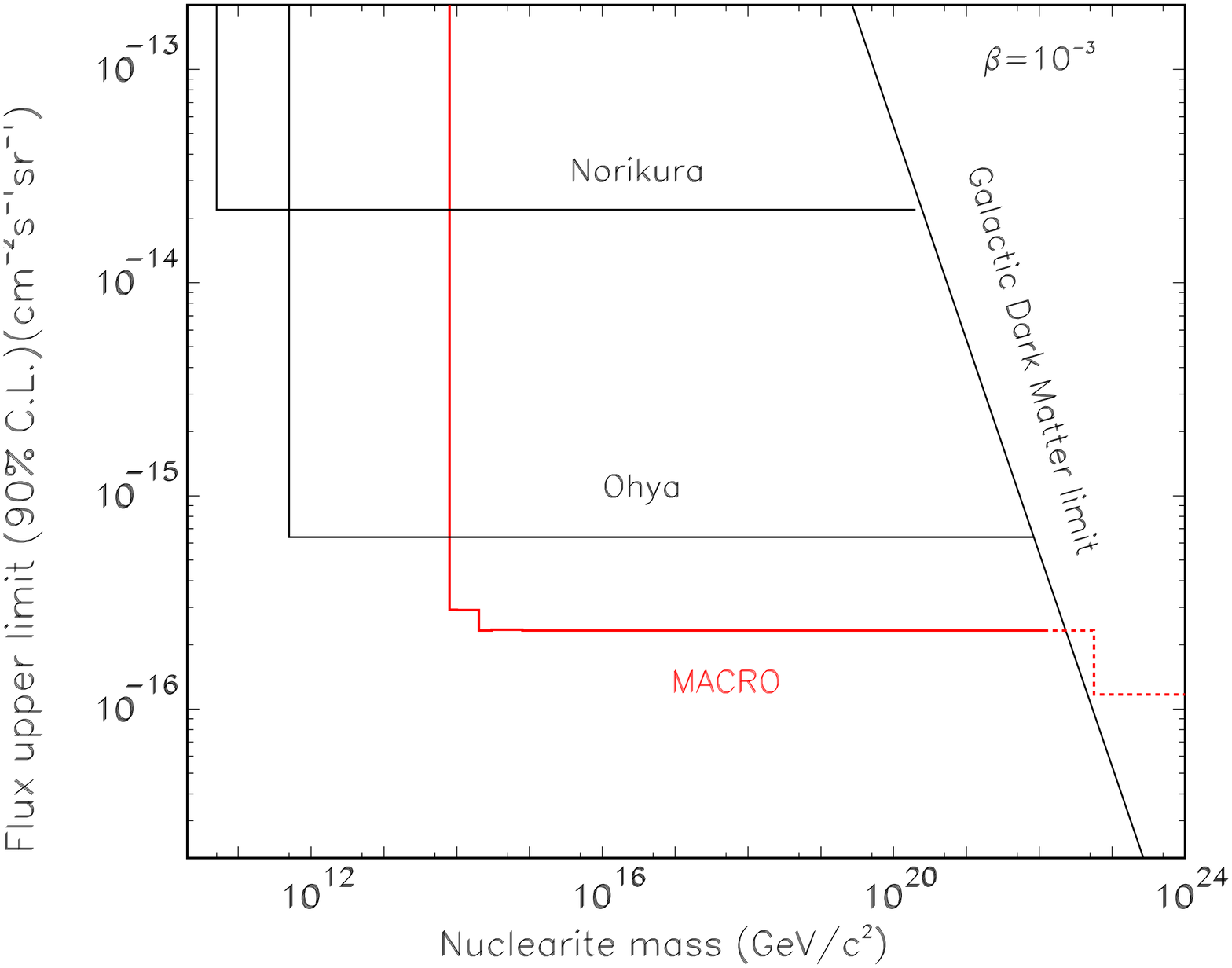,height=5.5cm,width=6cm}
\caption{$90\%$ C.L. upper limits versus mass for downgoing nuclearites
with $\beta=2 \cdot 10^{-3}$ at ground level. The MACRO limit for nuclearite 
masses larger than $5 \cdot 10^{22}$ GeV/c$^2$ has been extended and 
corresponds to an isotropic flux.}
\label{fig:nucl}
\end{minipage}
\end{figure*}

\section{\label{sec:exotic} Search for exotic particles}
\subsection{\label{sec:MM} Search for GUT magnetic monopoles (MMs)} 
\vspace{-0.3cm}
Supermassive magnetic monopoles predicted by Grand Unified Theories
(GUT) of the electroweak and strong interactions should have 
masses $m_M \sim 10^{17}$ GeV. 

MACRO was optimised to search for an isotropic flux of GUT MMs in the cosmic
radiation. The three sub-detectors had sensitivities in different $\beta$ 
regions, covering the velocity range $4 \cdot 10^{-5} < \beta < 1$.
They allowed multiple signatures of the same rare event candidate. No 
candidates were found by any of the three subdetectors. Fig. \ref{fig:mm} 
shows the $90\%$ C.L. flux upper limits for $g=g_D$ poles (one unit of Dirac 
magnetic charge) plotted versus $\beta$ \cite{mono} together with direct 
limits set by other experiments \cite{mono_altri}. The MACRO MM direct limits
 are by far the best existing over a very wide range of $\beta$.

The interaction of the GUT monopole core with a nucleon can lead to a 
reaction in which the nucleon decays, $M+p \rightarrow M + e^+ + \pi^0$. 
 MACRO dedicated an analysis procedure to detect nucleon decays induced 
by the passage of a GUT MM in the streamer tube system (a fast $e^+$
track from a slow ($\beta \sim 10^{-3}$) MM track). The $90\%$ C.L. flux 
upper limits established by this search are at the level of 
$\sim 3 \cdot 10^{-16}$ cm$^{-2}$ s$^{-1}$ sr$^{-1}$ for 
$10^{-4} \leq \beta \leq 0.5 \cdot 10^{-2}$; they are valid for catalysis 
cross sections $5\cdot 10^2 < \sigma_{cat} < 10^3$ mb \cite{pdecay}. 

\vspace{-0.3cm}
\subsection{\label{sec:nucl} Search for nuclearites, Q-balls and LIPs} 
\vspace{-0.35cm}
Strangelets should consist of aggregates of $u,~d$ and $s$ quarks in almost 
equal proportion \cite{nucl} and would have typical galactic velocities 
$\beta \sim 10^{-3}$. The MACRO $90\%$ C.L. upper limits for an isotropic 
flux of nuclearites with $10^{-5} \le \beta \le 1$ was at the level of 
$1.5 \cdot 10^{-16}$ cm$^{-2}$ s$^{-1}$ sr$^{-1}$ \cite{nuclea}.

MACRO searched also for charged Q-balls (aggregates of squarks, sleptons 
and Higgs fields) \cite{qballs}, giving an upper limit of 
$\sim 10^{-16}$ cm$^{-2}$ s$^{-1}$ sr$^{-1}$ \cite{qballs2}.

Fractionally charged particles could be expected in Grand Unified 
Theories as deconfined quarks; the expected charges range 
from $Q=e/5$ to $Q= 2/3e$. LIPs should release a fraction $(Q/e)^2$ of the
energy deposited by a muon traversing a medium. The $90\%$ C.L. flux upper 
limits for LIPs with charges $e/3,~2/3e$ and $e/5$ are at the level 
of $10^{-15}$ cm$^{-2}$ s$^{-1}$ sr$^{-1}$ \cite{lips}.

\section{\label{sec:conclu} Conclusions}
 Standard atmospheric $\nm$ oscillations: no-oscillation hypothesis 
ruled out at $5 \sigma$. \par
 VLI: $|\Delta_v|$ upper limits of the order of $10^{-25}$. \par
 MM search: upper flux limit of $1.4 \cdot 10^{-16}$ 
cm$^{-2}$ s$^{-1}$ sr$^{-1}$ for $4 \cdot 10^{-5} < \beta < 1$.  \par
 Nuclearite search: upper flux limit of $10^{-16}$  
cm$^{-2}$ s$^{-1}$ sr$^{-1}$ for $\beta \simeq 10^{-3}$. \par
 Charged Q-balls search: upper flux limit of $\sim 10^{-16}$ 
cm$^{-2}$ s$^{-1}$ sr$^{-1}$. \par
 WIMP search: upper flux limit of $\sim 10^{-14}$ cm$^{-2}$ s$^{-1}$ 
sr$^{-1}$. \par 
 LIP search: upper flux limit of $6.1 \cdot 10^{-16}$ 
cm$^{-2}$ s$^{-1}$ sr$^{-1}$.

\section*{Acknowledgements}
{\small 
I acknowledge the cooperation of the members of the MACRO Collaboration 
and, in particular, of the Bologna group.
}

\setcounter{section}{0}
\setcounter{subsection}{0}
\setcounter{figure}{0}
\setcounter{table}{0}
\setcounter{equation}{0}
\newpage

%%%%%%%%%%%%%%%%%%%%%%%%%%%%%%%%%%%%%%%%%%%%%%%%%%% Title, authors and addresses
\begin{frontmatter}


\title{Results with the Baksan Neutrino Telescope}

% use optional labels between square brackets to link authors explicitly to 
%addresses:% \author[label1,label2]{}
% If more than one author, keep a comma between the author tags

\author {M.M. Boliev},
\author {A.V. Butkevich},
\author {S.P. Mikheyev},
\author {O.V. Suvorova}

\address{ Institute for Nuclear Research, Russian Academy of Sciences;\\
    60th October Anniversary Avenue 7a; RU-117312 Moscow, Russia}

\begin{abstract}
The obtained exposure of the Baksan Underground Scintillation Telescope (BUST)
is now one of the best for astrophysical challenges. We present a performance
of the sample of 1102 upward through-going muons collected by the BUST during 21.15 years of
live time. We looked for an excess in arrival directions relatively to the expectations
from atmospheric neutrinos and for annihilation signal of Dark Matter weakly interacting massive particles (WIMPs) inside the Sun.

\end{abstract}

% \begin{keyword}

%\PACS
% \end{keyword}

\end{frontmatter}

%%%%%%%%%%%%%%%%%%%%%%%%%%%%%%%%%%%%%%%%%%%%%%%%%%%%%% MAIN TEXT
\section{\label{sec:intro} Introduction}

Nowadays searches for exotic physics serve mainly to resolve a "Double Dark"~\cite{Prim:06} picture
of the Universe evolution. The neutrino telescopes may contribute by
evidence of new neutrino properties or a new hypothetical heavy particle
either neutral or charged, like WIMPs, monopole, or Q-balls.

Recent data from WMAP~\cite{WMAP:06} as well as the combined observations of $\gamma$, 
and X-rays in various experiments~\cite{COSM:06} point out with high precision on the existing of 
non-luminous, invisible and unknown form of matter filling in all space of the flat universe with 
only 5$\%$  of barions.

We are focusing on searching for one of a favorite Dark Matter WIMP
candidate - the neutralino
(for a  review see ~\cite{Jungman:96}). 
The probabilities for WIMP to be gravitationally trapped by the Sun and by the Earth were 
calculated theoretically~\cite{Gould:87}. It has been shown that WIMPs can scatter
off a nucleus in the medium, sink to the central regions and accumulate there. 
During the evolution of the Universe the rate of neutralinos captured by the Sun could be large enough 
to provide a thermal equilibrium with their annihilation rate in the centre. Decays of generated 
fermions and bosons in neutralino annihilation channels produce high energy leptons from which 
only neutrinos survive on their way to detector. With benchmark points of mSUGRA we analyze 
their survival in comparison with our upper limits on annihilation rates.\\

\section{\label{sec:instructions} Experiment, Site and Triggers}

   The Baksan Underground Scintillation Telescope is known as high energy
neutrino telescope of early generation in the big underground laboratory
of Baksan Neutrino Observatory (BNO). The duration of continuous measurements
covers 26 years since 1978, while effective exposure amounts 
to $1.6\times10^{15} cm^2 s$.

   The BNO site is in North Caucasus mountains at 1700 m above sea level
and at the latitude of Mediterranean Sea:
$42^{\circ}41^{\prime}$E and $43^{\circ}16^{\prime}$N.
The BUST is located at effective depth of $850hg/cm^2$ which is
relevant to the $5\times10^3$ reduction of background from the downward-going
atmospheric muons. Trajectories of penetrating particles are reconstructed
by the positions of hit tanks, which put together a system of 3,150 liquid
scintillation counters of standard type $(70cm \times 70cm \times 30cm)$ in configuration 
of parallelepiped $(17m \times 17m \times 11m)$. The counters entirely cover all its sides and 
two horizontal planes inside at the distances 3.6 m and 7.2 m from the bottom. Each plane is 
separated from another one by concrete absorber of $160g/cm^2$.
The configuration provides $1.5^{\circ}$ of angular accuracy. The time 
resolution measured with downward-going muons is equal to 5 ns. 
The arrival direction is determined by time-of-flight method, so that the measured value 
of $\beta$ ~\cite{Baksan:91} is expected to be negative for upward-going particles. More details about the detector
can be found in ~\cite{Baksan:91}.

   While expected ratio of downward-going muon flux to upward-going one is about $10^7$,
there were two hardware triggers used to select upward-going muons. They reduce initial
rate of downward-going muons approximately by factor of $10^3$. Trigger 1 covers the zenith angle
range $95^{\circ} \div 180^{\circ}$ while trigger 2 selects horizontal muons in
the range $80^{\circ} \div 100^{\circ}$. The hardware trigger efficiency of
99\% has been measured with the flux of atmospheric muons. These two triggers
give about 1,800 events per day for further processing. In the year 2000 the
telescope data acquisition system was upgraded and allowed to simplify the trigger
system down to one general trigger with the rate of 17 Hz. All raw information
is than undergone further selection using off-line reconstruction code.

  There are two simple requirements in data selection to be satisfied: only
single trajectory of penetrating particle and a negative measured value
of $\beta$. However all events with negative values of $\beta$ have been
scanned by eye to check possible misinterpretation.

   It was found finally 1439 events survived these cuts and collected from December
of 1978 until April of 2006 with 21.15 years of live time.

\section{\label{sec:plots} Sample of upward through-going muons}

The reconstructed muon trajectories are required to have more than 7 m of measured length inside 
the telescope to
reject upward-going muons that could be mimicked by downward-going
atmospheric muon interactions or muon groups. So far we cut off particles 
with energy below 1 GeV \cite{Baksan:96}. In total 1102 events survived all cuts. 
From Fig.1 one can conclude about continuous and stable rate of measured upward through-going 
muons during all years of observations. The obtained zenith angular distribution
of these events is shown in Fig.2 in comparison with two hypotheses of simulated events:
with and without two-flavours oscillating neutrino.\par

  The complete Monte Carlo simulations of both atmospheric neutrino interactions and the detector 
response on induced upward through-going muons have been done on base of author codes. 
In part
of neutrino interactions we used the Bartol atmospheric neutrino flux \cite{ref:Bart95}, 
the GRV-94 parton density distributions \cite{ref:pdfGRV} and energy-loss parameterization 
\cite{ref:melLoh} for muon propagation in surrounding rock. Induced muons in the generated 
neutrino interactions have to pass the same hardware trigger requirements and the same set of cuts 
as real data. In the simulations of detector response the spread in threshold setting ($\approx 10\% $), 
PMT's gain ($\approx 10\%$), and time off-set of PMT's ($\approx 2ns$) as
well as the part of dead tanks ($\approx 1\%$) have been accounted. By varying the detector 
parameters relevant to observed upward-going muon events the systematic uncertainties have been 
evaluated.

Finally we have found the averaged number of 1268.5 events survived all cuts, 
corresponding to 22 runs of real data taking (in total of 460 years).
The comparison of observed and simulated distributions in number of hit tanks
as well as in thickness of absorber crossed by trajectories shows good agreement.
Mean energy of generated neutrinos for muons going out through the telescope is about 50 GeV.

  All-in-all, the ratio of observed total number of events to expected
one without oscillations is found
to be $0.87 \pm 0.03(stat.) \pm 0.05(syst.) \pm 0.15(theor.)$.

\begin{figure*}[t]
\begin{minipage}[t]{0.48\linewidth}
\centering\epsfig{file=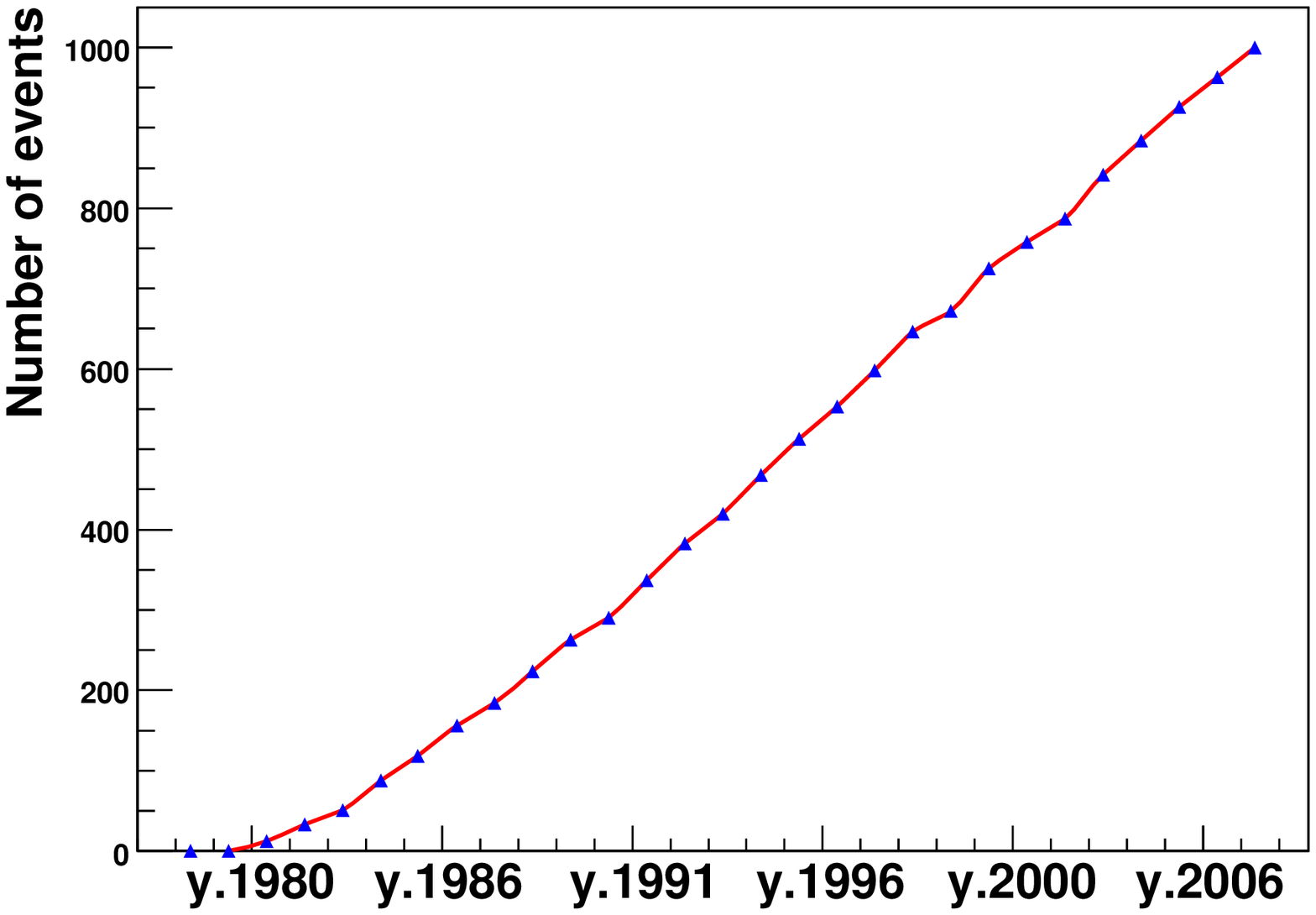,height=\linewidth,width=\linewidth}
\caption{Integral number of measured upward through-going muons versus years.}
\label{fig1bust}
\end{minipage}\hfill
\begin{minipage}[t]{0.48\linewidth}
\centering\epsfig{file=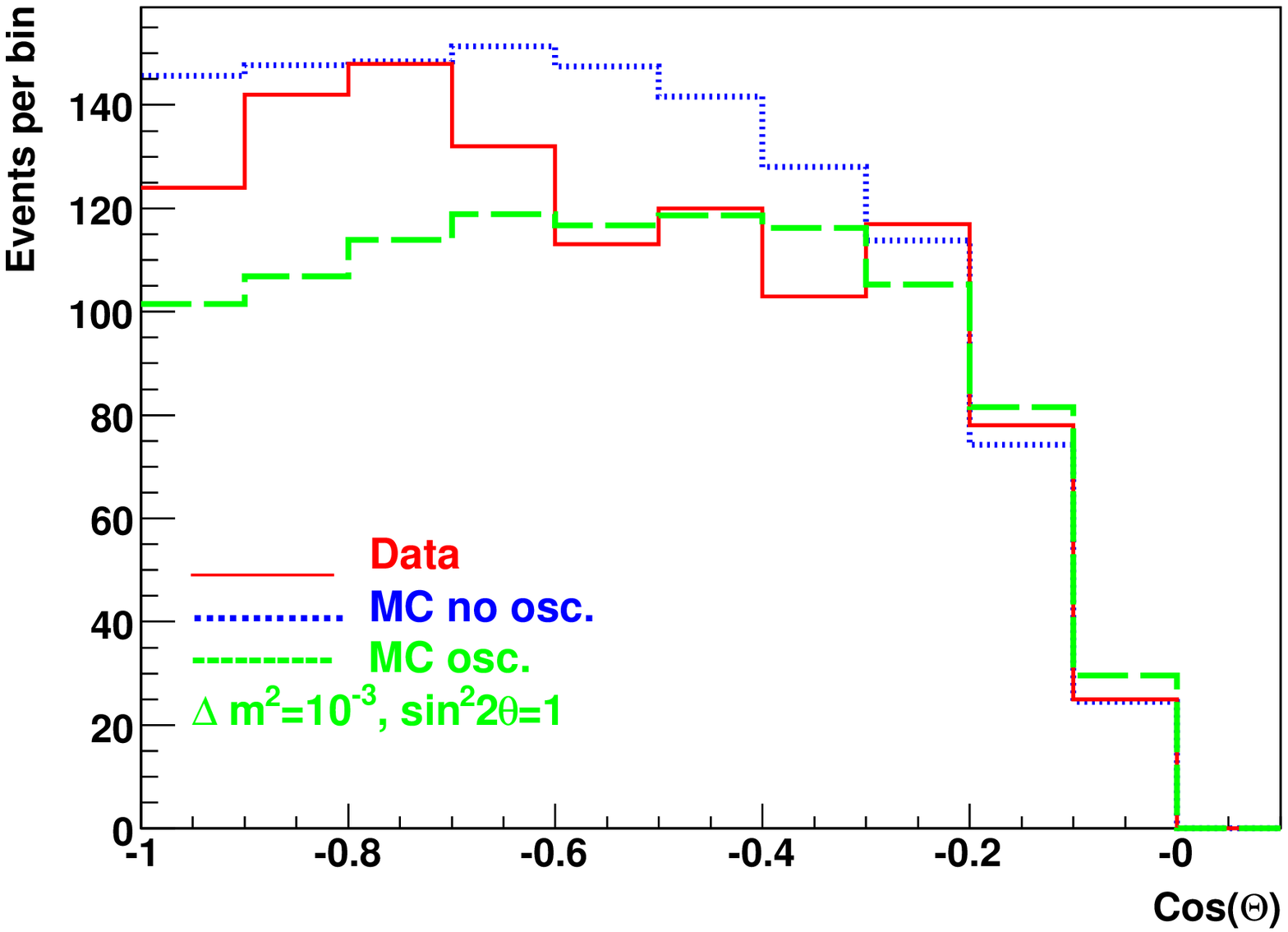,height=\linewidth,width=\linewidth}
\caption{Zenith distributions of upward through-going muons:
data (solid) and MC simulations with oscillations (dash)
and without ones (dot).}
\label{fig2bust}
\end{minipage}
\end{figure*}

\section{\label{sec:eqs} Performance, Astrophysics Aspects and Limits}

The analysis of upward through-going muons at the BUST for searching for high energy
neutrinos other than atmospheric origin has been done previously, when a live time
of observations were approximately twice less. Corresponding flux limits
have been obtained for astronomical point-like sources~\cite{Baksan:91}
and for WIMPs annihilations in the Sun and the Earth' core~\cite{Baksan:96}.
Now with selected upward through-going muons for 21.15 years of live time we have
analyzed arrival muon directions in sky map by bootstrap method in searching for
specified and unidentified sources and by optimal cone-size method for WIMP
search from the Sun.

For neutrino energy range of reconstructed events mentioned above the neutrino-induced muons
deviate from parent neutrino direction inside cone of 4 degree with $90\%$ efficiency.
In Fig.3 the event number distribution of upward through-going muons is shown,
where the sky map is in galactic coordinates and sizes of bin are $4^{\circ}\times4^{\circ}$.
To get a sky map in standard deviations for each point, we obtained the expected noise 
by swapping the event time and event direction
randomly and using  Poisson statistics with sigma of 4 degree per bin. The obtained sky map 
in sigma
deviations is shown in Fig.4. There is no bin with sigma exceeding the value of 3.6 and it is the
evidence of the absence of statistically significant excess in any direction at $90\%$ c.l.
We have set
new upper limit on muon flux from the Galactic Centre to be equal to
$6.3\times10^{-15}cm^{-2}s^{-1}$. The limits for some known sources among pulsars,
SNRs, AGNs, etc will be published later.

 According to latest theoretical estimates the Sun is most promising source in the search
for hidden processes of Dark Matter than others (the Earth or G.C.).
We have analyzed the sample of 1102 muon arrival directions in correlation with the
Sun position. The distribution is shown in Fig.5 together with averaged histogram 
of angles between of muon arrival directions and 5 fake suns. In absence of any features 
and excess toward the Sun we have set upper limits on number of muons at 90$\%$c.l. 
for each cone sizes $\le 16^{\circ}$ and improve our previous numbers by
factor of 1.5 approximately.

The optimal angular windows to collect $90\%$ of muons from annihilation source
for different neutralino masses are shown in Table 1 in~\cite{Baksan:96}. They have
been found by using the code wSUSY4~\cite{Baksan:96} to model expected
neutralinos, and by  further Monte Carlo simulations down to the detector response. 
The effects from oscillating 
neutrino propagations in absorption medium were not included yet. However we intend
to clarify their consequences on sizes of the optimal angular windows. 

Due to low energy threshold of the detector we can look for signatures from
WIMPs with masses starting from magnitude about 12 GeV e.g. much smaller than one from LEP
limits, if we suppose WIMP could be non supersymmetric particle.

\begin{figure*}[t]
\begin{minipage}[t]{0.48\linewidth}
\centering\epsfig{file=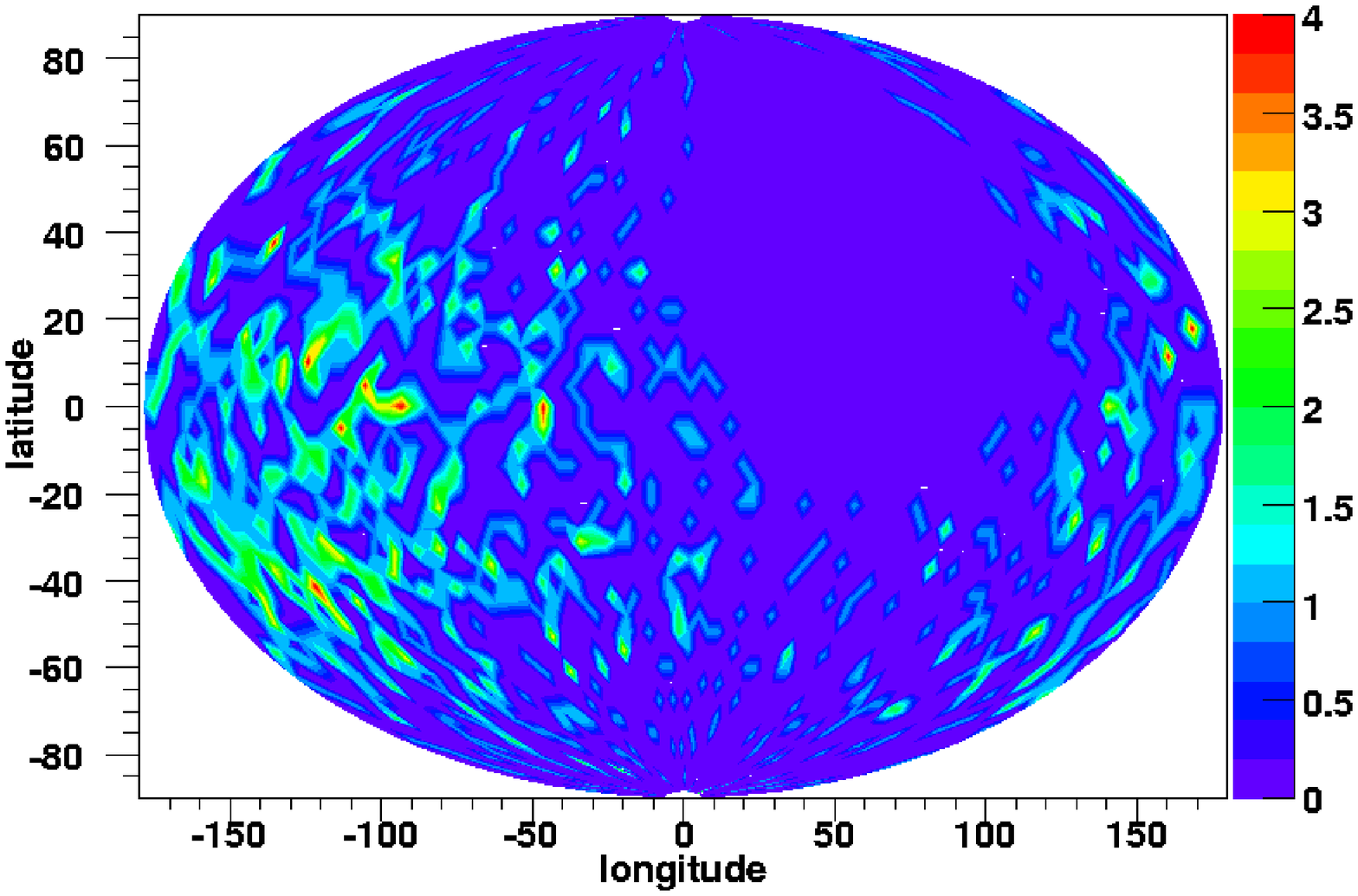,height=\linewidth,width=\linewidth}
\caption{Arrival directions in galactic coordinates with binning $4^{\circ}\times4^{\circ}$.}
\label{fig3bust}
\end{minipage}\hfill
\begin{minipage}[t]{0.48\linewidth}
\centering\epsfig{file=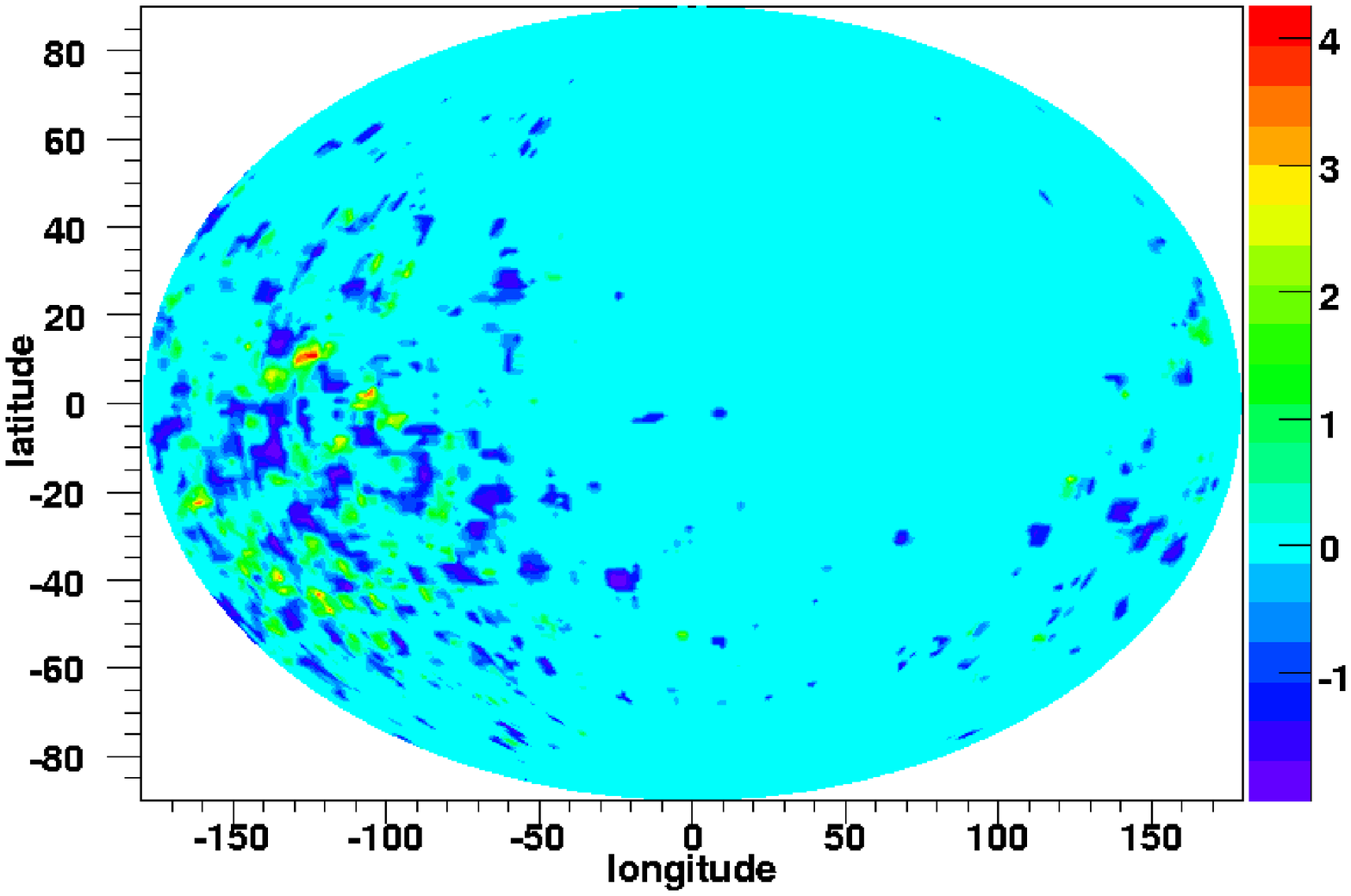,height=\linewidth,width=\linewidth}
\caption{Standard deviations in galactic coordinates with binning $4^{\circ}\times4^{\circ}$.}
\label{fig4bust}
\end{minipage}
\end{figure*}

We have compared the obtained upper limits at 90$\%$ c.l. on annihilation rates in the Sun with
DarkSusy ~\cite{DarkSusy:04} calculations for twelve known benchmark points of mSUGRA, and 
also the lightest neutralino mass model (~$53$ GeV) and declared phenomena signature in EGRET
analysis (~$64$ GeV) ~\cite{bench:03}. The results are plotted in Fig.6 showing that all these models
are still not excluded.\\

\begin{figure*}[t]
\begin{minipage}[t]{0.48\linewidth}
\centering\epsfig{file=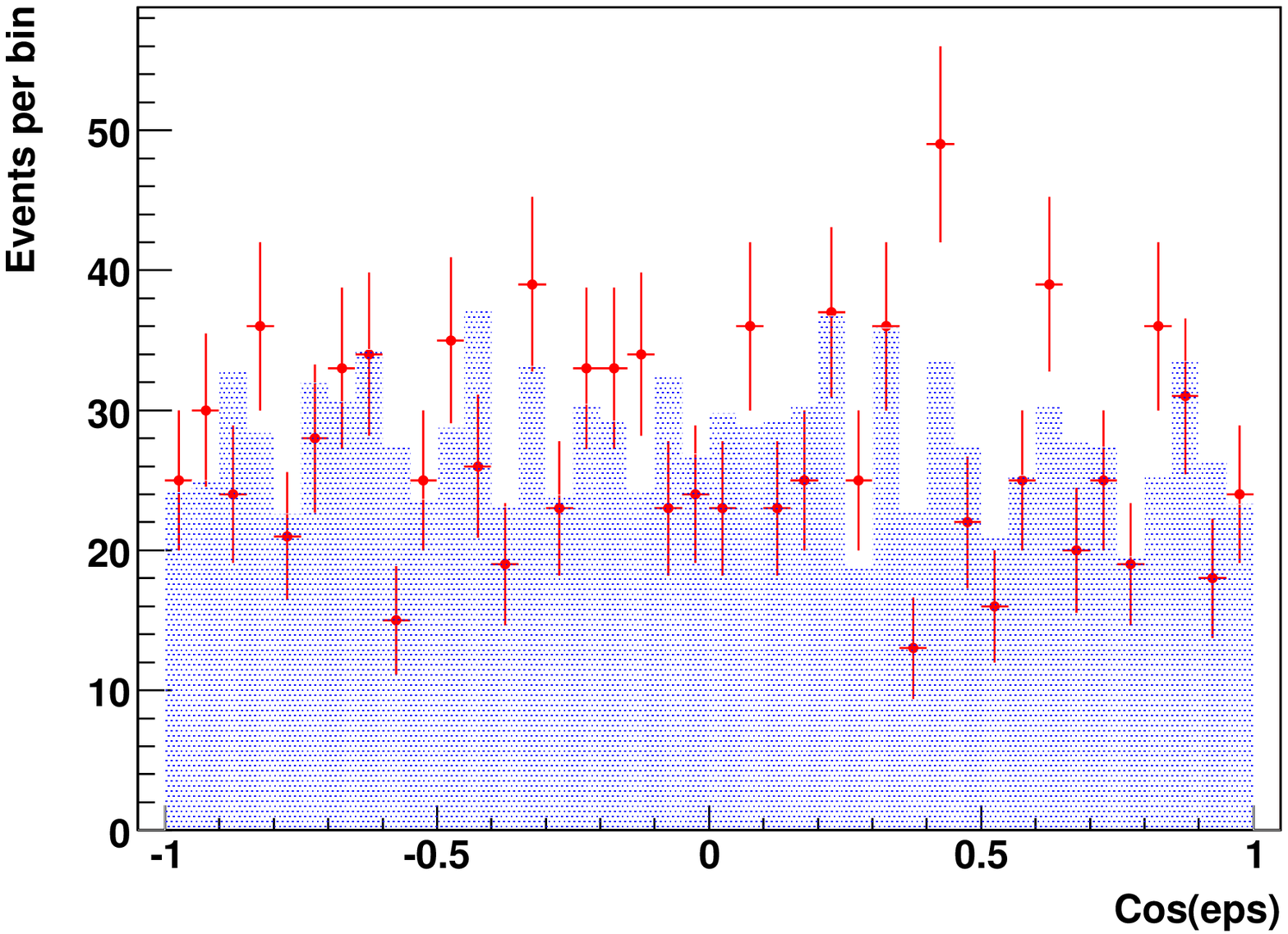,height=\linewidth,width=\linewidth}
\caption{Distribution of angle (eps) between upward through-going muons and the Sun.
Shaded histogram is averaged one for five fake positions of the Sun.}

\label{fig5bust}
\end{minipage}\hfill
\begin{minipage}[t]{0.48\linewidth}
\centering\epsfig{file=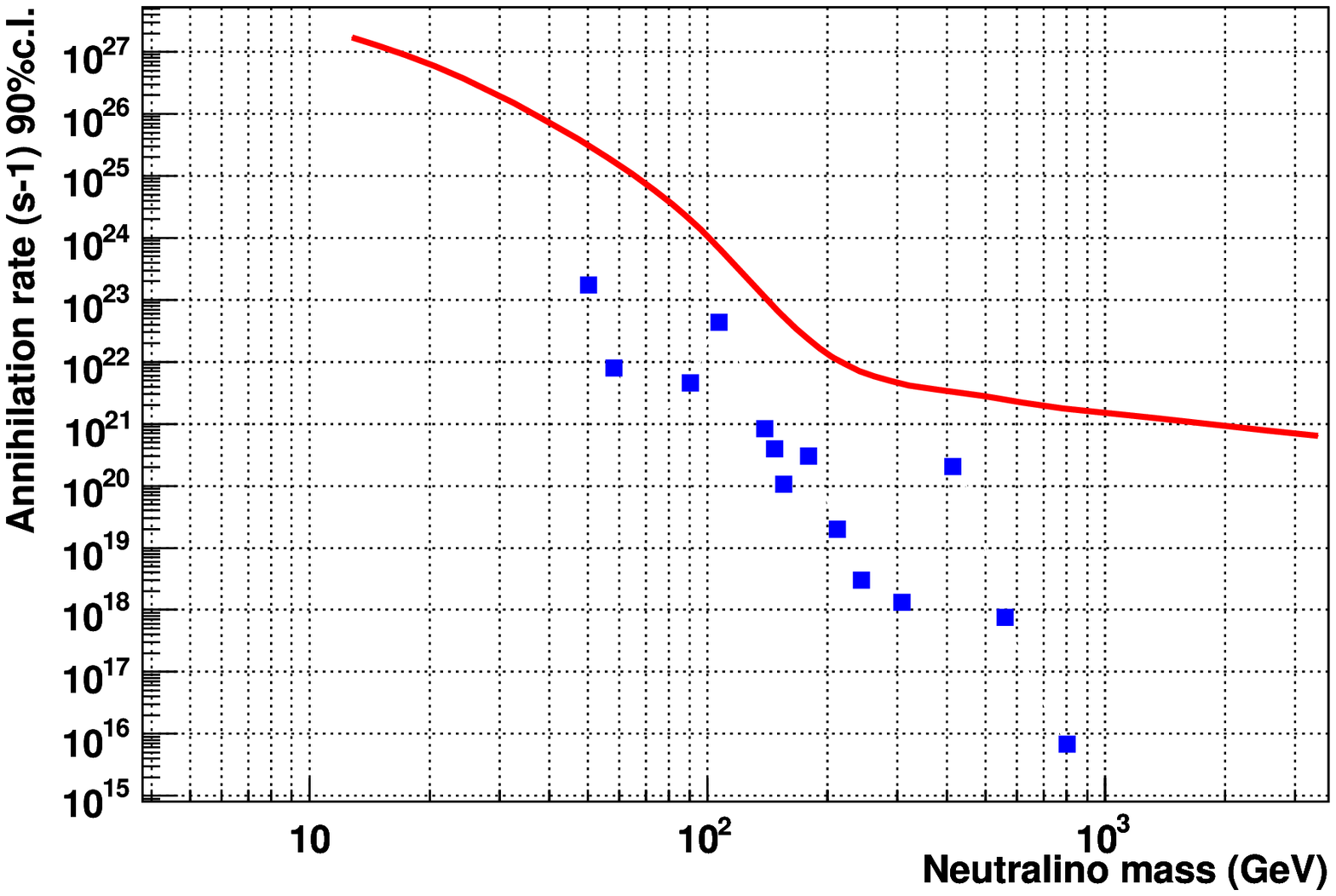,height=\linewidth,width=\linewidth}
\caption{Upper limits at $90\% c.l.$ on neutralino annihilation rate in the Sun as a function 
of neutralino mass. Dots are for mSUGRA benchmark points.}
\label{fig6bust}
\end{minipage}
\end{figure*}

In summary, the Baksan data do not show any excess in measured sky-binning muon fluxes,
including correlation with the Sun.  The reached exposure of $5\times10^{-3} km^2\times yr$
allows to improve our upper limits on muon fluxes and to get them down to limits from 
Super-Kamiokande ~\cite{SK:04}. Clearly, the exposure of several square kilometers
in year as well as decreasing of current uncertainties in nuclear physics and astrophysics
are required to produce adequate results compared to ones in standard cosmology.

\section*{Acknowledgments}
%% Keep the small font tag for the acknowledgements
{\small
O.V.S. thanks Carlos de los Heros for the invitation to give a talk in Uppsala University 
workshop EPNT06 and acknowledges partial support from Swedish Research Council. The work
was supported in part by Russian Foundation for Basic Research under grant No 0502-17196 and 
also by 
the Program of Fundamental Researches of Presidium of Russian Academy of Sciences "Neutrino 
Physics" 
and by Leading Scientific Schools of Russia under contract LSS-4580.2006.2.
}

%%%%%%%%%%%%%%%%%%%%%%%%%%%%%%%% reset.txt counters %%%%%%%%%%%%%%
%%
%%%%%%% do not change below here  %%%%%%%%%%%%%%%%%%%%%%%%%%%%%
%%
\setcounter{section}{0}
\setcounter{subsection}{0}
\setcounter{figure}{0}
\setcounter{table}{0}
\setcounter{equation}{0}
\newpage

%%%%%%%%%%%%%%%%%%%%%%%%%%%%%%%%%%%%%%%%%%%%%%%%%%% Title, authors and addresses
\begin{frontmatter}


\title{Discovery and Upper Limits in Search for Exotic Physics with Neutrino Telescopes}

% use optional labels between square brackets to link authors explicitly to addresses:
% If more than one author, keep a comma between the author tags

\author[address1]{J. Conrad}

\address[address1]{KTH Stockholm, Fysik, AlbaNova University Centre,
  SE-10691 Stockholm, Sweden}

\begin{abstract}
This note gives a short review of the statistical issues concerning upper limit calculation and claiming of discovery arising in the search for exotic physics with neutrino telescopes. Low sample sizes and significant instrumental uncertainties require special consideration. Methods for treating instrumental or theoretical uncertainties in the calculation of limits or discovery are described. Software implementing these methods is presented. The issue of optimization of analysis cuts and definition of sensitivity is briefly discussed.
\end{abstract}

% \begin{keyword}

%\PACS 
% \end{keyword}

\end{frontmatter}

%%%%%%%%%%%%%%%%%%%%%%%%%%%%%%%%%%%%%%%%%%%%%%%%%%%%%% MAIN TEXT
\section{\label{sec:intro} Introduction}
Searches for exotic physics in neutrino telescopes pose challenges to statistical inference. In neutrino telescopes often physicists have to deal with low event counts and significant instrumental uncertainties which need to be taken into account. The small sample size makes the use of asymptotic methods doubtful. Presence of significant instrumental or theoretical uncertainties introduce nuisance parameters which need to be considered in the estimation of physics parameter intervals or in claims for discovery.\par
Neutrino astronomy has so far been in an infant state, the main aim of existing experiments being proof of principle with little hope of detecting neutrinos of cosmic and/or exotic origin. Analyses have therefore often been optimized for setting the most stringent upper limit. With the advent of the IceCube detector this situation might change, in the sense that discovery should become more likely. Analysis optimization methods should therefore be reviewed.\par
In the next section, we remind the reader of methods to calculate confidence intervals and to claim discovery. In section \ref{sec:nuisance} we present two methods for including uncertainties, followed by a discussion on analysis optimization.\par
For simplicity, unless otherwise stated,  throughout the paper we will consider a statistical process with probability density function (PDF) $P(n|s)$, i.e. the probability to observe $n$ given the parameter $s$.

\section{\label{sec:discovery} Claiming discovery and calculating confidence intervals}
Though often mixed up \cite{Cranmer:2005,Conrad:2006}, claiming discovery and calculation of confidence intervals are generally presented as two different cases in statistical literature, see e.g. \cite{Eadie:1971} for an exhaustive introduction. 
\paragraph*{Claiming discovery}
Claiming discovery is an example of {\it hypothesis testing}. In the search of exotic physics the null hypothesis, $H_{0}$, is usually ``No signal present'' and the alternative hypothesis, $H_{1}$, is ``Signal of exotic physics present''.  Discovery is claimed if the measurement result would be very unlikely under the condition that the null hypothesis is true. This is quantified by calculating the {\it p-value}:
\begin{equation}
p = P(T \geq T_{obs} | H_{0})
\end{equation}
where $T$ denotes a {\it test statistics}, which is a function of the observation and the involved hypotheses (see below), and $P$ is a probability density function describing the distribution of the test statistics under the null hypothesis.
If the p-value is smaller than some predetermined threshold, $\alpha_{sign}$, then the null hypothesis is rejected. A particular common choice of the threshold is the probability corresponding to 5$\sigma$ in an equivalent Gaussian distribution, i.e.  $\alpha_{sign} \sim 3 \cdot 10^{-7}$.\par
The test statistics which in a statistical sense is most optimal in distinguishing two alternative hypotheses is according to the Neyman-Pearson lemma the likelihood ratio:
\begin{equation}
T = \frac{\mathcal{L}(n|H_{0})}{\mathcal{L}(n|H_{1})}
\end{equation}
One useful property of the likelihood ratio is that under the assumption that the null hypothesis is true, asymptotically the following statement holds:
\begin{equation}
-2 \ln{T} \sim \chi^2  
\end{equation}
i.e. minus twice the logarithm of the likelihood ratio follows a $\chi^2$ distribution. Though this property is often used in particle physics it is not always true. Except for normality, there are a series of other requirements, some of which are violated even for quite common cases \cite{Conrad:2006,Demortier:2006}. An experimenter intending to use the $\chi^2$ distribution should therefore be convinced that this is justified and in case of doubt, calculate the distribution of the test statistics under the null hypothesis for example by  using Monte Carlo  simulations.\par
The threshold probability in order to decide if the null hypothesis should be rejected is based on the number of false detection one is willing to accept and on the number of searches which are performed. For example, in the context of one of the large CERN experiments, there might be about 50000 independent channels \cite{Feldman:2005}, i.e. 500 channels, 1000 resolution elements each. This implies a false positive detection rate of about 1.5 \% if one requires a 5$\sigma$ detection. It is not obvious, that a similar reasoning would lead to the same requirement in case of searches with neutrino telescopes.

\paragraph*{Calculation of confidence intervals}
Consider $s$  a fixed (unknown) parameter. In frequentist statistics\footnote{the Bayesian calculation is somewhat simpler, since here the interval can be found by integrating in the probability density function $P(s|n)$, where the parameter is considered a random variable}, one defines a confidence interval [$s_1,s_2$] as a member of a set of intervals for which:
\begin{equation}
P(s \in [s_1,s_2]) = 1-\alpha \,\,\,\,\,\,,  \,\,\,\, for  \,\,\,\,\, all \,\,\,s
\label{eq:cov}
\end{equation}
These sets are to be constructed using only $P(n|s)$, $s$ being a fixed but unknown parameter and not $P(s|n)$. The latter would make the method Bayesian, since $s$ is treated as a random variable. The frequentist construction has been introduced by Neyman \cite{Neyman:1937a}. A special case, which is very common in high energy physics has been proposed by Feldman \& Cousins \cite{Feldman:1998a}. Other common approximative methods exist which make use only of the likelihood function, e.g. {\cite{Eadie:1971,Barlow:2003}.\par
All frequentist methods which fulfill the condition in equation \ref{eq:cov} are said to have {\it coverage}\footnote{To be more precise, in frequentist statistics in addition the probability has to be defined in terms of repeated identical experiments.}. Methods which do not have coverage are not valid or need correction\footnote{In general, physicists tend to accept over-coverage (meaning, being conservative), whereas under-coverage is considered unacceptable.}. In case of doubt, coverage has to be demonstrated for example by using Monte Carlo simulations.

\section{\label{sec:nuisance} Treatment of nuisance parameters}

{\it Nuisance parameters } are parameters which enter the data model, but which are not of prime interest. The probably most common example is the expected background in a Poisson process. In a usual physics experiment, only confidence intervals on the parameter of primary interest (for example signal flux or cross-section) are of interest. Thus, ways have to be found to marginalize the nuisance parameter. There are two common approaches: \par

In the first method, the PDF without uncertainty in nuisance parameters is replaced by one where there is an integration over all possible true values of the nuisance parameter {\it(integration method)}:
\begin{equation}
P(n|s,b_{true}) \longrightarrow \int_0^{\infty}{P(n|s,b_{true}) P(b_{true}|b_{est}) d\,b_{true}}
\label{eq:integration}
\end{equation}
Here $b_{true}$ is the true value of the nuisance parameter and $b_{est}$ is its estimate. Since the integrated PDF is describing the probability of the true value given its estimate (and not vice versa) this method is Bayesian. Some prior probability distribution of the true value of the nuisance parameter has to be assumed.\par
In the other common method, the PDF is replaced by one where for each $s$ the PDF is maximized with respect to the nuisance parameters {\it(profiling method)}
\begin{equation}
P(n|s,b_{true}) \longrightarrow \max_{b_{true}}{\mathcal{L}(n|s,b_{true})}
\label{eq:profile}
\end{equation}
with notation as above. This method is completely frequentist, since it never treats $b_{true}$ as a random variable. Therefore the argument of the maximisation is a likelihood function and not a PDF.}\par

Both approaches have recently been subjected to detailed studies regarding their coverage, i.e. with applications to confidence intervals \cite{Tegenfeldt:2004dk,Rolke:2004mj,Conrad:2005b}. For typical problems arising in neutrino telescopes they perform satisfactory. In the context of the LHC searches there are indications the Bayesian method under-covers badly, whereas the profiling method still seems to work fine \cite{Cranmer:2005}.\par
Software has been developed which implements the  {\it integration} and {\it profiling} method for typical problems arising in neutrino telescope analyses.\par

\noindent
\texttt{pole++} is a C++ library of classes which was developed based on the method presented in \cite{Conrad:2002a}. It allows calculating confidence intervals using the Feldman \& Cousins method with integration of nuisance parameters and coverage studies. The signal process considered is a Poisson with known (possibly uncertain) background with different models for nuisance parameters. Several experiments with correlated or uncorrelated uncertainties in the nuisance parameters can be combined. The pole++ library can be obtained from http://cern.ch/tegen/statistics.html \par
\texttt{TRolke} is class which is part of the \texttt{ROOT} analysis package \cite{ROOT}.  It treats a Poisson signal with background process with seven different models for nuisance parameters, which are marginalized using the profiling method. For extraction of the confidence intervals the likelihood function is used. For a description of the method see \cite{Rolke:2004mj}.\par
It should be noted that the \texttt{MINUIT} package \cite{James:1975dr} (in particular using the \texttt{MINOS} facility) also applies the profiling method and is completely general as long as the likelihood function can be written down. For the special case of a Poisson process \texttt{TRolke} applies some improvements.\par

Though profiling and integration are here presented in the context of confidence intervals, both approaches can also be used in hypothesis testing. The substitutions in equations \ref{eq:integration} and \ref{eq:profile} then have to be applied to the likelihoods. The statistical properties to be studied in this case would be the distribution of the test statistics under the null hypothesis as well as possibly the power function (see next section).

%\begin{figure*}[t]
%\begin{minipage}[t]{0.48\linewidth}
%\centering\epsfig{file=frame.eps,height=\linewidth,width=\linewidth}
%\caption{Caption to figure 1 }
%\label{fig1}
%\end{minipage}\hfill
%\begin{minipage}[t]{0.48\linewidth}
%\centering\epsfig{file=frame.eps,height=\linewidth,width=\linewidth}
%\caption{Caption to figure 2}
%\label{fig2}
%\end{minipage}
%\end{figure*}

\section{\label{sec:optimization} Analysis optimization}
Analyses are optimized defining some {\it figure of merit} (FOM), which will be maximized (or minimized) with respect to some cut value $t$\par
In searches with neutrino telescopes, it is often chosen to try to set the most stringent upper limit leading to the introduction of Model Rejection Factor \cite{Hill:2002nv}: 
\begin{equation}
MRF = \frac{<s_{1-\alpha}>}{n_s}
\end{equation}
Here, $s_{1-\alpha}$ denotes the (1-$\alpha$) confidence level upper limit on $s$. The mean is taken over the Poisson distribution with no signal. In case of presence of uncertainties in nuisance parameters the corresponding mean becomes:
\begin{equation}
<s_{1-\alpha}> = \int_0^{\infty} \int_0^{\infty}\,dn \, db_{est} s_{1-\alpha} P(n|b_{est}) P(b_{est}|b_{true}) 
\end{equation}
Unless the size of the uncertainties depends on the cut value or assymetric PDFs for the nuisance parameters have to be considered, the mean calculated using only the Poisson distribution should be sufficient for the calculation of the FOM.\par
In case a decision on whether to set a limit or claim discovery beforehand is not desired, the figure of merit will have to be based on a sensitivity region which is meaningful both if the experimenter would report a limit and if the experimenter wants to claim discovery. Punzi \cite{Punzi:2003} suggests to define the sensitivity region by:
\begin{equation}
1-\beta_{\alpha_{sign}}(s) > 1-\alpha_{CL}
\label{eq:Punzi}
\end{equation}
Where $\alpha_{sign}$ is the significance which the experimenter requires to claim discovery and $\alpha_{CL}$ is the confidence level required in case the experimenter wants to calculate a limit. $\beta (s)$ is the so called power function. Power is a concept arising in hypothesis testing. The power of a test is the probability to reject the null-hypothesis given the alternative hypothesis is true. The concept can also be applied to confidence intervals \cite{Conrad:2005b}. This definition of sensitivity fulfills several desirable properties: for example if $s$ is inside the region defined by equation \ref{eq:Punzi}, then there is a probability of at least $1-\alpha_{CL}$ that it will be discovered. From the sensitivity a FOM can be calculated. Simple expressions of the FOM for common problems can be found in \cite{Punzi:2003}.

\section{Conclusions}
Major statistical challenges to be faced in searches for new physics with neutrino telescopes are low statistics, large instrumental uncertainties and the desire to detect an unknown process while at the same time being able to set stringent upper limits.\par
In this note we point out and briefly discuss different solutions to the above challenges. There are several methods existing to calculate confidence intervals in the presence of instrumental uncertainties which have been tested and behave well also for the small sample sizes and relatively large uncertainties usually encountered in neutrino astronomy. Code performing the necessary calculations is readily available. Sensitivity regions linking the power of a hypothesis test with the confidence level of a confidence interval yield figure of merits which can be used to optimize analysis with respect to both discovery and stringent limits.

\section*{Acknowledgments}

%% Keep the small font tag for the acknowledegments
{\small 
I thank the organizers of the conference for the invitation to give this talk and Olga Botner and Allan Hallgren for reading the manuscript.

%for discovering his duty towards the audience just in time for my talk. 
}

%%%%%%%%%%%%%%%%%%%%%%%%%%%%%%%%%%%%%% reset.txt counters %%%%%%%%%%%%%%
%%
%%%%%%% do not change below here  %%%%%%%%%%%%%%%%%%%%%%%%%%%%%
\setcounter{section}{0}
\setcounter{subsection}{0}
\setcounter{figure}{0}
\setcounter{table}{0}
\setcounter{equation}{0}
\newpage

%%%%%%%%%%%%%%%%%%%%%%%%%%%%%%%%%%%%%%%%%%%%%%%%%%% Title, authors and addresses
\begin{frontmatter}

\title{Kaluza-Klein Dark Matter}

% use optional labels between square brackets to link authors explicitly to addresses:
% If more than one author, keep a comma between the author tags

\author[address1]{Dan Hooper},

\address[address1]{Theoretical Astrophysics, Fermi National
  Accelerator Laboratory, USA.}

\begin{abstract}
In this talk, I will discuss the potential for the direct and indirect detection of Kaluza-Klein dark matter found in models with universal extra dimensions. Although the prospects for direct detection are somewhat daunting, those for future kilometer-scale neutrino telescopes and satellite-based cosmic ray detectors are very encouraging. Experiments such as IceCube, PAMELA and AMS-02, will be quite sensitive to dark matter in this model. 
\end{abstract}

% \begin{keyword}

%\PACS 
% \end{keyword}

\end{frontmatter}

%%%%%%%%%%%%%%%%%%%%%%%%%%%%%%%%%%%%%%%%%%%%%%%%%%%%%% MAIN TEXT
\section{\label{sec:intro} Introduction}

Although there exists an enormous body of evidence for the existence of dark matter, its identity remains unknown~\cite{review}. Weakly Interacting Massive Particles (WIMPs) are, perhaps, the most well motivated class of candidates for dark matter. Among these, the lightest neutralino in models of supersymmetry is the most widely studied. 

Models with extra spatial dimensions can provide an alternative candidate for dark matter, however. In particular, in models in which all of the Standard Models fields are free to propagate in the bulk, called universal extra dimensions, the Lightest Kalzua-Klein Particle (LKP) may be stable and a potentially viable dark matter candidate~\cite{taitservant,fengmatchevcheng}. 

The most natural choice for the LKP is the first Kaluza-Klein excitation of the hypercharge gauge boson, $B^{(1)}$, sometimes called the ``KK photon". I will simply refer to this state as ``Kaluza-Klein Dark Matter" (or KKDM) throughout this talk. Previous studies of KKDM have found that the relic density predicted for such a state would naturally coincide with the measurements of WMAP for masses near about $m_{\rm{LKP}}\approx 900$ GeV if no other Kaluza-Klein states participate in the freeze-out process~\cite{taitservant}. If other states are light enough to significantly effect this process, however, the LKP can be substantially lighter or heavier why still generating the observed dark matter abundance~\cite{taitservant,relic1,relic2}.

In this talk, I will review some of aspects of the direct and indirect detection of KKDM. The prospects for indirect detection are particularly promising in this model.

\section{Direct Detection}

KKDM particles can elastically scatter via scalar couplings with nuclei through the exchange of Higgs bosons and KK quarks. These contributions lead to a spin-independent KKDM-nucleon elastic scattering cross section approximately given by~\cite{directsigma}:
\begin{eqnarray}
\sigma_{B^{(1)} N}   & \approx & \frac{g^4_1 \, m^4_N}{16 \pi m^2_{B^{(1)}}} \bigg[ \frac{1}{m^2_h} \bigg(f_{T_s}+\frac{6}{27}f_{TG}\bigg) + \frac{1}{r_{q^{(1)}}^2 m^2_{B^{(1)}}} \bigg(\frac{289}{81} f_{T_u}+ \frac{25}{81}f_{T_s}\bigg)\bigg]^2, \\ \nonumber & \sim &  1.2 \times 10^{-10} \, {\rm pb}\,\bigg(\frac{1\,\rm{TeV}}{m_{B^{(1)}}}\bigg)^2 \, \bigg[\bigg(\frac{100\, \rm{GeV}}{m_h}\bigg)^2 + 0.09 \, \bigg(\frac{1\,\rm{TeV}}{m_{B^{(1)}}}\bigg)^2 \bigg(\frac{0.1}{r_{q^{(1)}}}\bigg)^2\bigg]^2,
\label{sigmaueddirect}
\end{eqnarray}
where $r_{q^{(1)}} \equiv (m_{q^{(1)}}-m_{B^{(1)}})/m_{B^{(1)}}$ denotes the fractional mass splitting between the KK quarks and the LKP. The various quantities, $f$, account for the quark and gluon content of the nucleon.

Currently, the strongest constraints on the WIMP-nucleon elastic scattering are on the order of $10^{-6}$ pb~\cite{cdms:06}. The range of cross sections predicted for KKDM, in contrast, are far below the experimental sensitivity. Although future ton-scale detectors may be able to test such models, in the near future KKDM will go unexplored by direct detection experiments.

\section{Indirect Detection}

For the purposes of indirect detection, KKDM has several attractive phenomenological features. Firstly, approximately $60\%$ of KKDM annihilations are to charged lepton pairs ($20\%$ to each generation). $33\%$ of annihilations produce pairs of up-type quarks and $3.6\%$ produce neutrino pairs. The remaining fraction generate down type quarks and Higgs bosons. This is in stark contrast to neutralinos which do not annihilate efficiently to neutrinos, positrons, muons or other light fermions. 

Secondly, the total annihilation cross section for KKDM is given by
\begin{equation}
<\sigma v> = \frac{95 g_1^4}{324 \pi m^2_{LKP}} \simeq \frac{1.7 \times 10^{-26} \, \rm{cm^3}/\rm{s}}{m^2_{LKP}(\rm{TeV})}. 
\end{equation}
Notice that this consists entirely of an $a$-term in the expansion, $<\sigma v> =a + bv^2 + \mathcal{O}$$(v^4)$, thus the low velocity cross section is naturally the maximum possible for a thermal relic.

Furthermore, as we will see in the following section, KKDM's spin-dependent elastic scattering cross section with protons can be quite large, leading to the efficient capture of such particles in the Sun, and correspondingly large neutrino fluxes.

\subsection{Indirect Detection with Neutrino Telescopes}

Dark matter particles travelling through the Galactic halo can occasionally scatter and become trapped in deep gravitational wells, such as the Sun or Earth. Within these bodies, they accumulate and their annihilation rate is enhanced, potentially providing an observable flux of high-energy neutrinos~\cite{indirectneutrino}. The capture rate of KKDM particles in the Sun is given by~\cite{kkneu}
\begin{equation}
C^{\odot} \simeq 3.35 \times 10^{18} \rm{s}^{-1} \bigg(\frac{\sigma_{\rm{H, SD}}}{10^{-6}\, \rm{pb}}\bigg) \bigg(\frac{1000 \, \rm{GeV}}{m_{\rm{LKP}}}\bigg)^2,
\end{equation}
where $\sigma_{\rm{H, SD}}$ is the spin-dependent, elastic scattering cross section of KKDM off of hydrogen. This expression assumes a local dark matter density of 0.3 GeV/cm$^3$ and a RMS velocity of 270 km/s. The elastic scattering cross section is given by~\cite{taitservantdirect}
\begin{equation}
\sigma_{\mathrm{H,SD}} = \frac{g'^4 m_p^2}{648 \pi m_{\rm{LKP}}^4 r_{q^{(1)}}^2} 
\left( 4 \Delta_u^p + \Delta_d^p + \Delta_s^p \right)^2,
\end{equation}
where $r_{q^{(1)}}=(m_{q^{(1)}}-m_{\rm{LKP}})/m_{\rm{LKP}}$ is the fractional shift of the Kaluza-Klein quark masses over the LKP mass and the $\Delta^p_q$'s parameterize the fraction of spin carried by 
a constituent quark $q$. Inserting numerical values for the $\Delta^p_q$'s, one finds 
\begin{equation}
\sigma_{\rm{H, SD}} \approx 1.8 \times 10^{-6} \, \mathrm{pb} 
\left( \frac{1000 \, \mathrm{GeV}}{m_{\rm{LKP}}} \right)^4 
\left( \frac{0.1}{r_{q^{(1)}}} \right)^2 \; .
\end{equation}
For the annihilation and elastic scattering cross sections of KKDM, the annihilation rate in the Sun should reach (or nearly reach) equilibrium with the capture rate. These annihilations can produce neutrinos directly or in the decays of tau leptons or quarks~\cite{kkneu}.

\begin{figure*}[t]
%\begin{minipage}[t]{0.48\linewidth}
%\centering\epsfig{file=eventsx.ps,height=\linewidth,width=\linewidth}
%\label{fig3}
%\end{minipage}\hfill
%
\begin{minipage}[t]{\linewidth}
\centering\epsfig{file=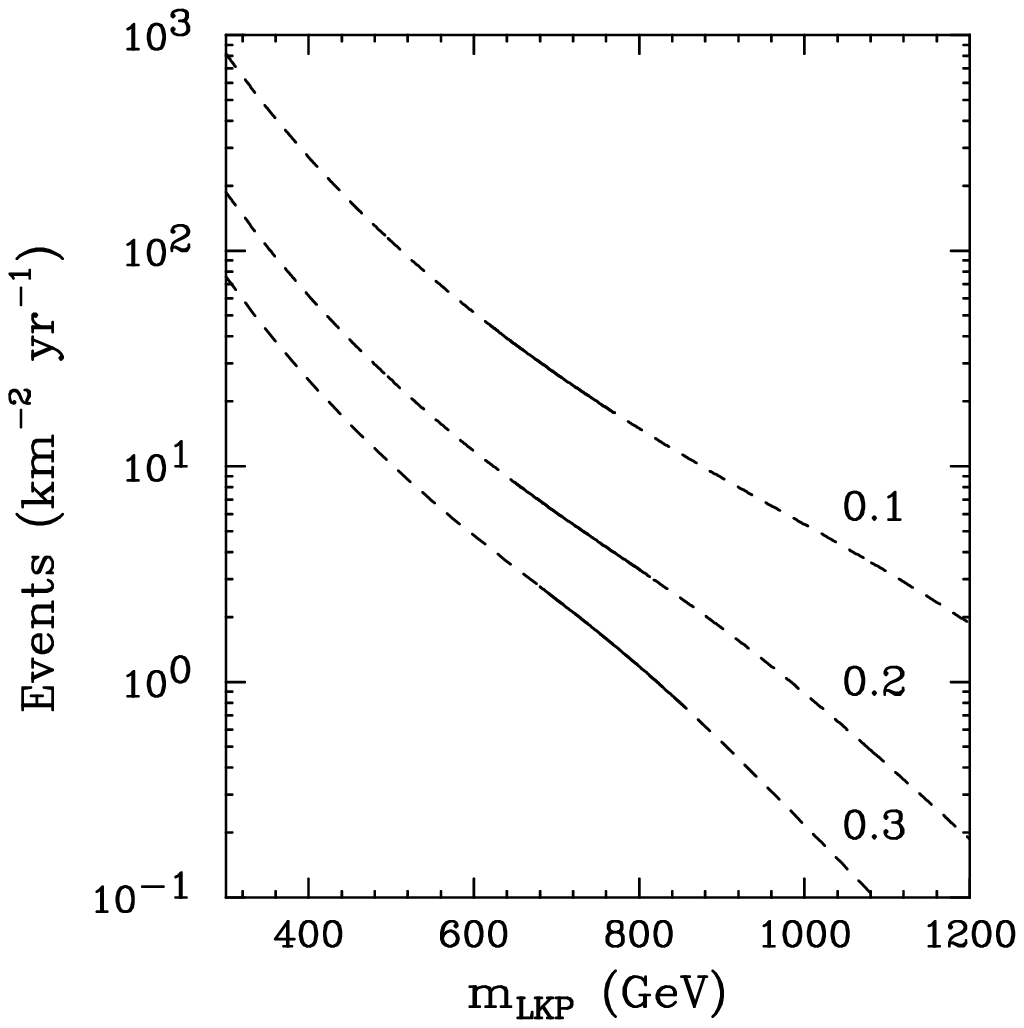,height=0.5\linewidth,width=0.5\linewidth}
\label{neutrinorate}
\end{minipage}
\caption{The rate of neutrino-induced muons above 50 GeV predicted in a kilometer-scale neutrino telescope, such as IceCube. Curves are shown for Kaluza-Klein quarks 10$\%$, 20$\%$ and 30$\%$ heavier than the LKP.}
\end{figure*}

Muon neutrinos which reach the Earth from the Sun can scatter in charged current interactions with nucleons to produce high-energy muons. These muons produce observable ``tracks'' as they propagate through the medium of a neutrino telescope, such as the Antarctic ice of the IceCube experiment. The rate of muon tracks generated in a kilometer-scale neutrino telescope from KKDM annihilations in the Sun in shown in figure~\ref{neutrinorate}. Notice that these results depend strongly on the LKP's mass and the mass of the Kaluza-Klein quarks. Calculations of the radiative corrections to the Kaluza-Klein spectrum estimate values of $r_{q^{(1)}}$ roughly in the range of 0.1 to 0.2~\cite{radiative}. For an 800 GeV LKP, about 5 to 50 events per year are predicted in IceCube over this range. For a lighter LKP of 500-600 GeV, up to 100 events could be observed.

\begin{figure*}[t]
\begin{minipage}[t]{0.48\linewidth}
\centering\epsfig{file=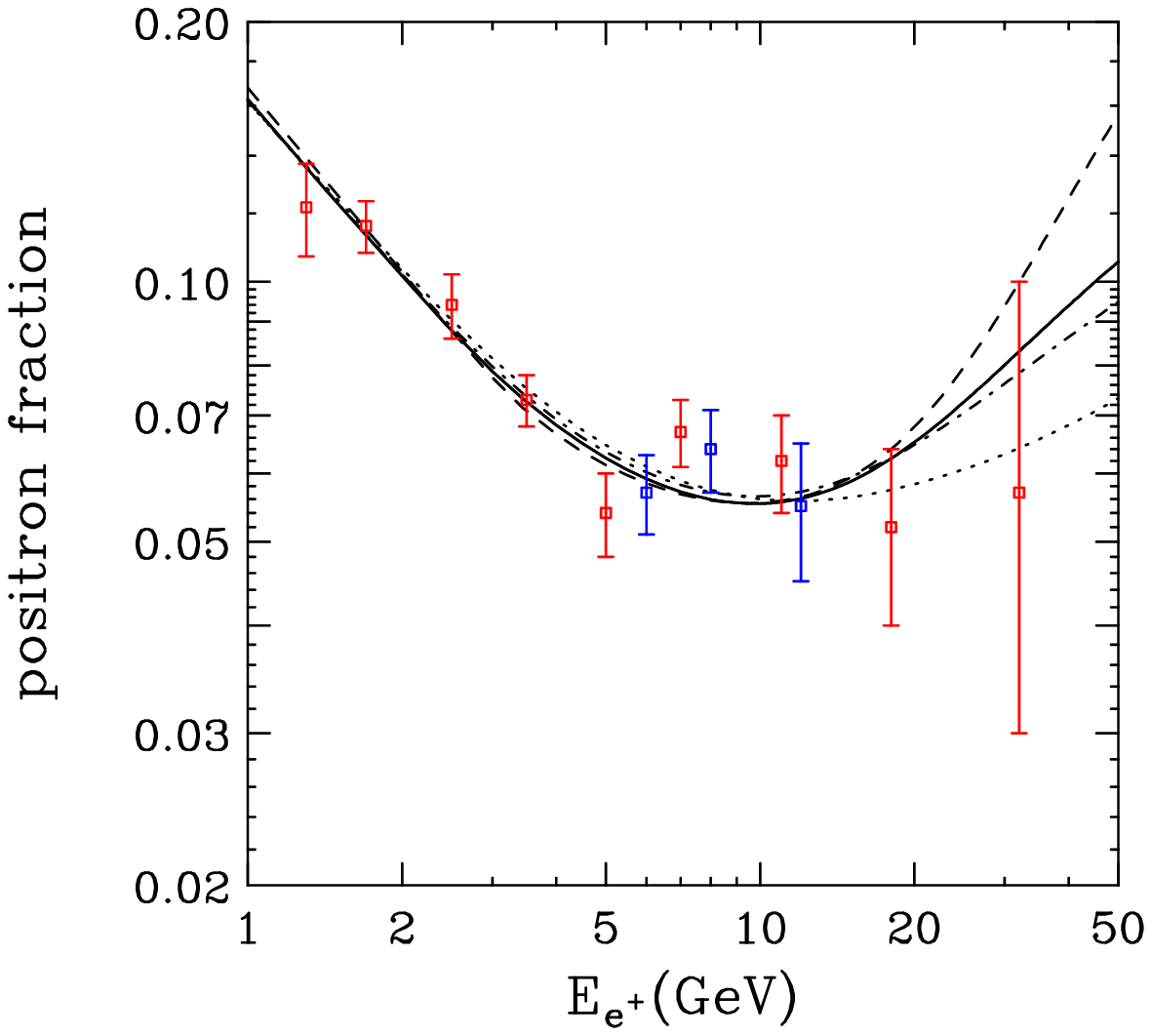,height=\linewidth,width=\linewidth}
\caption{The ratio of positrons to positrons plus electrons with a contribution from annihilating KKDM, compared to the HEAT data. Results for several choices of diffusion parameters are shown. The annihilation rate was normalized to the HEAT data.}
\label{heat}
\end{minipage}
\hfill
\begin{minipage}[t]{0.48\linewidth}
\centering\epsfig{file=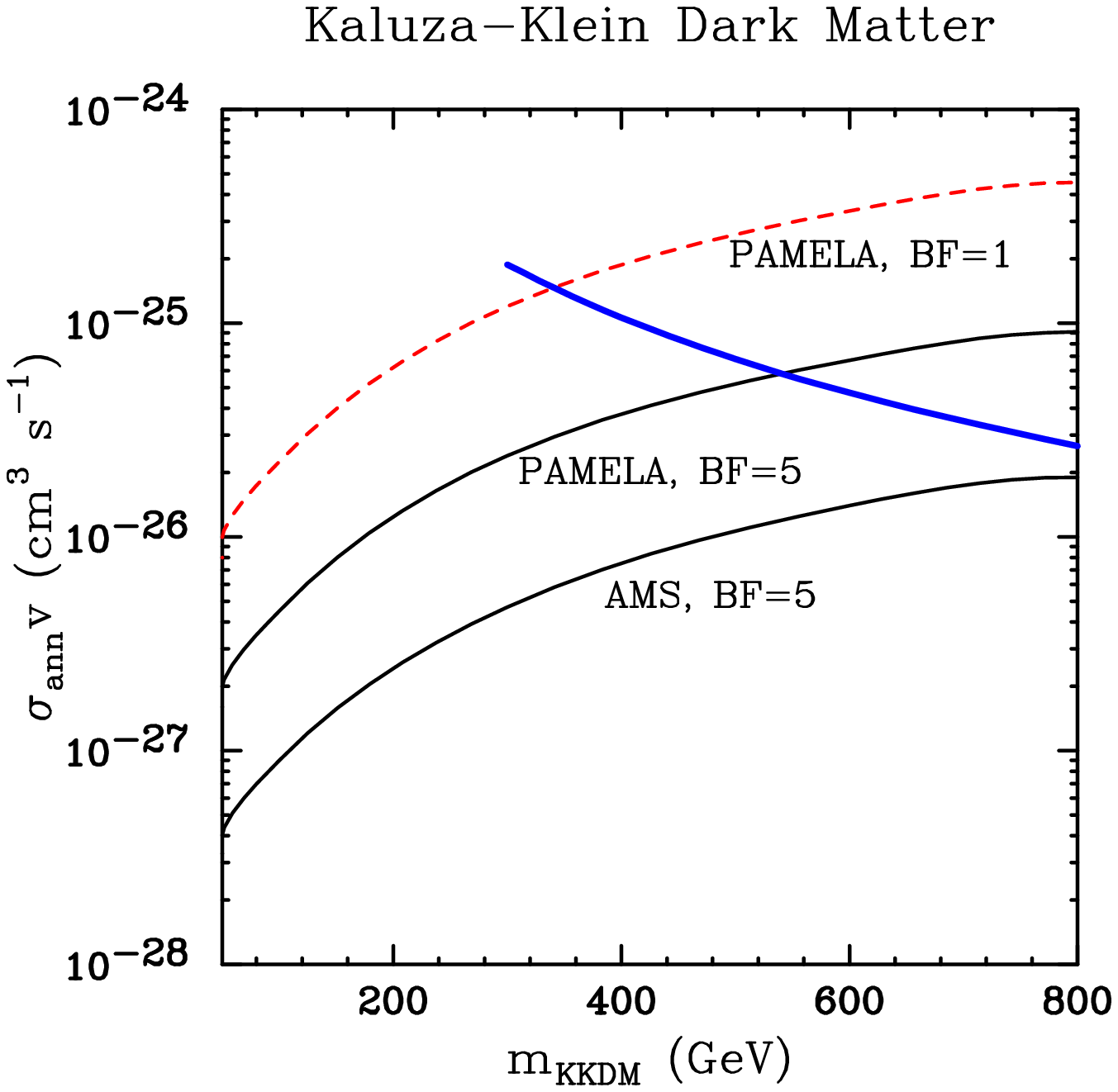,height=\linewidth,width=\linewidth}
\caption{The sensitivity of PAMELA and AMS-02 to cosmic positrons
  generated in KKDM annihilations. Contours are shown for a completely homogeneous distribution (BF=1) and a mildly clumped distribution (BF=5). The downward sloping curve represents the annihilation cross section predicted for KKDM.}
\label{future}
\end{minipage}
\end{figure*}

\subsection{Indirect Detection with Positron Experiments}

Dark matter annihilating in the Galactic halo can produce several potentially observable species of cosmic rays, including gamma-rays, anti-protons, anti-deuterons and positrons. Kaluza-Klein Dark Matter has characteristics which are favorable for detection with cosmic positrons: a large low-velocity annihilation cross section and a large fraction of annihilations which produce energetic positrons (such as the modes $e^+ e^-$, $\mu^+ \mu^-$ and $\tau^+ \tau^-$). 

To calculate the observed positron spectrum, the positrons injected in dark matter annihilations must be propagated through the Galactic magnetic fields, including scattering with starlight and the CMB. These effects are taken into account by solving the diffusion-loss equation:
\begin{equation}
\frac{\partial}{\partial t}\frac{dn_{e^{+}}}{dE_{e^{+}}} = \vec{\bigtriangledown} \cdot \bigg[K(E_{e^{+}},\vec{x})  \vec{\bigtriangledown} \frac{dn_{e^{+}}}{dE_{e^{+}}} \bigg] \frac{\partial}{\partial E_{e^{+}}} \bigg[b(E_{e^{+}},\vec{x})\frac{dn_{e^{+}}}{dE_{e^{+}}}  \bigg] + Q(E_{e^{+}},\vec{x}),
\end{equation}
where $K(E_{e^{+}},\vec{x})$ is the diffusion constant, $b(E_{e^{+}},\vec{x})$ is the energy loss rate and $Q(E_{e^{+}},\vec{x})$ is the source term.

The HEAT experiment, during balloon flights in 1994--95 and 2000, observed an excess in the cosmic positron flux when compared to the electron flux. This excess peaks near 8-10 GeV and extends to above 30 GeV where the detector's sensitivity falls off~\cite{heat}. It has been suggested that this excess could be the product of dark matter annihilations~\cite{positrons}. Neutralino dark matter, which does not annihilate significantly to light fermions, produces positrons inefficiently, however, and therefore requires a very high annihilation rate ({\it i.e.} a clumpy distribution) to account for the excess observed by HEAT~\cite{clumps}. KKDM, on the other hand, can produce such positron fluxes more naturally~\cite{kkpos}.

Satellite based cosmic ray experiments, such as PAMELA and AMS-02, will be capable of measuring the cosmic positron spectrum with much greater precision and to much higher energies than HEAT~\cite{poss2,futurepositron}. The first results from PAMELA, which began its mission in June of 2006, are eagerly awaited. Experiments such as these will be very sensitive to the presence of KKDM in our Galaxy~\cite{poss2}. In figure~\ref{future}, the sensitivity of these experiments to KKDM are shown. 

\section{Summary}

Kaluza-Klein particles within the context of models with universal extra dimensions constitute a viable candidate for dark matter with a number of attractive and interesting phenomenological features. Although the prospects for the direct detection of such a state, at least within the next few years, are not particularly encouraging, indirect detection techniques can be particularly powerful within such models.

The prospects for the indirect detection of Kaluza-Klein Dark Matter (KKDM) are very promising for both future neutrino telescopes and cosmic positron experiments. Experiments such as IceCube, PAMELA and AMS-02 will provide powerful probes of KKDM in the coming years. With PAMELA currently taking data, the near future may well indeed reveal to us new insights into the nature of dark matter.

\section*{Acknowledgments}
\enlargethispage{0.5cm}
%% Keep the small font tag for the acknowledegments
{\small 
This work has been supported in part by the Department of Energy and by NASA grant NAG5-10842.
}

\setcounter{section}{0}
\setcounter{subsection}{0}
\setcounter{figure}{0}
\setcounter{table}{0}
\setcounter{equation}{0}
\newpage

%%%%%%%%%%%%%%%%%%%%%%%%%%%%%%%%%%%%%%%%%%%%%%%%%%% Title, authors and addresses
\begin{frontmatter}


\title{ Nearly Vertical Muons From the Lower Hemisphere in the BAIKAL
Neutrino Experiment}

% use optional labels between square brackets to link authors explicitly to addresses:
% If more than one author, keep a comma between the author tags

\author[address1]{K. Antipin},
\author[address1]{V. Aynutdinov},
\author[address1]{V. Balkanov},
\author[address4]{I. Belolaptikov},
\author[address2]{N. Budnev},
\author[address1]{I. Danilchenko},
\author[address1]{G. Domogatsky},
\author[address1]{A. Doroshenko},
\author[address2]{A. Dyachok},
\author[address1]{Zh. Dzhilkibaev},
\author[address6]{S. Fialkovsky},
\author[address1]{O. Gaponenko},
\author[address4]{K. Golubkov},
\author[address2]{O. Gress},
\author[address2]{T. Gress},
\author[address2]{O. Grishin},
\author[address1]{A. Klabukov},
\author[address8]{A. Klimov},
\author[address2]{A. Kochanov},
\author[address4]{K. Konischev},
\author[address1]{A. Koshechkin},
\author[address6]{V. Kulepov},
\author[address3]{L. Kuzmichev},
\author[address5]{E. Middell},
\author[address1]{S. Mikheyev},
\author[address5]{T. Mikolajski},
\author[address6]{M. Milenin},
\author[address2]{R. Mirgazov},
\author[address3]{E. Osipova},
\author[address2]{G. Pan'kov},
\author[address2]{L. Pan'kov},
\author[address1]{A. Panfilov},
\author[address1]{D. Petukhov},
\author[address4]{E. Pliskovsky},
\author[address1]{P. Pokhil},
\author[address1]{V. Polecshuk},
\author[address3]{E. Popova},
\author[address3]{V. Prosin},
\author[address7]{M. Rosanov},
\author[address2]{V. Rubtzov},
\author[address4]{B. Shaibonov},
\author[address1]{A. Sheifler},
\author[address3]{A. Shirokov},
\author[address5]{Ch. Spiering},
\author[address2]{B. Tarashansky},
\author[address5]{R. Wischnewski},
\author[address3]{I. Yashin},
\author[address1]{V. Zhukov}

\address[address1]{Institute for Nuclear Research, Moscow, Russia }
\address[address2]{Irkutsk State University, Irkutsk, Russia}
\address[address3]{Skobeltsyn Institute of Nuclear Physics  MSU, Moscow, Russia}
\address[address4]{Joint Institute for Nuclear Research, Dubna, Russia}
\address[address5]{DESY, Zeuthen, Germany}
\address[address6]{Nizhni Novgorod State Technical University, Nizhni Novgorod, Russia}
\address[address7]{St Petersburg State Marine University, St Petersburg, Russia}
\address[address8]{Kurchatov Institute, Moscow, Russia}

\begin{abstract}
We present the results of a search for nearly vertically upward going neutrino-induced
muons with 1038 live days data from the neutrino telescope NT200 taken over the five year
period 1998-2002. No excess from WIMP annihilation
in the Earth above the expected atmospheric neutrino background has been found.
Upper limits at 90\% confidence level have been derived on the moun flux
induced by neutrinos from WIMP annihilation.
\end{abstract}

% \begin{keyword}

%\PACS 
% \end{keyword}

\end{frontmatter}

%%%%%%%%%%%%%%%%%%%%%%%%%%%%%%%%%%%%%%%%%%%%%%%%%%%%%% MAIN TEXT
\section{\label{sec:intro} Introduction}

\begin{figure}[t]
\includegraphics[width=.48\textwidth]{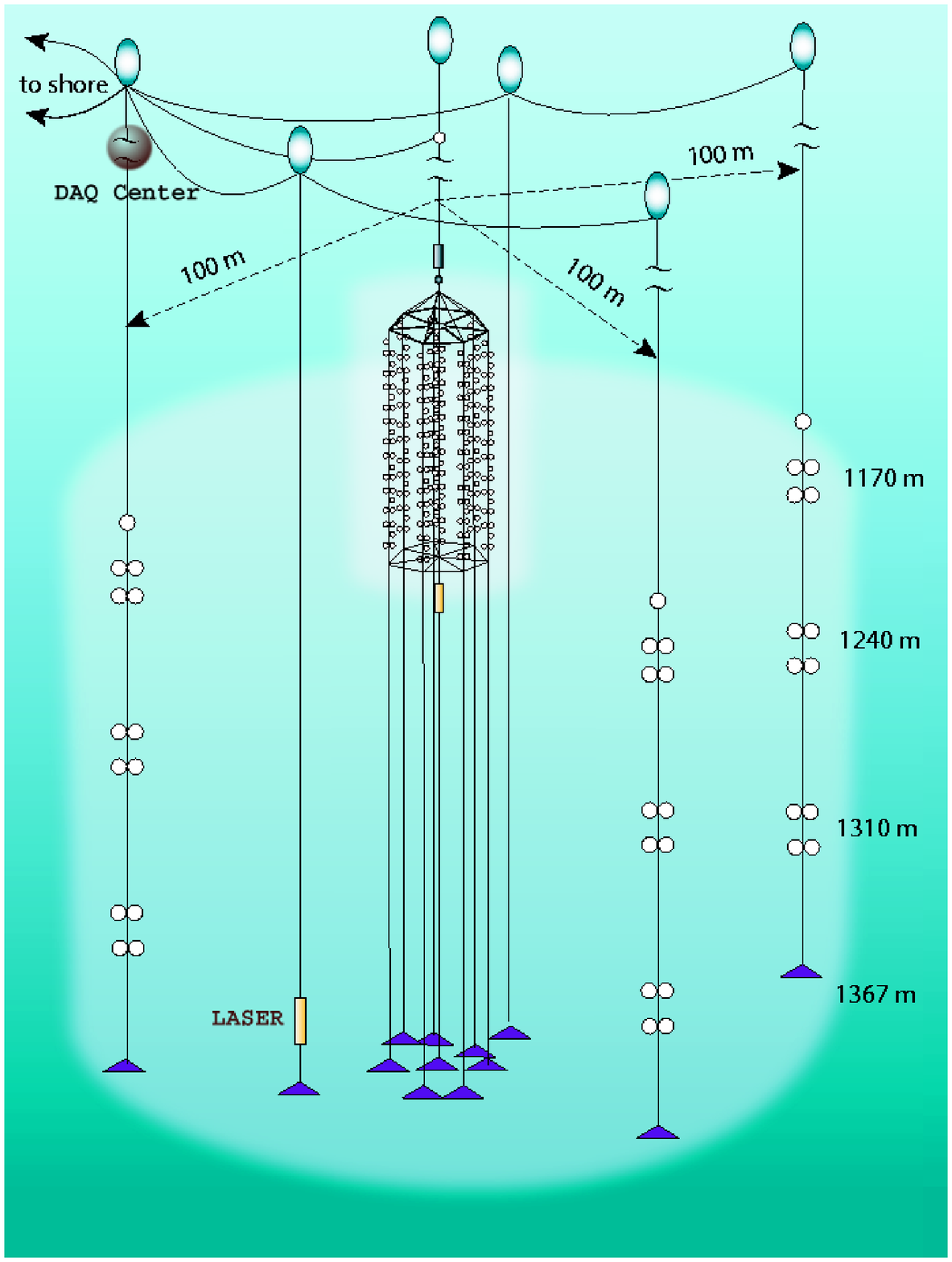}
\hfill
\includegraphics[width=.48\textwidth]{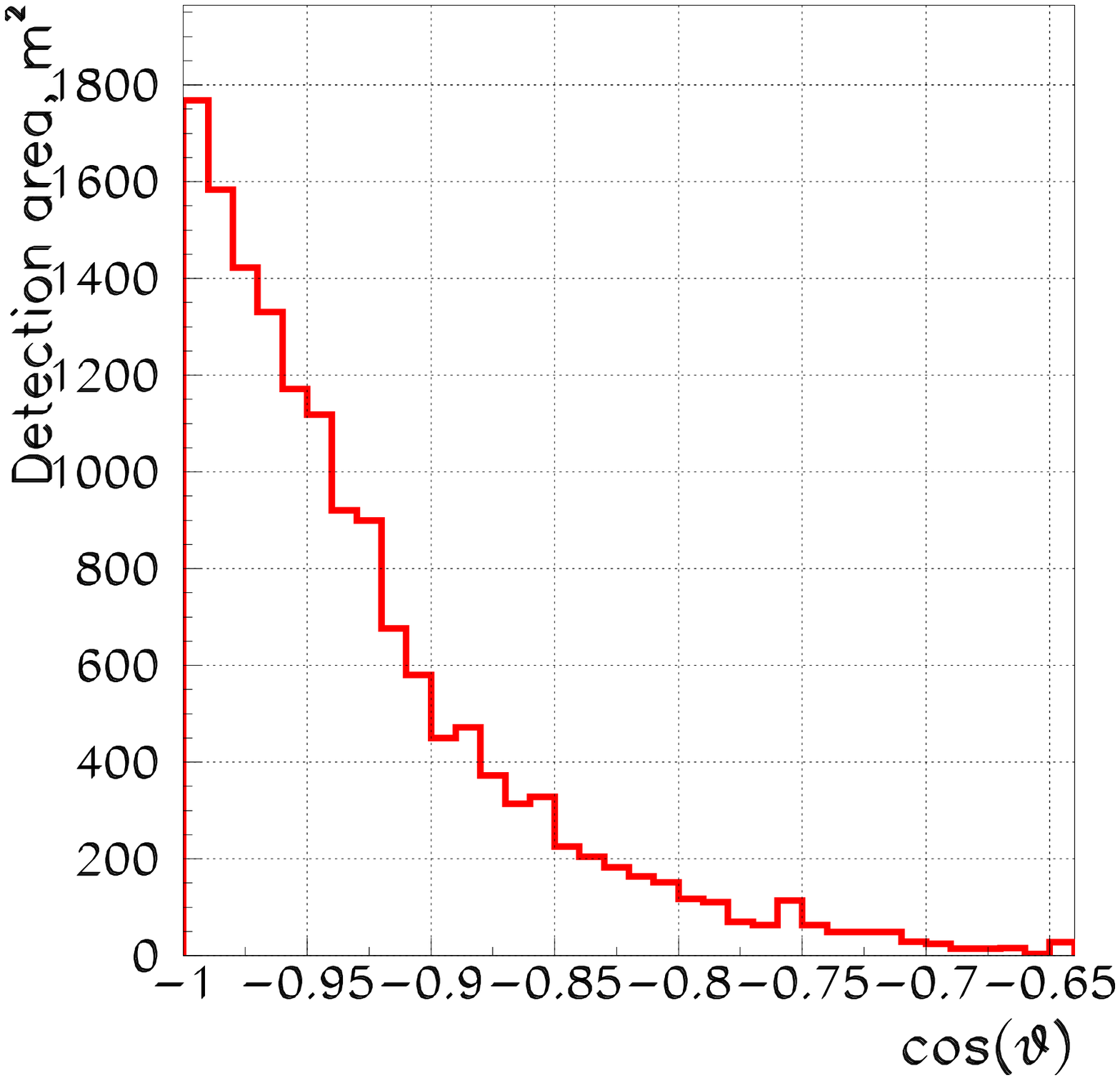}
\\
  \caption{Left: The upgraded Baikal Telescope NT200+ : 
The old NT200 surrounded
by three external long strings at 100\,m radius from the center.
Also indicated: external laser and DAQ center.
  Right: Detection area after WIMP cuts as a 
function of zenith angle. 
}
\label{FIG_NT200}
\end{figure}

The Baikal Neutrino Telescope  is operated in Lake 
Baikal, Siberia,  at a depth of \mbox{1.1 km}. 
Lake Baikal deep water is characterized by an
absorption length of $L_{abs}$(480 nm)=20$\div $24 m, a scattering
length of $L_s=$30$\div $70 m and a strongly anisotropic scattering
function $f(\theta)$ with a mean cosine of the scattering angle 
$\overline{\cos}(\theta)=0.85 \div 0.9$.
The first stage telescope configuration  NT200 
\cite{APP1} was put into permanent operation 
on April 6th, 1998 and consists of 192
optical modules (OMs). 
An umbrella-like frame carries  8 strings,
each with 24 pairwise arranged OMs (see central part of 
Fig. \ref{FIG_NT200} left panel).
All OMs face downward, with the exception of 
the second and eleventh pairs on each string which face
upward. Three underwater electrical cables 
connect the detector with the shore station. 
Each OM contains a 37-cm diameter {\it QUASAR} - photo multiplier (PM), 
which has been developed specially for our project \cite{OM2}. 
The PMs record the Cherenkov light produced by charged particles
in water. The two PMs of a pair are switched in coincidence in order to 
suppress background from bioluminescence and PM noise. 
A pair of OMs defines a {\it channel}. 
The light arrival time assigned to a channel is the response
time of the OM with the earliest hit. The amplitude assigned
to a channel is that recorded by one pre-selected PM of the two
PMs in a pair. 

A {\it trigger}
is formed by the requirement of \mbox{$\geq N$ {\it hits}}
(with {\it hit} referring to a channel) within \mbox{500 ns}.
$N$ is typically set to 
\mbox{3 or 4.} For  these events, amplitude and time of all fired
channels are digitized and sent to shore. 

%%%%%%%%%%%%%%%%%%%%%%%%%%%%%%%%
%%%%%%%%%%%%%%%%%%%%%%%%%%%%%%%

Two nitrogen lasers are used for the calibration of the detector.
The first one ({\it fiber laser})
is mounted just above the array. Its light is guided via optical
fibers of equal length to each OM pair. The fiber laser provides
the OMs with simultaneous light signals in order to determine the offset
for each channel. The second laser ({\it water laser}) is arranged 20 m
below the array. Its light propagates through the water.
This laser serves to monitor the water quality, in addition to
dedicated environmental devices located along a separate string.
A full cycle of detector calibration running both lasers over a wide range of
intensities is repeated every third day.

The upgraded telescope NT200+ was put into operation on April 9th, 2005
\cite{NT200+}. 
This configuration consists of the old NT200 telescope,
surrounded by three new external strings (see Fig.\ref{FIG_NT200}).
The external strings are 140~m long and are placed at 100~meter
distance from the  center of NT200. Each string contains 12 OMs grouped 
in pairs like in NT200. The upper 
pairs are at approximately the same level as the bottom OMs of NT200.
For NT200+, calibration is done with a powerful external
laser light source
with up to $5\times10^{13}$ photons per pulse and nsec-pulse duration,
which is 
located between two outer strings, see Fig.\ref{FIG_NT200}.

Our earlier limits on muon flux induced by WIMPs annihilation at the center
of the Earth, which have been obtained with NT200 using data collected during 1998-1999,
were presented elsewhere \cite{NEUTRINO2004}.
Here we present the new results of a search for nearly vertically upward going neutrino-induced
muons with 1038 live days data from detector NT200 taken over the five year
period 1998-2002.

\section{Search for Neutrinos from WIMP Annihilation}

The search for WIMPs with the Baikal
neutrino telescope is based on a possible signal of
nearly vertically upward going muons, exceeding
the flux of atmospheric neutrinos. 
The method of event selection relies on 
the application of a series of cuts which are tailored to the response
of the telescope to nearly vertically upward moving muons. 
The cuts remove muon events far away from the opposite
zenith as well as  background events which are mostly due to
pair and bremsstrahlung showers below the array and to bare downward
moving atmospheric muons with zenith angles close to the horizon
($\theta>60^{\circ}$). The candidates identified by the cuts are 
afterwards fitted in order to determine their zenith angles.

%%%%%%%%%%%%%%%%%%%%%%%%%%%%%%%%%%
\begin{figure*}[t]
\begin{minipage}[t]{0.48\linewidth}
\includegraphics[width=15pc,height=14pc]{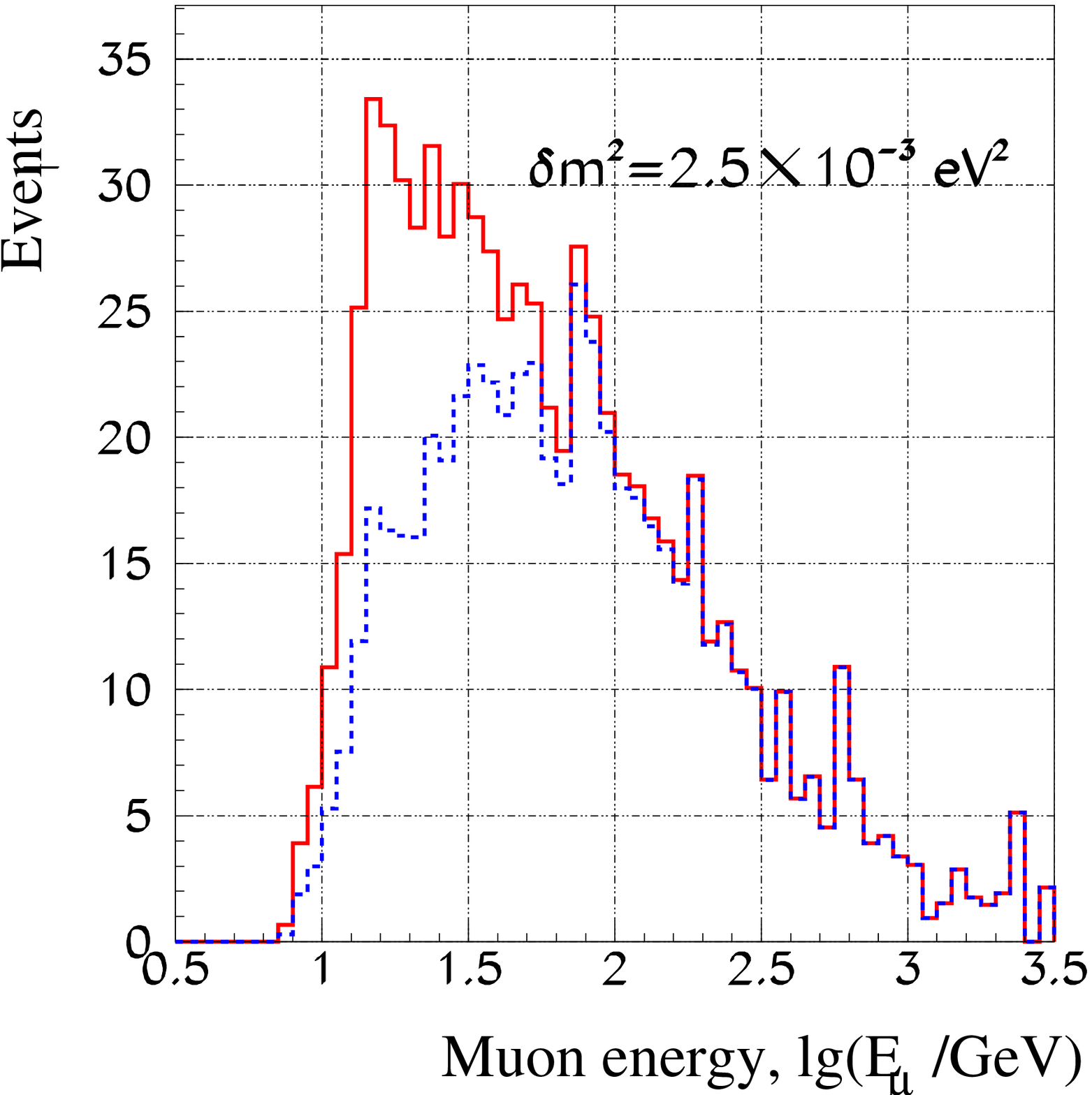}
\end{minipage}\hfill
\begin{minipage}[t]{0.48\linewidth}
\includegraphics[width=15pc,height=16pc]{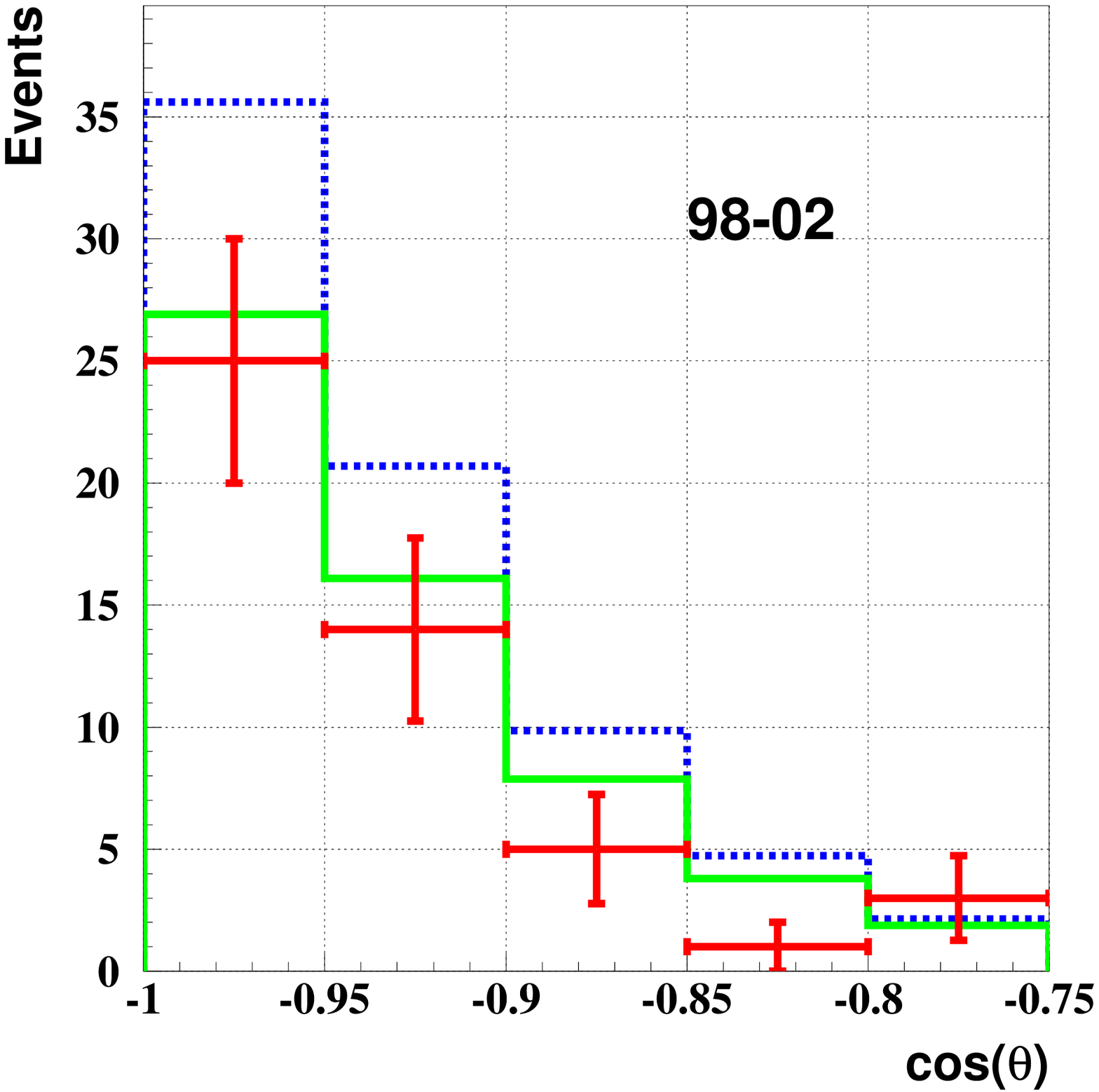}
\end{minipage}
 \caption{Left: Energy distribution of muons 
produced by atmospheric neutrinos, after cuts. 
Dashed/full histogram include/neglect oscillations. 
Right: Angular distributions of selected neutrino
candidates as well as expected distributions in case
of and without oscillations (solid and dashed curves
respectively). 
}
\end{figure*}
%****************************************

For the present analysis
we included all events with $\ge$5 hit channels, out of which 
$\ge$4 hits are along at least one of all hit strings.
To this sample, a series of 5 cuts is applied. Firstly,
the time differences of hit channels along each individual  string
have to be compatible with a particle close to the opposite zenith (1). 
The event length should be large enough (2), and 
the center of gravity of hit channels should not be close to
the detector bottom (3). The latter two cuts reject
efficiently brems showers from downward muons.
Finally,  also time differences of hits along {\it different}
strings have to correspond to a nearly vertical muon (cuts 4 and 5).

Fig. \ref{FIG_NT200} (right panel) shows the dependence of the detection area 
on the cosine of the zenith angle  $\theta$.
The applied cuts select muons with -1$< \cos(\theta) <$-0.65 
and result in a detection area of about 1800 m$^2$ for vertically 
upward going muons.

The expected (normalized) energy spectrum of muons produced 
by atmospheric neutrinos (Bartol flux \cite{Bartol}), 
which survive all cuts, 
%and of their parent neutrinos 
with oscillations (using Super-Kamiokande parameter set 
$\delta m^2 = $2.5$\cdot$10$^{-3}$ eV$^2$ 
with full mixing, $\theta_m\approx \pi/$4 \cite{SK1})
and without oscillations is shown in Fig. 2 (left panel). 
The energy threshold for this analysis is $E_{\mbox{thr}}\sim 10$ GeV.
We expect a muon event suppression of (25-30)\% due to neutrino oscillations.  

From 1038 days of effective data taking 
between April 1998 and February 2003,
48 events with -1$< \cos(\theta) <$-0.75 have been selected 
as neutrino candidates, comparing to 56.6 events expected from atmospheric neutrinos
in case of oscillations and 73.1 without oscillations. 
The angular distribution of these events
as well as the MC - predicted distributions 
without (dashed curve) and with (solid curve) oscillations are shown in
\mbox{Fig. 2} (right panel). 
Within 1$\sigma$ statistical uncertainties the experimental
angular distribution is  consistent with the prediction
including neutrino oscillations.
Note, also, that the theoretical uncertainty in the atmospheric 
neutrino flux calculation is estimated to be about 20\% \cite{Gaisser}. 

%%%%%%%%%%%%%%%%%%%%%%%%%%%%%%%%%%
\begin{figure*}[t]
\begin{minipage}[t]{0.48\linewidth}
\includegraphics[width=15pc,height=16pc]{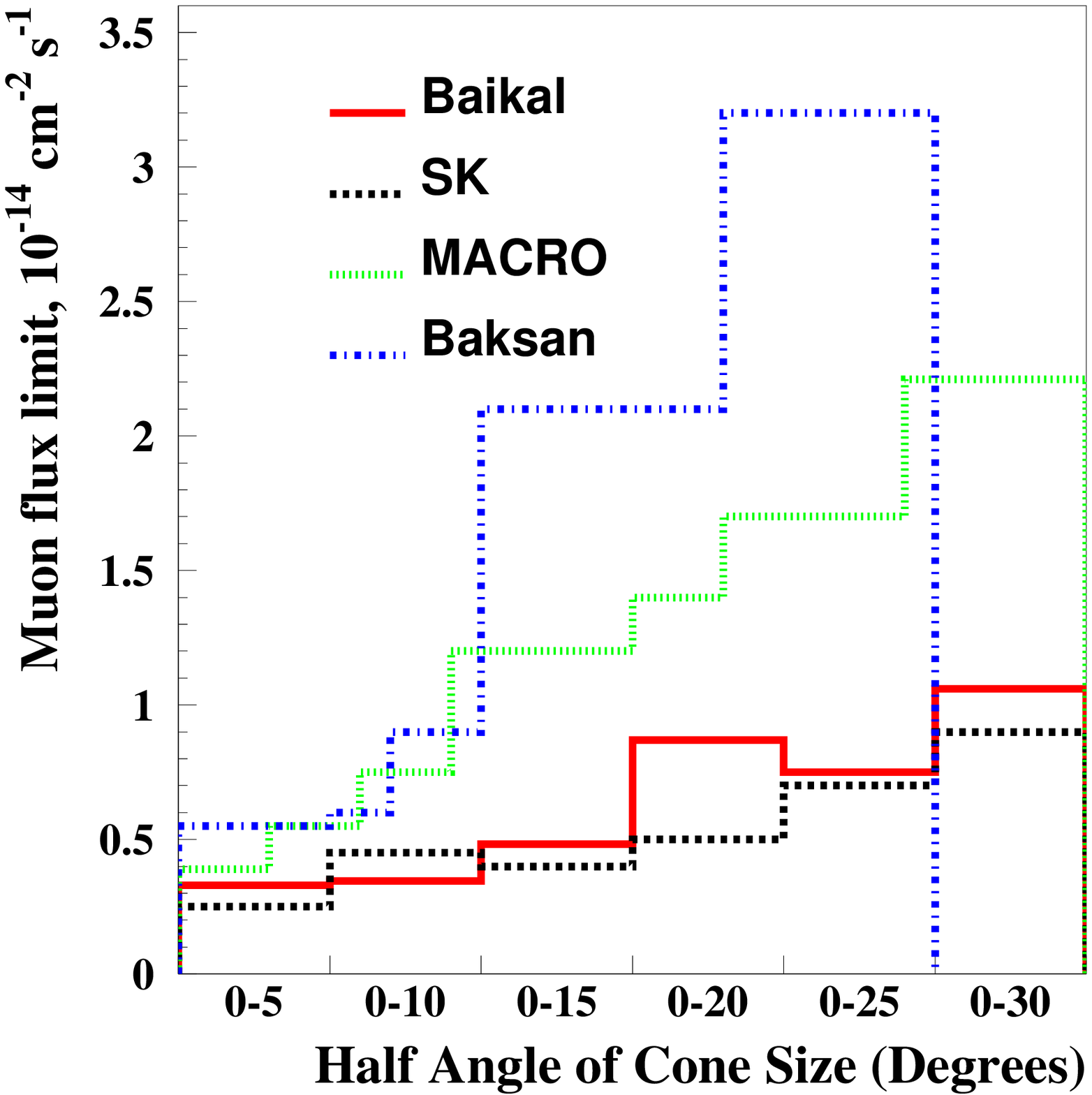}
\end{minipage}\hfill
\begin{minipage}[t]{0.48\linewidth}
\includegraphics[width=16pc,height=16pc]{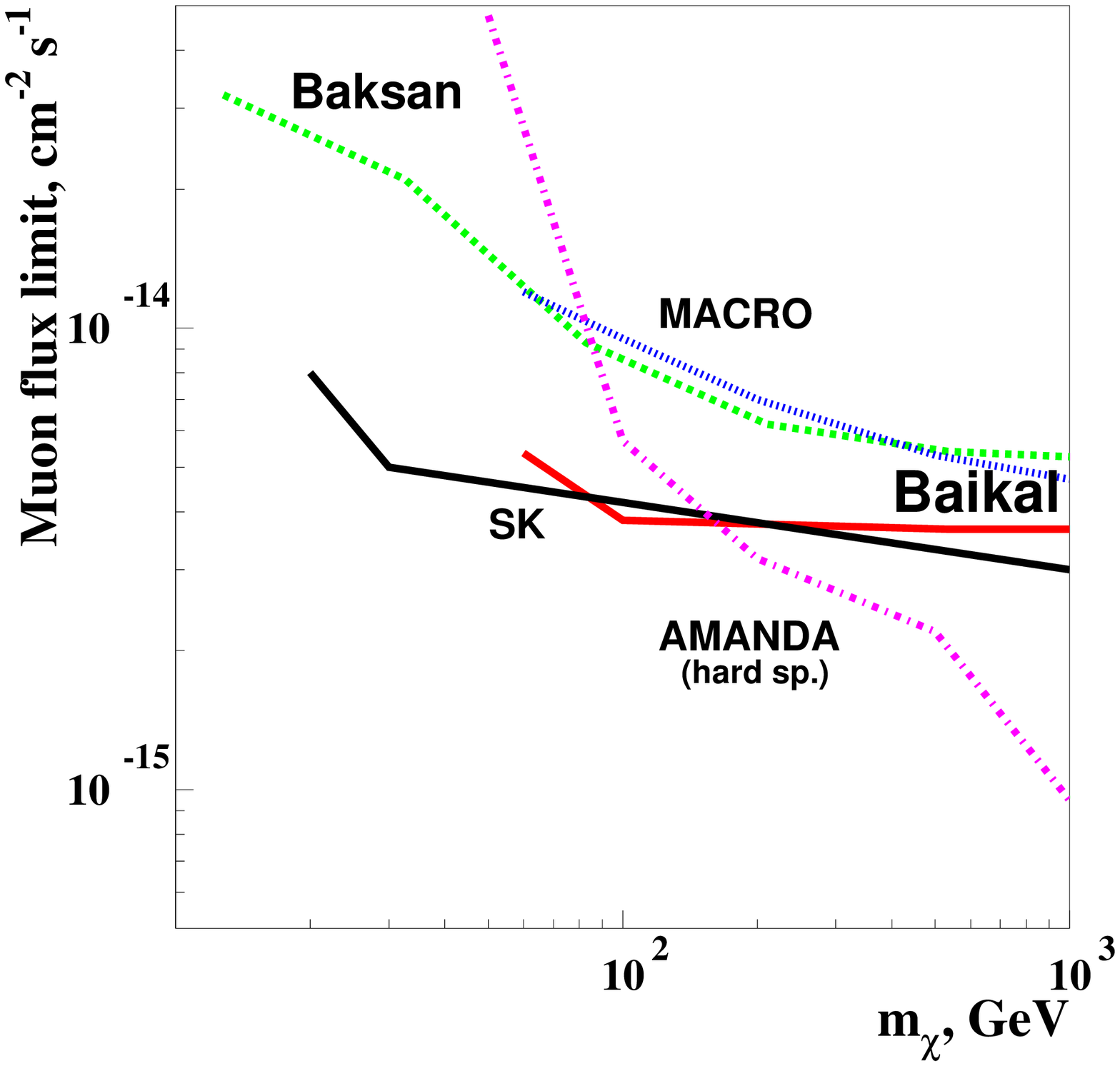}
\label{fig4}
\end{minipage}
 \caption{Left: Limits on the excess muon flux from the center of 
the Earth versus half-cone of the search angle. 
Right: Limits on the excess muon flux from the center of 
the Earth as a function of WIMP mass.
\label{fig3}
}
\end{figure*}
%****************************************

\subsection{Limit on muon excess  from WIMP annihilation}

Regarding the 48 detected events as being induced by atmospheric 
neutrinos, one can derive an upper limit on the additional flux of muons 
from the center of the Earth due to annihilation of neutralinos - the 
favored candidate for cold dark matter. The 90\% C.L. muon flux limits 
for six cones around the opposite zenith obtained with NT200 
($E_{\mbox{thr}}>$10 GeV) in 1998-2002 are shown in Fig. 3 (left panel).
It was shown \cite{BAKSANWIMP,MACROWIMP,SKWIMP} that the size of a cone 
which contains 90\% of signal strongly depends on neutralino mass.
The 90\% C.L. flux limits are calculated as a function of neutralino 
mass using cones which collect 90\% of the expected signal \cite{BAKSANWIMP}
and  are corrected for the 90\% collection efficiency due to cone size.
These limits are shown in Fig. 3 (right panel). Also shown in Fig. 3
are limits obtained by 
Baksan \cite{BAKSANWIMP}, MACRO \cite{MACROWIMP}, 
Super-Kamiokande \cite{SKWIMP} and AMANDA \cite{AMANDAWIMP}.
 
\section{Conclusion}

The Baikal neutrino telescope NT200 is 
taking data since April 1998. 
A search has been performed for nearly vertically
upward going muons based on NT200 telescope data from
the five years 1998-2002. The number of detected events,
as well as their angular distribution, is compatible with expectation
from atmospheric neutrinos. No statistically significant excess
of  nearly vertically  upward going muons over atmospheric neutrino
background was seen. 

Limits on an excess of the muon flux due to WIMP annihilation in the center
of the Earth for various cone angles around nadir 
as well as muon flux limits for different neutralino masses
have been derived, and compared with previous estimates by other experiments.

\section*{Acknowledgments}
{\small 
 This work was supported by the Russian Ministry of Education and Science, the 
  German Ministry of Education and Research and the Russian Fund of Basic Research
  (grants 05-02-17476, 04-02-17289, 04-02-16171,
  05-02-16593, 06-02-31012, 06-02-31005), by the Grant of
  the President of Russia NSh-4580.2006.2 and by NATO-Grant
  NIG-9811707 (2005).
}

\setcounter{section}{0}
\setcounter{subsection}{0}
\setcounter{figure}{0}
\setcounter{table}{0}
\setcounter{equation}{0}
\newpage

%%%%%%%%%
\hyphenation{sub-sample location cosmic energies atmo-sphere atmo-spheric con-taminate eli-minate electro-weak neu-tra-li-no energy matter products array showers sample under re-duction angles azi-muthal ge-ne-ra-ted filter totaling griest akerib limits anni-hilate anni-hilation samples possible}

\begin{frontmatter}
\title{Neutralino searches with AMANDA and IceCube\\
	-- past, present and future --}

\author[daan]{D. Hubert}
\author{for the IceCube Collaboration}
\address[daan]{Vrije Universiteit Brussel, Dienst ELEM, Pleinlaan 2, B-1050 Brussels, Belgium}
\begin{center}{\small (dhubert@vub.ac.be)}\end{center}

\begin{abstract}
Data taken with the AMANDA high energy neutrino telescope can be used in a search for an indirect dark matter signal.  If non-baryonic dark matter exists in the form of neutralinos a neutrino flux at energies lower than about 1 TeV is expected from the decay of neutralino pair annihilation products.  We present published results from searches for neutralinos accumulated in the Earth and the Sun, using AMANDA data from 1997-1999 and 2001 respectively.  Current efforts to improve neutralino trigger and selection efficiencies in analyses of multi-year AMANDA data samples are discussed.  Finally, expected sensitivities for a best-case 1 km$^{3}$ IceCube scenario are given.
\end{abstract}
\end{frontmatter}
%%%%%%%%%

\vspace{-0.5pc}
\section{Introduction}
%\vspace{-0.5pc}
Cosmological observations have long suggested the presence of non-baryonic dark matter on all distance scales.  The WMAP results \cite{wmap} confirmed our current understanding of the Universe, summarized in the concordance model.  In this model the Universe contains about 23\% non-baryonic cold dark matter, but nothing is predicted about the nature of this dark matter.  
A massive, weakly interacting and stable particle appears in Minimally Supersymmetric extensions to the Standard Model that assume R-parity conservation.  Indeed, the supersymmetric partners of the neutral electroweak and Higgs Standard Model bosons mix into an interesting dark matter candidate, the neutralino, whose mass is expected in the GeV-TeV range \cite{neutralinoDM}.  
On their trajectory through the Universe these particles will scatter weakly on normal matter and lose energy.  Eventually, the dark matter particles will be trapped in the gravitational field of heavy celestial objects, like the Earth and the Sun \cite{trap}.  
The particles accumulated in the center of these bodies will annihilate pairwise.  The neutrinos produced in the decays of the Standard Model annihilation products can then be detected with a high energy neutrino detector as an excess over the expected atmospheric neutrino flux.
In this paper we present the published results of searches with the Antarctic Muon And Neutrino Detector Array$\;$(AMANDA) for neutralino dark matter accumulated in the Earth (1997-1999 data set) and the Sun (2001 data set).  We also summarize current improvements and show preliminary results from ongoing analyses on higher statistics data samples accumulated during recent years (2001-2003).

The AMANDA detector \cite{amanda} at the South Pole uses the polar ice cap as a Cherenkov medium for the detection of relativistic charged leptons produced in high energy neutrino interactions with nuclei.  The detector was constructed between 1996 and 2000.  Now totaling 677 light sensitive devices distributed on 19 strings, the detector is triggered when at least 24 detector modules are hit within a sliding 2.5$~\mu$s window.
Before 2000, the detector configuration consisted of between 10 and 13 strings and consequently a lower multiplicity trigger condition was able to cope with the high rate of events produced by cosmic ray interactions with the atmosphere.
Reconstruction of muons, with their long range, offers the angular resolution required to reject the atmospheric background and search for a neutralino-induced signal, which, due to the geographic location of AMANDA, yields vertical upward-going (Earth) or horizontal (Sun) tracks in the instrumented volume.  Indeed, it is possible to eliminate the dominant background, downward-going atmospheric muons.  However, upward-going atmospheric neutrinos will always contaminate the final, selected data sample.
IceCube \cite{icecube}, the 1 km$^3$ successor of AMANDA, is currently being deployed and will continue searching for dark matter in the coming 10-15 years.

%\vspace{-0.5pc}
\section{Signal and background simulation}
%\vspace{-0.5pc}
We have used the DARKSUSY program \cite{darksusy:04a} to generate dark matter induced events for seven neutralino masses $m_\chi$ between 50$~$GeV and 5000$~$GeV, and two annihilation channels for each mass: the $W^+W^-$ channel produces a hard neutrino energy spectrum ($\tau^+\tau^-$ for $m_\chi < m_W$), while $b\bar{b}$ yields a soft spectrum.
The cosmic ray showers in the atmosphere, in which downward-going muons are created, are generated with CORSIKA \cite{corsika:98} with a primary spectral index of $\gamma$=2.7 and energies between 600$~$GeV and 10$^{11}~$GeV.  The atmospheric neutrinos are produced with NUSIM \cite{nusim} with energies between 10$~$GeV and 10$^8~$GeV and zenith angles above$~$80$^\circ$.

%\vspace{-0.5pc}
\section{\label{earth_section}Search for neutralino annihilations in the center of the Earth}
%\vspace{-0.5pc}
\begin{figure}[t]
\vspace{-1.pc}
\begin{center}
%%% old
%\includegraphics*[width=0.41\textwidth,angle=0,clip=true]{earth_efficiencies9799.eps}
%\includegraphics*[width=0.42\textwidth,angle=0,viewport=0 15 400 400,clip=true]{earth_limits_comp_all.eps}
%%% new
\includegraphics*[width=0.47\textwidth,height=0.255\textheight,angle=0,clip=true]{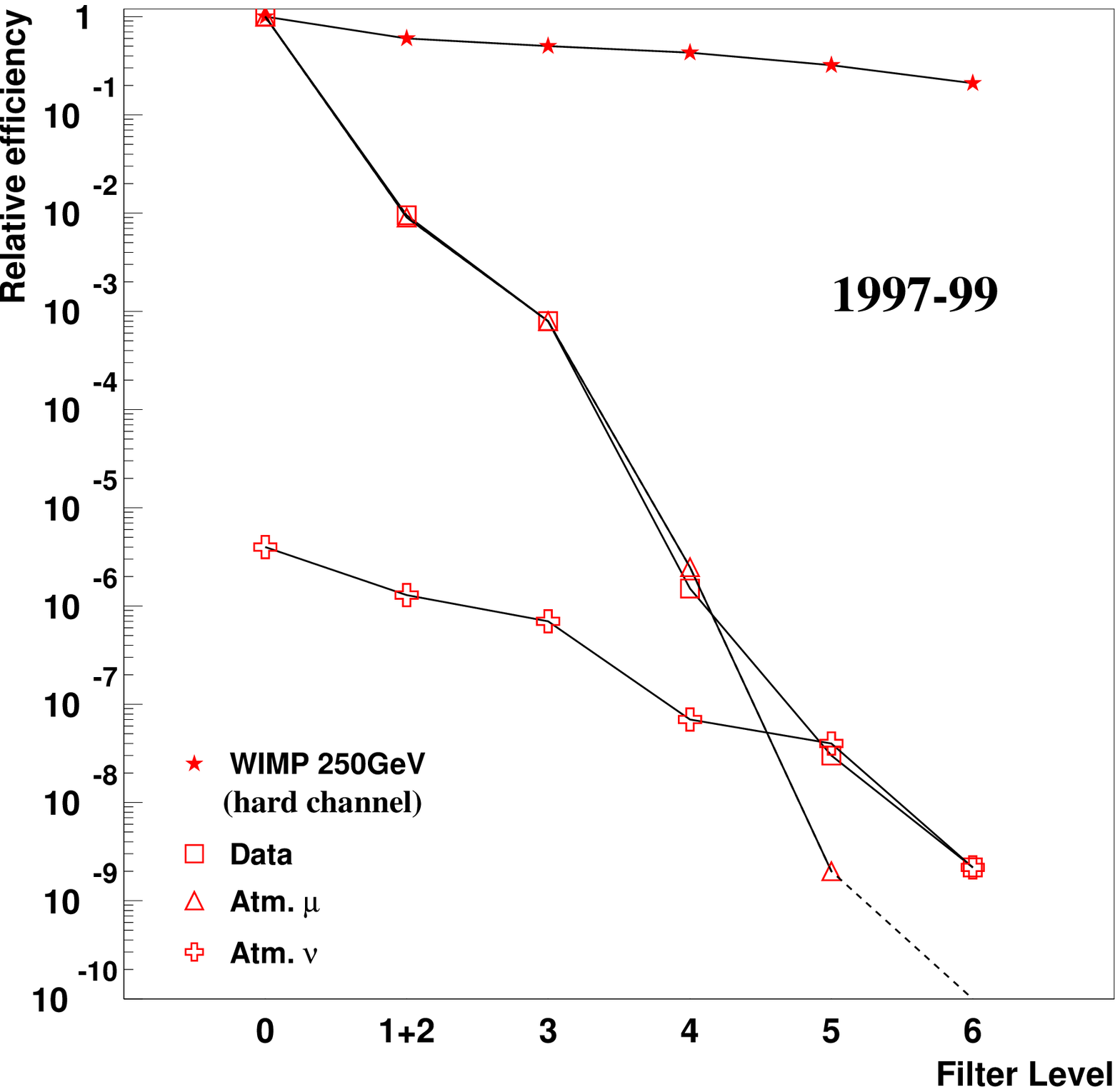}
\includegraphics*[width=0.48\textwidth,height=0.25\textheight,angle=0,viewport=0 15 400 400,clip=true]{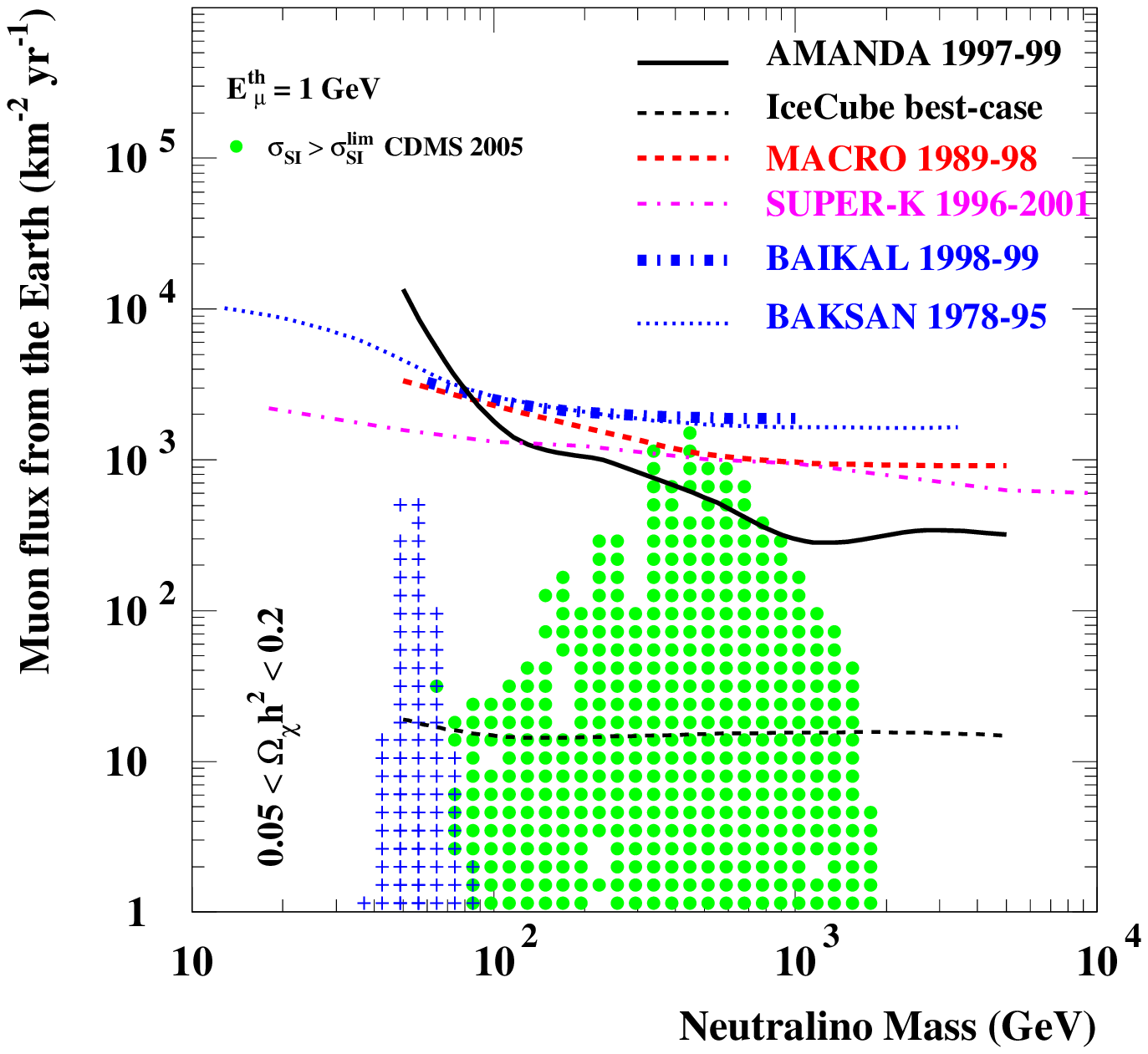}
%\vspace{-2.5pc}
\caption{\label{earth_fig} (a) Detection efficiencies relative to trigger level for the different filter levels in the terrestrial neutralino analysis ($m_{\chi}$=250$~$GeV, hard spectrum) for 1997-1999 data, neutralino signal, atmospheric muons and neutrinos, see Ref. \cite{newearth}.  (b) As a function of neutralino mass, the 90\% confidence level upper limit on the muon flux coming from hard neutralino annihilations in the center of the Earth compared to the limits of other indirect experiments \cite{indirect} and the sensitivity estimated for a best-case IceCube scenario.  Markers show predictions for cosmologically relevant MSSM models, the dots represent parameter space excluded by CDMS~\cite{cdms}.}
\end{center}
%\vspace{-2.5pc}
\end{figure}
A neutralino-induced signal from the center of the Earth was searched for in AMANDA data collected between 1997 and 1999, with a total effective livetime of 536.3 days.  To reduce the risk of experimenter bias, the complete data set of 5.0$\times$10$^9$ events was divided in a 20\% subsample, used for optimisation of the selection procedure, and a remaining 80\% sample, on which the selection was applied and final results calculated.  Similarly, the sets of simulated events were divided in two samples: the first for use in the selection optimisation and the second for the selection efficiency calculations.  The simulated atmospheric muon sample contains 4.2$\times$10$^9$ triggered events (equivalent to an effective livetime of 649.6 days).  The sample of atmospheric$~$neutrinos totals 1.2$\times$10$^8$ events, which corresponds to 2.2$\times$10$^4$ triggers when scaled to the livetime of the analysis.

First, we try to suppress the dominant atmospheric muon background which is about 10$^6$ times more abundant than the atmospheric neutrino background.  This is partially done by selecting the events that are reconstructed as upward-going and that satisfy a cut correlated with reconstruction quality (filter level 3).  However, only a 10$^{-3}$ reduction of the atmospheric muons is obtained this way (Fig. \ref{earth_fig}a) and more elaborate selection criteria are needed to reject downward-going muon tracks misreconstructed as upward-going.
Depending on the detector configuration and the neutralino model under study, the characteristics of the signal differ, which influences selection efficiencies significantly at this point.  Therefore, all further cuts are fine-tuned separately for each neutralino model and year of data taking.
At filter level 4, a neural network selection is applied, reaching another 10$^{-3}$ reduction.  Filter level 5 cuts sequentially on observables, with the goal of removing downward-going muons that resemble signal events.

At filter level 5 the data sample is dominated by atmospheric neutrinos (see Fig. \ref{earth_fig}a).  With no significant excess of vertical tracks observed, the final selection on reconstructed zenith angle (filter level 6) was optimized for the average most stringent 90\% confidence level upper limit on the muon flux.  From the number of observed events and the amount of (simulated) background in the final angular search bin, we infer the 90\% confidence level upper limit on the number of signal events.  Combined with the effective volume at the final cut level and the livetime of the collected data, this yields an upper limit on the neutrino-to-muon conversion rate, which can then be related to the muon flux \cite{newearth} (see Fig. \ref{earth_fig}b).

%\vspace{-0.5pc}
\section{Search for neutralino annihilations in the Sun}
%\vspace{-0.5pc}
\begin{figure}[t] 
\vspace{-1.pc}
\begin{center}
%%% old
%\includegraphics*[width=0.43\textwidth,angle=0,clip]{sun_efficiencies.eps}
%\includegraphics*[width=0.44\textwidth,angle=0,clip,viewport=0 15 400 400]{sun_limits_comp_all.eps}
%%% new
\includegraphics*[width=0.47\textwidth,height=0.255\textheight,angle=0,clip]{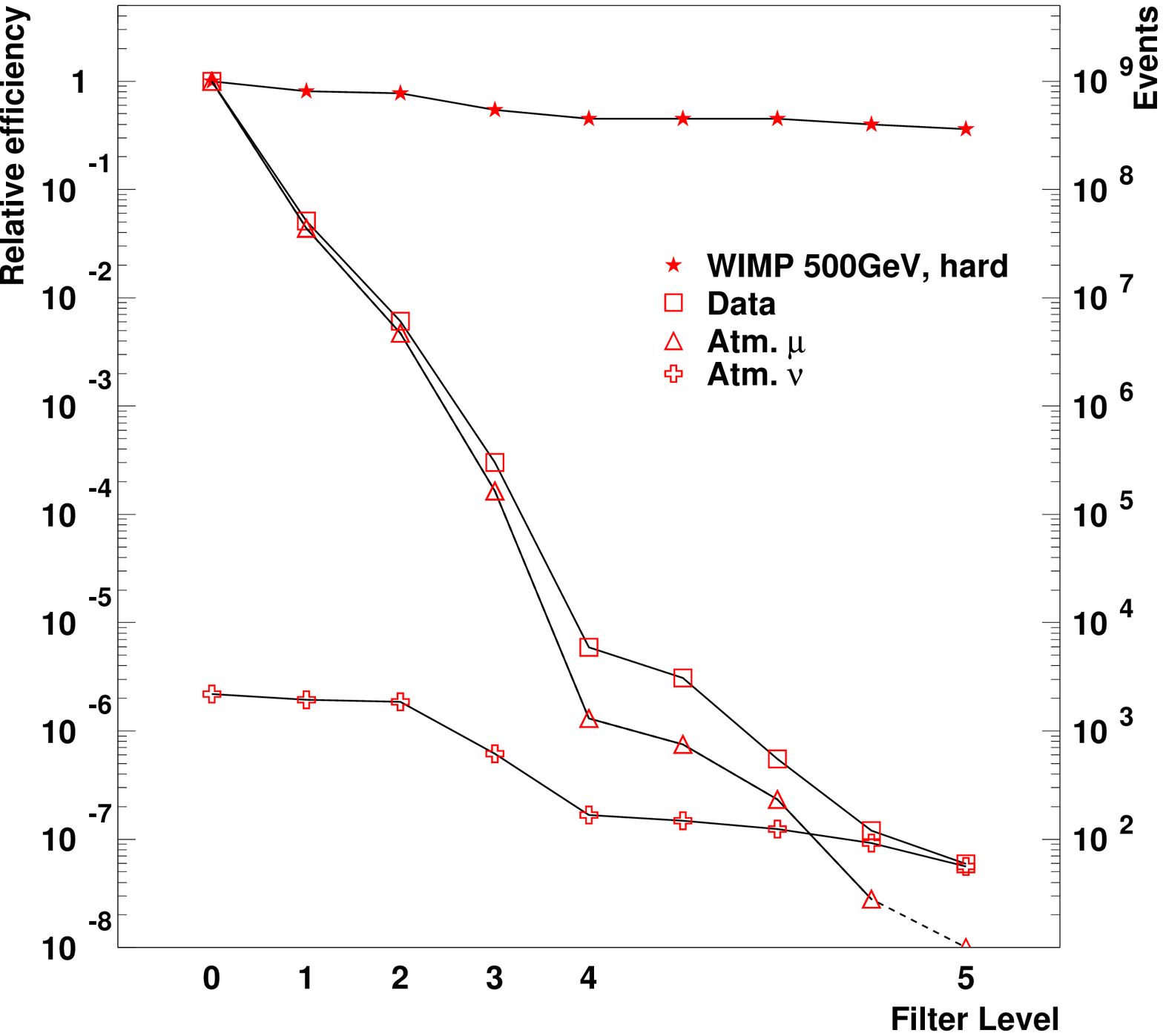}
\includegraphics*[width=0.48\textwidth,height=0.25\textheight,angle=0,clip,viewport=0 15 400 400]{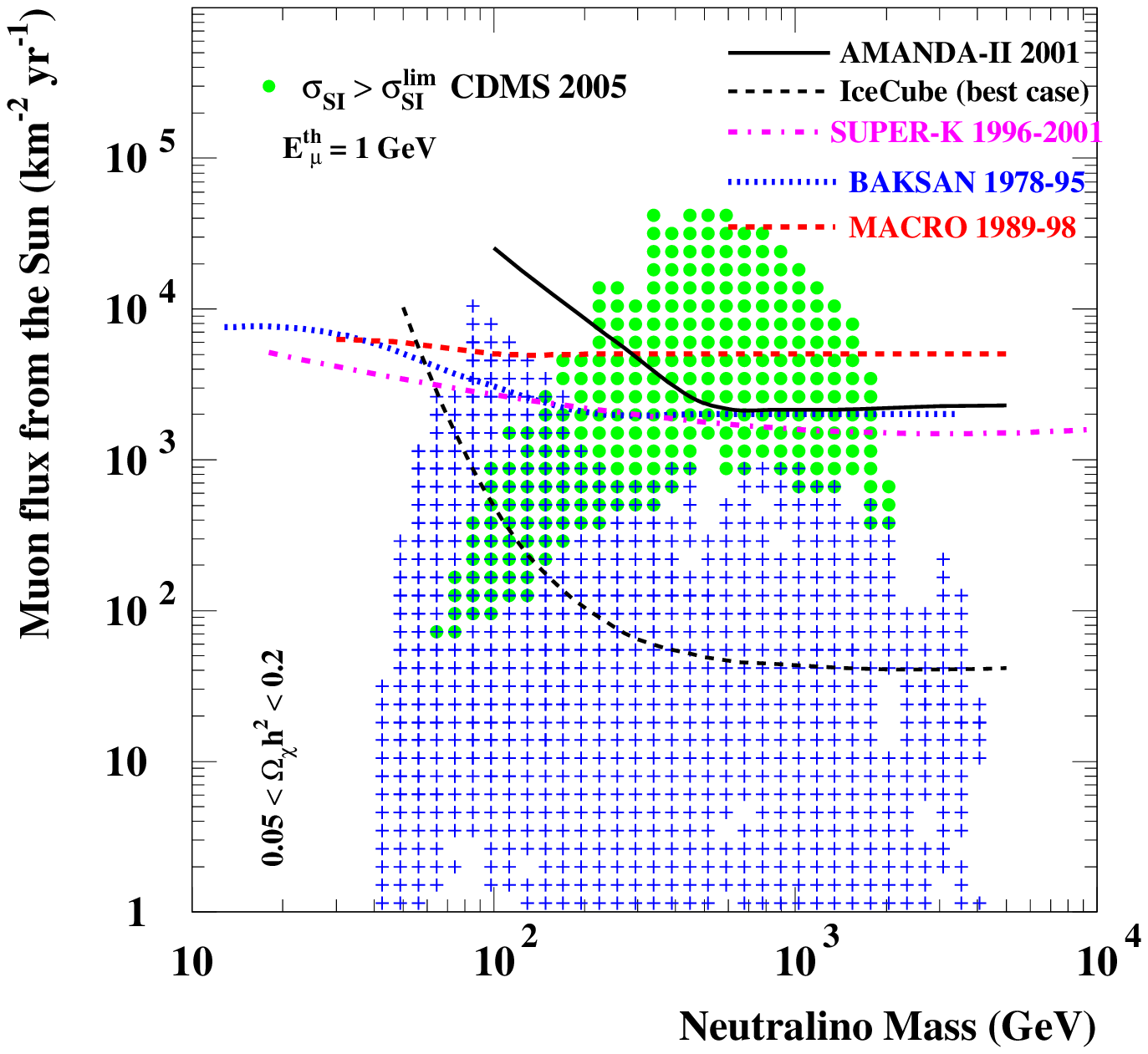}
%\vspace{-2.5pc}
\caption{\label{sun_fig} (a) Detection efficiencies relative to trigger level for the different filter levels in the solar neutralino analysis ($m_{\chi}$=500$~$GeV, hard spectrum) for 2001 data, neutralino signal, atmospheric muons and neutrinos.  (b) As a function of neutralino mass, the 90\% confidence level upper limit on the muon flux coming from hard neutralino annihilations in the center of the Sun compared to limits of other indirect experiments \cite{indirect} and the sensitivity estimated for a best-case IceCube scenario.  Markers show predictions for cosmologically relevant MSSM models, the dots represent parameter space excluded by CDMS~\cite{cdms}.}
\end{center}
%\vspace{-2.5pc}
\end{figure}
The AMANDA data used in the search for solar neutralinos consists of 8.7$\times$10$^8$ events, corresponding to 143.7 days of effective livetime, collected in 2001.  In contrast to the search in Section \ref{earth_section}, reducing the risk of experimenter bias in this analysis can be achieved by randomizing the azimuthal angles of the data.  The advantage of this procedure is that it allows the use of the full data set for cut optimization.  The azimuthal angles are restored once the optimization is finalized and results are calculated.  The simulated atmospheric background sample at trigger level totals 1.6$\times$10$^8$ muons (equivalent to 32.5 days of effective livetime) and$~$1.9$\times$10$^4~$neutrinos.

The solar neutralino analysis suffers the same backgrounds as the terrestrial neutralinos, but the signal is expected from a direction near the horizon, due to the trajectory of the Sun at the South Pole.  This analysis was only possible after completion of the AMANDA-II detector, whose 200 m diameter size provides enough lever arm for robust reconstruction of horizontal tracks. 

A similar analysis strategy as in Section \ref{earth_section} was adopted.  First, events were selected with well-reconstructed horizontal or upward-going tracks.  The remaining events are passed through a neural network that was trained separately for the neutralino models under study and used data as background.  Although a data reduction of $\sim$10$^{-5}$ compared to trigger level is achieved, the data sample is still dominated by misreconstructed downward-going muons.  Finally, these are removed by cuts on observables related to reconstruction quality.

There is no sign of a significant excess of tracks from the direction of the Sun in the final data sample.  The expected background in the final search bin around the Sun is estimated from off-source data in the same declination band, which eliminates the effects of uncertainties in background simulation.  Combining this with the number of observed events, the effective volume and the detector livetime, we obtain 90\% confidence level upper limits on the muon flux coming from annihilations in the Sun for each considered neutralino mass \cite{newsun}, as shown in Fig. \ref{sun_fig}b.

\vspace{-0.5pc}
\section{\label{discussion}Discussion and outlook}
%\vspace{-0.5pc}
Figures \ref{earth_fig}b and \ref{sun_fig}b present the AMANDA upper limits on the muon flux from neutralino annihilations into a hard channel
%\footnote{$W^{+}W^{-}$ for $m_\chi > m_W$, $\tau^+\tau^-$ for smaller masses.}
in the Earth and the Sun respectively, together with the results from other indirect searches \cite{indirect}.  Limits have been rescaled to a common muon threshold of 1$~$GeV using the known energy spectrum of the neutralinos.  Also shown are the cosmologically relevant MSSM models allowed (crosses) and disfavoured (dots) by the direct search from CDMS \cite{cdms}.  Compared to our search for a terrestrial neutralino signal in 1997 AMANDA data \cite{earth97}, the limit has been improved by at least a factor of 2, more than expected from additional statistics alone.  This is due mainly to the separate cut optimization for each neutralino mass, which exploits the characteristic muon energy spectrum of each model.

% amanda efforts
In 2001 an additional trigger was installed that lowered the energy threshold of the detector.  This trigger takes into account spatio-temporal correlations in the event hit pattern.  A preliminary analysis with data taken between 2001 and 2003 shows an improvement of a factor of about 30 in the effective volume in the search for 50$~$GeV neutralinos (soft annihilation channel) from the Earth with respect to the analysis presented in this note.  We are currently performing searches for a dark matter signal both from the Earth and the Sun with data taken from 2000 and later.  The increased detector exposure combined with improved reconstruction techniques and the new trigger setting will result in improved sensitivities (a 4-year exposure alone would already give an improvement of a factor of two).

% icecube
As of February 2006 the first nine IceCube instrumented strings are operational, monitoring a volume as big as that of AMANDA (0.022 km$^3$).  First results show that the performance of the system meets or exceeds design requirements \cite{icecube_perf}.  The final IceCube detector, to be completed around 2011, will encompass an approximately hexagonal instrumented volume of 1 km$^3$.  The sensitivity to dark matter signals will not only be improved by the factor 50 increase in fiducial volume, but also (especially for solar neutralinos) by the greater horizontal extension compared to AMANDA.  Furthermore, the smaller trigger efficiencies for low energy muons ($E_{\mu}<$100$~$GeV), due to sparse instrumentation, will be partially compensated for by incorporating the denser AMANDA array in IceCube.
Assuming a best-case scenario (10 and 5 km$^3$yr exposure for Earth and Sun respectively, no misreconstructed atmospheric background and without incorporating systematic uncertainties) we estimated the IceCube sensitivity for a terrestrial and solar neutralino-induced muon flux, shown in Fig. \ref{earth_fig}b and \ref{sun_fig}b.  The IceCube results will be complementary to future direct detection experiments, and even have an advantage over these for low mass solar neutralinos with large spin-dependent scattering cross sections.

\setcounter{section}{0}
\setcounter{subsection}{0}
\setcounter{figure}{0}
\setcounter{table}{0}
\setcounter{equation}{0}
\newpage

\begin{frontmatter}

\title{DARKSUSY: a dark matter Monte Carlo}

\author[address1]{Joakim Edsj\"o}

\address[address1]{Department of Physics, AlbaNova, Stockholm
  University, Sweden.}

\end{frontmatter}
\vspace*{5cm}
\begin{center}
 No contribution received
\end{center}
\setcounter{section}{0}
\setcounter{subsection}{0}
\setcounter{figure}{0}
\setcounter{table}{0}
\setcounter{equation}{0}
\newpage

%%%%%%%%%%%%%%%%%%%%%%%%%%%%%%%%%%%%%%%%%%%%%%%%%%% Title, authors and addresses
\begin{frontmatter}


\title{Direct Detection of Physics Beyond the Standard Model}

% use optional labels between square brackets to link authors explicitly to addresses:
% If more than one author, keep a comma between the author tags

\author[address1]{Ivone F. M. Albuquerque}

\address[address1]{Instituto de F\'{i}sica, Universidade de S\~{a}o Paulo, Brasil}

\begin{abstract}
In supersymmetric theories where the lightest supersymmetric particle is the
gravitino the next to lightest supersymmetric particle (NLSP) is typically a 
long lived charged slepton. These NLSPs can be produced by
high energy neutrinos interactions with nucleons in the Earth and be detected
by km$^3$ neutrino telescopes. The signal, consists of two parallel charged 
tracks separated by a few hundred
meters. This is compared to the main background, coming from
direct di-muon production. The distance between the background tracks is much
smaller and allows for a clean separation from the NLSP signal.
We conclude that neutrino telescopes
will complement collider searches in the determination of the supersymmetry 
breaking
scale, and may even provide the first evidence for supersymmetry at the weak scale.
\end{abstract}

% \begin{keyword}
% neutrinos \sep supersymmetry \sep gravitino

% \PACS 11.30.pb \sep 13.15+g \sep 12.60.jv \sep 95.30.Cq 
% \end{keyword}

\end{frontmatter}

%%%%%%%%%%%%%%%%%%%%%%%%%%%%%%%%%%%%%%%%%%%%%%%%%%%%%% MAIN TEXT
\section{\label{sec:intro} Introduction}
One of the most attractive candidate theory for the extension of the
standard model of particle physics is weak scale supersymmetry (susy).
The particle spectrum of this model is determined by the susy breaking
mechanism ($\sqrt{F}$). When susy is broken at high scales
such that $\sqrt{F} \gtrsim 10^{10}$GeV the LSP is typically the 
neutralino. If however susy is broken at lower scales, 
$\sqrt{F} \lesssim 10^{10}$GeV, the LSP
tends to be the gravitino and the Next to
Lightest Supersymmetric Particle (NLSP) is usually a charged slepton, typically the
right-handed stau. Within these models, if $\sqrt{F}$ is much
larger than a TeV the stau NLSP decay is suppressed \cite{abcl}
and its lifetime can be very large.

It was recently proposed \cite{abcl} that the diffuse flux of high
energy neutrinos colliding with the Earth can produce pairs of slepton NLSPs.
The energy loss of these particles is very small and they travel for long distances
before stopping. This compensates their small production cross section 
since they can be detected far away from their production point.

The NLSP production typically has a 
%The neutrino-nucleon reaction can produce a pair of
%supersymmetric particles that will promptly decay to a pair of NLSPs and standard
%model (SM) particles. This process results in NLSPs which typically have a very 
high boost and they will not decay inside the
Earth provided the susy breaking scale is $\sqrt{F} > 10^7$GeV.
Since the NLSP is charged, its upward going tracks can be
detected in large ice or water Cerenkov detectors, such as IceCube.  
%This is in analogy with the standard model charged current interaction muon
%production,
%the primary signal in neutrino telescopes. 
%%The high boost and large range of the
%%NLSPs implies that the tracks are parallel and well separated, enabling them to be
%%distinguished from background events.

We performed a Monte Carlo simulation of the NLSP production and
propagation through the Earth and determined their signature in km$^3$ 
neutrino telescope 
\cite{abc2}. The main  background (di-muon events) was also simulated and 
compared to the NLSP signal.

\section{\label{sec:prod} NLSP Production and Propagation Through the Earth}

The susy processes for NLSP production is analogous to the
standard model (SM) charged current (CC) interaction and involves a t-channel
production of a left-handed slepton ($\tilde l_L$) and a squark ($\tilde q$) 
through a gaugino exchange. In the dominant process the gaugino is a chargino 
and in the subdominant process a neutralino.
The $\tilde l_L$ and $\tilde q$ prompt decay results in two lighter
right-handed sleptons (NLSPs) plus SM particles. 
The NLSP will always be produced in pairs and is typically 
the stau ($\tilde\tau_R$). 

The energy threshold
for the NLSP production is given by the $\tilde l_L$ and $\tilde q$
masses. We take the typical values of 250 GeV for both chargino and
$\tilde l_L$ masses, 150 GeV for the NLSP mass and set the $\tilde q$  mass to
three values of 300, 600 and 900 GeV.

We compute the NLSP cross section including both dominant and subdominant
processes. The susy cross section is about three orders of magnitude
lower than the SM production cross section \cite{abc2}. 

For the high energy neutrino flux reaching the Earth, we take the 
Waxman-Bahcall~\cite{wb} limit

\begin{equation}
\left(\frac{d\phi_\nu}{dE}\right)_{\rm WB} = \frac{(1-4) \times
10^{-8}}{E^2} {\rm GeV~ cm^{-2} s^{-1} ~sr{-1}}~.
\label{eq:wblimit}
\end{equation}
Since the NLSP production
is independent of the neutrino flavor, the initial
neutrino flux contains both $\nu_\mu$ and $\nu_e$ in a 2 : 1 ratio.

The NLSP propagation through the Earth depends on its energy loss which at 
high energies is dominated by radiative losses.
The main processes are bremsstrahlung, pair production and photo-nuclear
interactions~\cite{abc2,ina}. As a result the NLSP will travel distances
much greater than a muon which will compensate for the lower cross section.

\section{Simulation of NLSP Production, Propagation and Detection}

In order to analyze the NLSP production, propagation and detection
we developed a Monte Carlo simulation. Assuming an isotropic incoming
neutrino flux we generated 30K NLSP events for each $\tilde q$ mass.
The neutrino incoming energy is distributed in steps ranging from the stau
production threshold to $10^{11}$ GeV. 

The survival probability ($P_S$) for a neutrino traveling through a 
path $dl$ is given by $P_S = \exp(\int n \sigma dl)$
where n is the Earth number density \cite{gqrs} and $\sigma$ is the
interaction cross section. The probability of interaction is $1 - P_S$.

To simulate the NLSP production an interaction point is randomly chosen
from an interaction probability distribution. If the interaction point falls
within a distance from the detector that is smaller than the NLSP range
the event is accepted. If this distance is greater than the NLSP range
the event will only count for the normalization.

The center of mass (CM) angular distribution was chosen based on the 
differential cross section distribution. We assume that the CM angular
distribution of the right-handed slepton pair (NLSPs) is the same as
the CM angular distribution of the $\tilde l_L$ and $\tilde q$ which is
a good approximation~\cite{abc2}. The 4-momentum of the NLSP pair is determined 
in the CM from the angular distribution and boosted to
the lab frame. The lab angular ($\Theta_{lab}$) distribution is determined 
from the lab 4-momentum.

With this procedure we can determine not only the rate of events at the
detector as well as the two NLSP track separation in the detector.
The separation is simply $\Theta_{lab}$ times the distance
from the neutrino interaction point and the detector.

\section{Background}

Muons produced in the Earth could be a potential source of background
for the NLSPs events. They can however be eliminated by requiring two
charged tracks in the detector. The remaining important background is
di-muon events originated from charm decay. Charm is produced
from high energy neutrino interactions through the following process: 
$\nu N \rightarrow \mu^- H_c \rightarrow \mu^- \mu^+ H_x \nu$,
%\begin{equation}
%\nu N \rightarrow \mu^- H_c \rightarrow \mu^- \mu^+ H_x \nu
%\end{equation}
where the charm hadron decays according to $H_c \rightarrow H_x \mu^+\nu$
and $H_x$ can be a strange or non-strange quark.

We computed the charm production cross section 
$\sigma(\nu N \rightarrow cX)$ from a d or s quark~\cite{abc2}.
%The di-muon production cross section is shown in Figure~\ref{fig:xs}.
Although this cross section is around an order of magnitude greater than
the one for NLSP production~\cite{abc2}, muons lose much more energy while 
traveling through the Earth
and therefore have to be produced very close to the detector. For this
reason their track separation in the detector will also be much smaller.

We performed a Monte Carlo simulation of the di-muon production, propagation 
and detection using the same procedure as for the NLSP.

\section{Results}

\begin{figure}[t]
\centering
\epsfxsize=200pt \epsfbox{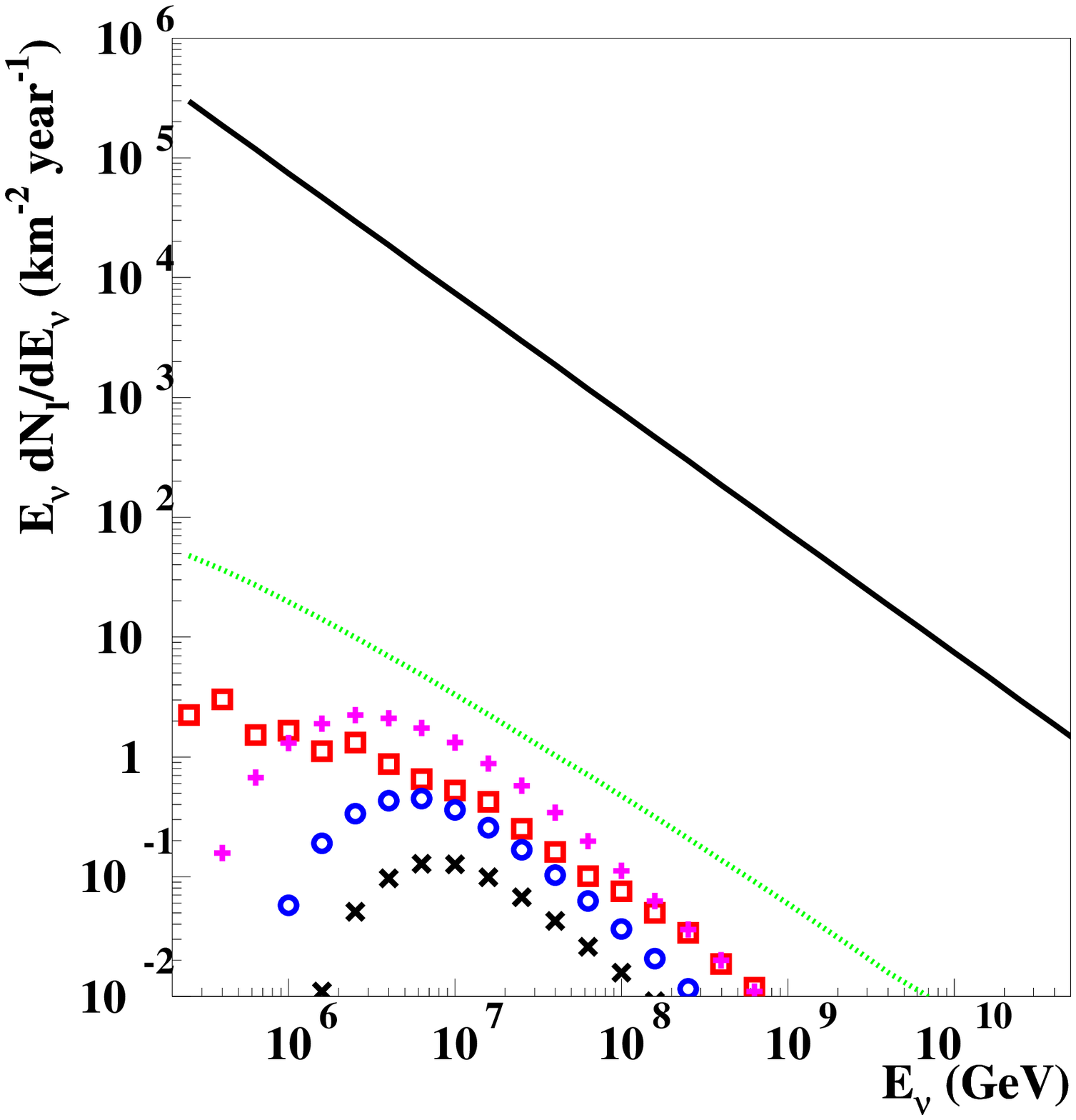}
\\*[-1.5cm]
\caption{Energy distribution of $\tilde\ell_R$ pair events per 
km$^2$, per year, at the detector. 
Curves that do not reach y axis; from top to bottom: 
$m_{\tilde q}=300$, ~$600$ and $900$~GeV. 
Here, $m_{\tilde\ell_R}=150$~GeV and
$m_{\tilde w}=250$~GeV. 
Also shown are the neutrino flux at earth and the $\mu$ and
the di-muon flux through the detector (curves that reach y axis; from top to bottom
respectively). In all cases we make use of the WB limit 
for the neutrino flux. 
}
\label{fig:rate}
%\end{flushright}
%
\end{figure}

The energy distribution of NLSPs and di-muon background are shown in
Figure~\ref{fig:rate}. The number of events per km$^2$ per year is 
shown in Table~\ref{tab:nev} where not only a neutrino
incoming rate equal to the WB limit is assumed but also one
equal to the Mannheim, Protheroe and Rachen (MPR) limit~\cite{mpr}.

Figure~\ref{fig:rate} and Table~\ref{tab:nev} show that km$^3$
neutrino telescopes are sensitive to NLSP detection. Although the
number of di-muon events is larger than the NLSP rate there are
many ways to reduce this background.

\begin{table}
\begin{center}
\begin{tabular}{l|cccc}
\hline
\hline
 & $\mu^+\mu^-$ & $m_{\tilde{q}}=300$ & $600$ & $900$ (GeV) \\
\hline
WB & 30 & 6 & 1 & 0.3 \\
MPR & 1412 & 21 & 3 & 1 \\
\hline
\hline
\end{tabular}
\vspace*{.5cm}
\caption{\scriptsize Number of events per km$^2$ per year assuming the WB and MPR limits. The first column refers to 
di-muon events. The last three columns correspond to NLSP pair events, 
for three different choices of squark masses: $300$~GeV, $600$~GeV and $900$~GeV.
The number of di-muon events are given for energies
above $10^3$ GeV and of NLSPs above threshold for production.}
\label{tab:nev}
\end{center}
\end{table}

\begin{figure}[t]
\begin{center}
\epsfig{file=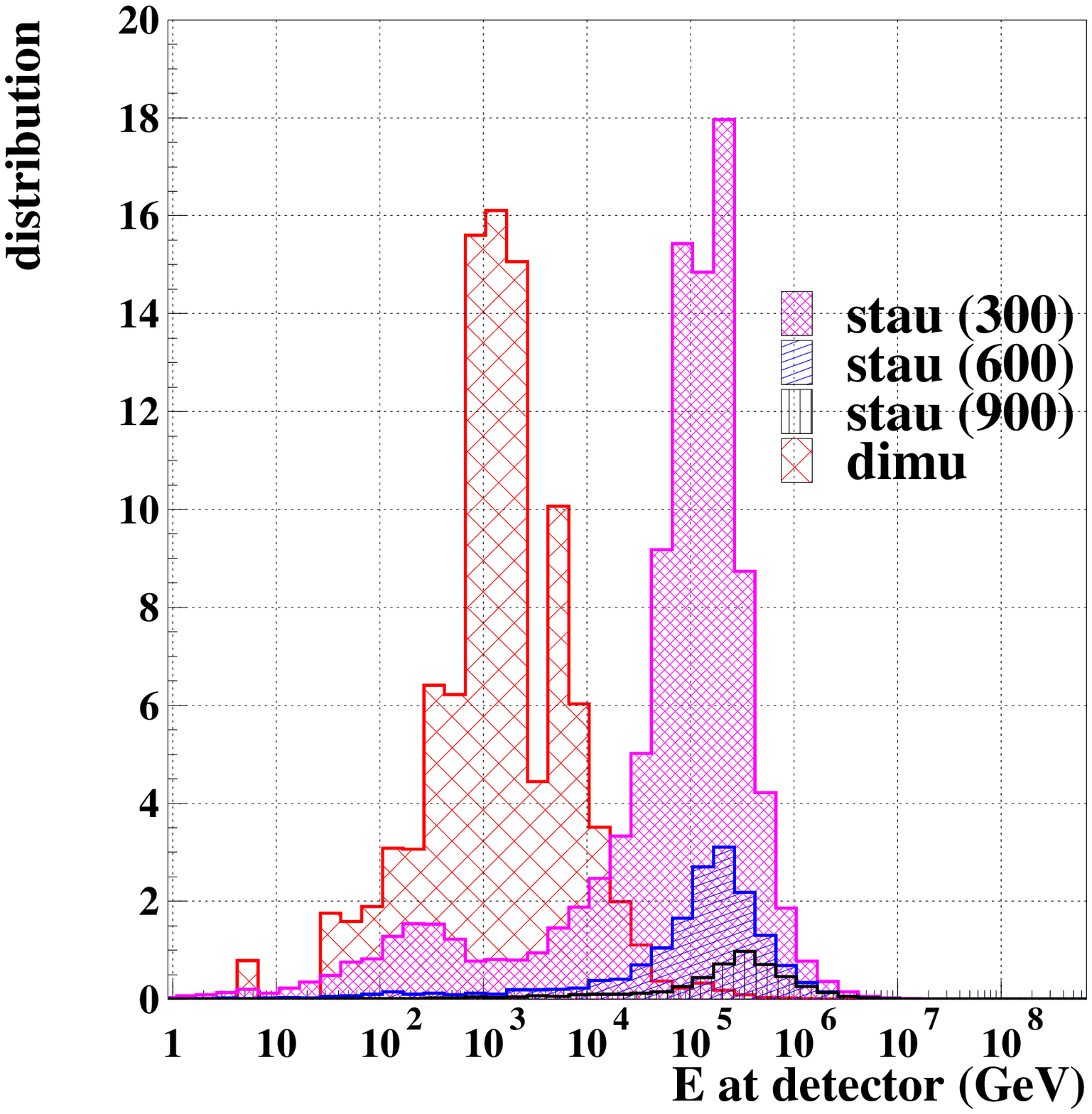,width=9.cm,height=8.cm,angle=0}
\\*[-1.cm]
\caption{Arrival energy distribution of the $\tilde\ell_R$ at the 
detector and for $m_{\tilde q}=300,~600 {\rm ~and~} 900$~GeV..
Here, $m_{\tilde\ell_R}=150$~GeV and
$m_{\tilde w}=250$~GeV. 
Also shown is the arrival distribution for the di-muon
background. The energy {\em deposited} in the detector by a stau 
traveling the average track length of $800~$m~\cite{als} is
$E_{\tilde{\ell}_R}^{\rm dep}= 150~$GeV, approximately the same for
all the masses considered here.
}
\label{fig:endist}
\end{center}
\end{figure}

First of all the lower integration limit to determine the number of events shown 
in Table~\ref{tab:nev} is different for the NLSP and the background. It is the
NLSP production threshold energy for the number of NLSPs and $10^3$ GeV
for the di-muon background.
The reason for this is that as the NLSPs lose less energy than
muons, they will look like lower energy muons in the detector. The NLSP
energy deposition in the detector is 150 GeV (see figure~( in \cite{abc2}).
For the same reason, one can cut higher energy events removing events
that deposit more than 300 GeV and reduce most of the dimuon
background. Figure~\ref{fig:endist} shows the arrival energy at the
detector for both NLSP and di-muon background where the efficiency of such
a cut is clear.

\begin{figure}[t]
%
%\centering
\begin{center}
\epsfig{file=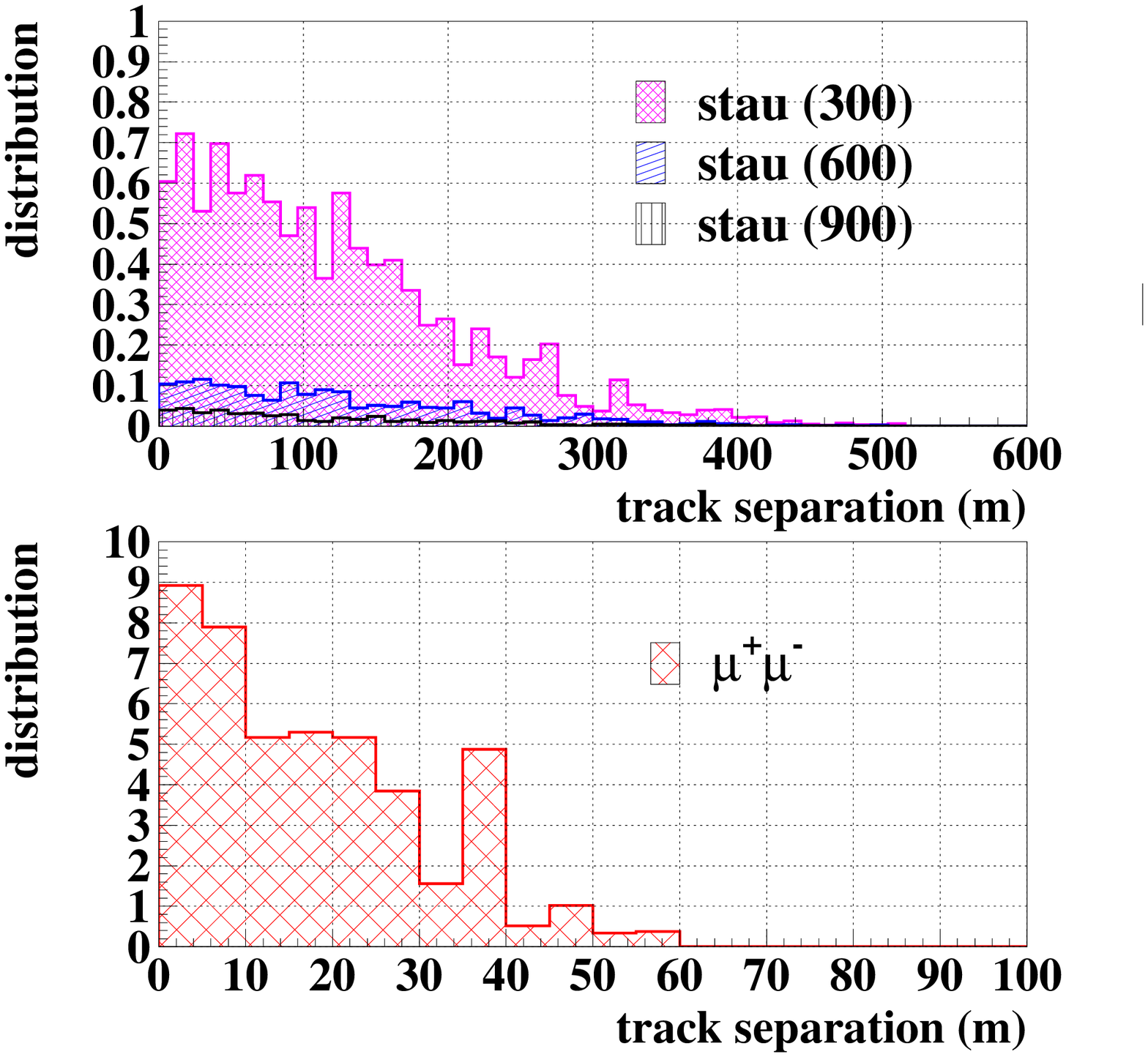,width=7.2cm,height=9cm,angle=0}
\\*[-1.cm]
\caption{Top panel: Track separation distribution of $\tilde\ell_R$
pair events. 
Bottom panel: The track separation distribution of the di-muon
background.  The relative 
normalization corresponds to the relative number of events for signal 
an background. Note the different horizontal scales, as well as
different binning between the two figures.
}
\label{fig:sep}
\end{center}
\end{figure}

Another powerful way for background reduction is the track separation between
2 NLSPs in the detector. This is shown in Figure~\ref{fig:sep}. While
a good fraction of the NLSP events are more than 50 m separated the dimuons
are less than 50 m separated. 

\section{Conclusions}
\enlargethispage{0.5cm}
We conclude that Km$^3$ neutrino telescopes have the potential to discover
the NLSP if susy models that predict the gravitino as the LSP and
a charged slepton as the NLSP are correct. This will be an indirect 
determination of the dark matter. If the NLSP is observed it will constitute
a direct probe of the susy breaking scale. This search is complementary to
the LHC.

\section*{Acknowledgements}
{\small I.F.M.A participation in the EPNT workshop was partially supported
by the Brazilian National Counsel for Technological and Scientific 
Development (CNPq).}

%%%%%%%%%%%%%%%%%%%%%%%%%%%%%%%%%%%%%% reset.txt counters %%%%%%%%%%%%%%
%%
%%%%%%% do not change below here  %%%%%%%%%%%%%%%%%%%%%%%%%%%%%
%%
\setcounter{section}{0}
\setcounter{subsection}{0}
\setcounter{figure}{0}
\setcounter{table}{0}
\setcounter{equation}{0}
\newpage

%%%%%%%%%%%%%%%%%%%%%%%%%%%%%%%%%%%%%%%%%%%%%%%%%%% Title, authors and addresses
\begin{frontmatter}


\title{Dark Matter Seaches with Balloon Experiments and Gamma-Ray Telescopes}

% use optional labels between square brackets to link authors explicitly to addresses:
% If more than one author, keep a comma between the author tags

\author[address1]{U.~Schwanke}
%\author[address2]{C. O. Author}

\address[address1]{Humboldt University Berlin, Department of Physics, Newtonstrasse 15, 12489~Berlin, Germany}
%\address[address2]{Department of Something Else, Univesity2}

\begin{abstract}
A number of signals involving charged cosmic rays and high-energy photons 
have been interpreted as being due to annihilating dark 
matter. This article provides an overview of the experimental evidence 
and discusses in particular detections of antiprotons and positrons
in the cosmic radiation, the diffuse $\gamma$-ray emission between 10\,MeV and 100\,GeV
from the Milky Way, and the 511\,keV annihilation radiation and 
the flux of very high-energy photons ($>100$\,GeV) from the Galactic Centre.
\end{abstract}

% \begin{keyword}

%\PACS 
% \end{keyword}

\end{frontmatter}

%%%%%%%%%%%%%%%%%%%%%%%%%%%%%%%%%%%%%%%%%%%%%%%%%%%%%% MAIN TEXT
\section{Introduction}

In contrast to laboratory-based dark matter searches (see \cite{sumner} for a review
of direct methods), indirect searches do not aim at the dark matter particles as such, but try to
measure their annihilation or decay products that arrive from outer space. 
Various experiments have searched for antiprotons and positrons at the boundary 
of the Earth's atmosphere since those antiparticles should stick out among the wealth of ordinary particles 
in the cosmic radiation. Gamma-rays as annihilation products have the 
additional advantage that their arrival direction can be correlated with
the dark matter density, and signals recorded by both space-borne and
ground-based $\gamma$-ray detectors have been searched for indications for
dark matter.
%We review here searches for excess antiprotons, positrons 
%and photons reported by various instruments.

\section{Antiproton and Positron Searches with Balloon Experiments}

Balloons carry particle spectro\-meters to heights of around 40\,km
above sea level but the reached exposure is somewhat limited since these experiments
%(like BESS\,\cite{bess}, CAPRICE\,\cite{caprice}, HEAT\,\cite{heat}, IMAX\,\cite{imax})
(like BESS or HEAT)
are flown typically once a year for a duration of roughly 10 hours. 
The BESS experiment\,\cite{bessmitchell}, for example, recorded around 2000 antiprotons
in eight flights which took place in northern Canada over a period of 
11 years. The small fraction of antimatter in the cosmic radiation ($\bar{p}/p\approx 10^{-5}$)
poses a particular challenge to the balloon-borne spectrometers, and the large
background of ordinary particles (i.e.\ protons and electrons)
requires high precision tracking and excellent particle identification 
capabilities. Some of the associated experimental uncertainties were 
maybe not so well known in early experiments whose data suggested a 
much larger number of antiprotons and positrons than expected from purely secondary 
production. Meanwhile, both the understanding of the measurements and of the 
expected background of secondary antiprotons have improved, resulting in
less room for additional sources of antimatter. 

\begin{figure*}[t]
\begin{minipage}[t]{0.48\linewidth}
\centering\epsfig{file=posfrac0alt.eps,height=\linewidth,width=\linewidth}
\caption{
Positron fraction measured by HEAT in 1994/95 and 2000.
The lines indicate a contribution from annihilating Kaluza-Klein dark matter 
which has been normalized to the data. The various line styles correspond
to different choices for the positron diffusion parameters\,\cite{heathooper}.
}
\label{fig:heat}
\end{minipage}\hfill
\begin{minipage}[t]{0.48\linewidth}  
\centering\epsfig{file=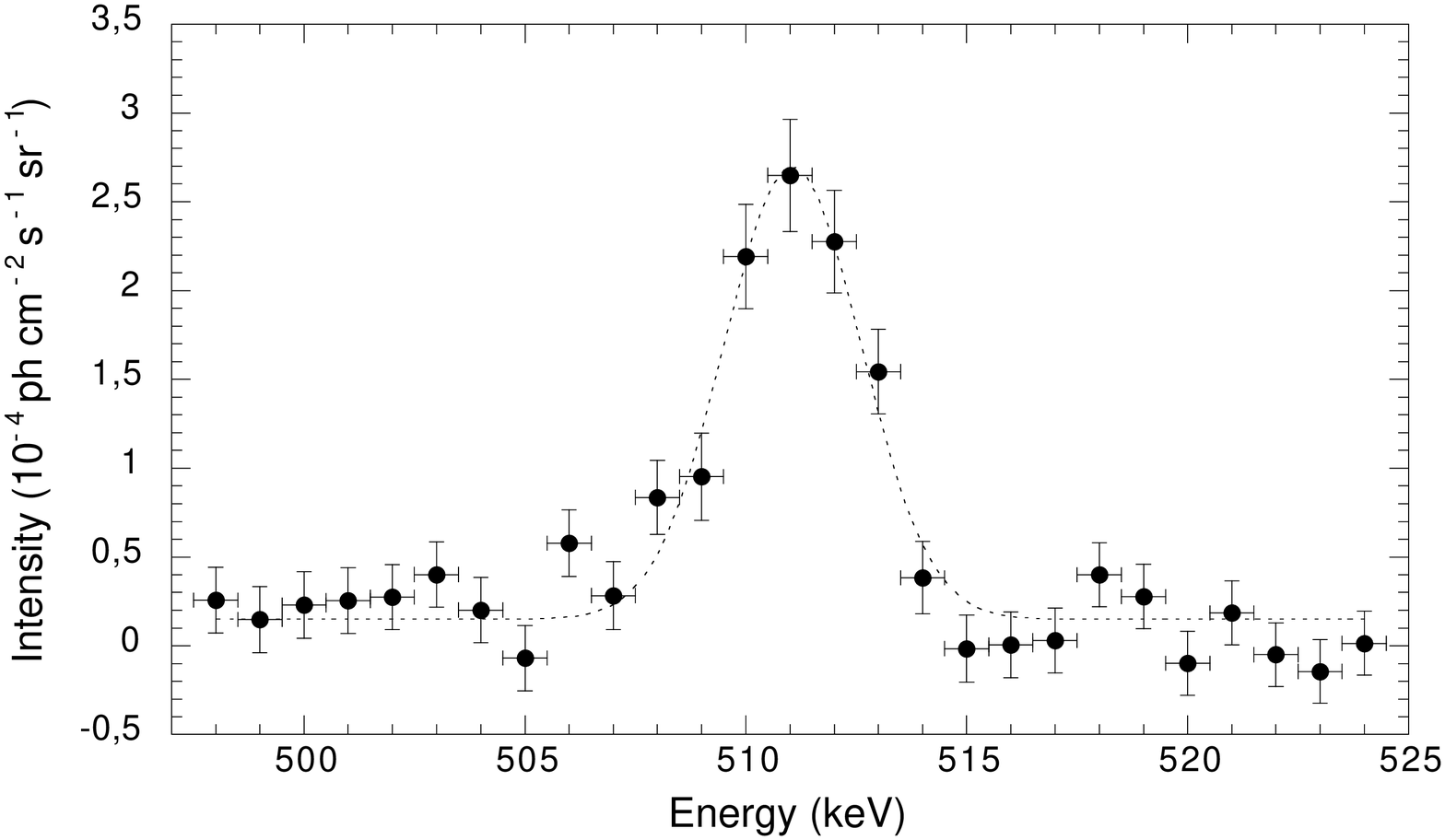,height=\linewidth,width=\linewidth}
\caption{
SPI 511\,keV flux spectrum that is obtained when modelling the Galactic
Centre emission region as a Gaussian with a FWHM of $10^\circ$\,\cite{spijean}.
}
\label{fig:spi}
\end{minipage}
\end{figure*}

For antiprotons, both the absolute flux and the energy dependence of
the antipro\-ton-to-proton ratio seem to be broadly consistent with a picture
where no dark matter contribution is needed to account for the observations.
The BESS collaboration reported a possible excess of antiprotons\,\cite{bessmitchell}
at energies below 1\,GeV, but the effect seems small given the limited 
statistics of the data and the uncertainties related to the proper
description of solar modulation in this energy range. In December 2004, the BESS collaboration
managed for the first time to conduct a 9-day flight in Antarctica that should have doubled the
available antiproton statistics and might give new insight into the
reality of the observed effect.

Positron measurements were conducted as well and analysed in terms of the 
so-called {\em positron fraction} (the fraction of positrons among all
detected electrons and positrons). The energy dependence of the positron fraction has been 
measured between 1 and 30\,GeV, and features of the 
observed spectrum have been interpreted as being due to dark matter.
In particular the data of the HEAT experiment (High Energy Antimatter Telescope)
have been discussed in this context. The HEAT spectrometer has been
flown in two different detector configurations (HEAT-$e^\pm$ and
HEAT-pbar). The flights took place near solar maximum (1994 and 1995, HEAT-$e^\pm$) 
and near solar minimum (2000, HEAT-pbar), and at different geomagnetic
cutoffs (around 1\,GeV in 1995, and around 4\,GeV in 1994 and 2000).
The energy dependence of the positron fraction measured during the 1994/95 flights 
agrees with what has been found in 2000\,\cite{heatdata1,heatdata2} (Fig.~\ref{fig:heat}),
and the HEAT collaboration finally concluded that the intensity of positrons
is consistent with a purely secondary origin due to nuclear interactions
in interstellar space\,\cite{heatdata2}.

Given the experimental and model-related uncertainties, a contribution from 
dark matter cannot be ruled out and this possibility has been the 
subject of theoretical studies. 
A popular dark matter candidate motivated by elementary particle physics
(the supersymmetric neutralino) couples in proportion to the mass of
the final state particles and is thus an inefficient electron-positron
source. In this situation, particles expected in Kaluza-Klein theories ofter a viable
alternative and have been studied in detail\,\cite{heathooper} (Fig.~\ref{fig:heat}).  

\section{Electron-Positron Annihilation Radiation from the Galactic Centre}

Another way of tracing positrons is the observation of the 511\,keV 
radiation that is created when electrons and positrons annihilate at rest
somewhere in the Milky Way. 
Energetic positrons suffer rapid ionization losses and can annihilate with 
electrons depending on the atom and electron density in the interstellar medium.
Since the stopping distance of positrons is typically much smaller than their mean
free path for annihilation, positrons do indeed thermalise before
annihilation. The annihilation process itself seems to be dominated
by positronium formation\,\cite{posfrac} creating a narrow line (25\,\% of the 
time) or a three-photon continuum (75\,\% of the time).

Electron-positron annihilation radiation from the Galactic Centre was detected for the first time
in the 70s\,\cite{posdetect,posdetect2,posident}. After initial claims of
a flux variability in time, consistent results for the absolute photon flux and the line shape
were obtained by a number of experiments\,\cite{pos1,pos2,pos3,pos4}.
There was much less agreement on the morphology of the emission region. Some
measurements suggested both a contribution from the bulge and the disk of the Milky
Way, and an additional component from a location at 
positive galactic latitude (i.e.~above the galactic plane) was also 
discussed.

The most recent high-statistics observations of the 511\,keV annihilation
radiation have been obtained using the SPI detector onboard 
the INTEGRAL satellite. SPI is sensitive 
for photon energies between 20\,keV and 10\,MeV and has an energy resolution of 2\,keV (at 1\,MeV).
The arrival direction of photons is reconstructed with an angular
resolution of around 2$^\circ$.

\begin{figure*}[t]
\begin{minipage}[t]{0.48\linewidth}
\centering\epsfig{file=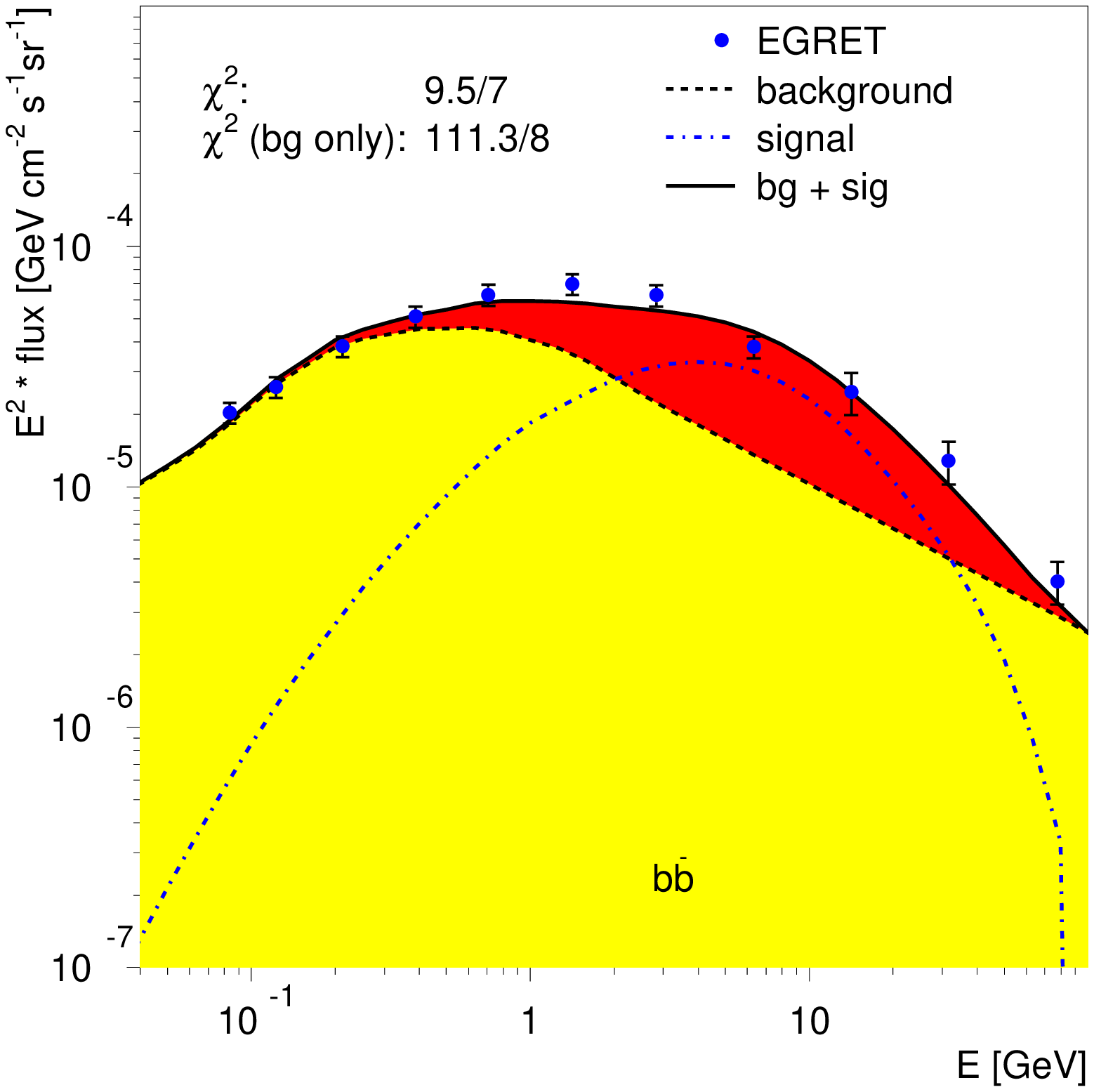,height=\linewidth,width=\linewidth}
\caption{
Diffuse $\gamma$-ray energy spectrum of the Inner Galaxy
(longitude $330-30^\circ$, latitude $|b|<5^\circ$). The EGRET data are shown as dots.
The dashed line denotes the photon contribution from hadronic interactions
in the interstellar gas, the dot-dashed line denotes the photon spectrum
expected from neutralino annihilations into $b\bar{b}$ pairs. The solid line
is the sum of the two components\,\protect\cite{wimdeboer}.
}
\label{fig:egret}
\end{minipage}\hfill
\begin{minipage}[t]{0.48\linewidth}  
\centering\epsfig{file=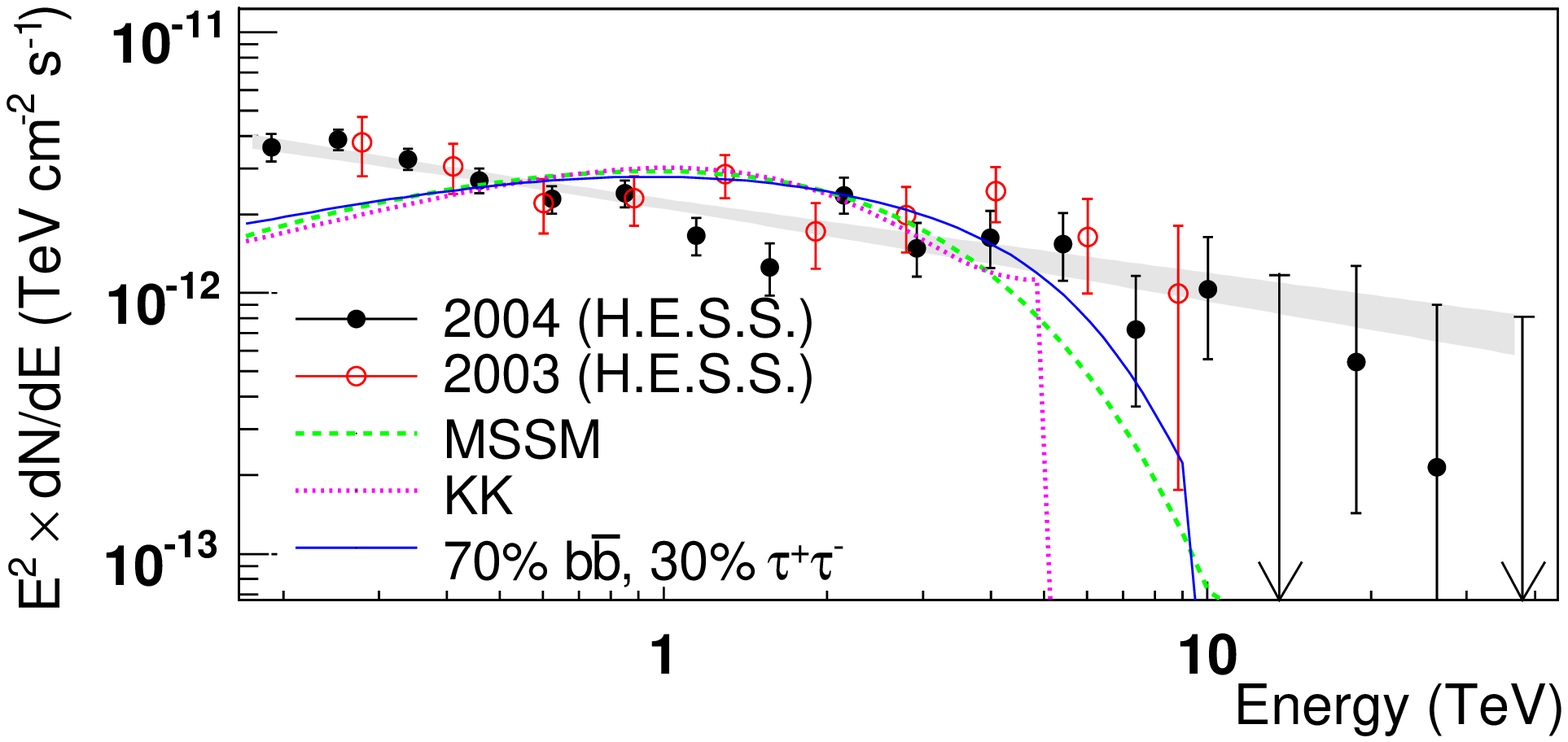,height=\linewidth,width=\linewidth}
\caption{
  Spectral energy density of $\gamma$-rays from observation of the Galactic Centre
  with H.E.S.S.~in 2003 (red circles) and 2004 (black dots). The shaded area
  is the result of a power-law fit $dN/dE\propto E^{-\Gamma}$. The lines 
  show the spectra expected from a MSSM neutralino with a mass of 14\,TeV (dashed),
  from Kaluza-Klein DM with a mass of 5\,TeV, and a 10\,TeV DM particle (solid)
  annihilating into $\tau^+\tau^-$ (30\,\%) and $b\bar{b}$ (70\,\%).(From \cite{gcprl}.)
}
\label{fig:hess}
\end{minipage}
\end{figure*}

The SPI data \cite{spijean,spiknoedl,spiweiden}  cover galactic longitudes of $\pm 30^\circ$ and galactic latitudes of $\pm 10^\circ$.
The data show a strong signal (Fig.~\ref{fig:spi})
with a significance exceeding 20 standard deviations. The measured flux 
and the line width are consistent with earlier measurements. Despite significant
exposure out to galactic longitudes of $\pm 30^\circ$ the morphology of the emission
region is accurately modelled by a spherical distribution which is centred at the Galactic Centre
and exhibits a Gaussian radial profile with a FWHM of $\approx 10^\circ$. Upper limits
on the flux contribution from the galactic disk were derived.

The restriction of the annihilation signal to the Galactic Centre region seems difficult 
to explain in the framework of conventional astrophysical positron sources. 
Most of the discussed sources (like supernovae, micro-quasars, pulsars, stellar flares, Wolf-Rayet stars, cosmic-ray interactions
with the interstellar medium and others, see for example \cite{wang,prantzos,guessoum}) would 
probably imply a non-negligible contribution from the galactic disk. To avoid 
such a contribution is has even been speculated that the large-scale magnetic field
in the galaxy could be such that positrons are channeled from the disk into the bulge\,\cite{prantzos}.

The SPI measurement suggests the galactic bulge as the positron source
and this agrees with the notion that the density of dark matter particles should
be highest in the Galactic Centre region, while a contribution from the galactic disk
is not necessarily expected. This has resulted in attempts to model the SPI data as being
due to dark matter annihilation\,\cite{boehm} and it has been shown that the right
relic density can be obtained. To evade observational constraints 
the dark matter should be light (1--100\,MeV) to create
the $e^+e^-$ final state almost exclusively. The exchange particle in the 
annihilation process could be a fermion (with suppressed $Z$ 
couplings to evade collider constraints) or a new type of gauge boson ($U$ boson). 
This dark matter interpretation has been checked against various experimental 
constraints, and it has been shown for example that already the 
presence of Bremsstrahlung photons in the annihilation process would 
violate COMPTEL data if the mass of the dark matter particle exceeds 20\,MeV\,\cite{beacom}.

\section{Diffuse $\gamma$-ray Emission from the Milky Way}

Photons from the decay cascade of particles created by annihilating dark matter
can potentially make a significant contribution to the diffuse $\gamma$-ray emission
from the Milky Way and this possibility has lately been investigated\,\cite{wimdeboer}
using EGRET data. The EGRET (Energetic Gamma Ray Experiment Telescope) instrument was
a detector sensitive to photons with energies between 20\,MeV and 30\,GeV that was
flown between 1991 and 2000 aboard NASA's Compton Gamma-Ray Observatory. 
It allowed to reconstruct photon energies with an accuracy of 20\,\%, and its
angular resolution varied between 1.3$^\circ$ (at 1\,GeV) and 0.4$^\circ$ (at 10\,GeV).

The EGRET observations discovered several hundred $\gamma$-ray sources. Diffuse
emission was searched for by subtracting those sources from the overall photon
flux, and it was noted already in 1997 that the energy spectra of the diffuse 
component show a 60\,\% excess compared with model predictions\,\cite{hunter}. 
This excess was found at energies above several GeV, while the EGRET data
below 1\,GeV were well accounted for by models that describe the photon spectrum
as entirely due to secondary interactions of charged cosmic rays. The excess
was found with a very similar spectral shape for all sky directions, i.e.\ it 
does not only occur when observing, for example, the Galactic Centre region.

An excess of $\gamma$-rays from all directions is a signature that could 
be provided by dark matter annihilations taking place throughout the Milky
Way. The shape of the 
energy spectrum of the excess photons is determined by the mass and 
the couplings of the dark matter particle and should indeed be independent of
the observation direction. An analysis of the total EGRET data set has been 
performed to investigate possible dark matter contributions to the 
diffuse photon flux\,\cite{wimdeboer}. The analysis extracts the background 
shape from a model of cosmic ray propagation and $\gamma$-ray 
production in the Galaxy (GALPROP\,\cite{model2})
with a particular sets of parameters, while the signal shape is based
on expectations from supersymmetry\,\cite{darksusy:04b} combined with the known 
fragmentation of quarks. The absolute normalizations of signal and 
background are allowed to vary, and a fit of the energy spectra is used
to limit the mass of the dark matter particle to the 50--100\,GeV range (Fig.~\ref{fig:egret}).
The spatial distribution of the detected photons is used to extract
very detailed information about the dark matter distribution in the Galaxy.
It is found that the EGRET data can be described with a contribution
from annihilating sypersymmetric dark matter when the annihilation rate
is enhanced by a 'boost factor' of 20 which could be possible due to clumping on small
scales.

The proper modelling of the sizable fraction of background photons 
created by interactions of charged cosmic rays is a key element when
looking for dark matter induced photons. The GALPROP model combined with  
cosmic ray measurements (i.e.\ local (anti)proton and electron spectra) 
has been used to describe the EGRET data on diffuse photon emission 
without any dark matter contribution. This alternative analysis\,\cite{smreimer} 
uses the diffuse emission itself to adjust the electron injection spectrum
and also scales up the initial proton spectum using antiproton data. 
It arrives at a different GALPROP parameter set and somewhat relaxes the 
requirement on the agreement of locally measured and predicted 
electron and proton spectra. This is not implausible since details
of the galactic structure are anyway not taken into account in the
background model. Recently, it has been pointed out that the 
description of the EGRET excess with the help of dark matter 
as proposed by \cite{wimdeboer} would result in an antiproton 
flux higher than measured by balloon experiments\,\cite{bergstromantip}.

\section{Very High-Energy $\gamma$-rays from the Galactic Centre}

The Galactic Centre region is considered as a promising target for
dark matter searches since an increased dark matter density is 
expected there. The dynamical centre of the Galaxy is dominated 
by Sagittarius $A^\ast$, a putative black hole of $\approx 3\,\times 10^6$ solar masses\,\cite{schoedel}.
Sgr $A^\ast$ was first discovered as compact radio source and
exhibits a very low bolometric luminosity. It was found to be 
variable on sub-hour timescales in the infrared band and in X-rays\,\cite{genzel,x1,x2,x3} 
suggesting that the observed radiation is created close 
to the event horizon of the black hole. Sgr $A^\ast$ is surrounded
by a supernova remnant (Sgr $A$ East) and a region with a high density
of ionized hydrogen (Sgr $A$ West). Recently, also an energetic pulsar
candidate has been discovered close to Sgr $A^\ast$\,\cite{gcplerion}.

In 2004, results of observations of the Galactic Centre in very high-energy $\gamma$-rays ($>100\,\mbox{GeV}$) 
have been published by several experiments\,\cite{gcwhipple,gccangaroo,gchess} using the imaging atmospheric Cherenkov technique.
The most significant and precise $\gamma$-ray signal was reported by the H.E.S.S.\ experiment above
an energy threshold of 160\,GeV\,\cite{gchess}. The detected source appears point-like compared with the
point spread function of the instrument. Within errors, the source position is compatible
with Sgr $A^\ast$, but also with the supernova remnant Sgr $A$ East and the energetic 
pulsar candidate. The energy spectrum measured by H.E.S.S.\ is well described by a power law $E^{-\Gamma}$
with a spectral index $\Gamma$ of 2.25 (symbols in Fig.~\ref{fig:hess}), and 
extends well beyond to 20\,TeV. No indications
for flux variability were found in the data on time scales down to 10 minutes.

The $\gamma$-ray signal from the Galactic Centre has been interpreted 
both as dark matter\,\cite{gchooper,gcprl} and in terms of 'ordinary' astrophysical processes.
When ascribing the total photon signal to dark matter annihilations 
the H.E.S.S.\ energy spectrum implies dark matter masses in the order of
10\,TeV (Fig.~\ref{fig:hess}) regardless whether the dark matter candidate
comes supersymmetric or Kaluza-Klein theories. The power-law shape of
the spectrum does not fit to the cutoff that is in general expected from the 
annihilation spectra of dark matter particles. It has also been tried to
search for a dark matter annihilation spectrum that is hidden under an
astrophysical background with a power-law shape, but no significant 
dark matter component was found\,\cite{gcprl}.

Astrophysical explanations of the Galactic Centre signal have been
studied as well and could be viable alternatives. Accretion onto the black hole Sgr $A^\ast$ 
and associated acceleration processes can produce high-energy protons and electrons
that create secondary $\gamma$-rays due to synchrotron and curvature radiation
or due to interactions with ambient matter and photon fields. The low 
luminosity of Sgr $A^\ast$ implies that it appears transparent for TeV $\gamma$-rays,
and it has been demonstrated\,\cite{aha} that the H.E.S.S.\ data can be 
explained assuming populations of high-energy protons or electrons
associated with Sgr $A^\ast$. On the other hand, the lack of temporal 
variability makes an association of the TeV $\gamma$-ray signal with
Sgr $A^\ast$ less likely.

Other astrophysical interpretations focus on the supernova
remnant Sgr $A$ East\,\cite{crocker}. The spectral index measured by H.E.S.S.\ is close
to the value expected for particle acceleration in the shock front of 
supernova explosions. The observed $\gamma$-rays would then be 
created by protons and/or electrons accelerated to energies of
around 100\,TeV. This explanation appears not unmotivated 
since Sgr $A$ East is a very powerful supernova remnant with an
above-average explosion energy, and the presence of a 100\,TeV-particles 
has lately been demonstrated for other supernova remnants\,\cite{rxj,velajr}. 
It has also been argued that the energetic young pulsar would 
be luminous enough to explain the TeV $\gamma$-radiation.

\section{Summary}

A number of signals recorded by balloon experiments, satellites and imaging 
Cherenkov telescopes have been interpreted as being due to dark matter. 
Dark matter explanations, often of the same signal, invoke various theoretical 
frameworks, and there seems to be no scenario that can explain more than 
experimental signature. In most cases, astrophysical sources of the 
observed signals offer a reasonable alternative.

%%%%%%%%%%%%%%%%%%%%%%%%%%%%%%%%%%%%%% reset.txt counters %%%%%%%%%%%%%%
%%
%%%%%%% do not change below here  %%%%%%%%%%%%%%%%%%%%%%%%%%%%%
%%
\setcounter{section}{0}
\setcounter{subsection}{0}
\setcounter{figure}{0}
\setcounter{table}{0}
\setcounter{equation}{0}
\newpage

%%%%%%%%%%%%%%%%%%%%%%%%%%%%%%%%%%%%%%%%%%%%%%%%%%% Title, authors and
%%%%%%%%%%%%%%%%%%%%%%%%%%%%%%%%%%%%%%%%%%%%%%%%%%% addresses
\begin{frontmatter}
\title{ Strangelets, Nuclearites, Q-balls---A Brief Overview}
\author{Jes Madsen}
\address{Department of Physics and Astronomy, University of Aarhus, Denmark }
\begin{abstract}
Astrophysical bounds on the properties and abundances of primordial quark nuggets
and cosmic ray strangelets are reviewed. New experiments to search for cosmic
ray strangelets in lunar soil and from the International Space Station
are described. Analogies with baryonic and supersymmetric
Q-balls are briefly mentioned, as are prospects for strangelets as ultra-high energy
cosmic rays.
\end{abstract}
\end{frontmatter}

%%%%%%%%%%%%%%%%%%%%%%%%%%%%%%%%%%%%%%%%%%%%%%%%%%%%%% MAIN TEXT

\vspace{-1.5pc}
\section{ Introduction}

Quark nuggets, nuclearites and strangelets are different names for lumps of a
hypothetical phase of absolutely stable quark matter, so-called strange quark
matter because of the admixture of slightly less than one-third strange quarks
with the up and down quarks. Whether strange quark matter is absolutely stable
is a question yet to be decided by experiment or astrophysical observation
(see \cite{Weber:2004kj,Madsen:1998uh} for reviews), 
but if it is the case, then strange quark matter
objects may exist with baryon numbers ranging from ordinary nuclei to a
maximum of order $2\times10^{57}$ corresponding to gravitational instability
of strange stars.

Truly macroscopic quark matter lumps surviving from the cosmological
quark-hadron phase transition are often referred to as quark nuggets, and if
they hit the Earth they are sometimes dubbed nuclearites. Strangelets are
smaller lumps (baryon number $A<10^{7}$) 
where the electron cloud neutralizing the slightly
positive quark charge mainly resides outside the quark core. Strangelets are
unlikely to survive from the early Universe, but may form as a result of strange
star binary collisions and/or acceleration from the surface of pulsars.
The nomenclature is not strictly defined, and in the following the word
strangelet will be used as a general name except when discussing leftovers
from the early Universe.

Q-balls are non-topological solitons suggested from various origins in the
early Universe and as such have nothing to do with strange quark matter (their
origin and general properties were described by Kusenko at this workshop).
However for some classes of Q-balls (baryonic and supersymmetric Q-balls), the
astrophysical bounds that can be derived on strangelets are easily generalized
to these creatures as well and are therefore of interest in the context of
this Workshop on Extreme Physics with Neutrino Telescopes.

In the following I shall discuss the (unlikely) survival of cosmological quark
nuggets, the more optimistic prospects for cosmic rays strangelets from
strange stars, and two new experimental efforts to search for them. Due to
space limitations I will not go into details with the Q-ball analogies but
refer the reader to Kusenko's contribution and to \cite{madqball}.

\vspace{-0.8pc}
\section{Sources, sizes, and fluxes}
\subsection{Primordial quark nuggets}
\vspace{-0.5pc}
In a first order cosmological quark-hadron phase transition at $T\approx
100\,$MeV, supercooling may lead to concentration of baryon number inside
shrinking bubbles of quark phase, that may reach nuclear matter density and
form quark nuggets. The baryon number inside the horizon during the cosmic
quark-hadron phase transition (an upper limit for causal formation of quark
nuggets) is $A_{\mathrm{hor}}\approx10^{49}.$ Witten \cite{wit84} predicted
that typical nuggets would be somewhat smaller than this and argued that quark
nuggets might explain the cosmological dark matter problem. Quark nuggets
would decouple from the radiation bath very early in the history of the
Universe and behave as cold dark matter in the context of galaxy formation.
Today the nuggets would move with typical galactic halo velocities of a few
hundred kilometers per second through the Milky Way.

Later studies showed that the hot environment made cosmological nuggets
unstable against surface evaporation \cite{alcfar85}
and boiling \cite{alcoli89}, effectively destroying nuggets
with $A$ below $10^{39-46}$ depending on assumptions. Small traces of
primordial nuggets with lower baryon numbers could also be left over from the
destruction processes in the early Universe. Even such traces may in fact be
\textquotedblleft observed\textquotedblright\ using the astrophysical
detectors discussed below. Let $v\equiv250\mathrm{km\,s}^{-1}v_{250}$ and
$\rho\equiv10^{-24}\mathrm{g\,cm}^{-3}\rho_{24}$ be the typical speed and mass
density of nuggets in the galactic halo. The speed is given by the depth of
the gravitational potential of our galaxy, whereas $\rho_{24}\approx1$
corresponds to the density of dark matter. Then the number of nuggets hitting
the Earth is
\begin{equation}
\mathcal{F}\approx 6.0\times10^{5}A^{-1}\rho_{24}v_{250}\,\mathrm{cm}^{-2}%
\,\mathrm{s}^{-1}\,\mathrm{sr}^{-1}.
\end{equation}

Quark nuggets have a positive electrostatic surface potential (several MeV) of
the quark phase because quarks are stronger bound than electrons, so
during Big Bang nucleosynthesis ($T\leq1$ MeV), nuggets absorb neutrons but
not protons. This reduces the neutron-to-proton ratio, thereby lowering the
production of $^{4}$He. The helium-production is very sensitive to the total
amount of nugget-area present, and in order not to ruin the concordance with
observations, one finds \cite{madrii85} that only nuggets with
$A>A_{\mathrm{BBN}}\approx10^{22}\Omega_{\mathrm{nug}}^{3}f_{n}^{3}$ are
allowed during nucleosynthesis. Here $\Omega_{\mathrm{nug}}$ is the
present-day nugget contribution to the cosmic density in units of the critical
density, and $f_{n}\leq1$ is the penetrability of the nugget surface. 

In spite
of carrying baryon number, primordial quark nuggets do \textit{not\/}
contribute to the usual nucleosynthesis limit on $\Omega_{\mathrm{baryon}}$.
The baryon number is \textquotedblleft hidden\textquotedblright\ in quark
nuggets long before Big Bang nucleosynthesis begins, and the nuggets only
influence nucleosynthesis via the neutron absorption just described.
The same is true for baryon number carrying Q-balls.

While primordial quark nuggets remain a possibility within the 
tight restrictions
mentioned, the main problem with this scenario is the need for a first order
quark-hadron phase transition which is currently not favored in lattice QCD
studies at zero chemical potential.

\subsection{Strangelets from compact stars}
\vspace{-0.7pc}
If strange quark matter is absolutely stable all compact stars are likely to
be strange stars (see the following Section), and therefore the galactic
coalescence rate estimated for neutron star binaries that inspiral due to loss
of orbital energy by emission of gravitational radiation, believed to be of
order one collision in our Galaxy every 10,000 years
\cite{Kalogera:2003tn}, is really the rate of
strange star collisions. Each event involves a phase of tidal disruption as
the stars approach each other before the final collision. During this stage
small fractions of the total mass may be released from the binary system in
the form of strange quark matter. No realistic simulation of collisions
involving two strange stars has been performed. Simulations of binary neutron
star collisions, depending on orbital and other parameters, lead to the
release of anywhere from $10^{-5}-10^{-2}M_{\odot}$
($M_{\odot}$ is the solar mass), corresponding to a total
mass release in the Galaxy of $10^{-10}-10^{-6}M_{\odot}$ per year. The
equation of state for strange quark matter is stiff, so strange star
collisions probably lie in the low end of the mass release range. A
conservative estimate of the galactic production rate of strangelets is
$10^{-10}M_{\odot}$yr$^{-1}.$

Quark matter lumps released by tidal forces are macroscopic 
\cite{Madsen:2001bw},
but subsequent collisions lead to fragmentation, and if the collision energy
compensates for the surface energy involved in making smaller strangelets, a
significant fraction of the mass released from binary strange star collisions
might end up in the form of strangelets with $A\approx10^{2}-10^{4}$
\cite{Madsen:2001bw}. Incidentally, a similar range of strangelet masses is expected if the surface
of strange stars consists of a layer with strangelets embedded in an electron
gas rather than pure quark matter all the way to the surface 
\cite{Jaikumar:2005ne}.

Assuming that strangelets from binary collisions are accelerated
and propagate like cosmic ray nuclei in our Galaxy, taking
proper account of their small charge-to-mass ratio, as well as energy loss,
spallation, escape from the Galaxy, etc., it was shown in 
\cite{Madsen:2004vw} that the
expected flux for color-flavor locked strangelets (charge $Z=0.3A^{2/3}$
\cite{Madsen:2001fu}) near Earth is
\begin{equation}
\mathcal{F}\approx10^{-6}A^{-1.47}\mathrm{cm}^{-2}\,\mathrm{s}^{-1}%
\,\mathrm{sr}^{-1}.
\label{appflux}
\end{equation}
Most of these nuggets have rigidities (momentum divided by charge) of a few
GV, but with a powerlaw tail at higher rigidity. Apart from the slightly
different $A$-dependence this is some 12 orders of magnitude smaller than the
flux estimate for dark matter nuggets, which is not unreasonable because the
total strangelet mass originating from binary collisions over the age of the
Galaxy is around 1 $M_{\odot}$, compared to $10^{12}$ $M_{\odot}$ of dark matter.

Another possible cosmic ray strangelet source is extraction from the
surface of pulsars and acceleration in the strong pulsar electric
fields. A measurable flux is predicted in \cite{Cheng:2006ak} 
within the scenario where the strange star surface consists of
strangelets embedded in an electron gas \cite{Jaikumar:2005ne}.
Formation and acceleration in
supernova explosions has also been suggested \cite{Benvenuto:1989hw}.

\section{Detection of cosmic ray strangelets (or why either all or no compact
stars are strange)}
De R\'{u}jula and Glashow \cite{dergla84} argued that quark nuggets hitting
the Earth would show up as unusual meteor-events, earth-quakes, etched tracks
in old mica, in meteorites and in cosmic-ray detectors. A negative search for
tracks in ancient mica corresponded to a lower nugget flux limit of
$8\times10^{-19}\,\mathrm{cm}^{-2}\,\mathrm{s}^{-1}\,\mathrm{sr}^{-1}$ for
nuggets with $A>1.4\times10^{14}$ (smaller nuggets being trapped in layers
above the mica samples studied).

Later investigations have improved these flux limits by a few orders of
magnitude and extended them to lower $A$, though with higher flux limits (see
other contributions to these proceedings for examples). This has excluded
quark nuggets with $3\times10^{7}<A<5\times10^{25}$ as dark matter, but a low
flux from the Big Bang or from collision of strange stars cannot be ruled out.

Neutron stars and their stellar progenitors\ may be thought of
as alternative large
surface area, long integration time detectors leading to much tighter flux
limits \cite{mad88}, see also \cite{calfri91}. 
The presence of a single quark nugget in
the interior of a neutron star is sufficient to initiate a transformation of
the star into a strange star \cite{wit84,alcfar86a}. The time-scale
for the transformation is short, between seconds and minutes, so observed
pulsars would have been converted long ago if their stellar progenitors ever
captured a quark nugget, or if neutron stars absorbed one after formation.

To convert a neutron star into strange matter a quark nugget should not only
hit a supernova progenitor but also be caught in the core \cite{mad88}. A main
sequence star is capable of capturing non-relativistic quark nuggets with
baryon numbers below $A_{\mathrm{STOP}}\approx10^{31}$, which are braked by
inertia, i.\ e.\ they are slowed down by electrostatic scatterings after
plowing through a column of mass similar to their own, and settle in the
stellar core. Relativistic nuggets may be destroyed after collisions with
nuclei in the star, but even a
tiny fraction of a nugget surviving such an event and settling in the star is
sufficient to convert the neutron star to a strange star, so the
non-relativistic flux limits may still apply.

For nuggets with $A<A_{\mathrm{STOP}}$ the sensitivity of main sequence stars
as detectors is given by the limit of one nugget hitting the surface of the
supernova progenitor in its main sequence lifetime. Converted into a flux,
$\mathcal{F}$, of nuggets hitting the Earth it corresponds to
\begin{equation}
\mathcal{F}\approx 4\times10^{-42}v_{250}^{2}\mathrm{cm}^{-2}\,\mathrm{s}%
^{-1}\,\mathrm{sr}^{-1}.
\end{equation}
This is a factor of $10^{20}$--$10^{40}$ more sensitive 
than direct detection experiments!

\textit{If} it is possible to prove that some neutron stars are indeed neutron
stars rather than strange stars, the sensitivity of the astrophysical
detectors rules out quark nuggets as dark matter for $A<10^{34-38}$. And it
questions the whole idea of stable strange quark matter, since it is
impossible to avoid polluting the interstellar medium with nuggets from
strange star collisions or supernova explosions at fluxes many orders of
magnitude above the limit measurable in this way.

\textit{If} on the other hand strange quark matter
is stable, then all neutron stars are likely
to be strange stars, again because some pollution can not be avoided.

\vspace{-0.5pc}
\section{Experiments underway}

Several experiments have searched for strangelets in cosmic rays. While some
interesting events have been found that are consistent with the predictions
for strangelets, none of these have been claimed as real discoveries.
Interpreted as flux limits rather than detections these results are consistent
with the flux estimates given above. For discussions see
\cite{Sandweiss:2004bu,Finch:2006pq}.

If the interesting events were actual measurements, two experiments that are
currently underway will reach sensitivities that would provide real statistics.

\textbf{AMS-02: }The Alpha Magnetic Spectrometer (AMS) is a space-based
particle physics experiment involving several hundred physicists from more
than 50 institutions in 16 countries, led by Nobel laureate Samuel Ting of
MIT. A prototype (AMS-01) flew in June 1998 aboard the Space Shuttle Discovery
\cite{Aguilar:2002ad}, and AMS-02 is currently scheduled to fly to the
International Space Station (ISS) in 2009. Once on the ISS AMS-02 will remain
active for at least three years. AMS-02 will provide data with unprecedented
accuracy on cosmic ray electrons, positrons, protons, nuclei, anti-nuclei and
gammas in the GV-TV range and probe issues such as antimatter, dark matter,
cosmic ray formation and propagation. In addition it will be uniquely suited
to discover strangelets characterized by extreme rigidities for a given
velocity compared to nuclei \cite{Sandweiss:2004bu,Finch:2006pq}. 
AMS-02 will have
excellent charge resolution up to $Z\approx30$, and should be able to probe a
large mass range for strangelets. A reanalysis of data from the AMS-01
mission has given hints of some interesting events, such as one with
$Z=2,A=16$ \cite{Choutko:2003} and another with $Z=8,$ but with the larger
AMS-02 detector running for 3 years or more, real statistics is achievable.

\textbf{LSSS: }The Lunar Soil Strangelet Search (LSSS) is a search for $Z=8$
strangelets using the tandem accelerator at the Wright Nuclear Structure
Laboratory at Yale \cite{Finch:2006pq,Han:2006}. 
The experiment involves a dozen people from Yale, MIT, and
\AA rhus, led by Jack Sandweiss of Yale. The experiment which is about to
begin its real data taking phase, studies a sample of 15 grams of lunar soil
from Apollo 11. It will reach a sensitivity of $10^{-17}$ over a wide mass
range, sufficient to provide detection according to Eq.~\ref{appflux}
if strangelets have been trapped in the lunar surface layer, which has an
effective cosmic ray exposure time of around 500 million years and an
effective mixing depth due to micrometeorite impacts of only around one
meter, in contrast to the deep geological and oceanic mixing on Earth.
Combined with the fact that the Moon has no shielding magnetic field,
this results in an expected strangelet concentration in lunar soil
which is at least four orders of magnitude larger than the corresponding
concentration on Earth.

\vspace{-0.8pc}
\section{Strangelets as ultra-high energy cosmic rays}

The existence of cosmic rays with energies well beyond $10^{19}\mathrm{eV}$,
with measured energies as high as $3\times10^{20}\mathrm{eV}$, is one of the
most interesting puzzles in cosmic ray physics
\cite{Greisen:1966jv0}. It is almost impossible to find a
mechanism to accelerate cosmic rays to these energies. Furthermore ultra-high
energy cosmic rays lose energy in interactions with cosmic microwave
background photons, and only cosmic rays from nearby (unidentified) sources
would be able to reach us with the energies measured. Strangelets circumvent
both problems \cite{Madsen:2002iw},
and therefore provide a possible mechanism for cosmic rays
beyond the so-called GZK-cutoff.

\textbf{Acceleration: }All astrophysical \textquotedblleft
accelerators\textquotedblright\ involve electromagnetic fields, and the
maximal energy of a charged particle is proportional to its charge. The charge
of massive strangelets has no upper bound in contrast to nuclei, so highly
charged strangelets are capable of reaching energies much higher than those of
cosmic ray protons or nuclei using the same \textquotedblleft
accelerator\textquotedblright\ \cite{Madsen:2002iw}.

\textbf{The GZK-cutoff} is a consequence of ultra-relativistic cosmic rays
hitting a $2.7\mathrm{K}$ background photon with a Lorentz-factor $\gamma$
large enough to boost the $7\times10^{-4}\,\mathrm{eV}$ photon to energies
beyond the threshold of energy loss processes, such as photo-pion production
or photo-disintegration. The threshold for such a process has a fixed energy,
$E_{\mathrm{Thr}}$, in the frame of the cosmic ray, e.g., $E_{\mathrm{Thr}%
}\approx10\mathrm{MeV}$ for photo-disintegration of a nucleus or a strangelet,
corresponding to $\gamma_{\mathrm{Thr}}=E_{\mathrm{Thr}}/E_{2.7\mathrm{K}%
}\approx10^{10},$ or a cosmic ray total energy
\begin{equation}
E_{\mathrm{Total}}=\gamma_{\mathrm{Thr}}Am_{0}c^{2}\approx10^{19}%
A~\mathrm{eV.}%
\end{equation}
Since strangelets can have much higher $A$-values than nuclei, this pushes the
GZK-cutoff energy well beyond the current observational limits for ultra-high
energy cosmic rays \cite{Madsen:2002iw,Rybczynski:2001bw}.

\vspace{-0.8pc}
\section{Conclusion}

Lumps of strange quark matter (quark nuggets, nuclearites, strangelets) may
form in a first-order cosmological quark-hadron phase transition (unlikely),
or in processes related to compact stars (more likely). Flux estimates for
lumps reaching our neighborhood of the Galaxy as cosmic rays are in a range
that makes it realistic to either detect them in upcoming experiments like
AMS-02 or LSSS, or place severe limits on the existence of stable strange
quark matter. A similar line of reasoning applies to Q-balls.

\section*{Acknowledgments}

{\small This work was supported by the Danish Natural Science Research
Council. }\\[-0.6cm]

%%%%%%%%%%%%%%%%%%%%%%%%%%%%%%%%%%%%%% reset.txt counters %%%%%%%%%%%%%%
%%%%%%% do not change below here  %%%%%%%%%%%%%%%%%%%%%%%%%%%%%
\setcounter{section}{0}
\setcounter{subsection}{0}
\setcounter{figure}{0}
\setcounter{table}{0}
\setcounter{equation}{0}
\newpage

%%%%%%%%%%%%%%%%%%%%%%%%%%%%%%%%%%%%%%%%%%%%%%%%%%% Title, authors and addresses
\begin{frontmatter}


\title{Properties and signatures of supersymmetric Q-balls
}

% use optional labels between square brackets to link authors explicitly to addresses:
% If more than one author, keep a comma between the author tags

\author{Alexander Kusenko}

\address{Department of Physics and Astronomy, University of
California, Los Angeles,~CA~90095,~USA }

\begin{abstract}
Supersymmetric extensions of the Standard Model predict the existence of
Q-balls with baryon and lepton numbers.  Stable Q-balls can form at the end of
inflation from the fragmentation of the Affleck--Dine condensate and can exist
as dark matter.  The best current limits come from Super-Kamiokande and MACRO.
 The search beyond these limits can be conducted using the future water
Cherenkov detectors.  

\end{abstract}

% \begin{keyword}

%\PACS 
% \end{keyword}

\end{frontmatter}

%%%%%%%%%%%%%%%%%%%%%%%%%%%%%%%%%%%%%%%%%%%%%%%%%%%%%% MAIN TEXT

\section{Introduction}

Supersymmetry is a theoretically appealing possibility for physics beyond the
Standard Model.  It predicts the existence of new particles and extended
objects, Q-balls~\cite{ak_mssm,dk}.   Both the Q-balls and the lightest
supersymmetric particle are candidates for dark matter.  Here we will briefly
review the former possibility.

\section{Q-balls from Supersymmetry} 

In a class of theories with interacting scalar fields $\phi$ that carry
some conserved global charge, the ground state is a
Q-ball~\cite{q,nts_review}, a lump of coherent scalar condensate that can
be described semiclassically as a non-topological soliton of the form
\begin{equation}
\phi(x,t) = e^{i \omega t} \bar{\phi}(x).
\label{q}
\end{equation}
Q-balls exist whenever the scalar potential satisfies certain conditions
that were first derived for a single scalar degree of freedom~\cite{q} with
some abelian global charge and were later generalized to a theory of many
scalar fields with different charges~\cite{ak_mssm}.  Non-abelian global
symmetries~\cite{nonabelian} and abelian local symmetries~\cite{gauge} can
also yield Q-balls.

For a simple example, let us consider a field theory with a scalar
potential $U(\vf) $  that has a global minimum $U(0)=0$ at $\vf=0$.
Let $U(\vf)$ have an unbroken global\footnote{
Q-balls associated with a local symmetry have been constructed 
\cite{gauge}.  An important qualitative difference is that, in the case of a
local symmetry, there is an upper limit on the charge of a stable Q-ball.} 
U(1) symmetry at the origin, $\vf=0$.  And let the scalar field $\vf$ have
a unit charge with respect to this U(1).

The charge of some field configuration $\vf(x,t)$ is  
\beq
Q= \frac{1}{2i} \int \vf^* \stackrel{\leftrightarrow}{\partial}_t  
\vf \, d^3x . 
\label{Qt}
\eeq
Since a  trivial configuration $\vf(x)\equiv 0$ has zero charge, the
solution that minimizes the energy, 
\beq
E=\int d^3x \ \left [ \frac{1}{2} |\dot{\vf}|^2+
\frac{1}{2} |\nabla \vf|^2 
+U(\vf) \right], 
\label{e}
\eeq
and has a given charge $Q>0$, must differ from zero in some (finite)
domain.  This is a Q-ball.   It is a time-dependent solution, more
precisely it has a time-dependent phase. However, all physical quantities
are time-independent.  Of course, we have not proved that such a 
``lump'' is finite, or that it has a lesser energy than the collection of
free particles with the same charge; neither is true for a general
potential.  A finite-size  Q-ball is a minimum of energy and is
stable with respect to decay into free $\vf$-particles if 
\beq
U(\vf) \left/ \vf^2 \right. = {\rm min},
\ \ {\rm for} \ 
\vf=\vf_0>0 .
\label{condmin}
\eeq

One can show that the equations of motion for a Q-ball in 3+1 dimensions
are equivalent to those for the bounce associated with tunneling 
in 3 Euclidean dimensions in an effective potential $\hat{U}_\omega
(\vf)= U(\vf) - (1/2) \omega^2 \vf^2$, where $\omega$ is such that it
extremizes~\cite{ak_qb}
\beq
{\sf E}_\omega = S_3(\omega) +\omega Q. 
\label{Ew}
\eeq
Here $S_3(\omega)$ is the three-dimensional Euclidean action of the bounce
in the potential $\hat{U}_\omega (\vf)$.  The Q-ball
solution has the form (\ref{q}), 
%
%\beq
%\vf(x,t) = e^{i\omega t} \bar{\vf}(x),
%\eeq
where $\bar{\vf}(x)$ is the bounce. 

The analogy with tunneling clarifies the meaning of condition 
(\ref{condmin}), which simply requires that there exist a value of 
$\omega$, for which $\hat{U}_\omega (\vf)$ is negative for some value of 
$\vf=\vf_0 \neq 0$ separated from the false vacuum by a barrier. 
This condition ensures the  existence of a bounce.  (Clearly, the
bounce does not exist if $\hat{U}_\omega (\vf) \ge 0$ for all $\vf$ because
there is nowhere to tunnel.)  

In the true vacuum, there is a minimal value $\omega_0$, so that only for
$\omega>\omega_0$, $\hat{U}_\omega (\vf)$ is somewhere negative.  If one
considers a Q-ball in a metastable false vacuum, then $\omega_0=0$.  The
mass of the $\vf$ particle is the upper bound on $\omega$ in either
case. Large values of $\omega$ correspond to small charges~\cite{ak_qb}.
As $Q \rightarrow \infty$, $\omega \rightarrow \omega_0$.  In this case,
the effective potential $\hat{U}_\omega (\vf)$ has two nearly-degenerate
minima; and one can apply the thin-wall approximation to calculate the
Q-ball energy~\cite{q}.  For smaller charges, the thin-wall approximation
breaks down, and one has to resort to other methods~\cite{ak_qb}.

The above discussion can be generalized to the case of several fields, 
$\vf_k$, with different charges, $q_k$~\cite{ak_mssm}.  Then the Q-ball is
a solution of the form 
\beq
\vf_k(x,t) = e^{iq_k \omega t} \vf_k(x),
\label{tsol}
\eeq
where $\vf(x)$ is again a three-dimensional bounce associated with
tunneling in the potential 
\beq
\hat{U}_\omega (\vf) = U(\vf)\ - \ \frac{1}{2} \omega^2 \, 
\sum_k q_k^2 \, |\vf_k|^2. 
\label{Uhat}
\eeq
As before, the value of $\omega$ is found by minimizing ${\sf E}_\omega$
in equation  (\ref{Ew}).  The bounce, and, therefore, the Q-ball, exists if 
\begin{eqnarray}
\mu^2 & = & 
2 U(\vf) \left/ \left (\sum_k q_k \vf_{k,0}^2 \right ) \right. = {\rm min},
\ \nonumber \\ & & {\rm for} \ |\vec{\vf}_0|^2 > 0.
\label{condmin1}
\end{eqnarray}

The soliton mass can be calculated by extremizing ${\sf E}_\omega$ in
equation (\ref{Ew}).  If $ |\vec{\vf}_0|^2 $ defined by equation
(\ref{condmin1}) is finite, then the mass of a soliton $M(Q)$ is
proportional to the first power of $Q$: 
\beq
M(Q) = \tilde{\mu} Q, \ \ {\rm if} \ |\vec{\vf}_0|^2 \neq \infty. 
\label{MQ}
\eeq In particular, if $Q\rightarrow \infty$, $\tilde{\mu}\rightarrow \mu$
(thin-wall limit)~\cite{q,nts_review}.  For smaller values of $Q$,
$\tilde{\mu}$ was computed in~\cite{ak_qb}.  In any case, $\tilde{\mu}$ is
less than the mass of the $\phi$ particle by definition (\ref{condmin1}).

However, if the scalar potential grows slower than the second power of
$\phi$, then $|\vec{\vf}_0|^2 = \infty$, and the Q-ball never reaches the
thin-wall regime, even if Q is large.  The value of $\phi$ inside the
soliton extends as far as the gradient terms allow, and the mass of a
Q-ball is proportional to $Q^{p}$, $p<1$.  In particular, if the scalar
potential has a flat plateau $U(\phi) \sim m $ at large $\phi$, then the
mass of a Q-ball is~\cite{dks} 
\beq
M(Q) \sim m Q^{3/4}.
\label{MQflat}
\eeq
This is the case for the stable baryonic Q-balls in the MSSM discussed
below. 

It turns out that all phenomenologically viable supersymmetric extensions
of the Standard Model predict the existence of non-topological
solitons~\cite{ak_mssm} associated with the conservation of baryon and
lepton number. If the physics beyond the standard model reveals some
additional global symmetries, this will further enrich the spectrum of
Q-balls~\cite{Demir}.  The MSSM admits a large number of different Q-balls,
characterized by (i) the quantum numbers of the fields that form a
spatially-inhomogeneous ground state and (ii) the net global charge of this
state.

First, there is a class of Q-balls associated with the tri-linear
interactions that are inevitably present in the MSSM~\cite{ak_mssm}.  The
masses of such Q-balls grow linearly with their global charge, which can be
an arbitrary integer number~\cite{ak_qb}.  Baryonic and leptonic Q-balls of
this variety are, in general, unstable with respect to their decay into
fermions.  However, they could form in the early universe through the
accretion of global charge~\cite{gk,ak_pt} or, possibly, in a first-order
phase transition~\cite{s_gen}.

The second class~\cite{dks} of solitons comprises the Q-balls whose VEVs
are aligned with some flat directions of the MSSM.  The scalar field inside
such a Q-ball is a gauge-singlet~\cite{kst} combination of squarks and
sleptons with a non-zero baryon or lepton number.  The potential along a
flat direction is lifted by some soft super\-sy\-mmetry-breaking terms that
originate in a ``hidden sector'' of the theory at some scale $\Lambda_{_S}$
and are communicated to the observable sector by some interaction with a
coupling $g$, so that $g \Lambda \sim 100$~GeV.  Depending on the strength
of the mediating interaction, the scale $\Lambda_{_S}$ can be as low as a
few TeV (as in the case of gauge-mediated SUSY breaking), or it can be some
intermediate scale if the mediating interaction is weaker (for instance,
$g\sim \Lambda_{_S}/m_{_{Planck}}$ and $\Lambda_{_S}\sim 10^{10}$~GeV in
the case of gravity-mediated SUSY breaking).  For the lack of a definitive
scenario, one can regard $\Lambda_{_S}$ as a free parameter.  Below
$\Lambda_{_S}$ the mass terms are generated for all the scalar degrees of
freedom, including those that parameterize the flat direction.  At the
energy scales larger than $\Lambda_{_S}$, the mass terms turn off and the
potential is ``flat'' up to some logarithmic corrections.  If the Q-ball
VEV extends beyond $\Lambda_{_S}$, the mass of a soliton~\cite{dks,ks} is
no longer proportional to its global charge $Q$, but rather to $Q^{3/4}$.
A hybrid of the two types is yet another possibility~\cite{hybrid}. 

This allows for the existence of some entirely stable Q-balls with a large
baryon number $B$ (B-balls).  Indeed, if the mass of a B-ball is $M_{_B} \sim
({\rm 1~TeV}) \times B^{3/4}$, then the energy per baryon number
$(M_{_B}/B)\sim ({\rm 1~TeV}) \times B^{-1/4}$ is less than 1~GeV for $B >
10^{12}$.  Such large B-balls cannot  dissociate into protons and neutrons
and are entirely stable thanks to the conservation of energy and the baryon
number.  If they were  produced in the early universe, they would exist at
present as a form of dark matter~\cite{ks}.

\section{Fragmentation of Affleck--Dine Condensate into Q-balls}
Several mechanisms could lead to formation of B-balls and L-balls in the
early universe. First, they can be produced in the course of a phase
transition~\cite{s_gen}.  Second, thermal fluctuations of a baryonic and
leptonic charge can, under some conditions, form a Q-ball.  Finally, a
process of a gradual charge accretion, similar to nucleosynthesis, can take
place~\cite{gk,ak_pt,dew}.  However, it seems that the only process that can
lead to a copious production of very large, and, hence, stable, B-balls, is
fragmentation of the Affleck-Dine condensate~\cite{ks}. 

At the end of inflation, the scalar fields of the MSSM develop some large
expectation values along the flat directions, some of which have a non-zero
baryon number~\cite{ad}. Initially, the scalar condensate has the form
given in eq.~(\ref{q}) with $\bar{\phi}(x)= const$ over the length scales
greater than a horizon size. One can think of it as a universe filled with
Q-matter.  The relaxation of this condensate to the potential minimum is
the basis of the Affleck--Dine (AD) scenario for baryogenesis.

It was often assumed that the condensate remains spatially homogeneous from
the time of formation until its decay into the matter baryons.  This
assumption is, in general, incorrect.  In fact, the initially homogeneous
condensate can become unstable~\cite{ks} and break up into Q-balls whose
size is determined by the potential and the rate of expansion of the
Universe.  B-balls with $12 < \log_{10} B < 30$ can form naturally
from the breakdown of the AD condensate.  These are entirely
stable if the flat direction is ``sufficiently flat'', that is if the
potential grows slower than $\phi^2$ on the scales or the order of
$\bar{\phi}(0)$.   The evolution of the primordial condensate can be
summarized as follows: 

\vspace{3mm}
\psfig{figure=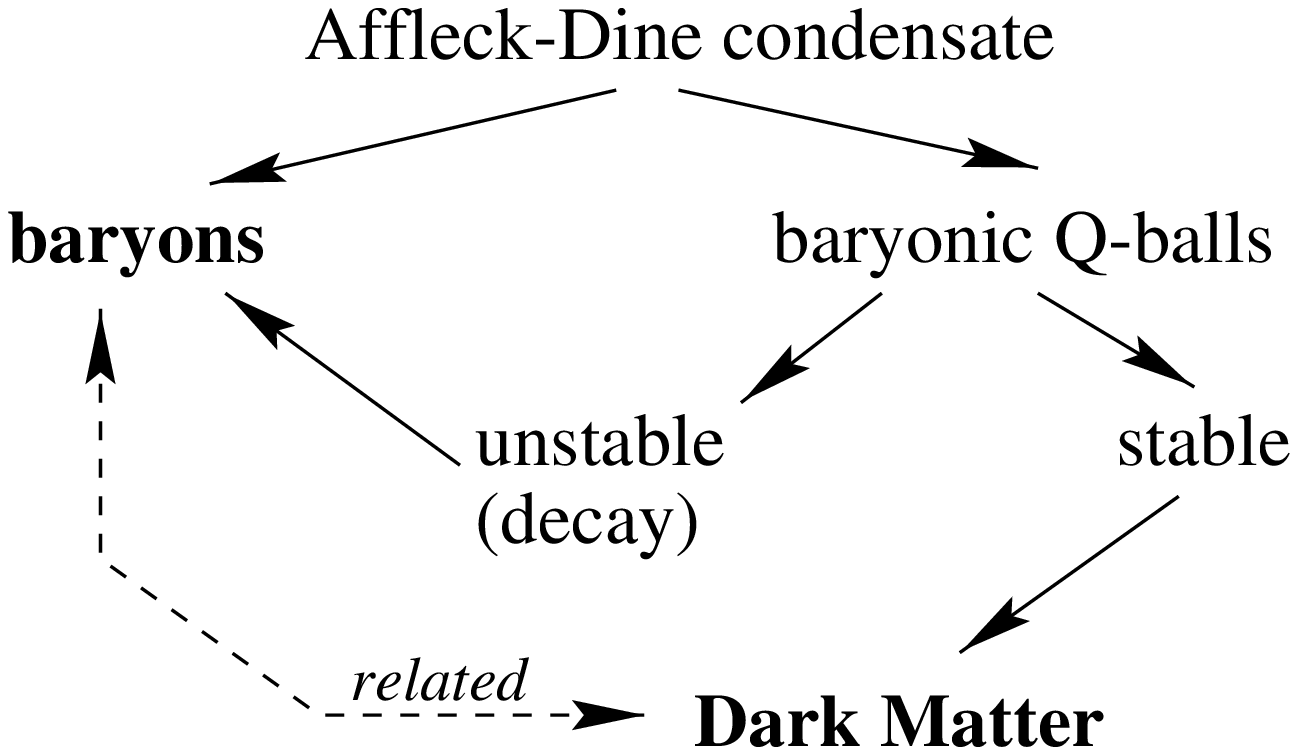,height=2.1in,width=4.5in}

This process has been analyzed analytically~\cite{ks,em} in the linear
approximation.  Recently, some impressive numerical simulations of Q-ball
formation have been performed~\cite{kasuya}; they confirm that the
fragmentation of the condensate into Q-balls occurs in some Affleck-Dine
models.  The global charges of Q-balls that form this way are model
dependent.  The subsequent collisions~\cite{ks,Axenides:1999hs} can further
modify the distribution of soliton sizes.

% \begin{figure}
% \setlength{\epsfxsize}{3.4in}
% %%%%%%%%%%%%%%%%%%%%%%%%%%%\setlength{\epsfysize}{}
% \centerline{\epsfbox{figure_r_3d.eps}}
% \caption{ 
% The charge density per comoving volume in (1+1) dimensions for a sample
% potential analyzed numerically during the fragmentation of the condensate
% into Q-balls. 
% }
% \label{fig_charge}
% \end{figure}

In supersymmetric extensions of the Standard Model, Q-ball formation occurs
along flat directions of a certain type, which appear to be generic in the
MSSM~\cite{Enqvist:2000gq}.

\section{SUSY Q-balls as Dark Matter} 

Conceivably, the cold dark matter in the Universe can be made up entirely
of SUSY Q-balls.  Since the baryonic matter and the dark matter share the
same origin in this scenario, their contributions to the mass density of
the Universe are related.  Most of dark-matter scenarios offer no
explanation as to why the
observations find $\Omega_{_{DARK}} \sim \Omega_{B} $ within an order of
magnitude.  This fact is extremely difficult to explain in models that
invoke a dark-matter candidate whose present-day abundance is determined by
the process of freeze-out, independent of baryogenesis.  
If one doesn't want to accept this equality
as fortuitous, one is forced to hypothesize some {\it ad hoc}
symmetries~\cite{kaplan} that could relate the two quantities.  In the MSSM
with AD baryogenesis, the amounts of dark-matter Q-balls and the ordinary
matter baryons are related~\cite{ks}; one
predicts~\cite{lsh} $\Omega_{_{DARK}} \sim 10 \, \Omega_{B} $ for
B-balls with $B \sim 10^{26}$. However, the size of Q-balls depends on the
supersymmetry breaking terms that lift the flat direction.  
The required size is in the middle of the range of
Q-ball sizes that can form in the Affleck--Dine
scenario~\cite{ks,em,kasuya}.  Diffusion effects may force the Q-balls
sizes to be somewhat smaller, $B\sim 10^{22} - 
10^{24}$~\cite{Banerjee:2000mb}.

% \begin{figure}
% \setlength{\epsfxsize}{4.5in}
% %%%%%%%%%%%%%%%%%%%%%%%%%%%\setlength{\epsfysize}{}
% \epsfclipon
% \centerline{\epsfbox{QMs.eps}}
% \caption{ The resent limits on the baryon numbers of electrically neural
% dark-matter Q-balls from a paper by J.~Arafune {\em et al.}~\cite{arafune}.
% }
% \epsfclipoff
% \label{fig_limits}
% \end{figure}

The value $B\sim 10^{26}$ is well within the present experimental limits 
on the baryon number of an average relic B-ball, under the assumption
that all or most of cold dark matter is made up of Q-balls.  On their 
passage through matter, the electrically neutral baryonic SUSY Q-balls can
cause a proton decay, while the electrically charged B-balls produce 
massive ionization.  Although the condensate inside a Q-ball is
electrically neutral~\cite{kst}, it may pick up some electric charge
through its interaction with matter~\cite{kkst}.  Regardless of its ability
to retain electric charge, the Q-ball would produce a straight track in a
detector and would release the energy of, roughly, 10 GeV/mm.  The present
limits~\cite{kkst,exp,arafune,Super-K,MACRO} constrain the baryon number of a
relic dark-matter B-ball to be greater than $10^{22}$.  Future experiments are
expected to improve these limits.  It would take a detector with the area of
several square kilometers to cover the entire interesting range $B\sim
10^{22} ... 10^{30}$.

\section{Interactions with matter, experimental and astrophysical bounds}  

A Dirac fermion scattering off a Q-ball can convert into an antifermion, as
long as the fermion number is spontaneously broken inside the Q-ball by the
scalar condensate~\cite{Loveridge_int}.  The baryonic Q-balls can, therefore,
interact with matter nuclei converting them into pions whose decay can produce
a signal in various detectors,
including Super-Kamiokande~\cite{kkst,arafune,Super-K,Super-K_talk}.

Dark-matter superballs pass through the ordinary stars and planets with a
negligible change in their velocity.  However, Q-balls can stop in
the neutron stars and accumulate there~\cite{sw}.  As soon as the first
Q-ball is captured by a neutron star, it sinks to the center and begins to
absorb the baryons into the condensate.  High baryon density inside a
neutron star makes this absorption very efficient, and the B-ball grows at
the rate that increases with time due to the gradual increase in the
surface area.  After some time, the additional dark-matter Q-balls that
fall onto the neutron star make only a negligible contribution to the
growth of the central Q-ball~\cite{sw}.  So, the fate of the neutron star
is sealed when it captures the first superball.

Baryonic Q-balls can destroy neutron stars by stimulated nucleon decay in
nuclear matter~\cite{sw,Loveridge_int,Loveridge_astro}.
Neutron stars are stable in some range of masses.  In particular, there is
a minimal mass (about 0.18 solar mass), below which the force of gravity is
not strong enough to prevent the neutrons from decaying into protons and
electrons.  While the star is being consumed by a superball, its mass
gradually decreases, reaching the critical value eventually.  Neutron stars
are known to exist for as long as 10 Gyr, which sets a bound on the rate of
absorption by SUSY Q-balls. 

The astrophysical limits on SUSY Q-balls depend on the nature of the flat
direction and the non-renormalizable operators that ``lift'' it for a
sufficiently large VEV.  The generic lifting terms can be written in the
form
\beq
V^{(m,n)}(\phi)_{\mbox{\tiny lifting}}\approx  
\lambda_{mn} M^4\left(\frac{\phi}{M}\right)^{n-1+m}
\left(\frac{\phi^*}{M}\right)^{n-1-m},
\label{lift}
\eeq
where $M$ is the characteristic high scale, presumably, of the order of the
Planck scale.  The possible lifting terms \cite{flat_directions} may preserve
the baryon number ($m=0$), or they may cause a violation of the baryon number
for a large scalar VEV ($m\neq 0$). If the leading lifting terms
in eq.~(\ref{lift}) have $m=0$, the baryon number is conserved even for
very large values of the VEV. The flat directions of this kind generate Q-balls
that can grow rapidly inside a neutron star, and stringent constraints exists
on such relic Q-balls~\cite{Loveridge_astro}.  In contrast, if the flat
direction is lifted by terms with $m\neq 0$, the corresponding Q-balls cannot
grow beyond certain size, and the limits from neutron stars or white dwarfs do
not apply~\cite{Loveridge_astro}.   In the MSSM, the flat directions which
are not constrained by the stability of neutron stars include $QLe$, $QLd$,
$Lude$, $QLde$, $QLud$, $QLude$, in the notation of
Ref.~\cite{flat_directions}.

\section{Conclusion}

Supersymmetric models of physics beyond the weak scale offer two plausible 
candidates for cold dark matter: the lightest supersymmetric
particle and a stable non-topological soliton, or Q-ball, carrying some
baryonic charge. 

SUSY Q-balls make an appealing dark-matter candidate because their
formation is a natural outcome of the Affleck--Dine baryogenesis.  The basic
assumptions are supersymmetry and inflation.  The search beyond the current
limits can be conducted using future water Cherenkov detectors. 

This work was supported in part by the DOE grant DE-FG03-91ER40662 and the NASA
ATP grants NAG~5-10842 and NAG~5-13399.

%%%%%%%%%%%%%%%%%%%%%%%%%%%%%%%%%%%%%% reset.txt counters %%%%%%%%%%%%%%
%%
%%%%%%% do not change below here  %%%%%%%%%%%%%%%%%%%%%%%%%%%%%
%%
\setcounter{section}{0}
\setcounter{subsection}{0}
\setcounter{figure}{0}
\setcounter{table}{0}
\setcounter{equation}{0}
\newpage

%%%%%%%%%%%%%%%%%%%%%%%%%%%%%%%%%%%%%%%%%%%%%%%%%%% Title, authors and addresses
\begin{frontmatter}


\title{Search for Neutral Q-balls in Super-Kamiokande II}

% use optional labels between square brackets to link authors explicitly to addresses:
% If more than one author, keep a comma between the author tags

\author[ncen]{Y. Takenaga}

\address[ncen]{Research Center for Cosmic Neutrinos, Institute for Cosmic Ray Research, University of Tokyo, Kashiwa, Chiba 277-8582, Japan}
\collab{The Super-Kamiokande Collaboration}

\begin{abstract}
 A search for Q-balls has been carried out in Super-Kamiokande II with 541.7\,days of live time. A neutral Q-ball can interact with nuclei to produce pions, generating a signal of successive contained pion events along a track. No candidate for successive contained event groups has been found, and upper limits on the flux of such Q-balls are obtained.
\end{abstract}

% \begin{keyword}

%\PACS 
% \end{keyword}

\end{frontmatter}

%%%%%%%%%%%%%%%%%%%%%%%%%%%%%%%%%%%%%%%%%%%%%%%%%%%%%% MAIN TEXT
\section{\label{sec:intro} Introduction}

 In the framework of some supersymmetric models, stable non-topological solitons called Q-balls~\cite{Coleman:1985ki} can be produced in the early universe and contribute to the dark matter~\cite{Affleck:1984fy,Kusenko:1997vp,Kasuya:1999wu,Enqvist:2003gh}. If Q-balls are formed in the early universe and have survived until the present time, they would be part of the dark matter located in the halo of Galaxy. Q-balls are solitons with theoretically appealing roles in some dark matter scenarios and in explanations of the baryon asymmetry~\cite{Enqvist:1997si}. 

 In this model~\cite{Kusenko:1997vp}, the Q-ball mass and radius are given by
\begin{equation}
M_{Q} = \frac{4\pi \sqrt{2}}{3} {M_S} Q^{\frac{3}{4}},\qquad R_Q = \frac{1}{\sqrt{2}} {M_S}^{-1} Q^{\frac{1}{4}}
  \label{eq:qball}
\end{equation}
where $M_S$ is the energy scale of SUSY breaking and Q is the Q-ball's baryon number.

Q-balls are classified into two groups according to the properties of their interactions with matter: supersymmetric electrically charged solitons and supersymmetric electrically neutral solitons.
Scintillator detectors may be more suitable for the detection of charged Q-balls with large energy loss in matter. Thus, we will confine our attention here to neutral Q-balls.

 When a neutral Q-ball collides with a nucleon it absorbs its baryonic charge and induces the dissociation of the nucleon into free quarks. In this process, about 1\,GeV energy is released by the emission of typically two or three pions~\cite{Bakari:2000dq,Kusenko:2004yw}. 

 The cross section of the interaction between neutral Q-balls and the matter is roughly estimated to be the geometrical size of the Q-ball,
\begin{equation}
  \sigma = \pi {R_Q}^2.
  \label{eq:qball_sigma}
\end{equation}
 For 100\,mb cross section, the energy loss of Q-ball is estimated to be about 60\,MeV/cm. The mass of a Q-ball with this cross section is about $10^{19}$\,MeV/$c^2$. Therefore the effect of the energy loss in the Earth is expected to be negligible compared to the kinetic energy of a Q-ball. 

\section{\label{sec:detector} Super-Kamiokande detector}
 Super-Kamiokande is a cylindrical 50\,kton water Cherenkov detector located
 at a depth of 2,700\,m water equivalent. The water tank is optically separated into two regions; the inner detector~(ID) instrumented with inward facing 20\,inch diameter photomutiplier tubes~(PMT) and the outer detector~(OD) instrumented with outward facing 8\,inch PMTs.
 Super-Kamiokande II (SK-II) is the second phase of Super-Kamiokande experiment and started in December 2002 with 5,182\,PMTs in the ID and 1,885\,PMTs in the OD. 
 See~\cite{Fukuda:2002uc} for more details on the detector. The inner PMTs are instrumented with acrylic and FRP covers to avoid a chain reaction implosion.
 The OD is used to veto entering cosmic ray muons and to tag exiting charged particles. 

In this paper we report the results from a search for neutral Q-balls using 541.7\,live days of data for the SK-II period. 
 A Q-ball interacting in the detector will produce several pions which emit Cherenkov light. Thus, we can observe these pion events along the Q-ball track in the detector.

\section{\label{sec:simulation} Simulation of Q-ball events}
 A Monte Carlo simulation is developed to estimate the detection efficiency of Q-ball events. 
The direction of a Q-ball is randomly generated, and the distance between the center of the detector and a Q-ball track is assumed to be uniformly distributed within 50\,m.
The Q-ball velocity is generated assuming a Maxwellian distribution with 
a mean value of 270\,km/sec---a typical galactic velocity. Therefore, the velocity of Q-balls observed on the earth is determined by superposing the velocity of solar system 220\,km/sec on the Q-ball velocity in the halo~\cite{Smith:1988kw}.
 The pions produced in the interactions with a Q-ball and a nucleon can be determined using KNO scaling~\cite{Koba:1972mr}. The total energy of the released pions is equal to the nucleon mass. The charges of the pions are randomly generated within the constraint of the charge conservation. The kinematics of generated pions are determined by assuming the Q-ball and nucleus scatter into the Q-ball and the several pions.

\section{\label{sec:reduction} Data Reduction}
 In this work we look for a Q-ball signature consisting of at least two successive events in which 1\,GeV of energy is released in the form of pions in the ID. A Q-ball with $\sigma$=10\,mb could interact about 18\,times within 100\,$\mu$sec, since the interaction length of a Q-ball with $\beta$=10$^{-3}$ is 30\,m. The details of the event reconstruction in Super-Kamiokande can be found in~\cite{Shiozawa:1999}.

\begin{figure*}[t]
\begin{minipage}[t]{0.48\linewidth}
\centering\epsfig{file=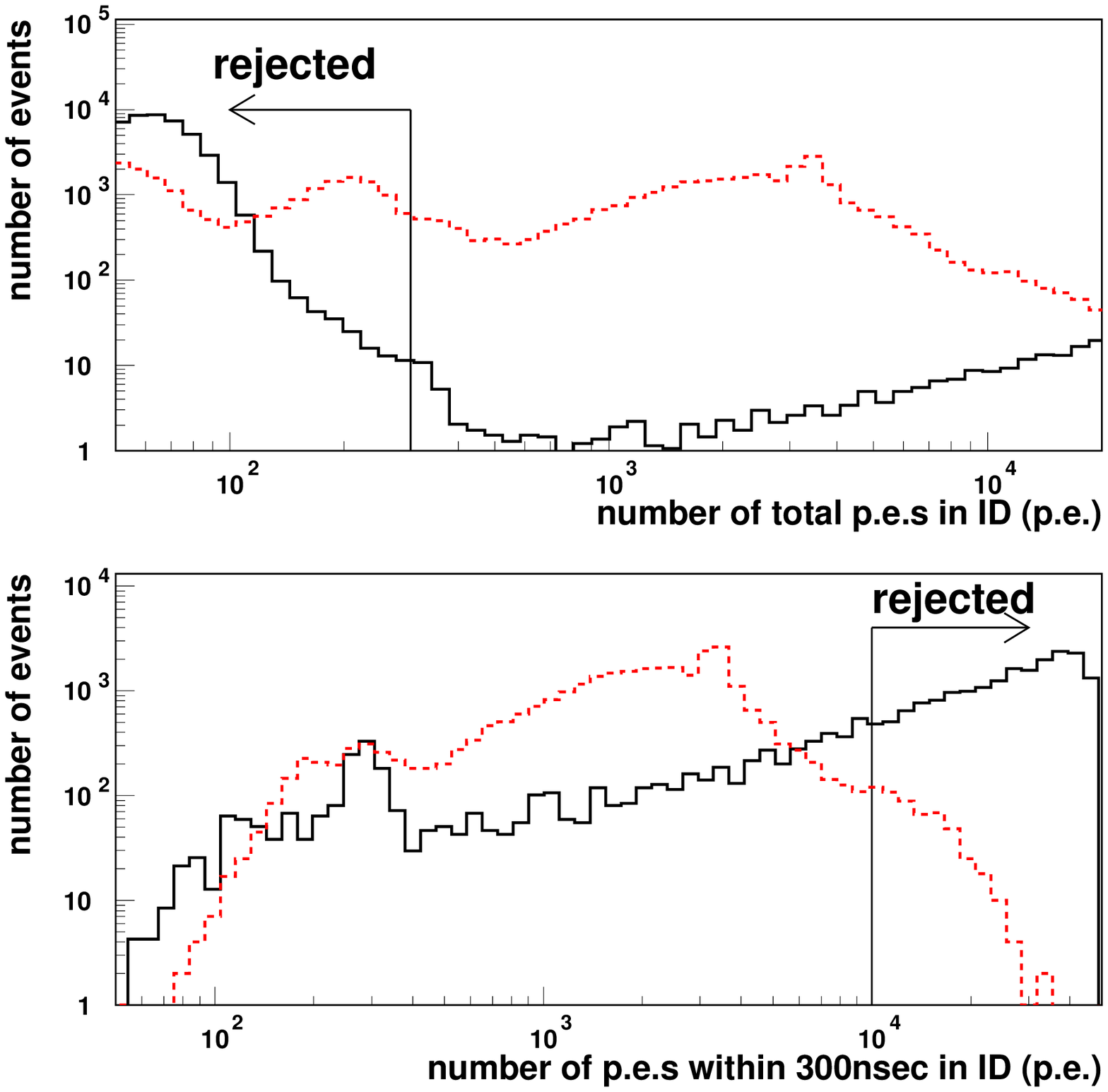,height=\linewidth,width=\linewidth}
\caption{The charge distributions for the data and Q-ball Monte Carlo events. The solid black line indicates the data and the dotted red line indicates the Monte Carlo events with $\sigma$=10\,mb.}
\label{takenaga_fig1}
\end{minipage}\hfill
\begin{minipage}[t]{0.48\linewidth}
\centering\epsfig{file=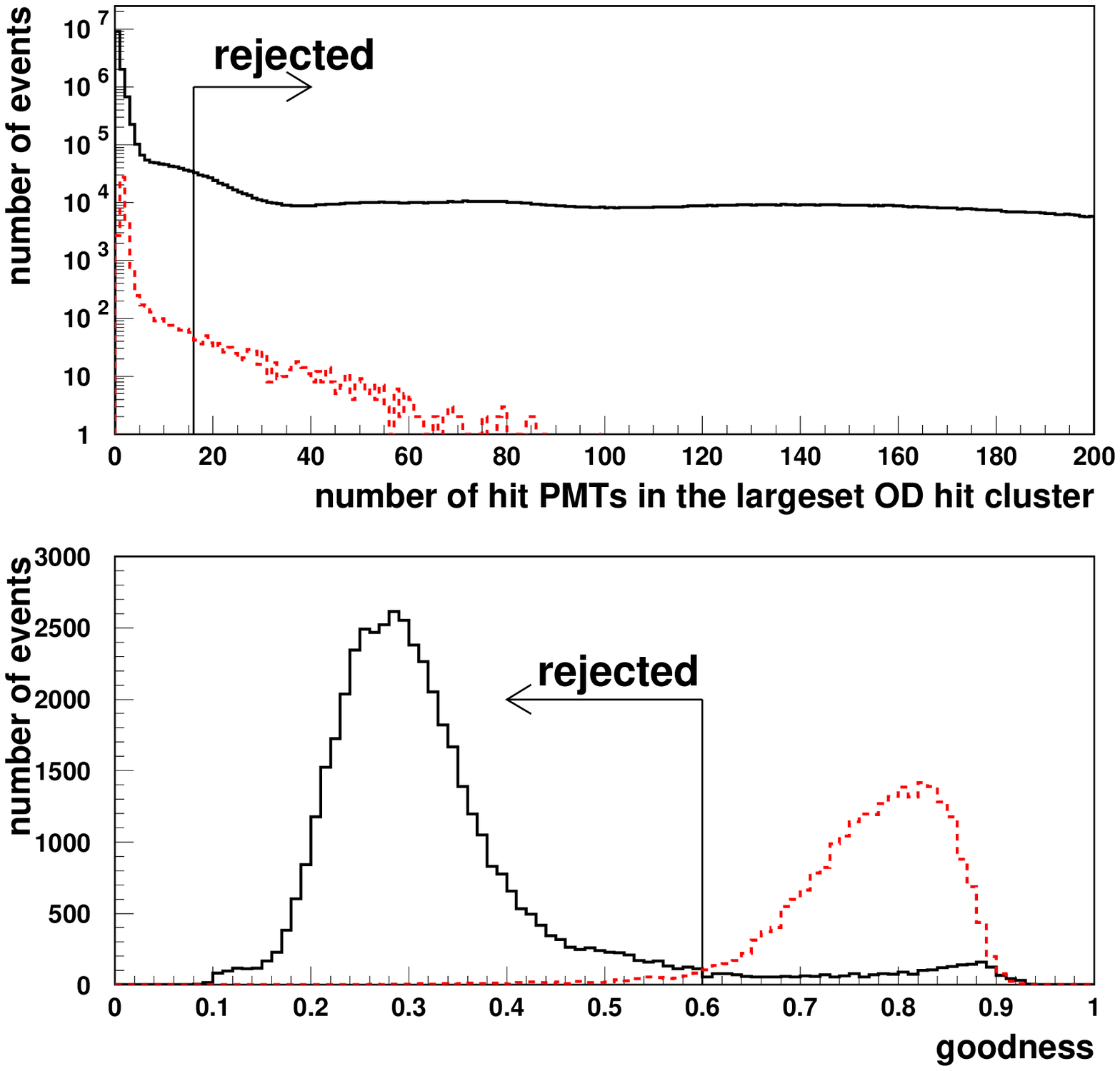,height=\linewidth,width=\linewidth}
\caption{The top figure shows the distribution of the number of hits in the largest OD cluster for the data and Q-ball Monte Carlo events. The bottom figure shows the distribution of the goodness of the vertex fit. The lines are same as in Fig.~\ref{takenaga_fig1}.}
\label{takenaga_fig2}
\end{minipage}
\end{figure*}

The first selection criteria are: 

 1-1) The number of total photoelectrons (p.e.'s) in the ID should be larger than 300~(the top of Fig.~\ref{takenaga_fig1}) and the maximum number of p.e.'s in the ID within sliding 300\,nsec time window should be less than 10,000~(the bottom of Fig.~\ref{takenaga_fig1}), which corresponds to about 60\,MeV and 2\,GeV energy for electrons, respectively. These cuts reject electron events from decays of cosmic ray muons, high energy through-going muon events and electrical noise events.

1-2) At least one delayed event should exist within a 100\,$\mu$sec window after the event selected by 1-1). This cut rejects single events.

 All events selected by 1-1) and 1-2) are considered to be an event group. The event selected by criterion 1-1) is regarded as the first event in an event group.

 After the first selection, there are still background event groups which are caused by cosmic ray muons, electrons from a decay of cosmic ray muons and electrical noise caused by large muon pulses. Background event groups still survive the above criteria with an approximate rate of 0.06\,/sec.
In the second reduction, we require at least two events which are entirely contained in the ID.

2-1) The number of hit PMTs in the largest OD hit cluster should be less than 16, in order to remove primarily cosmic ray muon events as shown in the top of Fig.~\ref{takenaga_fig2}.

2-2) The number of p.e.'s in the ID within 300\,nsec time window should be larger than 500 and the goodness of vertex fit should be larger than 0.6 (the bottom of Fig.~\ref{takenaga_fig2}) to better reject events associated with electrical noise.

2-3) In an event group two or more events should satisfy both 2-1) and 2-2). 

\begin{figure*}[t]
\begin{minipage}[t]{0.48\linewidth}
\centering\epsfig{file=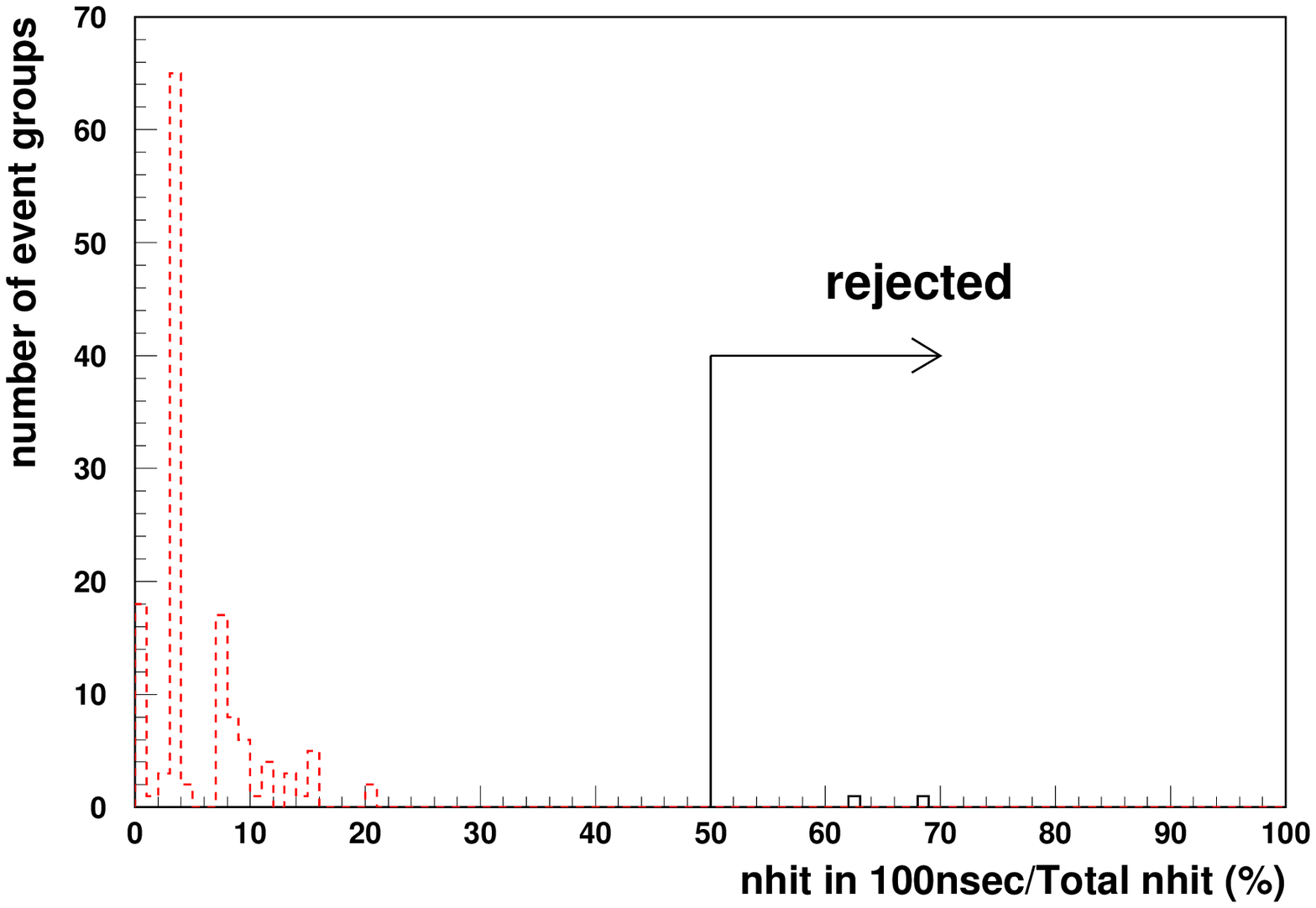,height=\linewidth,width=\linewidth}
\caption{The distribution of the ratio of the number of hit ID PMTs in a 100\,nsec time window at the end of the trigger gate to the total hit ID PMTs for the data and the Monte Carlo events.}
\label{takenaga_fig4}
\end{minipage}\hfill
\begin{minipage}[t]{0.48\linewidth}
\centering\epsfig{file=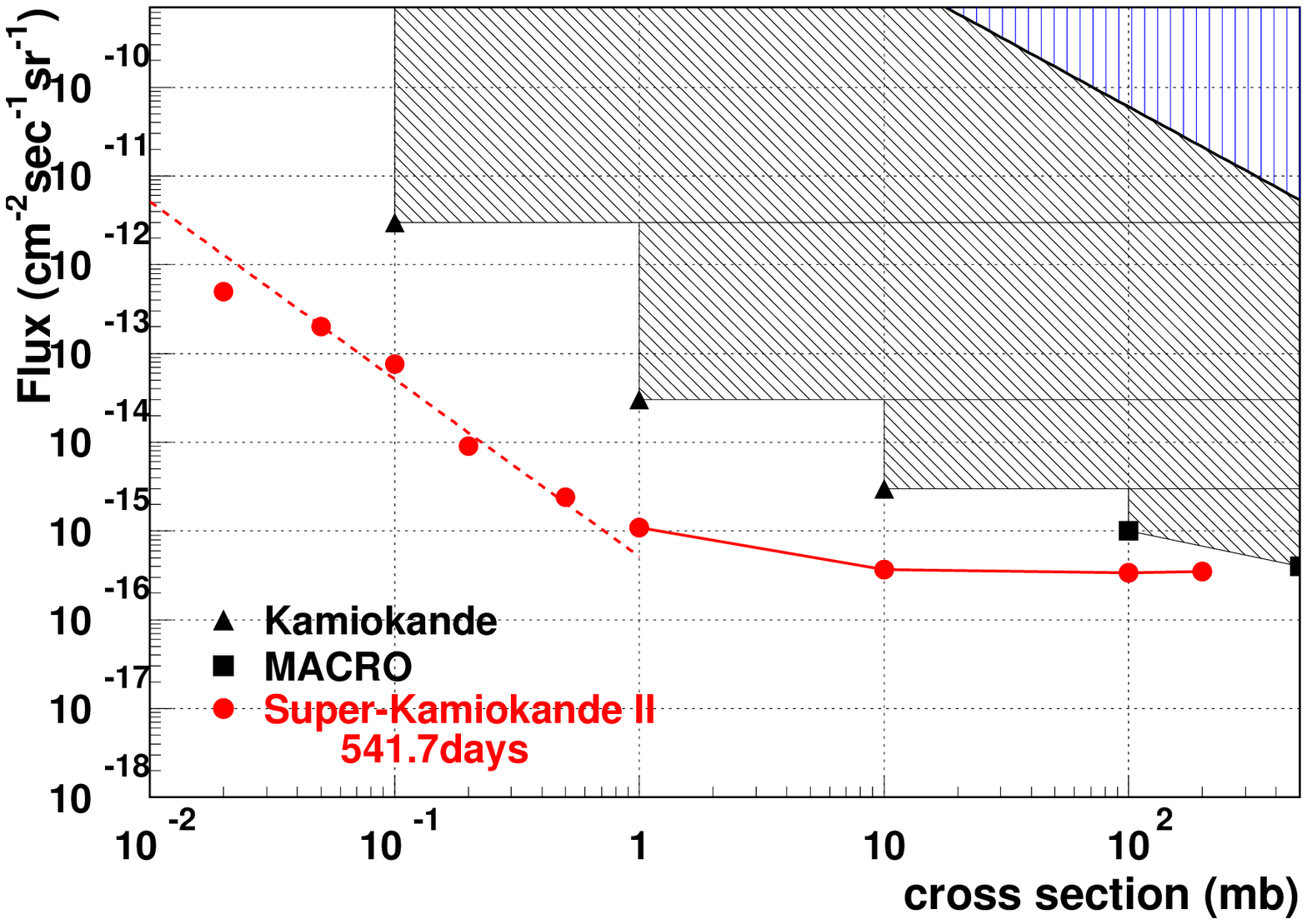,height=\linewidth,width=\linewidth}
\caption{90\,\% C.L. upper flux limits on the neutral Q-ball flux as a function of the Q-ball-nucleon interaction cross section. The circles show our results and the dashed line in the region below 1\,mb cross section is obtained assuming that the flux limits should be in inverse proportion to the square of cross section. }
\label{takenaga_fig5}
\end{minipage}
\end{figure*}

Two event groups have been found, both containing two successive events. However, they are accidental coincidence events consisting of a low energy event and a cosmic ray muon within the 1.3\,$\mu$sec trigger gate. The cosmic ray muon is separately recorded over two events since the cosmic ray muon hits at the end of the trigger gate.  These event groups are not rejected in the former reduction stages because of the absence of hit OD PMTs and the large number of total p.e.'s in the ID due to the muons.

We design the following selection criteria to reject the accidental coincidence events without losing any Q-ball Monte Carlo events.

3) If an event group contains two events, for the first event, the ratio of the number of hit ID PMTs in a 100\,nsec time window at the end of the trigger gate to the total number of hit ID PMTs should be less than 50\,$\%$ (see in Fig.~\ref{takenaga_fig4}).

 No group of events has survived after applying all of the selection criteria. The number of background event groups due to the accidental coincidence of 2 or more atmospheric neutrino events within 100\,$\mu$sec during the 541.7\,day exposure is estimated to be 8.6$\times$10$^{-5}$.

\section{\label{sec:results} Results}
 Since there is no candidate event group, the upper limits on the Q-ball flux for various mean interaction lengths can be calculated from the following formula,
\begin{equation}
 Flux < \frac{N_0}{S_{eff} \Omega T}\ (90\,\% C.L.),
  \label{eq:flux}
\end{equation}
where $N_{0}=2.3$ for no observed event group, T is the detector live time (541.7\,days), and 
$S_{eff} \Omega$, the effective aperture, which is defined as:
\begin{equation}
S_{eff} \Omega = \epsilon \pi r^2 \times 4 \pi \ \textrm{ [m$^2$ sr]},
\end{equation}
where $\epsilon$ is the detection efficiency estimated by Monte Carlo events and {\it r} is the impact parameter, which is defined to be 50\,m in this analysis.

In another theory, Q-balls could convert matter into antimatter on their surfaces~\cite{Kusenko:2004yw}. In this case, up to 2\,GeV energy can be released. However, the present selection criteria is almost equally efficient for Q-balls from this theory. We also find that varying the multiplicity of the generated pions from two to four does not change the detection efficiency.
 The velocity of the neutral Q-ball has uncertainty of as much as $\pm$10\,$\%$~\cite{Smith:1988kw}. The efficiency is not affected by this uncertainty.
Table~\ref{flux} summarizes the detection efficiencies and the obtained flux limits. 

\begin{table*}[t]
\caption{\label{flux} Summary of flux upper limits on neutral Q-balls for several Q-ball cross-sections.}
\centering\begin{tabular}{|ccc|}  
\hline
Cross section~(mb) & Detection efficiency~($\%$) & Flux upper limits~(cm$^{-2}$sec$^{-1}$sr$^{-1}$)\\
\hline
0.02 & 0.01 & 5.0$\times 10^{-13}$ \\
0.05 & 0.03 & 2.0$\times 10^{-13}$ \\
0.1 & 0.07 & 7.7$\times 10^{-14}$ \\
0.2 & 0.6 & 9.1$\times 10^{-15}$ \\
0.5 & 2.1 & 2.4$\times 10^{-15}$ \\
1   & 4.5 & 1.1$\times 10^{-15}$ \\
10  & 13.6 & 3.7$\times 10^{-16}$ \\
100 & 14.5 & 3.4$\times 10^{-16}$ \\
200 & 14.4 & 3.5$\times 10^{-16}$ \\
\hline
\end{tabular}
\end{table*}

Fig.~\ref{takenaga_fig5} shows the 90\,$\%$ C.L. upper limits on the neutral Q-ball flux as a function of the Q-ball cross section. The upper right hatched region is excluded because the Q-ball mass density exceeds the Galactic mass density~\cite{Arafune:2000yv}. 

The hatched region is previously excluded by the experiments which searched for monopole-catalyzed nucleon decays. The triangle points are the results from Kamiokande~\cite{Kajita:1986nb} and the square point is obtained by MACRO~\cite{Ambrosio:2002qu}. The present results, shown in circles, are the most stringent limits on neutral Q-ball flux for Q-ball cross section below 200\,mb.

In summary, we report on a search for groups of successive contained events at Super-Kamiokande II. We have found no evidence for the successive contained event groups in 541.7\,days of SK-II data. New upper limits at 90\,$\%$ C.L. for neutral Q-ball flux are obtained. These limits are the most stringent bounds on Q-ball flux for Q-ball cross section below 200\,mb.

\section*{Acknowledgments}
%% Keep the small font tag for the acknowledegments
{\small 
We gratefully acknowledge the cooperation of the Kamioka Mining and Smelting Company. The Super-K experiment has been built and operated from funding by the Japanese Ministry of Education, Culture, Sports, Science and Technology, the United States Department of Energy, and the U.S. National Science Foundation, and the National Natural Science Foundation of China. 
}

%%%%%%%%%%%%%%%%%%%%%%%%%%%%%%%%%%%%%% reset.txt counters %%%%%%%%%%%%%%
%%
%%%%%%% do not change below here  %%%%%%%%%%%%%%%%%%%%%%%%%%%%%
\setcounter{section}{0}
\setcounter{subsection}{0}
\setcounter{figure}{0}
\setcounter{table}{0}
\setcounter{equation}{0}
\newpage

%%%%%%%%%%%%%%%%%%%%%%%%%%%%%%%%%%%%%%%%%%%%%%%%%%% Title, authors and addresses
\begin{frontmatter}

\title{Status of Searches for Magnetic Monopoles, Q-Balls and Nuclearites with the AMANDA-II Detector}

\author[address1]{A. Pohl},
\author[address2]{H. Wissing}

\address[address1]{Dept. of Chemistry and Biomedical Sciences, Kalmar University, S-39182 Kalmar, Sweden}
\address[address2]{III Physikalisches Institut, RWTH Aachen University, D-52056 Aachen, Germany}

\vspace{-1.pc}
\begin{abstract}
Neutrino telescopes are sensitive to a variety of 
hypothetical super-heavy exotic particles.
We review the status of current searches for magnetic monopoles, 
nuclearites, and Q-balls using data taken with the AMANDA-II detector. 
\end{abstract}

\end{frontmatter}

%%%%%%%%%%%%%%%%%%%%%%%%%%%%%%%%%%%%%%%%%%%%%%%%%%%%%% MAIN TEXT
\section{\label{sec:intro} Introduction}
The existence of magnetic monopoles is mandatory 
in a large class of Grand Unified Theories (GUTs). 
Monopoles are supposed to be copiously 
produced as topological defects in 
symmetry breaking phase transitions in the early universe, but their density 
will have been strongly diluted by inflation. 
GUTs predict monopole masses to range from 10$^8$ - 10$^{17}$\,GeV, depending 
on the symmetry group and unification scale of the underlying theory \cite{tHooftPolyakov74}. 
The monopole magnetic charge will be an integer multiple of  
the {\it Dirac charge}  $g_D=e/(2\alpha)$, where $e$ is the electric elementary charge and 
$\alpha=1/137$ is the fine structure constant. 
Since magnetic monopoles are topologically stable, 
they  should still be present in
today's universe and can be searched for in cosmic radiation.   
Once created, monopoles can efficiently be accelerated in large scale magnetic fields or by gravitation. 
Monopoles with masses below $\sim 10^{14}$\,GeV can have been accelerated to 
relativistic velocities \cite{Wick03} and could be detected in neutrino telescopes
through their direct  Cherenkov emissions.
In some GUTs, monopoles can act as ``catalysts'' in interactions that 
violate baryon number conservation \cite{rubakov81}. In a neutrino telescope, 
the signature 
of these ``catalyzing'' monopoles would be 
a series of closely spaced light bursts from nucleon decay products 
produced along the monopole trajectory.   
Other massive particles have also been hypothesized to exist in cosmic radiation:  
Nuclearites (nuggets of strange dark matter) and Q-balls (supersymmetric coherent states of squarks, 
sleptons and Higgs fields, predicted by supersymmetric generalizations of the standard model).
Electrically neutral Q-balls will have the same experimental signature in a neutrino telescope as
catalyzing monopoles,
although the underlying process \cite{kusenko98b} is different 
from nucleon decay catalysis. 
Nuclearites and charged Q-balls might also be detectable, as,  traveling through matter, they 
would generate a thermal shock wave which emits blackbody radiation 
at visible wavelengths \cite{derujula84,arafune00}.

\section{\label{sec:detector} The AMANDA-II Neutrino Telescope}
AMANDA-II is a neutrino telescope  located  
at a depth between 1500 and 2000\,m under the ice at the geographic South Pole. 
A cylindrical volume of roughly 200\,m diameter of the Polar ice was 
instrumented with a total of 677 optical modules (OMs), 
consisting of a photomultiplier tube (PMT) and supporting electronics enclosed in  
a transparent pressure sphere. The OMs were deployed on 19 vertical strings, which  
are arranged in three concentric circles (see figure \ref{topview}). 
\begin{floatingfigure}[hr]{0.33\textwidth}
\centering
\includegraphics[width=0.3\textwidth]{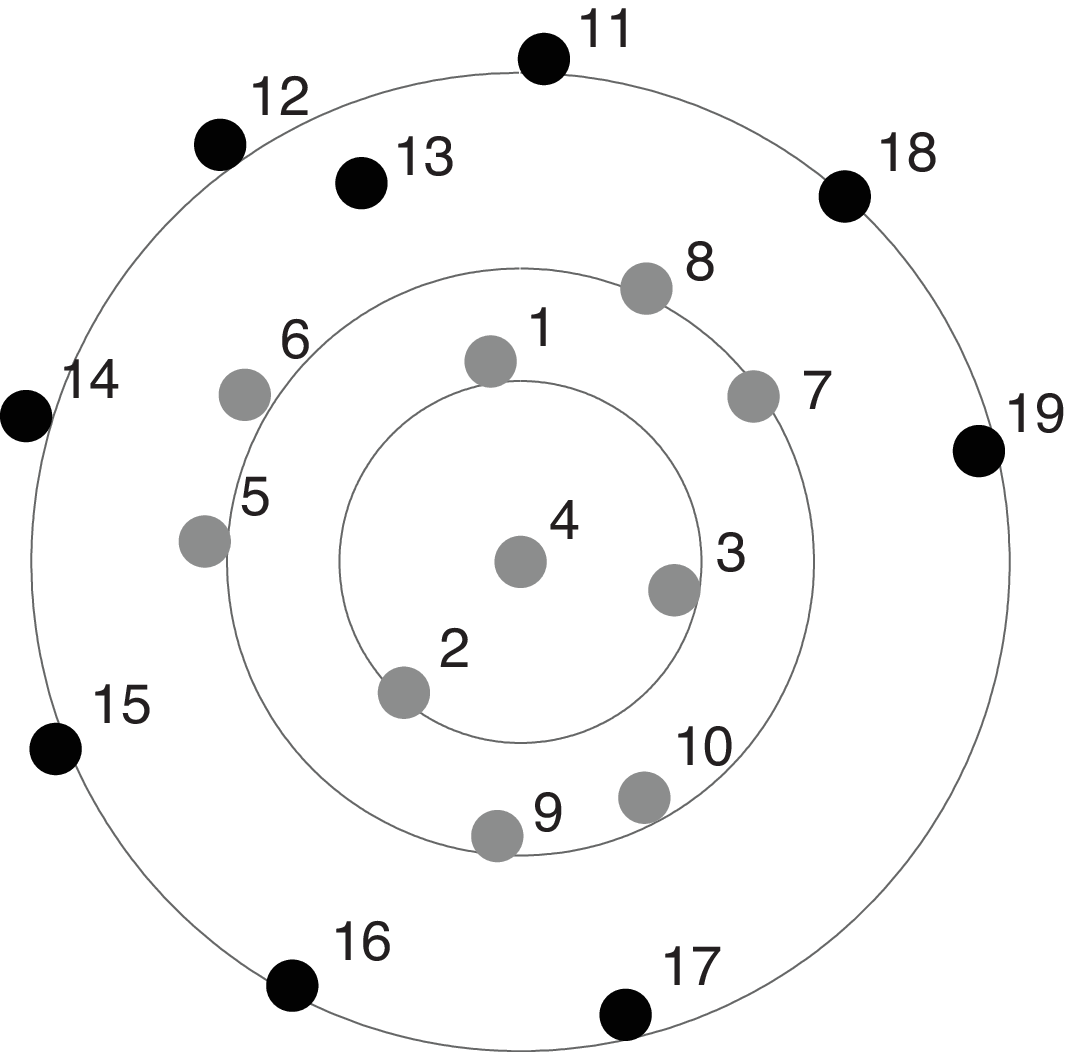}
\caption{Arrangement of the 19 strings of the AMANDA-II detector in the horizontal plane.}
\label{topview}
\end{floatingfigure}
The detector was built in three stages, each  incorporating step-wise improvements to the 
PMT signal transmission techniques. The inner ten strings of the detector   
are read out electrically via coaxial or twisted-pair cables, while 
the outermost strings  
use optical fiber transmission.
The detector is operated with a  variety 
of different triggers, two of which are relevant to the analyses presented here:
First, the 24-fold multiplicity trigger requiring a minimum of 24 OMs hit within a fixed 
coincidence window of $2.5\,\mu$s, and second, a so-called correlation trigger,
requiring $n$ OMs to be hit in any group of $m$ adjacent OMs on the same string. 
For each triggered event, PMT pulse data is recorded 
over a time window of $\sim33\,\mu$s. 
For each OM,
up to eight subsequent pulses (or \emph{hits}) can be recorded. 
The vast majority of triggers are due to down-going atmospheric muons, yielding an 
average event rate of roughly 90\,Hz.

\section{\label{sec:rel} Search for Relativistic Magnetic Monopoles}

The number of Cherenkov photons $N_{\gamma}$  emitted  per path length $dx$ 
and photon wavelength $d\lambda$ radiated from a relativistic 
magnetic monopole carrying one Dirac charge passing 
through matter with index of refraction $n$ is \cite{Tompkins65}
\begin{equation} 
\frac{dN_{\gamma}}{dxd\lambda} =
\frac{2\pi\alpha}{\lambda^2}\left(\frac{g_D n}{e}\right)^2 
\left( 1-\frac{1}{\beta^2n^2} \right), 
\end{equation} 
where $\beta$ is the velocity of the monopole as a fraction of the velocity of light. 
The Cherenkov light intensity in ice radiated from the monopole is  
 enhanced by a factor $(g_D \cdot n/e)^2=8300$ compared to the intensity 
radiated from a particle with electric charge $e$ and the same velocity.
Thus, in a neutrino telescope, a relativistic magnetic monopole will
stand out as an extremely bright event relative to the background of atmospheric muons. 

We are presently searching for the signal of relativistic magnetic monopoles in 
data taken with AMANDA-II during the year 2000. 
This data set corresponds to 194 days of effective livetime.
Only events that fulfill the 24-fold multiplicity trigger are included. 
The analysis is ``blind'', meaning that the entire data selection chain is optimized 
exclusively on Monte Carlo simulations,
and only a small subset of the experimental data is used to verify the detector simulation.
 
We have simulated the detector response to relativistic magnetic 
monopoles carrying one unit Dirac charge passing the 
detector's sensitive volume with
four different velocities $\beta = 0.76, 0.8, 0.9$, and $\beta = 1$. The background of down-going 
atmospheric muon bundles was simulated with the air-shower simulation 
package \texttt{CORSIKA} \cite{corsika:00}.
In several data filtering steps we reject the bulk of low-energy atmospheric muons.
We select events in which a large number of photons are detected.   
Observables like the number of hit OMs, the total number of 
PMT pulses recorded, and the pulse amplitudes are 
measures for the light deposition in the detector. 
We use these observables 
either as one dimensional cut parameters or as input variables to
a discriminant analysis \cite{Fisher:1936et}. 
By selecting events with large light deposition alone, 
we can identify relativistic magnetic monopoles
among the background of down-going atmospheric muons, even if they 
were arriving from above the horizon.  % ACTIVE
In this document, however, we present the search of up-going monopoles only.
The mass range over which this search is sensitive is 
limited, since only monopoles with masses larger than $10^{11}$\,GeV 
can penetrate the entire Earth and still be relativistic upon reaching 
the detector \cite{derkaoui}.        

\begin{figure}[tb]
\begin{center}
\includegraphics[width=0.495\textwidth]{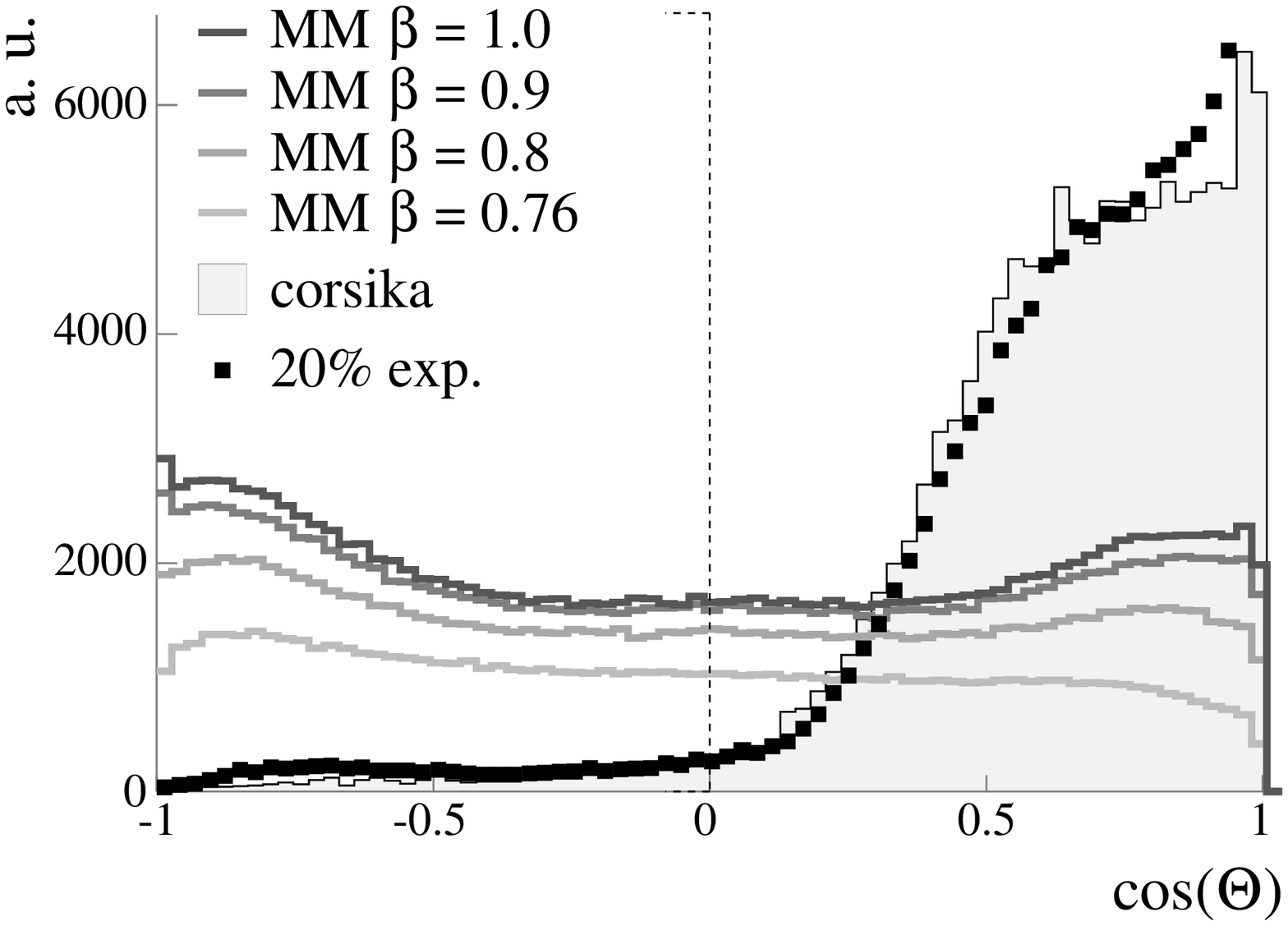}
\includegraphics[width=0.495\textwidth]{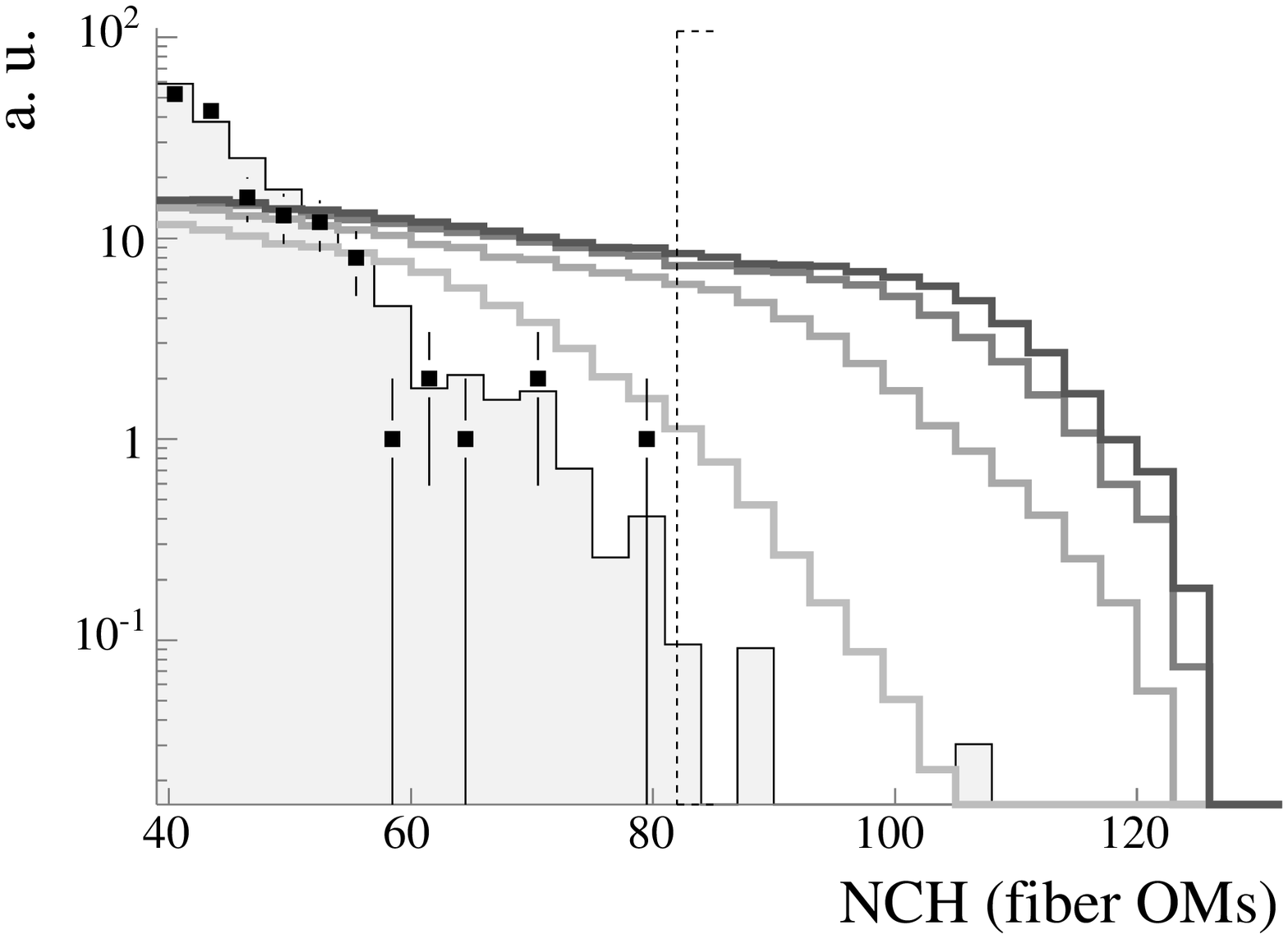}
\caption{Left: Zenith angle distribution from a likelihood track reconstruction.
The region $\cos(\Theta) < 0$ corresponds to events in which the particle direction
is reconstructed as up-going.
Right: Number of hit fiber OMs for the ``up-going'' event sample
($\cos(\Theta) < 0$). This observable serves as final cut parameter.
In both figures, the full histograms represent the simulated atmospheric muon background.
Black markers represent 20\% of experimental data taken during the year 2000, and 
the four open histograms represent the simulated signal of relativistic magnetic monopoles
with four different velocities.
}
\label{zenithNCH}
\end{center}
\end{figure}

We reconstruct the particle track using a likelihood method 
\cite{reco_paper} and consider only events for which the zenith angle of the track is 
bigger than $90^{\circ}$, corresponding to a particle entering from below the horizon.
Some of the down-going atmospheric muon bundles are mis-reconstructed as
up-going particles, 
posing a background to the search for up-going monopoles  (see figure \ref{zenithNCH}).
We reject this remaining background with a final cut on
the number of hit OMs with fiber readout.
We use only fiber OMs for this last cut. Located at the outermost strings, 
these OMs define the detector's surface area and are consequently 
most crucial to the acceptance of extremely bright signals.
% establish most of the s. acceptance
% are the detrming factor to .. acceptance
% contribute most 
Also, their response to large amounts of light is 
simulated most accurately by the detector simulation (which is essential in
a ``blind'' analysis). 
We optimize the final cut such that we achieve the optimum ``sensitivity''. In this context, 
``sensitivity'' means the 90\% C.L. flux upper limit that we expect to obtain 
for a given background prediction, if no true signal were present (see \cite{hillrawlins},
and references therein). The final cut parameter is shown in figure \ref{zenithNCH}.    
The signal acceptance, and hence the sensitivity of this analysis, 
depends on the monopole velocity.
We expect the analysis of 194 days of experimental data to yield sensitivities 
between $\sim 5\times10^{-17}$ ($\beta = 1$) and $\sim 1\times10^{-15}$ cm$^{-2}$s$^{-1}$sr$^{-1}$
($\beta=0.76$). This sensitivity range does not yet include the effect of systematic uncertainties. 
As of this writing, our sensitivity is comparable to the best upper 
limit on the flux of relativistic magnetic 
monopoles published by the BAIKAL Collaboration \cite{baikal}.

%%%%%%%%%%%%%%%%%%%%%%%%%%%%%%%%%%%%%%%%%%%%%%%%%%%%%%%%%%%%%%%%%%%%%%%%%%%%%%%

\section{\label{sec:subrel} Search for Subrelativistic Particles}

In the search for sub-relativistic particles, two hypothesized mechanisms for light production are exploited.
Magnetic monopoles have been suggested to catalyze nucleon decay with a cross section 
which is large enough to make them detectable but uncertain with several orders of magnitude 
\cite{rubakov81}. Here only protons are considered, 
which have several possible catalyzed decay channels. The channel 
$p\rightarrow e^+\pi^0$ creates an electromagnetic shower with energy close to the
proton mass. Other channels lose some of their shower energy to neutrinos.
                                                                                
Neutral Q-balls dissociate baryons in a fundamentally different way, but the experimental 
signature is quite similar \cite{kusenko98b}. The cross section is their geometric size. 
By limitations given in \cite{bakari01}, it ranges from $\sim 10^{-26}\mathrm{cm^2}$ and many orders of magnitude upwards. 
                                                                               
The luminosity of thermal shock waves from nuclearites and charged Q-balls, as given by \cite{derujula84,arafune00}, is also determined by their geometric size. Their luminosity would exceed that of neutral Q-balls by several orders of magnitude.

All particles are simulated with isotropic directions and with speed $\beta=10^{-2}$. 
In the simulations, the luminosity is expressed as the mean distance $\lambda$ between two 
electromagnetic showers, whose energy is the proton mass. The only value for $\lambda$ used so far is 
2\,cm, and only hydrogen decays in the decay channel 
$e^+\pi^0$ with a branching ratio of 0.9 are considered \cite{bais83}. The corresponding cross section of 
monopole catalyzed decay is $9\cdot 10^{-24}\mathrm{cm}^2$, which is at the upper edge of 
what appears to be allowed by the theoretical uncertainties. For neutral Q-balls, oxygen nucleon 
decay is considered, making $\lambda$ correspond to a cross section of $9\cdot 10^{-25}\mathrm{cm}^2$. For nuclearites and charged Q-balls, the chosen $\lambda$ corresponds to a much lower luminosity than that given by \cite{derujula84,arafune00}.

\begin{figure}[t]
\begin{center}
\includegraphics[width=0.49\textwidth]{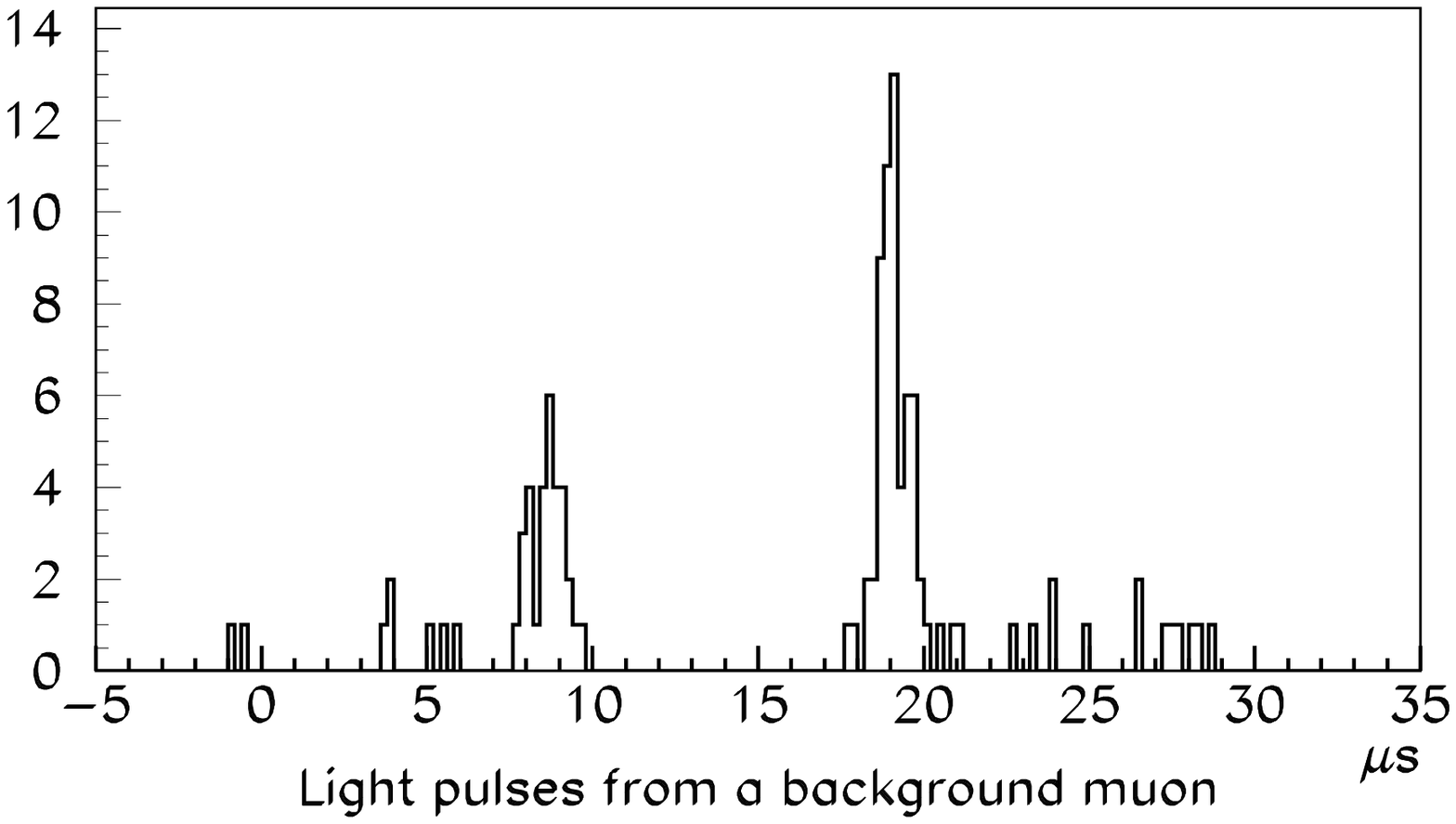}
\includegraphics[width=0.49\textwidth]{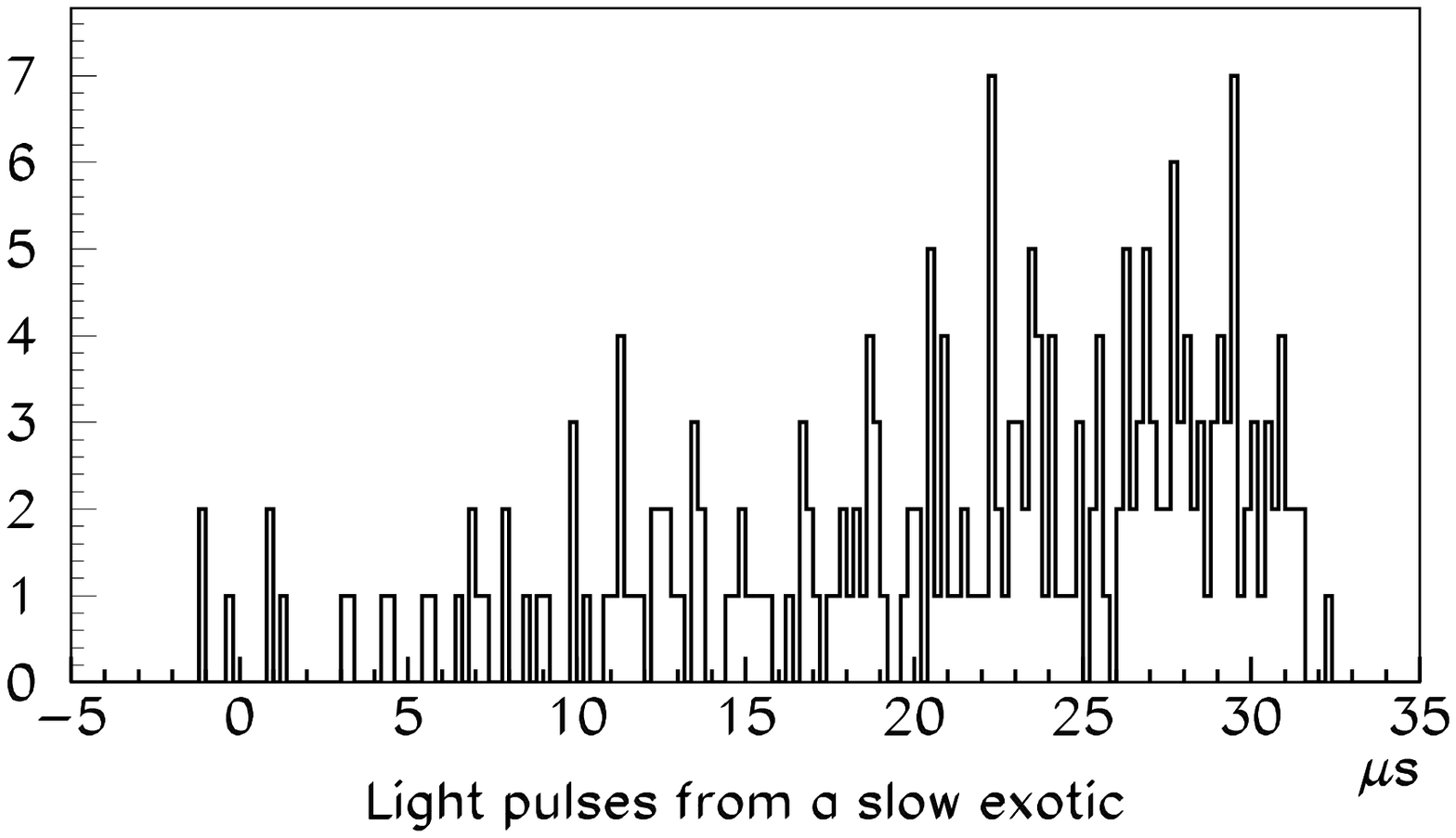}
\caption{Left: a background event with a triggering and a non-triggering muon. 
Right: a simulated signal 
event from a particle traveling at speed $\beta=0.01$.}
\label{signature}
\end{center}
\end{figure}
                                                                                
%The correlation trigger used in AMANDA during part of 2001
%requires a multiplicity of 6 within any 9 adjacent OMs for the 4 inner-most strings and a multiplicity of 7 within 
%any 11 adjacent OMs for the outer strings. 
Simulations show that the correlation trigger is substantially more sensitive to this type of signal 
than the multiplicity trigger. 
In each triggered event, hits are collected during a time window of $33\mathrm{\mu s}$. Background 
relativistic muons emit light during $\sim3\mathrm{\mu s}$, whereas slow particles emit during a large 
fraction of the time window. A comparison is shown in fig. \ref{signature}. The left picture shows a background 
event with a triggering muon at time $19\mathrm{\mu s}$, and an accidental early non-triggering muon at 
$9\mathrm{\mu s}$. The right picture shows a simulated signal event. The signal separation from background 
is based on hits at times when no light from triggering muons is expected, 
the \textit{early and late hits} outside the interval $16-24 \mathrm{\mu s}$.

A period of 113 days in 2001 when a constant correlation trigger definition was used, is considered here. 
The background properties and a preliminary expected sensitivity is determined using 20\% of the data. 
A first filter reduces the data by 99\%, requiring a total of at 
least 14 early and late hits. The second filtering uses additional trigger cleaning and three 
additional cuts based on early and late hits.
\begin{floatingfigure}[t]{0.5\textwidth}
\centering
\includegraphics[width=0.49\textwidth]{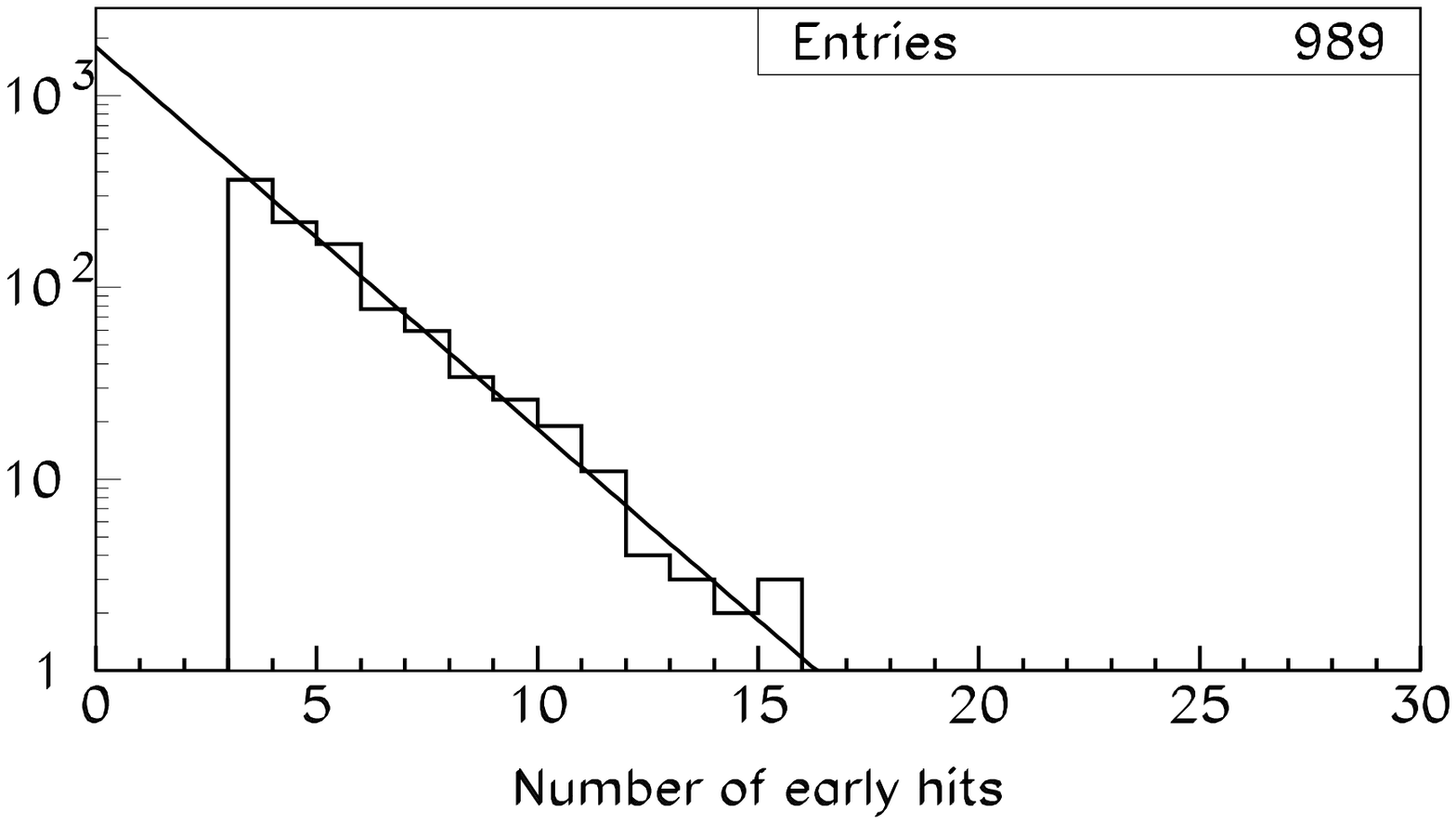}
\caption{An exponential fit parametrizes the background distribution in the number of early hits.}
\label{exp_tail}
\end{floatingfigure}  
The events passing the filter have an exponential distribution in the number of early hits. 
It is shown in fig. \ref{exp_tail}, along with a fit. 
Since about 80\% of any signal events would be expected above 20 (as seen from simulations), but none were found, fig. \ref{exp_tail} 
must be almost signal free. Thus, the fit parameters are suitable for background estimation. Using the same 
sensitivity optimization scheme as in the relativistic case, the optimal final cut for the 80\% sample requires $>24$ early hits. 
The corresponding expected sensitivity is $\sim 5\cdot 10^{-17}\mathrm{cm^{-2}s^{-1}sr^{-1}}$. 
This updates the oral presentation, where the correlation trigger had not been included in the simulations and data filtering.

\section{Discussion and Outlook}
The AMANDA neutrino telescope is an excellent instrument to search for 
several postulated super heavy exotic particles. In this document, we present first 
studies of the sensitivity of AMANDA to magnetic monopoles, Q-balls and nuclearites. 
The given sensitivities are still  preliminary. Specifically, systematic uncertainties 
are not yet included. So far, we have used relatively small sub-sets of the available AMANDA data 
in order to outline our analysis strategies. The sensitivity of the two 
analyses will improve substantially with more data.

%With the analysis of one year of 
%experimental data, we expect a reach a sensitivity 
%to reletivistic magnetic monopoles comparable to
%best existing limits.
%Our result is valid for monopoles with masses between roughly $10^{11}$ to 
%roughly $10^{14}$\,GeV. The lower bound corresponding to the mass above which 
%monopoles will be capable to cross the Earth and the upper bound 
%corresponding to the mass below which we can assume .monopoles to be relativistic. 
%An analysis dedicated to down-going magnetic monopoles is under way, extending the 
%mass range of monopoles to which AMANDA-II will be sensitive. 

%%%%%%%%%%%%%%%%%%%%%%%%%%%%%%%%%%%%%% reset.txt counters %%%%%%%%%%%%%%
%%
%%%%%%% do not change below here  %%%%%%%%%%%%%%%%%%%%%%%%%%%%%
%%
\setcounter{section}{0}
\setcounter{subsection}{0}
\setcounter{figure}{0}
\setcounter{table}{0}
\setcounter{equation}{0}
\newpage

%%%%%%%%%%%%%%%%%%%%%%%%%%%%%%%%%%%%%%%%%%%%%%%%%%% Title, authors and addresses
\begin{frontmatter}


\title{Search for Relativistic Magnetic Monopoles  with  the Baikal Neutrino Telescope}

% use optional labels between square brackets to link authors explicitly to addresses:
% If more than one author, keep a comma between the author tags

%\author[address1]{ K. Antipin^a  , V. Aynudinov^a , V. Balkanov^a ,}
%\author[address2]{I. Belolaptikov^d , D.Borshov^d }

%\address[address1]{a Institute for Nuclear Research Russian Academy of Science, Moscow, Russia }
%\address[address2]{ d Joint Institute for Nuclear Research, Dubna, Russia}

\author[address1]{K.Antipin},
\author[address1]{V.Aynudinov},
\author[address1]{V.Balkanov},
\author[address4]{I.Belolaptikov},
\author[address4]{D.Borschov},
\author[address2]{N.Budnev},
\author[address1]{I.Danilchenko},
\author[address1]{G.Domogatsky},
\author[address1]{A.Doroshenko},
\author[address2]{A.Dyachok}, 
\author[address1]{Zh.-A.Dzilkibaev},
\author[address5]{S.Fialkovsky},
\author[address1]{O.Gaponenko},
\author[address4]{K.Golubkov },
\author[address2]{O.Gress},
\author[address2]{T.Gress},
\author[address2]{O.Grishin},
\author[address1]{A.Klabukov},
\author[address8]{A.Klimov}, 
\author[address2]{A.Kochanov},
\author[address4]{K.Konishev}, 
\author[address1]{A.Koshechkin},
\author[address3]{L.Kuzmichev},
\author[address5]{V.Kulepov},
\author[address7]{E.Middell},
\author[address1]{S.Mikheyev},
\author[address7]{T.Mikolajski},
\author[address5]{M.Milenin},
\author[address2]{R.Mirgazov},
\author[address3]{E.Osipova},
\author[address2]{L.Pan'kov},
\author[address2]{G.Pan'kov},
\author[address1]{A.Panfilov},
\author[address1]{D.Petukhov}
\author[address4]{E.Pliskovsky}
\author[address1]{P.Pokhil},
\author[address1]{V.Poleschuk},
\author[address3]{E.Popova},
\author[address3]{V.Prosin},
\author[address6]{M.Rosanov},
\author[address2]{V.Rubtsov},
\author[address4]{B.Shaibonov},
\author[address3]{A.Shirikov},
\author[address1]{A.Sheifler},
\author[address7]{Ch.Spiering},
\author[address2]{B.Tarashansky},
\author[address4]{R.Vasiljev},
\author[address7]{R.Wischnewski}, 
\author[address3]{I.Yashin},
\author[address1]{V.Zhukov}

\address[address1]{  Institute for Nuclear Research, Russian Academy
  of Science,Moscow, Russia }
\address[address2]{  Irkutsk State University,Irkutsk,Russia}   
\address[address3]{  Skobeltsyn Institute of Nuclea Physics MSU, Moscow, Russia}
\address[address4]{  Joint Institute for Nuclear Research, Dubna,
  Russia}
\address[address5]{  Nizhni Novgorod State Technical University,
  Nizhni Novgorod, Russia}
\address[address6]{  St. Peterburg State Marine University,
 St. Peterbur,Russia} 
\address[address7]{  DESY, Zeuthen, Germany }
\address[address8]{  Kurchatov  Institute, Moscow, Russia }

\begin{abstract}
  We present  the results of a search for relativistic magnetic
  monopoles ($\beta>0.8$) with the Baikal underwater Cherenkov telescope NT200,
%RW   based on data taken in 1998-2003. The upper limit on the flux of
  based on data taken in 1998-2002. The upper limit on the flux of
  monopoles with $\beta>0.8$  from the lower hemisphere is the most
  stringent at the present time.
  
\end{abstract}

% \begin{keyword}

%\PACS 
% \end{keyword}

\end{frontmatter}

%%%%%%%%%%%%%%%%%%%%%%%%%%%%%%%%%%%%%%%%%%%%%%%%%%%%%% MAIN TEXT
\section{\label{sec:intro} Introduction}

  Fast magnetic monopoles with Dirac charge g=68.5e are attractive
 objects to search for with deep underwater neutrino telescopes. 
The intensity of Cherenkov radiation from a relativistic monopole in
 water is 8300 times higher than that of a muon. The light flux from a
 monopole
%RW is equvalent the light flux from a muon with an energy about 10 PeV.
is equivalent to the light flux from a muon with an energy of about 10 PeV.
An optical module (OM) of the Baikal experiment could detect such
 a bright object from  distances up to 100 m.

In 1974 Polyakov \cite{Polyakov}  and Hooft \cite{Hooft} 
discovered a monopole solution of the 
  SO3 Georgi-Glashow model. In a wide class of elementary particle 
 models the mass of monopoles may range between $10^7 -10^{14}$ GeV.
%RW  Propagating trough the Universe,  monopoles could be accelarated by
 Propagating through the Universe,  monopoles could be accelerated by
%RW  magnetic fields up to energy  $10^{12}-10^{15}$
 magnetic fields up to energies of  $10^{12}-10^{15}$
  GeV \cite{R.Beck}, \cite{D.Ryu}.
Hence  monopoles with 
 mass below $10^{14}$ GeV are expected to be relativistic. The monopole
energy loss is about 10 $\frac{GeV\cdot cm^2}{g}$
 for a Lorentz factor $<10^3$, but
%RW significantly rises with energy, by a factor 1000 for Lorentz factor
significantly rises with energy, by a factor of 1000 for a Lorentz factor
%RW $10^7 $ \cite{Wick}. Therefore light monopoles with mass less than $10^7$ GeV 
$10^7 $ \cite{Wick}. Therefore, light monopoles with mass less than $10^7$ GeV 
could not cross
the Earth. In order to suppress the huge background we search for
%RW  monopoles only from the lower hemisphere and supposing a mass range $10^7-10^{14}$
 monopoles only from the lower hemisphere and assume a mass range $10^7-10^{14}$
 GeV.
  In this paper we present the result of a search for relativistic
  monopoles with the Baikal neutrino telescope NT200, based on data
%RW  taken in the years 1998 - 2003.     
  taken in the years 1998 - 2002 (April 1998 - February 2003).

\section{\label{sec:instructions} Detector and search strategy}
%RW     The present stage of the telescope NT200  \cite{Balkanov} takes data
    The neutrino telescope NT200  \cite{Balkanov} takes data
   since April 1998 and consists of 192 OMs in 8 strings. 
%RW   A pair of  OMs  defines a channel. 
   The OMs are arranged in pairs, and each pair defines a channel. 
   A trigger is formed by the requirement
%RW    of $N_{hit}>4$ fired channels within 500 ns. For such events
    of $N_{hit}>4$ fired channels within 500 ns. For such events, the 
    amplitude and time of all fired channels within a time window of 2000ns
    are digitized and sent to the shore. 
        The space-time pattern of the 
    light recorded from a monopole depends on the water
    characteristics. The maximal absorption length of Baikal deep water
    is $L_{abs}(480nm)=22\div 24$\,m and varies slightly during years.
      Scattering in Baikal water is characterized by a strongly
    anisotropic function $f(\theta)$ with a mean cosine
    $cos(\theta)=0.85\div0.9$ and  a scattering length $L_s =30\div70$\,m. 
    Both scattering length and  $f(\theta)$ vary considerably
    during the year. Our Monte-Carlo simulations indicate that the uncertainty 
    of the scattering length from 15 to 30 m leads to an uncertainty of
    the effective area of nearly 20\%  for fast monopole registration.

%^ ************************************************

\begin{figure*}[t]
\begin{minipage}[t]{0.50\linewidth}
\centering\epsfig{file=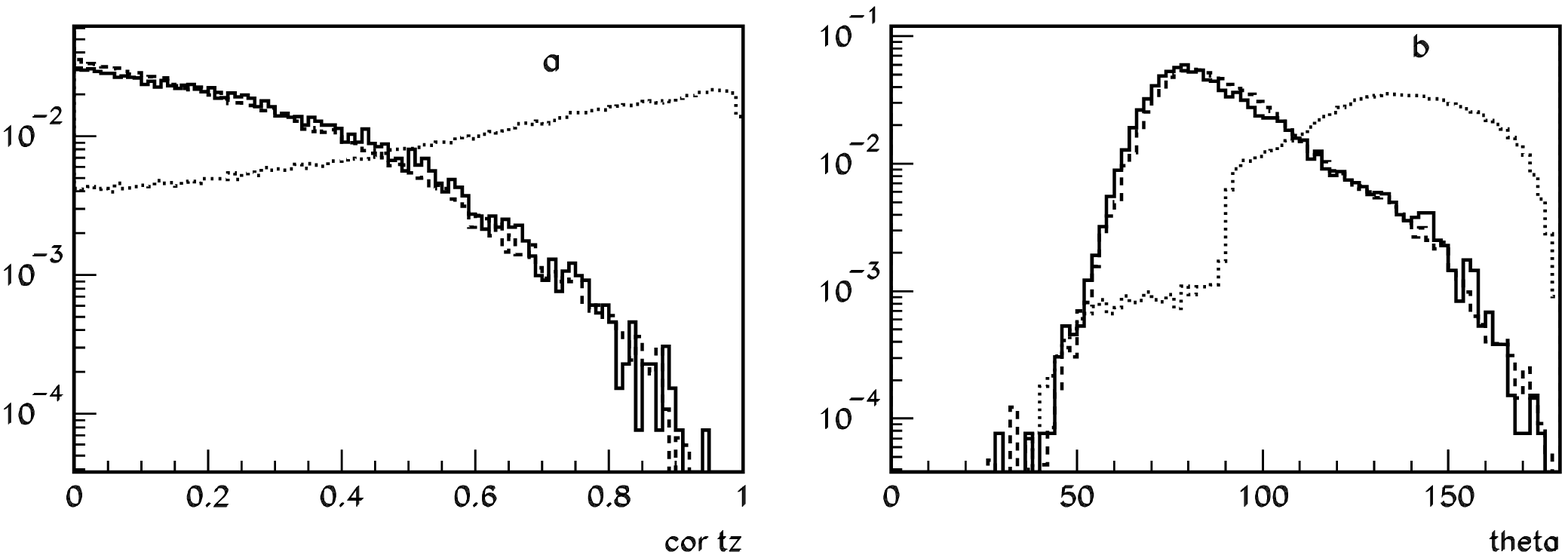,height=\linewidth,width=\linewidth}
\end{minipage}\hfill
\begin{minipage}[t]{0.50\linewidth}
\centering\epsfig{file=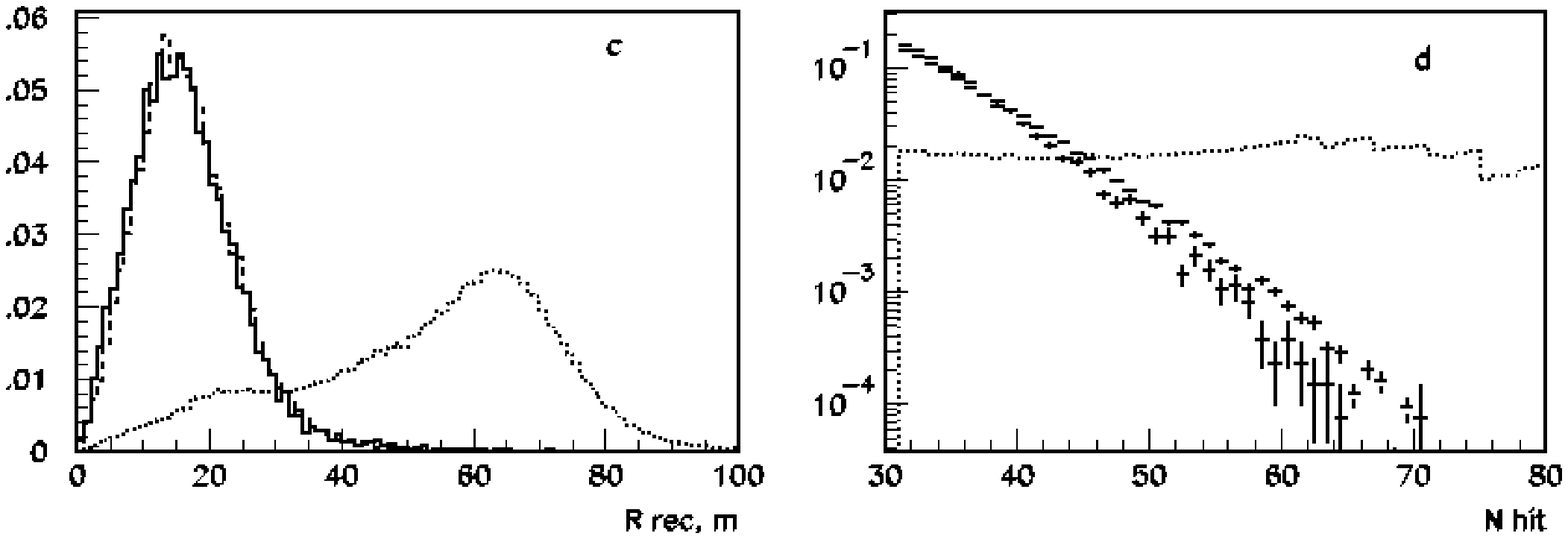,height=\linewidth,width=\linewidth}
\label{osipova_fig4}
\end{minipage}
%RW \caption{Distributions of experimental data(solid line), MC atmospheric
%RW  muons(dashed line) and MC from monopole(dotted line) on
%RW  a- $cor_{TZ}$, b- $\theta$, c- $R_{rec}$, d- $N_{hit}$}
\caption{Distributions of experimental data (solid line), MC atmospheric
  muons (dashed line) and MC from monopole (dotted line) for
  $cor_{TZ}$ (a), $\theta$ (b), $R_{rec}$ (c), $N_{hit}$ (d).}

\end{figure*}

%RW    The most natural way to search for such bright object as fast
    The most natural way to search for such bright objects as fast
    monopoles is to select events 
with a large number of fired channels, $N_{hit}>30$.
    To reduce the background from atmospheric muons we  
    select only up-going particles. For such particles the times of the hit channels
    increase with rising $z$-coordinates from bottom to top of the 
%RW    detector.  A cut on the value of the $z$--time correlation has been
    detector.  A cut on the value of the $z$--time correlation, $cor_{TZ}$, has been
    applied to suppress downward-moving particles: 

\begin{equation}
cor_{TZ}= \frac{ \sum_{i=1}^{N_{hit}}(t_i- \overline T )(z_i- \overline
  z)} {N_{hit} \sigma_t \sigma_z}>0
\end{equation}
  where $t_i$ and $z_i$ are time and $z$-coordinate of a fired channel, 
$\overline T$ and $\overline z$ are mean values for times and
  $z$-coordinates of the event, $ \sigma_t$ and $ \sigma_z$ the root mean
  square errors for time and z-coordinates.
  The first cut (CUT1) was $N_{hit}>30$ and  $cor_{TZ}>0$.
%RW This requirements  leave $1.5\cdot10^{-4}$ of events that satisfy
These requirements  leave $1.5\cdot10^{-4}$ of those events that satisfy 
the trigger $N_{hit}\ge6$ on $\ge$ 3 strings. CUT1 reduces the effective 
area for monopoles with $\beta=1$ by about a factor of 2.

The main background for the fast monopole signature is provided by  muon bundles, 
high energy muons and cascades from muons. 
The simulation chain of 
atmospheric muons starts with cosmic ray air showers generated with
 {\texttt{CORSIKA }}\cite{corsika:92}.  
We use the {\texttt{QGSJET1 }\cite{qgsjet} model and a
 primary composition according to \cite{composition}. 
The   {\texttt{MUM }}\cite{mum} program is
 used for muon propagation through the water. 
%RW To obtain the detector response to muons passing the detector volume,
To obtain the detector response to muons passing through the detector volume,
%RW we simulate all muon energy loss processes, taking into
%RW account the characteristics of OM and measurement system.
 we simulate all muon energy loss processes, taking into
 account the characteristics of the OMs and the measurement system.
In fig. 1 we compare experimental data and simulated atmospheric muon
events with respect to all parameters which have been used for
background rejection. The expected distributions for fast ($\beta=1$)
monopoles are presented also. 
One can see that the simulation describes the 
%RW experimental data quite well, even on the level of those rare events.
experimental data quite well, even on the level of rare events.
 
%RW Within 1003 days live time 21240 events satisfy CUT1 .
Within 1003 days live time (between April, 1998 and February, 2003), 21240 events satisfy CUT1 .
%RW All events have been reconstructed. For further background suppression
All events have been reconstructed as single muon tracks. For further background suppression
we use additional cuts, which essentially reject background events
from atmospheric muons and at
the same time only slightly reduce the area for registration of fast
monopoles.

The final cuts were : 
 
CUT1: $N_{hit}>30$ and $cor_{TZ}>0$

CUT2: $N_{hit}>35$ and $cor_{TZ}>0$,  events should be reconstructable

CUT3: CUT2 and $\chi^2$ determined from reconstruction less than 3

%RW CUT4: CUT3 and $\theta>100^o$,  ($\theta$-reconstructed zenith angle)       
CUT4: CUT3 and $\theta>100^o$; $\theta$: reconstructed zenith angle       

%RW CUT5: CUT4 and $R_{rec}>10\div25$\,m , $R_{rec}$ distance
%RW from the center of NT200
CUT5: CUT4 and $R_{rec}>10\div25$\,m;  $R_{rec}$: distance
from the center of NT200

%RW CUT6:  CUT5 and $cor_{TZ}>0.25\div0.65$    
CUT6:  CUT5 and $cor_{TZ}>0.25\div0.65$ .   

%RW NT200 took data in various configurations, due to the different numbers
%RW of operating channels. The different cuts on $cor_{TZ}$ and  $R_{rec}$
NT200 took data in various configurations, due to different numbers 
of operating channels for different time periods. The range of cuts on $cor_{TZ}$ and  $R_{rec}$
correspond to different configurations of NT200.
%RW No events from the experimental sample pass CUTs 1-6.
%RW  !!!!!
No events from the experimental sample pass CUT 6.
Passing rates versus cut level for experimental and simulated events 
are shown on fig.2. The figure also shows the dependence of the 
effective area for monopoles as a function of cut level.

\begin{figure*}[t]
\begin{minipage}[t]{0.45\linewidth}
\centering\epsfig{file=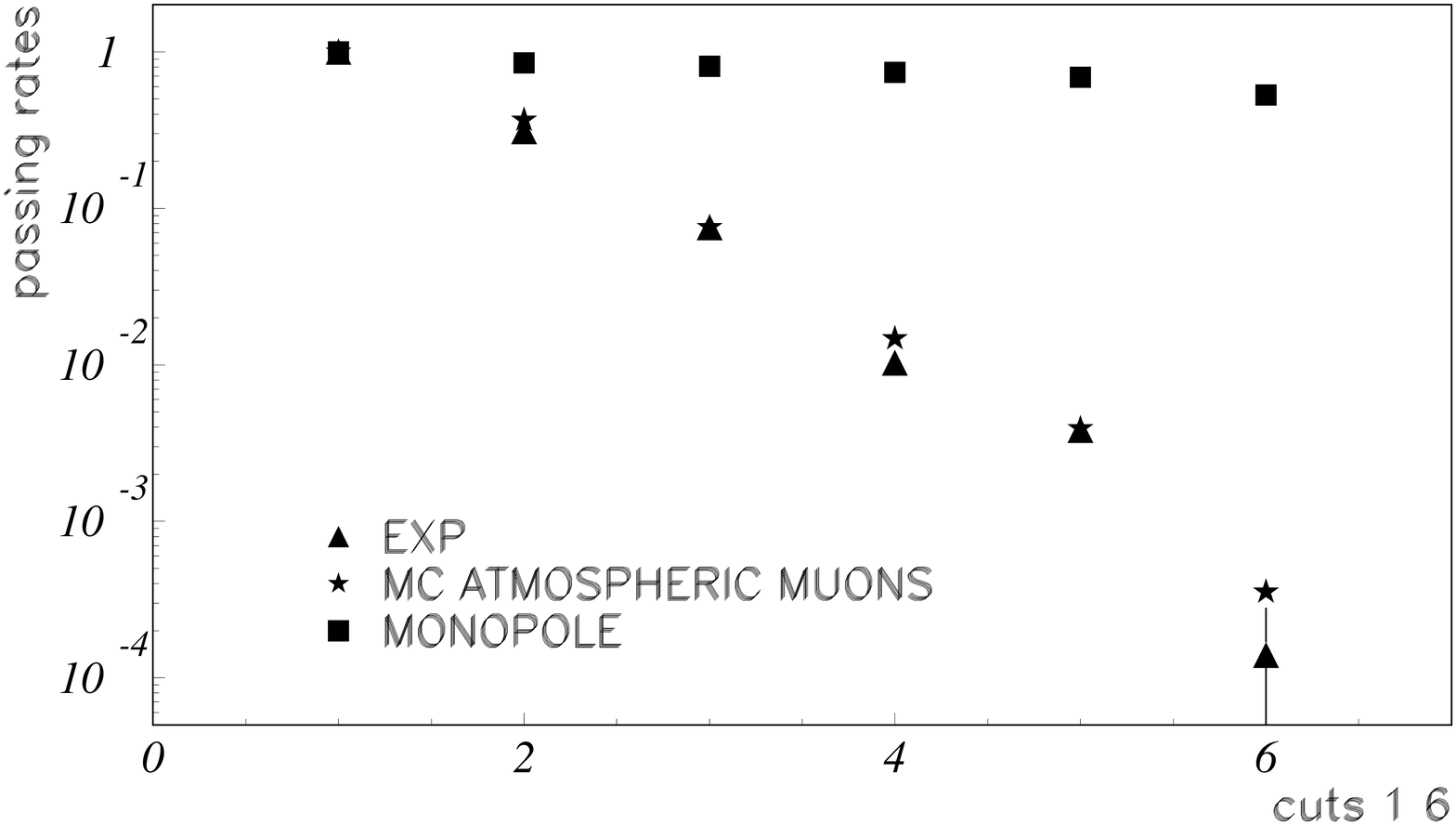,height=\linewidth,width=\linewidth}
\caption{Passing rates versus cut level for experimental
  data (triangles), MC for atmospheric muons (asterix), MC for
  monopoles (squares).}
\label{osipova_fig1}
\end{minipage}\hfill
\begin{minipage}[t]{0.45\linewidth}
\centering\epsfig{file=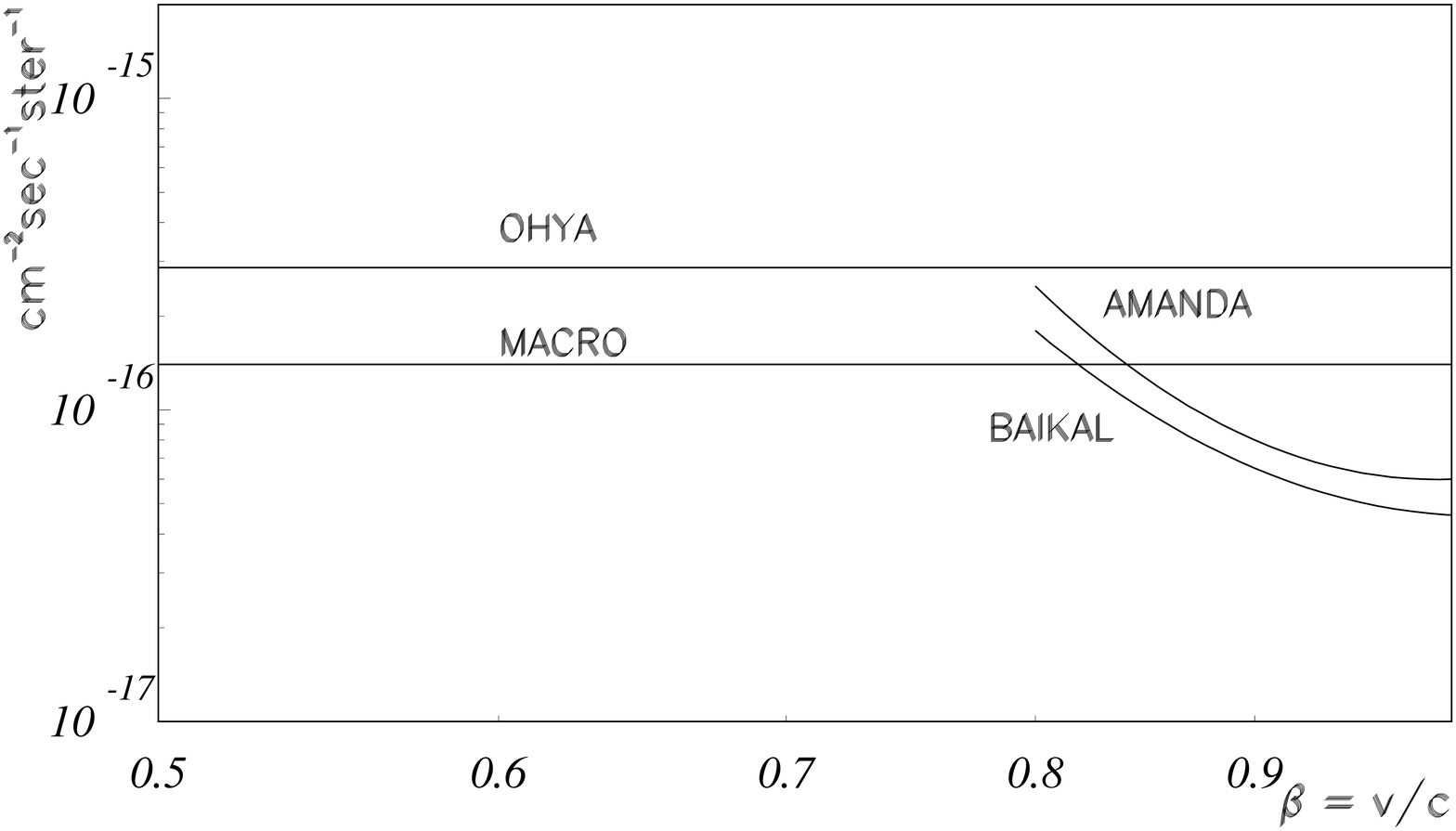,height=\linewidth,width=\linewidth}
%RW \caption{Upper limit (90\% C.L.) on the monopole flux}
\caption{Upper limit (90\% C.L.) on the monopole flux obtained in this work (BAIKAL), 
compared to other experiments.}
\label{osipova_fig2}
\end{minipage}
\end{figure*}

\section{\label{sec:plots} Limit on the flux of relativistic monopoles }
   The acceptance, $A_{eff}$, for monopoles with $\beta=1, 0.9, 0.8$ has 
   been calculated for each configuration of NT200 separately, depending
   on the number of operating channels and concrete values for 
   rejection cuts. 
   From the non-observation of candidate events in NT200 and the
%RW   earlier stages NT36, NT96, an upper limit on the flux of
   earlier stages NT36 and NT96 \cite {Belolaptikov}, an upper limit on the flux of
   relativistic $(\beta>0.8)$ monopoles at the 90\% confidence level is 
   obtained. 
The cumulative acceptances $A_{eff}\cdot T$ as well as the
   90\%C.L. upper limits are presented in Table 1.
These results are an update of an earlier analysis \cite{ayn}, 
mainly differing by a more realistic accounting for light scattering in 
Baikal water.

   In fig.3 the 90\% C.L. upper limit obtained with the Baikal neutrino
   telescope for an isotropic flux of fast monopoles is compared to
   the limits from the experiments Ohya \cite{ohya}  , MACRO
   \cite{macro} and 
    AMANDA \cite{AMANDA}. The search for fast magnetic monopole with the
   Baikal underwater detector NT200 results in the  strongest experimental
   bound on the flux of monopoles with  $\beta>0.8$.

\section{ Outlook}
 In order to increase the sensitivity for the searches of bright objects, 
NT200 was significantly upgraded. In April 2005, NT200+ was
put into  operation. It  consists of NT200 and 3 external strings
%RW spaced at 100 m distance from the center of NT200 \cite{Aynutdinov}.
spaced at 100 m distance from the center of NT200 \cite{Aynutdinov}. 
The main advantage of NT200+ is the possibility to identify high energy   
cascades which are the main source  of background for the monopole
search. It  permits to reject background by softer cuts on
$N_{hit}$ compared to the  NT200 analysis. We expect an increase of
the  effective area for fast
magnetic monopoles in NT200+ of about  1.5 times compared to NT200.

The construction  NT200+ can be considered as a first step in 
the construction of a neutrino telescope with an effective volume of about 1 km$^3$, 
%RW (gigavolume detector GVD).  The design of the new telescope GVD
(Gigaton Volume Detector, GVD).  The design of the new telescope GVD
has started recently \cite{V.Ayn}. 
According to preliminary  considerations, the telescope GVD will  consist
of 90-100 strings. 12-16 OM's will be placed on  each string and
the instrumented length of strings will be 300-350\,m. Following our pilot 
simulations we hope to reach a  sensitivity to fast monopoles on
the level of $\sim 5\cdot10^{-18}$ cm$^{-2}$ s$^{-1}$ sr$^{-1}$ after one year
%RW 
of data taking.

\begin{table}[t]   %[H] add [H] placement to break table across pages
\begin {center}
\caption{\label{crflux} $A_{eff}*T$ and 90\% C.L. upper limits on
%RW  the flux of fast monopoles} 
  the flux of fast monopoles.} 
 \end{center}

\begin{center}
\begin{tabular}{||c|c|c|c||} \hline \hline

   &   $\beta =1$  & $\beta=0.9$  & $\beta=0.8$       \\

                             \hline 

$A_{eff}\cdot T$ NT200    &  4.84$ \cdot$ 10$^{16}$  & 3.48$\cdot$ 10$^{16}$ &1.23$\cdot$10$^{16}$ \\
cm$^2$$\cdot$ sec$\cdot$ sr  &     &   & \\       
                              \hline    

 $A_{eff}\cdot T$ NT36+NT96       & 0.37 $\cdot$ 10$^{16}$ &   0.25
 $ \cdot$  10$^{16}$ & 0.094 $ \cdot$  10$^{16}$\\
cm$^2$$\cdot$ sec$\cdot$ sr  &     &     & \\  
                         \hline
  90 \% C.L. upper flux limit     & 0.46 $\cdot$ 10$^{-16} $ & 0.65
 $ \cdot$ 10$^{-16}$&1.83 $\cdot$ 10$^{-16}$\\ 
 cm$^{-2}$$\cdot$ sec$^{-1}$$ \cdot$ sr$^{-1}$      &    &     &           \\
                                   \hline

\end{tabular}
\end{center}
\end{table}

\section{Conclusion}
The neutrino telescope NT200 in Lake Baikal is taking data since April
1998. From the non-observation of candidate events in NT200 and the
early stages NT36 and
NT96, an upper limit on the flux of fast monopoles is
obtained. With 
$0.46\cdot 10^{-16}$ cm$^{-2}\cdot$ sec$^{-1}\cdot$
sr$^{-1}$ for $\beta=1$ (90\% C.L.) it is the most stringent
limit at present. 
In April 2005, the  upgraded array NT200+ was put into operation. It will have 
a sensitivity to fast monopoles of about 1.5 times that of
NT200. Our future plans are connected to the Gigaton Volume Detector. 
Its sensitivity to fast monopoles will be about  
$\sim 5\cdot10^{-18}$ cm$^{-2}$ sr$^{-1}$ sec$^{-1}$ for one year 
%RW 
of data taking. 

%$\backslash$section$\ast$\{Acknowledgements\}.\\
\section*{Acknowledgments}
%\section{\label{sec:plots} Acknowledgments}
{\small
This work was supported by the Russian Ministry of Education and
Science, the German Ministry of Education and Research and the Russian
Fund of Basic Research (grants 05-0217476 and 04-02-17289), by the
Grant of President of Russia NSh-1828.2003.2, and by NATO-Grant NIG-9811707(2005). }

%%%%%%%%%%%%%%%%%%%%%%%%%%%%%%%%%%%%%% reset.txt counters %%%%%%%%%%%%%%
%%
%%%%%%% do not change below here  %%%%%%%%%%%%%%%%%%%%%%%%%%%%%
%%
\setcounter{section}{0}
\setcounter{subsection}{0}
\setcounter{figure}{0}
\setcounter{table}{0}
\setcounter{equation}{0}
\newpage

\begin{frontmatter}

\title{Dark Matter searches with the ANTARES neutrino telescope}

\author{E. Castorina}, on behalf of the ANTARES collaboration
\address[address1]{Dipartimento di Fisica dell'Universit\`a e Sezione
  INFN, Pisa,~Italy}
\begin{center}{\small (ersilio.castorina@pi.infn.it)}\end{center}

\begin{abstract}
  The ANTARES collaboration is building an undersea neutrino telescope at 2400 m depth in the
Mediterranean Sea. The experiment aims to detect high-energy cosmic neutrinos using a 3D array of 900
photomultipliers (PMTs) arranged in 12 strings. The advantages of neutrinos as astrophysical and cosmic messengers are that they open a new window to observe known astrophysical objects as well as to look for new Physics, such as dark matter. In many supersymmetric  models, the favourite dark matter candidate is the lightest neutralino whose annihilation in the core of massive celestial objects can lead to the emission of neutrinos in the subsequent decay chains. The expected performance of ANTARES is discussed.
\end{abstract}

%  \begin{keyword}
% WIMPS \sep neutralino \sep neutrino telescope
%\PACS  
% \end{keyword}

\end{frontmatter}

%%%%%%%%%%%%%%%%%%%%%%%%%%%%%%%%%%%%%%%%%%%%%%%%%%%%%% MAIN TEXT
\section{\label{sec:intro} Introduction}

 Rapid advances and improvements in observational cosmology are leading to the beginning of a \textit{precision era} in the sense that many of the key cosmological parameters are being determined with greater and greater accuracy. A series of  observations have provided sound foundations for our current picture of the universe as mainly consisting of non-baryonic, non-luminous and non-absorbing matter, commonly referred to as Dark Matter, immersed in some form of energy featuring a negative pressure, named Dark Energy. Weakly Interacting Massive Particles or WIMPs can explain the growth of structures and are the natural candidate for dark matter particles.
The original WIMP candidate, and the only one known to exist, is the neutrino. However, only a tiny fraction of about 0.3\% of the total cosmic density can be attributed to neutrinos. Moreover the hot dark matter scenario, though not completely ruled out, seems less and less favourable since the simulations of the structure formation appear to be inconsistent with observations.
Unlike the neutrino, all the other candidates for cold WIMPs are hypothetical and most of them are associated with a new symmetry of nature known as \textbf{supersymmetry} (SUSY).
A hypothetical particle $\chi$ which is a good candidate for Dark Matter if it satisfies a number of constraints coming from current observational results:
\begin{itemize}
\item The cosmological abundance of $\chi$'s must be about $\Omega_{\chi}\simeq\Omega_{m}-\Omega_{b}\simeq 0.23$;
\item $\chi$ must not be a baryon;
\item  $\chi$ must be electrically and colour neutral, otherwise electromagnetically and strongly interacting $\chi$ particles would become bound by normal matter forming heavy isotopes: very sound upper limits exists for abundances of heavy nuclear isotopes which strongly support, over a wide mass range (1 GeV$\lesssim m_{\chi}\lesssim$1 TeV), the hypothesis of at most a weakly interacting particle. 
\end{itemize}
\section{\label{sec:ANTARES} The ANTARES detector}
The ANTARES collaboration is building a neutrino telescope in the Mediterranean Sea, at a depth of 2.4 km, located roughly 40 km off the southern coast of France at a latitude of 42${}^{\circ}$50~\cite{ant1,ant2}.
Neutrinos are detected when they interact with a nucleon $N$ via a charged current ($\nu_{l} N\rightarrow l X$) weak interaction. The detector operates by detecting the intensity and arrival time of the Cherenkov radiation emitted by charged particles produced in the neutrino interaction.
The detector will consist of 12 autonomous mooring lines called ``strings''. Each of them is equipped with 75 10-inch Hamamatsu PMTs, arranged in triplets (storeys), oriented at 45 degrees with respect to the downward vertical.
A series of LED and laser beacons, distributed across the storeys and at the bottom of the line, provide a tool for timing calibrations (Fig.~\ref{time}) while a system of acoustic transponders and receivers allows to perform precise position measurements (to within a precision of less than 20 cm). The first two lines of the detector have already been deployed and are currently being operated, while the complete 12-line detector will be operational by the end of 2007. As an example, two reconstructed muon tracks are shown in Fig.~\ref{track}.
\begin{figure*}[t]
\begin{minipage}[t]{0.48\linewidth}
\centering\epsfig{file=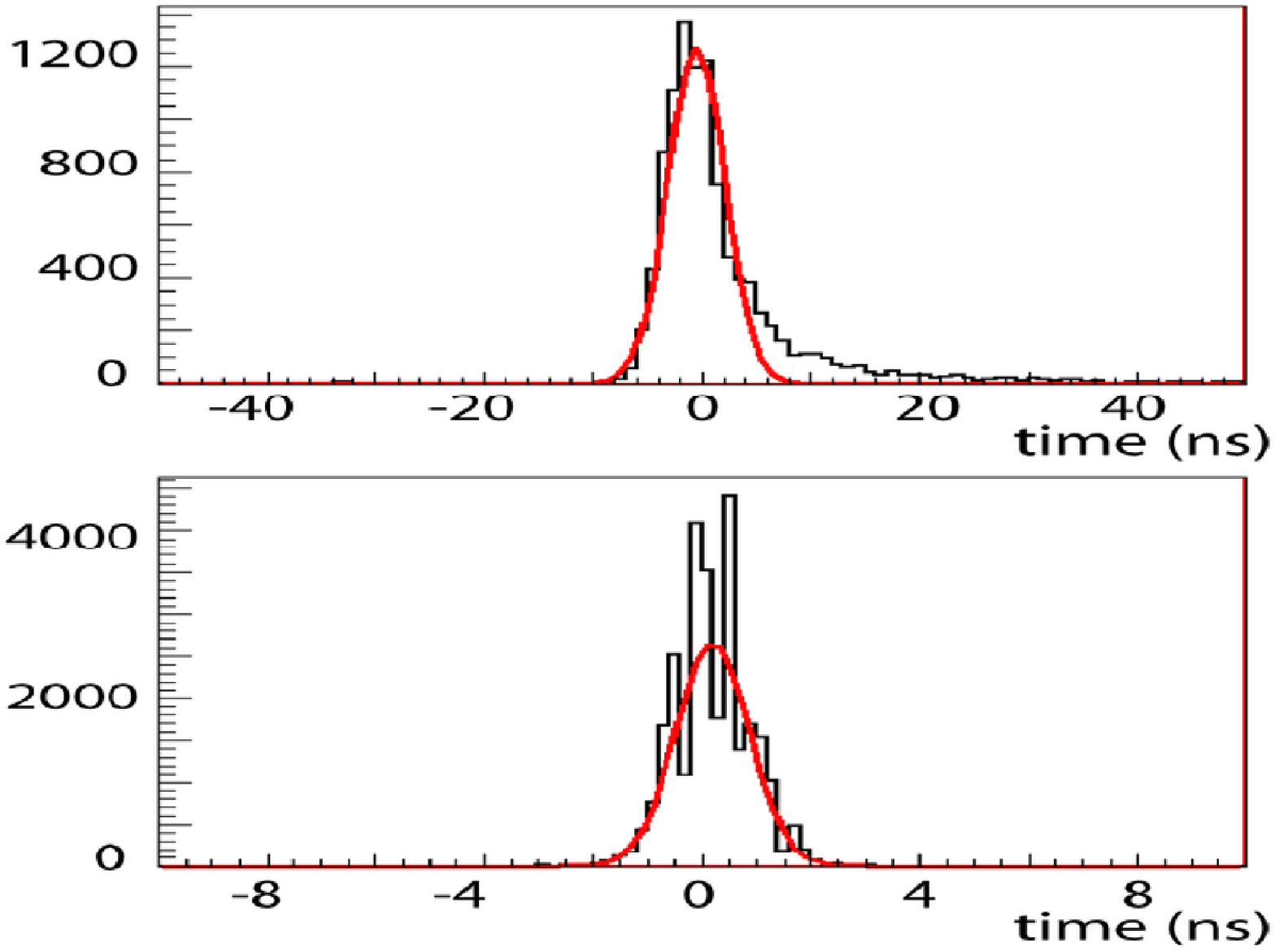,height=\linewidth,width=\linewidth}
\caption{Time differences between hits in a pair of OMs in storeys at different distances from the light source. The top distribution refers to OMs at 150 m from the emitter: the effect of light scattering is visible. The bottom plot corresponds to a 70 m distant storey: the distribution is Gaussian and has a width of 0.7 ns.}
\label{time}
\end{minipage}\hfill
\begin{minipage}[t]{0.48\linewidth}
\centering\epsfig{file=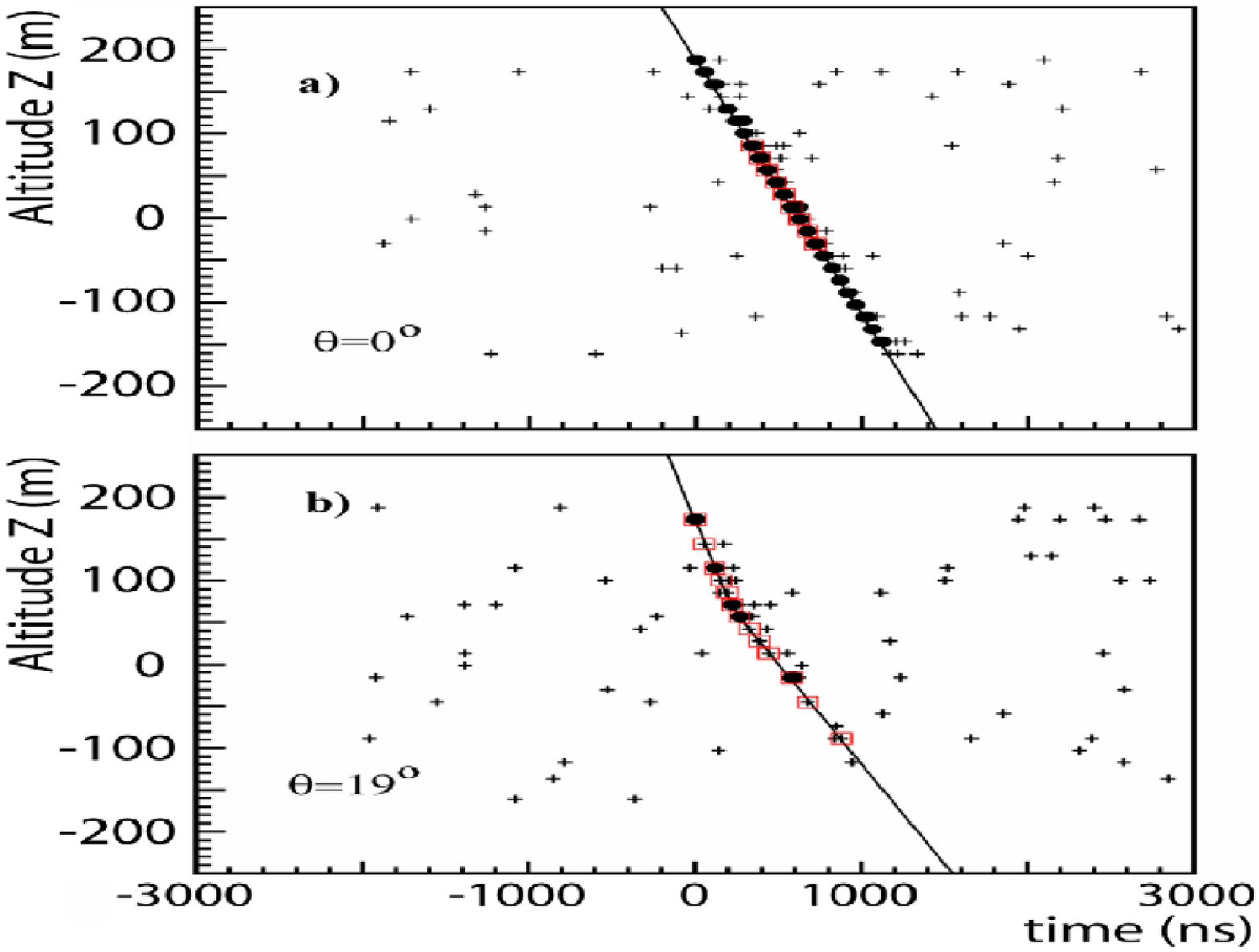,height=\linewidth,width=\linewidth}
\caption{Reconstructions of two events: height (Z) along the string vs. the measured hit times wrt an arbitrary zero. Crosses correspond to all the hits within a 1 $\mu$s time window, black dots to hits in coincidence while those used in the event fit are surrounded by a red square.}
\label{track}
\end{minipage}\hfill
\end{figure*}
\section{\label{sec:detection} Neutralino detection at Neutrino Telescopes}
The WIMPs in the galactic halo will pass through massive bodies in the galaxy and can lose energy by scattering off nuclei so as to become gravitationally bound. Over time, WIMPs concentrate near the centres of these bodies and annihilate producing Standard Model particles. The exact rate depends on the time the WIMPs have had to accumulate and the annihilation cross section. The products of these annihilations will possibly decay and produce neutrinos which will be able to escape from the centre of these bodies and would potentially be visible as a neutrino flux at the surface of the Earth.
 The advantage of this detection method, which is common to the case of gamma rays too (but different from antimatter searches), is that neutrinos do not interact in outer space, and therefore the direction from which they arrive points at the location where they were produced. A high-energy neutrino signal in the direction of the centre of the Sun or of the Earth is therefore an excellent experimental signature, which may stand up against what is the main limitation of the technique itself, namely the neutrino background generated by cosmic-ray interactions in the Earth's atmosphere.

Quantitatively, the neutrino flux from neutralino annihilations depends on the one hand on the particle physics setup, i.e. on the details of the decay chain of a neutralino of a given mass and composition; on the other hand, a crucial role is played by the capture versus annihilation balance in the core of the celestial bodies and on the physics of the propagation of the relevant SM decay products. The differential neutrino flux is given by
\begin{equation}
\frac{\ud N_\nu}{\ud E_\nu}=\frac{\Gamma_A}{4\pi D^2}\sum_fB^f_\chi\frac{\ud N_\nu^f}{\ud E_\nu}
\label{nuflux}
\end{equation}
where $\Gamma_A$ is the annihilation rate, $D$ is the distance of the detector from the source, $f$ is the neutralino pair annihilation final state and $B^f_\chi$ are the branching ratios into the final state, each giving rise to the energy distribution of neutrinos $\frac{\ud N_\nu^f}{\ud E_\nu}$
The differential flux of neutrino-induced muons reaching the detector is then provided by
\begin{equation}
\frac{\ud N_\mu}{\ud E_\mu}=\int_{E_\mu^{th}}^{+\infty}\!\ud E_\nu\int_0^{+\infty}\!\ud R\int_{E_\mu}^{E_\nu}\!\ud E_\mu^\prime P\left(E_\mu,E_\mu^\prime;R\right)\frac{\ud \sigma\left(E_\nu,E_\mu^\prime\right)}{\ud E_\mu^\prime}\frac{\ud N_\nu}{\ud E_\nu}
\label{muflux}
\end{equation}
%\,\:
where from right to left we can identify
\begin{itemize}
\item[-] $\frac{\ud N_\nu}{\ud E_\nu}$: the differential neutrino flux given by eq.\,\ref{nuflux};
\item[-] $\frac{\ud\sigma\left(E_\nu,E_\mu^\prime\right)}{\ud E_\mu^\prime}$: the production cross section of a $\mu$ of energy $E_\mu^\prime$ from a $\nu_\mu$ of energy $E_\nu$;
\item[-] $ P\left(E_\mu,E_\mu^\prime;R\right)$: the probability for a $\mu$ of initial energy  $E_\mu^\prime$ to end up with an energy  $E_\mu$ after traversing a path-length $R$;
\item[-]$R$ is the $\mu$ range in the material surrounding the detector.
\end{itemize}
At this point, the only missing ingredient for the computation is the annihilation rate $\Gamma_A$ which, in turn, depends on the details of the neutralino interactions. Some assumptions and models must then be made which can affect the final result of the computation. Moreover the Sun and the Earth compositions are quite different because the Sun is mainly consisting of hydrogen while the Earth is mostly composed of nuclei with zero spin. This in turn implies that the flux from the two sources are respectively dominated by spin-dependent and spin-independent processes.\\
\begin{figure*}[t]
\begin{minipage}[t]{0.48\linewidth}
\centering\epsfig{file=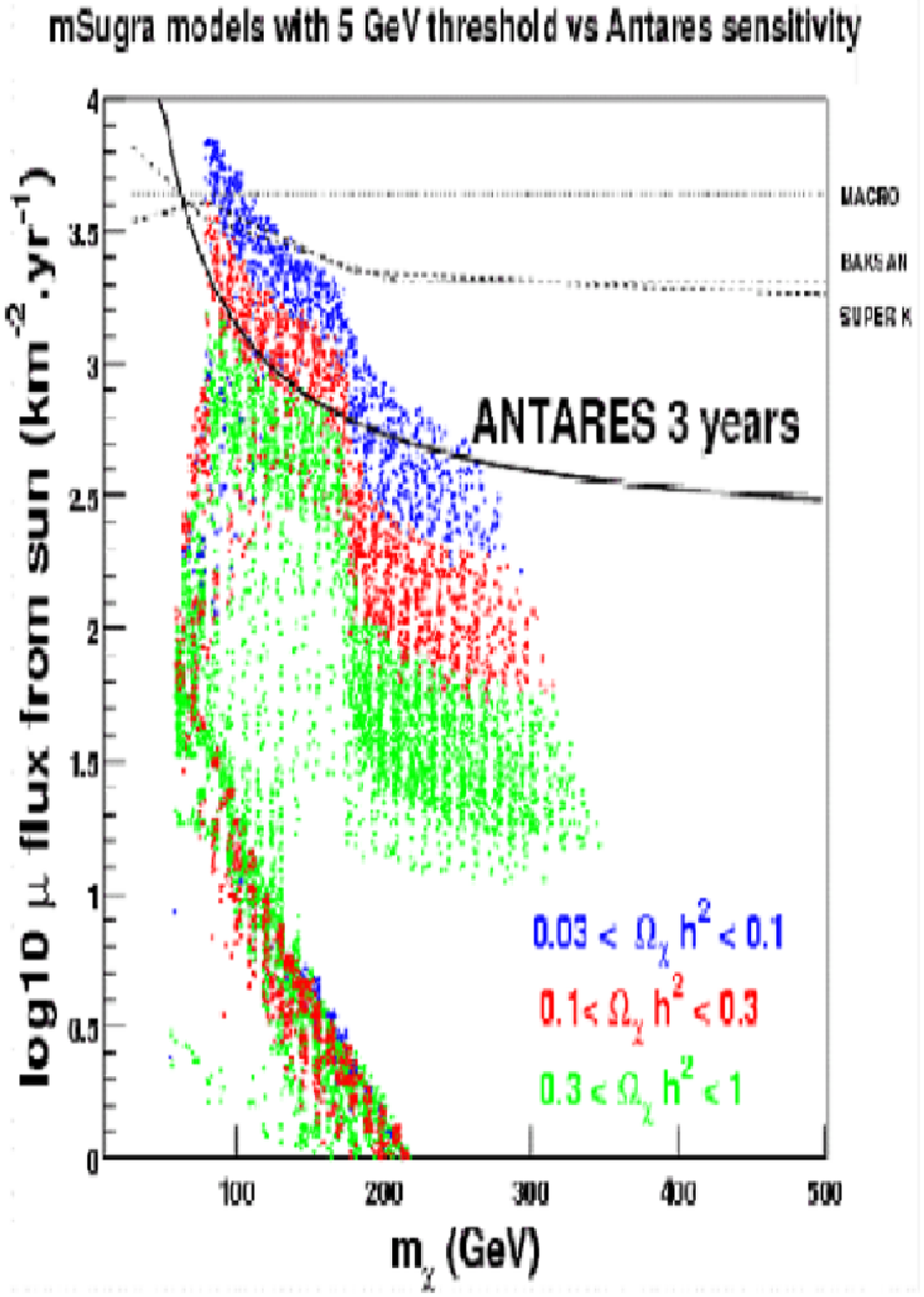,height=\linewidth,width=\linewidth}
\caption{ANTARES sensitivity to a muon flux from neutralinos in the Sun.}
\label{sensitiv}
\end{minipage}\hfill
\begin{minipage}[t]{0.48\linewidth}
\centering\epsfig{file=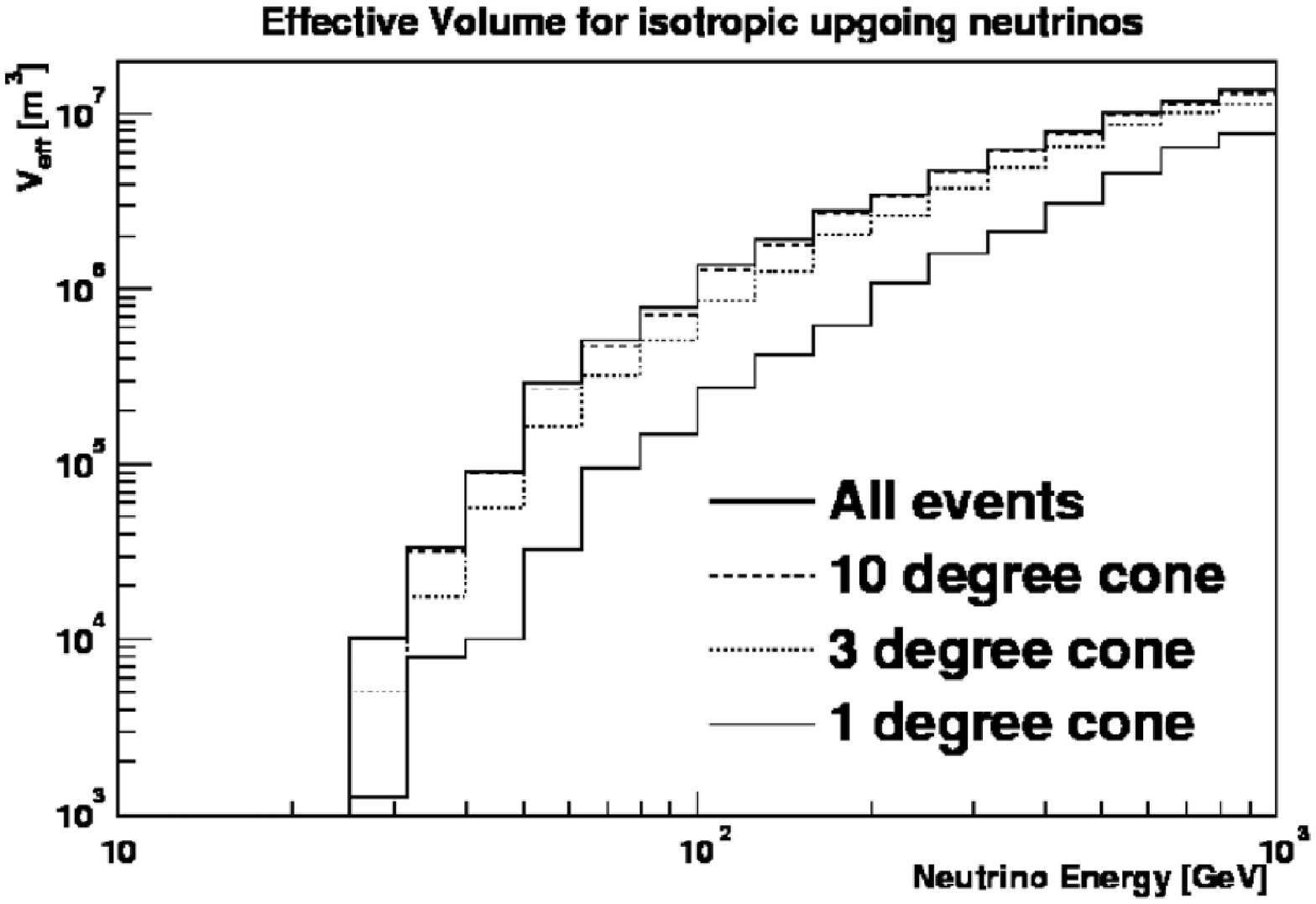,height=\linewidth,width=\linewidth}
\caption{Effective volume for isotropic upgoing neutrinos}
\label{eff_volume}
\end{minipage}\hfill

\end{figure*}
\section{\label{sec:sensitivity} ANTARES sensitivity to Neutralinos}
The sensitivity of ANTARES to neutrinos originating from Dark Matter annihilation in the context of mSUGRA~\cite{msugra} has been extensively studied through MonteCarlo simulations~\cite{bailey}. One of the results is shown in Fig.~\ref{sensitiv}. The effective volume (which is the detector sensitive volume including reconstruction efficiency) is calculated as a function of neutralino energy and optimised with respect to the angular size cut (Fig.~\ref{eff_volume}). For any assumed neutralino mass, the expected background inside the cone is calculated and the 90\% confidence level limit is set using the Feldman and Cousins technique \cite{FC} to yield a neutrino flux limit. This is in turn converted to a muon flux limit using the muon yield per neutrino provided by the \texttt{DarkSUSY}~\cite{darksusy} program. Superimposed on Fig.~\ref{sensitiv} are a number of points which correspond to the theoretical predictions within the mSUGRA framework. The fall-off in sensitivity at low energies arises from using a reconstruction package optimised for higher energies: dedicated studies to improve low energy reconstruction efficiencies and to evaluate the sensitivity to neutrino fluxes from neutralino annihilation in the Earth are under way.

%%%%%%%%%%%%%%%%%%%%%%%%%%%%%%%%%%%%%% reset.txt counters %%%%%%%%%%%%%%
%%
%%%%%%% do not change below here  %%%%%%%%%%%%%%%%%%%%%%%%%%%%%
\setcounter{section}{0}
\setcounter{subsection}{0}
\setcounter{figure}{0}
\setcounter{table}{0}
\setcounter{equation}{0}
\newpage

%%%%%%%%%%%%%%%%%%%%%%%%%%%%%%%%%%%%%%%%%%%%%%%%%%% Title, authors and addresses
\begin{frontmatter}


\title{Exotic Physics with IceCube}

% use optional labels between square brackets to link authors explicitly to addresses:
% If more than one author, keep a comma between the author tags

\author[address1]{David Hardtke}
\author[address2]{for the IceCube}collaboration

\address[address1]{Department of Physics, University of California, Berkeley,~CA~94720,~USA }
\address[address2]{http://www.icecube.wisc.edu/}

\begin{abstract}
The IceCube neutrino observatory, currently under construction at the South Pole, will be used to search for particles and interactions beyond the standard model.  These searches include both direct and indirect searches for dark matter, searches for non-standard neutrino interactions, searches for supersymmetric particles, and searches for both relativistic and non-relativistic monopoles. The expected sensitivity of IceCube to exotic physics will be discussed.  
\end{abstract}

% \begin{keyword}

%\PACS 
% \end{keyword}

\end{frontmatter}

%%%%%%%%%%%%%%%%%%%%%%%%%%%%%%%%%%%%%%%%%%%%%%%%%%%%%% MAIN TEXT
\section{\label{sec:intro} Introduction}

The IceCube neutrino observatory will be the largest volume instrumented detector ever constructed.  Nearly one cubic kilometer of South Pole ice will be instrumented with highly sensitive photomultiplier tubes and associated readout electronics.  The large size is driven by the desire to detect high-energy (TeV-PeV) extraterrestrial neutrinos.  A detector of this size can also open up new possibilities for exotic physics searches.

High energy neutrinos have many properties that make them ideal probes for physics beyond the standard model.  The small standard model neutrino interaction cross-section means that any new types of interactions will have pronounced effects on the observed neutrino spectra.  At this conference, we heard talks on several new types of neutrino interactions, including TeV scale gravity \cite{Illana}, micro black-holes\cite{Kowalski}, and supersymmetric stau production \cite{Albuquerque}.  The observation of these signatures requires the existence of a high-energy extra-terrestrial neutrino flux.   Once IceCube establishes the existence and makes quantitative measurements of a high energy extra-terrestrial neutrino flux we will use this lepton beam to probe physics beyond the reach of accelerators.  

IceCube will also be a unique instrument for dark matter searches. There is a long history of indirect dark matter searches in neutrino telescopes utilizing neutrinos produced in neutralino annihilations in the center of the Earth or Sun (for a review of such searches in AMANDA and IceCube, see the contribution by Hubert\cite{Hubert}).   
There are also dark matter 
candidates where IceCube has sensitivity for direct detection.  They
are generally very massive and therefore have extremely low fluxes due
to the dark matter bound, are expected to have virial 
velocities ($\beta\approx 10^{-3}$), and emit photons via non-Cherenkov mechanisms.
Three potential candidates are:
\begin{description}
\item[Supersymmetric Q-balls.] These are coherent states of squarks and Higgs fields that may be produced during the decay of the proposed Affleck-Dine condensate\cite{Kusenko}.  Q-balls carry large baryon number but would evade the non-baryonic constraint on dark matter from Big Bang Nucleosynthesis .  The observation 
of these objects would validate the Affleck-Dine mechanism as the origin of the baryon-antibaryon asymmetry in the universe.
These states are only stable for masses $>10^{15}$ GeV.  They produce light
via the catalysis of nucleon decay (which leads to pion production and finally
photons).
\item[Heavy Strangelets.] These objects are states with nearly equal 
numbers of up, down, and strange quarks \cite{Madsen}.  As they become very massive, they
have sizes of order atomic dimensions or above.  Objects this size passing through ice will trigger
a thermal shock and produce light via black-body radiation.  
\item[Massive Magnetic Monopoles.] The limits on massive monopoles with masses 
$>10^{16}$ GeV can be improved by IceCube.  These objects also register in the 
detector via the catalysis of nucleon decay. 
\end{description}
A search for slowly moving massive particles using AMANDA is in progress \cite{Pohl}. 

As with other Cerenkov neutrino telescopes, IceCube will be highly sensitive to relativistic monopoles.  A fast monopole with unit Dirac charge will produce $\sim$8300 times as much Cerenkov radiation as an identical particle with unit electric charge.  

\section{\label{sec:detector} Detector}

The possibility of using the South Pole Ice as a neutrino detector was first demonstrated by AMANDA \cite{AMANDA_nature}.  The South Pole has several obvious and non-obvious advantages as a location for high energy neutrino detection.  The ice thickness at the South Pole is $\sim$2800 m, which allows the active instrumentation to be shielded from the vast majority of cosmic-ray muons.   At depths below $\sim$1400m the trapped air bubbles in the ice convert to an air hydrate crystal form with a substantially increased scattering length.  Below 1500m, the effective scattering length varies between $\sim$15-45m for wavelengths between 300 and 600nm except near the dust peak at $\sim$2100m\cite{IcePaper}.   For wavelengths less than 400 nm the absorption length averages $\sim$110m.  

Although the optical properties of water (effective scattering lengths $\sim$200 m and average absorption length $\sim$50m) are superior to ice, South Pole Ice has several technical advantages for operation of a neutrino telescope.  For the instrumented depths (1450m to 2450m), the temperature of the ice varies from -40 C at the highest point to -15 C at the lowest point.   At these low temperatures, the dark noise rates in the photomultiplier tubes are reduced (averaging about 450 Hz for the IceCube DOMs).  The lack of bioluminescence allows the optical modules to deployed in isolation rather than in pairs.  While it is difficult to deploy the optical modules in ice, once the IceCube strings are installed they are mechanically stable and do not require constant alignment.  Several AMANDA strings have been operated for nearly 10 years with minor optical module failures.    

IceCube is both larger and technologically superior to AMANDA.  The instrumented volume of AMANDA is $\sim$0.015 km$^3$, while IceCube will eventually approach 1 km$^3$ once all 70-80 strings are installed (as of this conference, IceCube has 9 deployed strings).  IceCube, however, is more sparsely instrumented with a horizontal spacing between strings of  125m compared to 55-75 m for AMANDA.  Vertically, the modules are spaced every 17m in IceCube versus 10-20m in AMANDA.  In addition to the in ice optical modules, a surface array (IceTop) based on the same optical module technology will be installed.  The acceptance of the surface/deep ice joint array will approach $\sim$0.3km$^2$sr and will allow cosmic ray composition studies from below the knee up to 1 EeV.  

Both AMANDA and IceCube utilize large (10" in the case of IceCube) photomultiplier tubes encased in spherical pressure spheres as the active optical elements.  In the case of IceCube, however, each digital optical module (DOM) acts as an autonomous data collection unit.  Each DOM has its own high voltage power supply and waveform digitization electronics.  The digitized waveforms can be buffered in the DOMs, and are communicated digitally at a rate of up to 1 MBit/s over the cable.  In the case of AMANDA, HV was provided from the surface, and the photomultiplier signals were sent to the surface analog communications (with corresponding signal degradation).   

For the purposes of exotic physics studies, the in-ice signal digitization will give a improvement in the sensitivity of the detector.  The waveform digitization is done by three analog transient waveform digitizers (ATWDs) that run at three different gain settings.  This leads to a large increase in dynamic range and will reduce saturation for the brightest events (e.g. relativistic monopoles).   In addition to the ATWDs which record 128 waveform samplings with a sampling rate between 200 and 700 MHz, each DOM is equipped with a slow 40 MHz digitizer that records the waveform for up to 6.4$\mu$s.   This ability to record multiple photoelectrons over an extended period will allow for improved background rejection. 

In the case of slowly moving massive particle searches, the IceCube hardware opens up new possibilities as well.  The DOMs operate essentially dead-time free.  Particles moving at $10^{-3}$c  will spend a few milliseconds crossing the detector and the DOMs are able to record signals during this entire period.  The IceCube trigger is entirely software based and topological triggers for slow particles are under development.  The minimum velocity to which IceCube is sensitive is largely determined by the dark noise rates, and these are substantially lower than similar water based Cerenkov detectors. 

\begin{figure}[t]
\begin{center}
\includegraphics[width=3in]{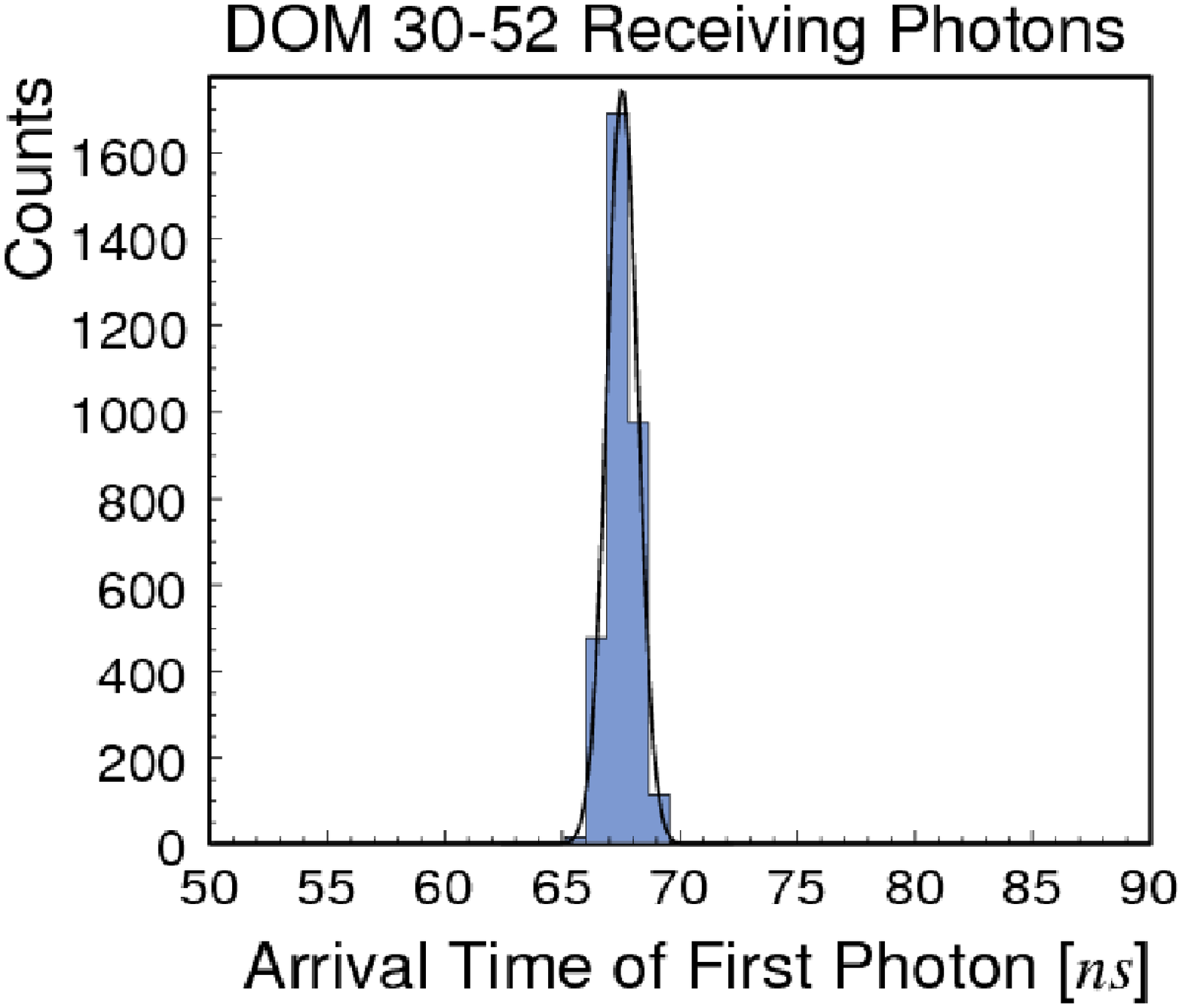}
%\vspace{-0.6in}
\caption{ Plot of the arrival time of the first photoelectron in the DOM 17m above an LED flasher.}
\label{Flasher}
\end{center}
\end{figure}

\section{\label{sec:Sensitivity} Detector Performance and Sensitivity Estimates} 

As of 2006, 9 IceCube strings have been deployed with 540 DOMs.  A total of 16 IceTop stations have been constructed with 64 DOMs.  Extensive testing has been done on the deployed hardware to verify the performance of the IceCube DOMs and data acquisition system\cite{IceCubePerformance}.   The detector has been used to reconstruct down-going muons as well as up-going atmospheric neutrinos.  Additionally, LEDs in each DOM have been used to test the performance of the array.  Figure \ref{Flasher} shows the arrival time distribution of the first photon recorded in the DOM 17m above a flashing LED.   This shows that the timing resolution of the deployed DOMs is better than 2ns.    

First simulations are underway to determine the sensitivity of the IceCube array to exotic physics signals.  The simulations use a detailed model of South Pole Ice, simulations of the photon propagation through the pressure sphere and glass, detailed modeling of the photomultiplier tube response and electronics, and a simulation of the detector trigger.   Figure \ref{monopole} shows a simulated event consisting of a M=1 PeV/c$^2$, $\gamma$ = 10 monopole passing through the 2006 IceCube configuration.  As monopole events are very bright, the large cross-sectional area of IceCube will allow even the present detector stage with 9 strings to perform monopole searches with increased sensitivity over existing limits.    It is expected that analysis of the 2006 IceCube data for will have a flux sensitivity of $\sim$$10^{-17}$ cm$^{-2}$s$^{-1}$sr$^{-1}$.  This will represent a factor of 3-4 improvement of the current best limits from Baikal\cite{Osipova}.  The full IceCube detector operated for several years will have a sensitivity to relativistic monopoles of $\sim$$10^{18}-10^{-19}$ cm$^{-2}$s$^{-1}$sr$^{-1}$.

\begin{figure}[t]
\begin{center}
\includegraphics[width=2.5in]{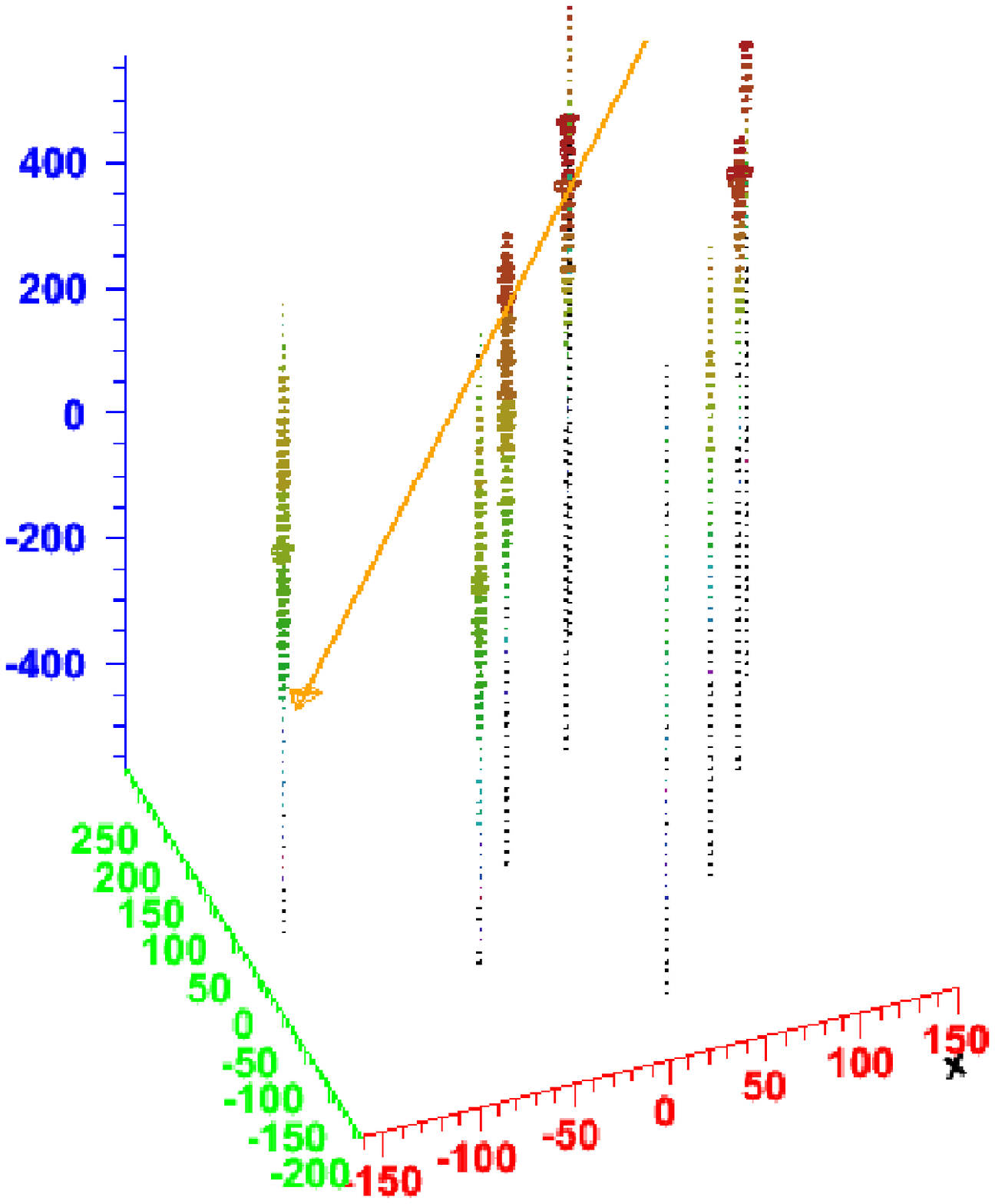}
\caption{Simulation of a mass =1 PeV/c$^2$ monopole with $\gamma=10$ passing through the 2006 nine-string IceCube detector.}
\label{monopole}
\end{center}
\end{figure}

IceCube will have a similar flux sensitivity to slowly moving massive particles.  Figure \ref{SMMP} shows the current limits as well as future IceCube sensitivity to massive strangelets, neutral Q-balls, and GUT monopoles that catalyze nucleon decay.  In this case the flux limit is shown versus the mass of the slowly moving massive particle.  The diagonal black line is the local dark matter limit ($\rho =$ 0.3 GeV/cm$^3$) assuming a velocity $\beta=10^{-3}$. The black line horizontal line (IceCube24) is for the expected 21-23 string 2007 IceCube configuration, and the blue line is for three years of full IceCube.  The strongest current limits on monopoles are from searches for nuclear tracks in ancient mica \cite{Price}.  IceCube should exceed that sensitivity level and will not be subject to the same theoretical uncertainties.

\begin{figure}[t]
\begin{center}
\includegraphics[width=4in,height=3in]{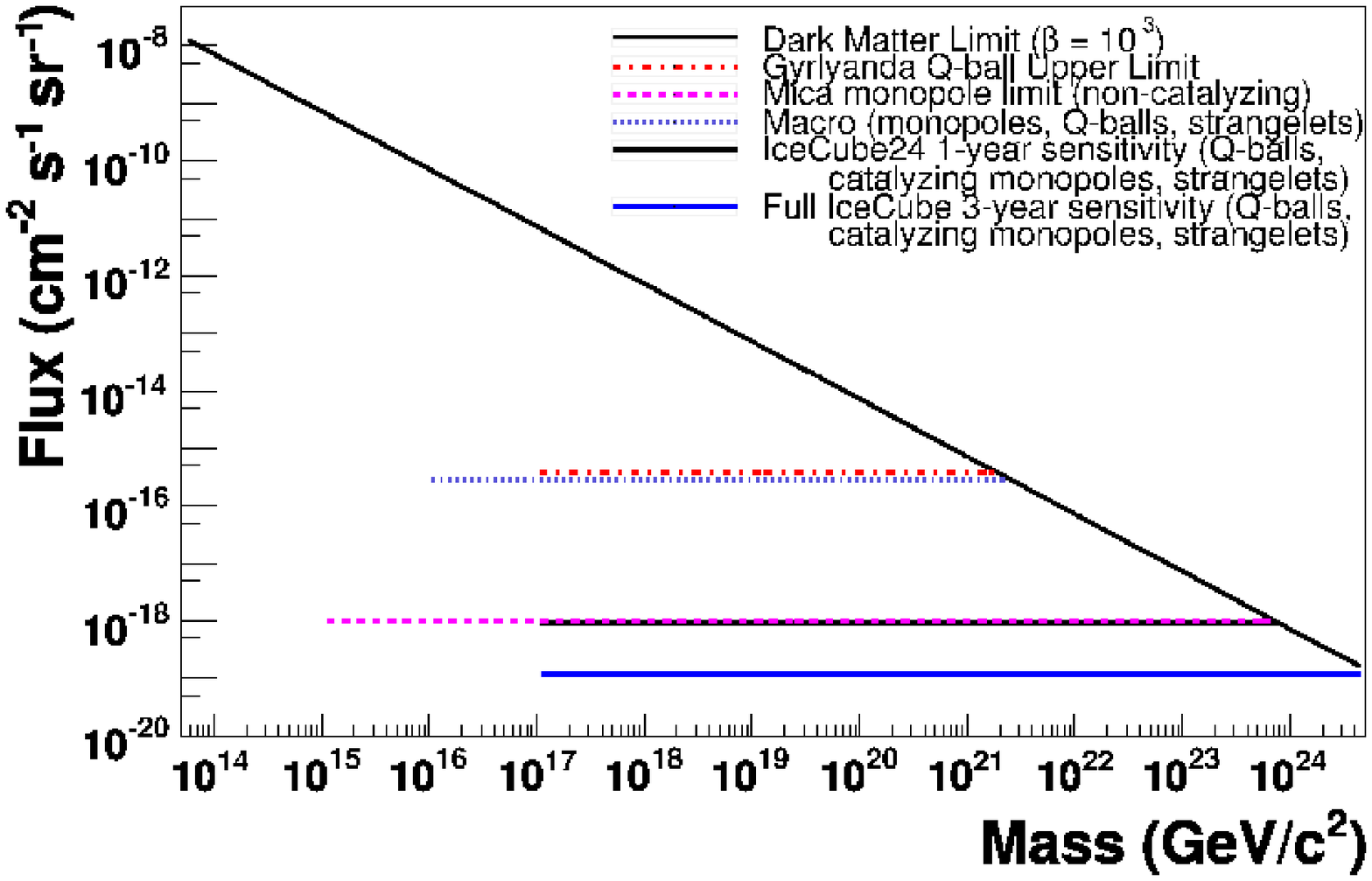}
\caption{ Current flux limits on $\beta=10^{-3}$ massive particles and predicted flux sensitivity for 2007 24 string IceCube configuration and full IceCube configuration. }
\label{SMMP}
\end{center}
\end{figure}

\section{Conclusions}

The IceCube neutrino observatory will be the most sensitive detector for many of the exotic physics topics discussed at this conference.  Extraterrestrial neutrino beams will probe particles and interactions beyond the standard model.  IceCube will conduct both indirect and direct searches for dark matter.    The initial experience with the IceCube hardware indicates that a rich exotic physics program is feasible.

%%%%%%%%%%%%%%%%%%%%%%%%%%%%%%%%%%%%%% reset.txt counters %%%%%%%%%%%%%%
%%
%%%%%%% do not change below here  %%%%%%%%%%%%%%%%%%%%%%%%%%%%%
\setcounter{section}{0}
\setcounter{subsection}{0}
\setcounter{figure}{0}
\setcounter{table}{0}
\setcounter{equation}{0}
\newpage

%%%%%%%%%%%%%%%%%%%%%%%%%%%%%%%%%%%%%%%%%%%%%%%%%%% Title, authors and addresses
\begin{frontmatter}


\title{NESTOR}

% use optional labels between square brackets to link authors explicitly to addresses:
% If more than one author, keep a comma between the author tags

\author[address1]{A. Belias}{\em for The NESTOR Collaboration},
\author[address1]{G. Aggouras},\author[address2]{E.~G. Anassontzis}, \author[address5]{A.~E. Ball}, \author[address7]{G. Burlis},
\author[address9]{W. Chinowsky}, \author[address8]{E. Fahrun},
\author[address6]{G. Grammatikakis},
\author[address8]{C. Green},
\author[address3]{P. Grieder},
\author[address10]{P. Katrivanos},
\author[address8]{P. Koske},
\author[address7]{A. Leisos}, 
\author[address9]{L. Ludvig},
\author[address1]{E. Markopoulos}, 
\author[address4]{P. Minkowsky},
\author[address9]{D. Nygren},
%\author[address1]{K. Papageorgiou}, 
\author[address9]{G. Przybylski},
\author[address10]{P. Rapidis},
\author[address1,address2]{L.~K. Resvanis}, 
\author[address10]{I. Siotis}, 
\author[address9]{J. Sopher},
\author[address1]{T. Staveris}, 
\author[address1]{S. Tsagli}, 
\author[address7]{A. Tsirigotis}, 
\author[address12]{V.~A. Zhukov}
\address[address1]{NESTOR Institute of Astroparticle Physics, National Observatory of Athens, Pylos, Greece}
\address[address2]{University of Athens, Physics Department, Greece}
\address[address3]{University of Bern, Physikalisches Institut, Switzerland}
\address[address4]{University of Bern, Institute for Theoretical Physics, Switzerland}
\address[address5]{CERN (European Organization for Nuclear Research), Geneva, Switzerland}
\address[address6]{University of Crete, Physics Department, Greece}
\address[address7]{Hellenic Open University, School of Science and Technology, Patra, Greece} 
\address[address8]{University of Kiel, Institute of Experimental and Applied Physics, Germany} 
\address[address9]{Lawrence Berkeley National Laboratory, Berkeley, CA, USA}
\address[address10]{NCSR Demokritos, Athens, Greece}
%\address[address11]{NESTOR Institute for Deep Sea Research, Technology and Neutrino Astroparticle Physics, Pylos, Greece}
\address[address12]{Institute for Nuclear Research, Russian Academy of Sciences, Moscow, Russia}

\begin{abstract}
 The NESTOR collaboration advances towards a large, deep-sea neutrino telescope in the Mediterranean Sea.
We highlight the achievements of operating the NESTOR neutrino telescope at a depth of about $4000m$ and outline the next phases of NESTOR for a neutrino telescope with large sensitive area for high energy astrophysics.    
\end{abstract}

% \begin{keyword}
%\PACS 
% \end{keyword}

\end{frontmatter}

%%%%%%%%%%%%%%%%%%%%%%%%%%%%%%%%%%%%%%%%%%%%%%%%%%%%%% MAIN TEXT
\section{\label{sec:NESTOR_intro} Introduction}
In attempting to detect intra- and extra-galactic High Energy Neutrinos on Earth
one is faced with a minuscule signal compared to the overwhelming backgrounds due to cosmic rays and atmospheric neutrinos. Furthermore, to observe as much as possible of the given neutrino fluxes from extraterrestrial sources, requires vast sensitive detection areas, in the order of $km^2$. 

The Neutrino Extended Submarine Telescope with Oceanographic Research, NES\-TOR, 
is a deep underwater neutrino detector designed to detect neutrino induced muons and showers of charged particles through their Cherenkov radiation produced in sea water~\cite{Resvanis:1,Resvanis:2}. 
After a description of the NESTOR site and detector, we highlight the main results of operations in deep sea, at about $4000m$ depth,~\cite{Anassontzis:3,Aggouras:1,Aggouras:2} and report on the NESTOR plans for future cubic kilometer size Neutrino telescopes.

\section{\label{sec:NESTOR_Detector} NESTOR Site and Detector}
Located off the South-West of the Peloponnese, in Greece, near the Ionian Sea with deepest waters, $5200m$, in the Mediterranean Sea, the NESTOR project, up to now makes use of a large horizontal plateau $8km \times 9km$ at a mean depth of $4000m$~\cite{Resvanis:1} about $7.5$ nautical miles from the shore. Several studies of the environmental properties,~\cite{Khanaev:1,Anassontzis:2,Trimonis:1}, show a water transmission length of $55m\pm 10m$ at a wavelength of $\lambda = 460nm$ and minimal underwater currents, few $cm/s$
 ~\cite{Demidova:1}. 

\begin{figure*}[t]
\begin{minipage}[t]{0.48\linewidth}
\centering\epsfig{file=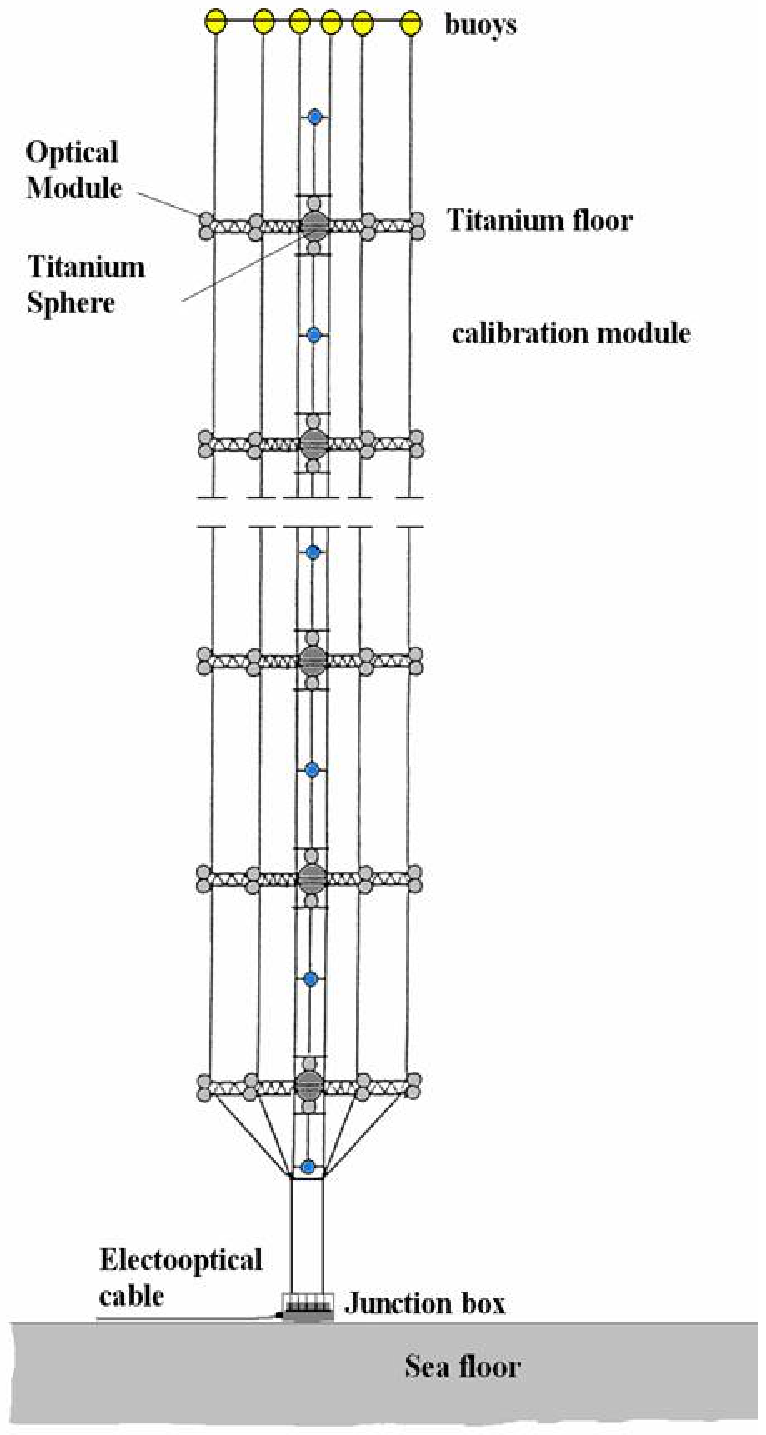,height=\linewidth,width=\linewidth}
\caption{A schematic drawing of a NESTOR tower made of vertically placed stars with Optical Modules at their apex, the readout electronics in the Titanium sphere and calibration modules in-between.}
\label{fig1}
\end{minipage}\hfill
\begin{minipage}[t]{0.48\linewidth}
\centering\epsfig{file=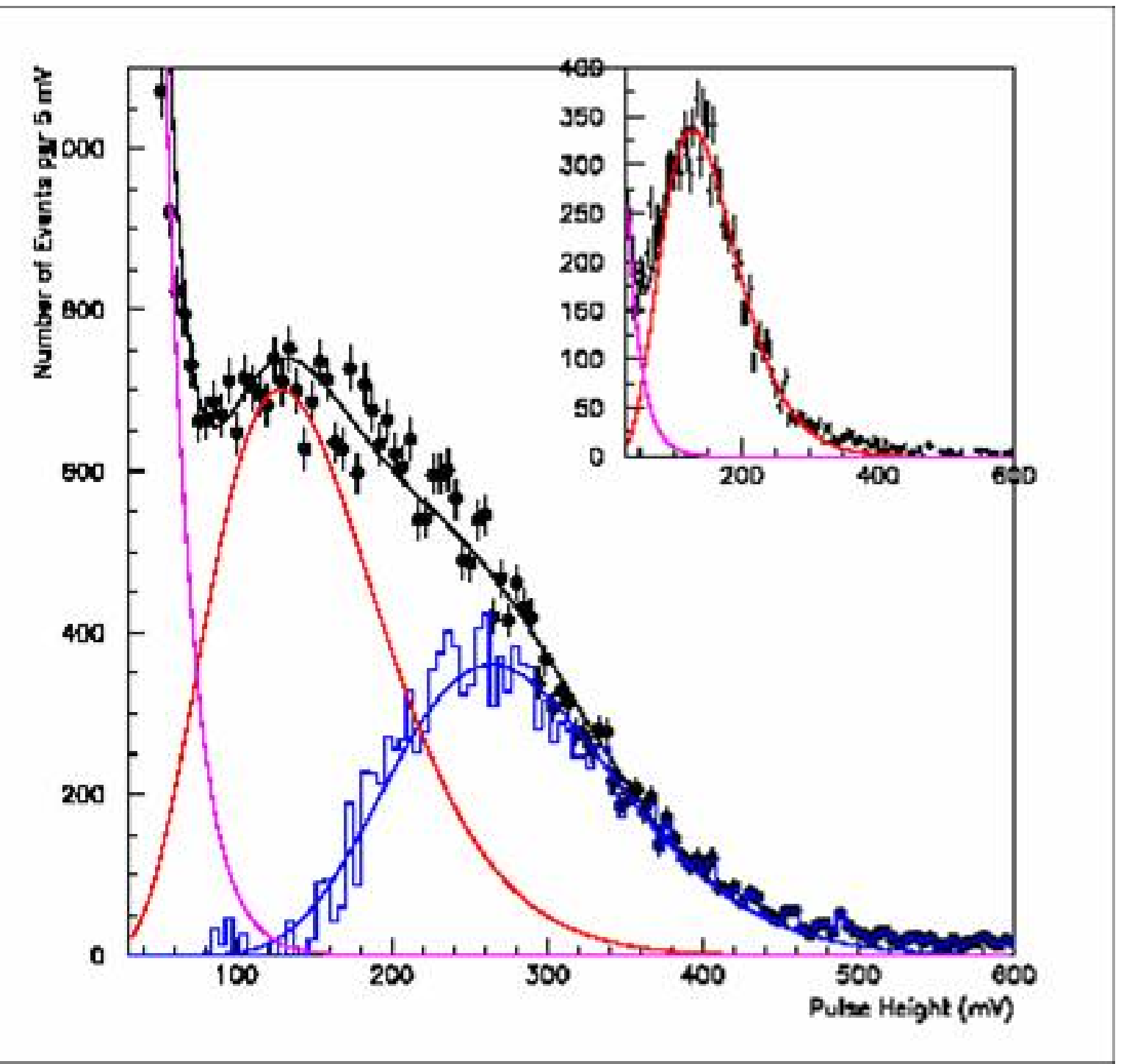,height=\linewidth,width=\linewidth}
\caption{Pulse height distribution of one PMT using data from 4000m depth (main plot) with three spectra, the PMT dark counts, and single and double photoelectron, compared to calibration data taken in the lab (insert plot).}
\label{fig2}
\end{minipage}
\end{figure*}

The NESTOR detector, is composed of rigid hexagonal stars which can be stacked vertically to form a tower (see figure~\ref{fig1}). 
The six-arm star (or floor) is made of Titanium latticed tubes with a diameter of typically $32m$. At each end of each arm two Optical Modules are mounted, each with a photomultiplier tube (PMT), with one photocathode 
facing down-wards and the other up-wards, providing $4\pi$ coverage.

The NESTOR Optical Module ~\cite{Anassontzis:1} is made of a Benthos glass sphere housing a $15$-inch Hamamatsu PMT.
Some important characteristics of these PMTs are large photocathode area, high quantum-efficiency ($20\%$) and small transit time jitter (FWHM $\approx 5.5ns$) at the single photoelectron level. Optical coupling between the glass housing and the PMT is improved by glycerine and shielding from the Earth's magnetic field is achieved with a mu-metal cage. 

At the centre of a star a Titanium sphere (1m diameter) hosts the electronics for voltage control of all PMTs, and the front-end readout electronics developed by the Lawrence Berkley National Laboratory~\cite{Joshua_Nygren_Arragain:1}, for signal acquisition, triggering, digitization and data transmission to shore. 
The electronics can apply a threshold to each PMT signal and a selectable majority trigger, allowing event selection by requiring a time coincidence for any subset of all PMTs on a star. When a trigger condition is satisfied the analogue signals of all PMTs on a star are digitized and transmitted to shore, to be monitored in the control room and recorded for offline analysis. 

Calibration modules, above and below a star, use bright blue LEDs with controlled light pulse sequences for in-situ calibration and monitoring of the PMTs. The relative timing between the PMTs was measured to be about 0.5ns.

The junction box mounted on the sea-bottom Anchor Unit connects the telescope to the shore through an electro-optical cable.

\section{\label{sec:NESTOR_results} Results from the deep-sea}
The NESTOR collaboration has deployed and operated environmental sensors attached to the Anchor Unit 
at a depth of $4000m$ in the Mediterranean Sea (project LAERTIS, Laboratory in the Abyss of Europe with Real-Time data Transfer to shore for Interdisciplinary Studies) in January 2002~\cite{Anassontzis:3} and a telescope module at $3800m$ depth in March 2003~\cite{Aggouras:1}. 
The event samples collected have been used to compare the data and Monte Carlo predictions of the cosmic ray muon flux~\cite{Aggouras:2}. 
The simulation tools and methods,~\cite{Leisos:1} and the reconstruction,~\cite{Tsirigotis:1} are described in detail elsewhere; here we highlight the main results.\par

\begin{figure*}[t]
\begin{minipage}[t]{0.48\linewidth}
\centering\epsfig{file=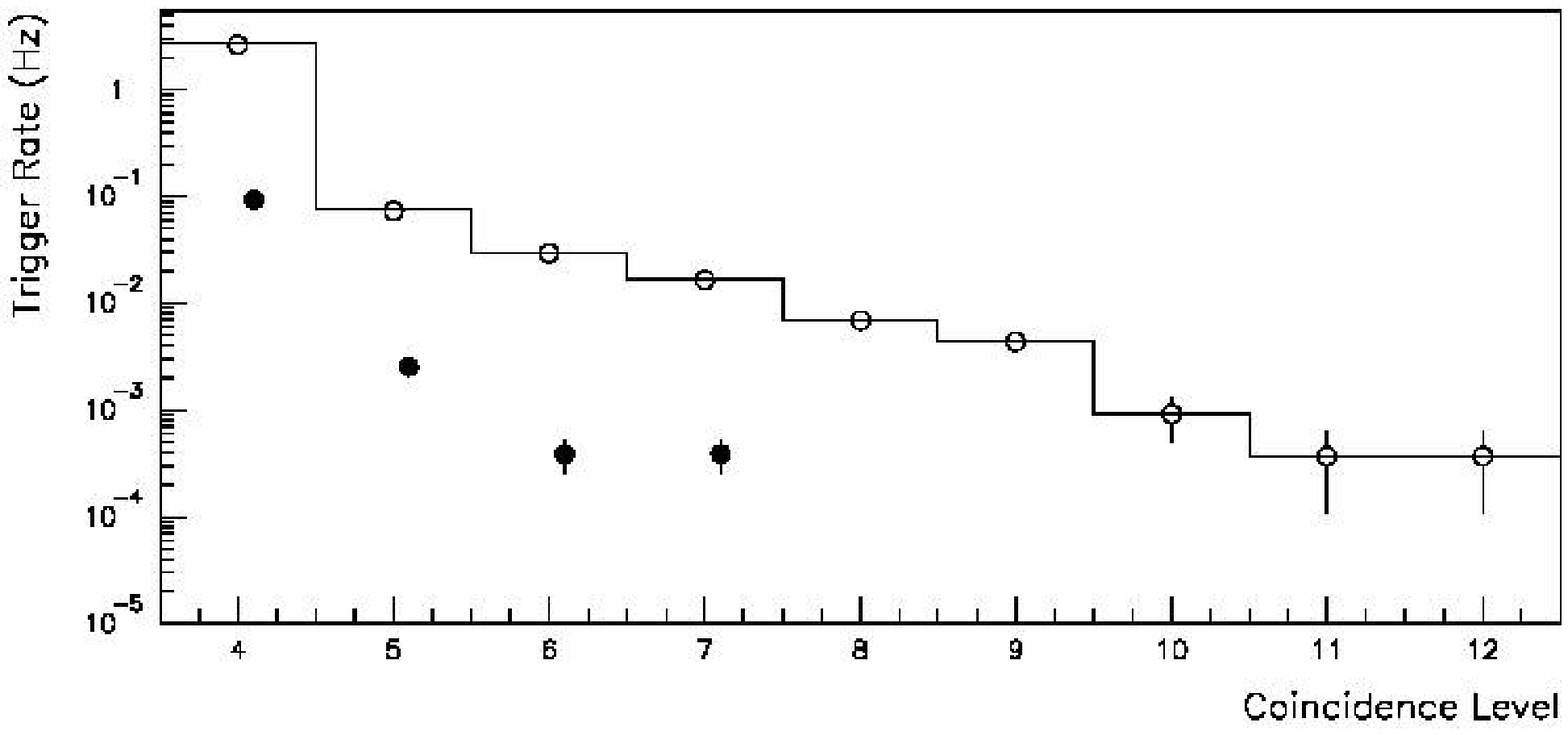,height=0.8\linewidth,width=\linewidth}
\caption{Total trigger rates (solid line), bioluminescence contribution to total trigger rates (open circle), 
experimental trigger rates from periods without bioluminescence (closed circle). Bioluminescence occurs only for $1.1\%\pm 0.1\%$ of the active experimental time.}
\label{belias_fig3}
\end{minipage}\hfill
\begin{minipage}[t]{0.48\linewidth}
\centering\epsfig{file=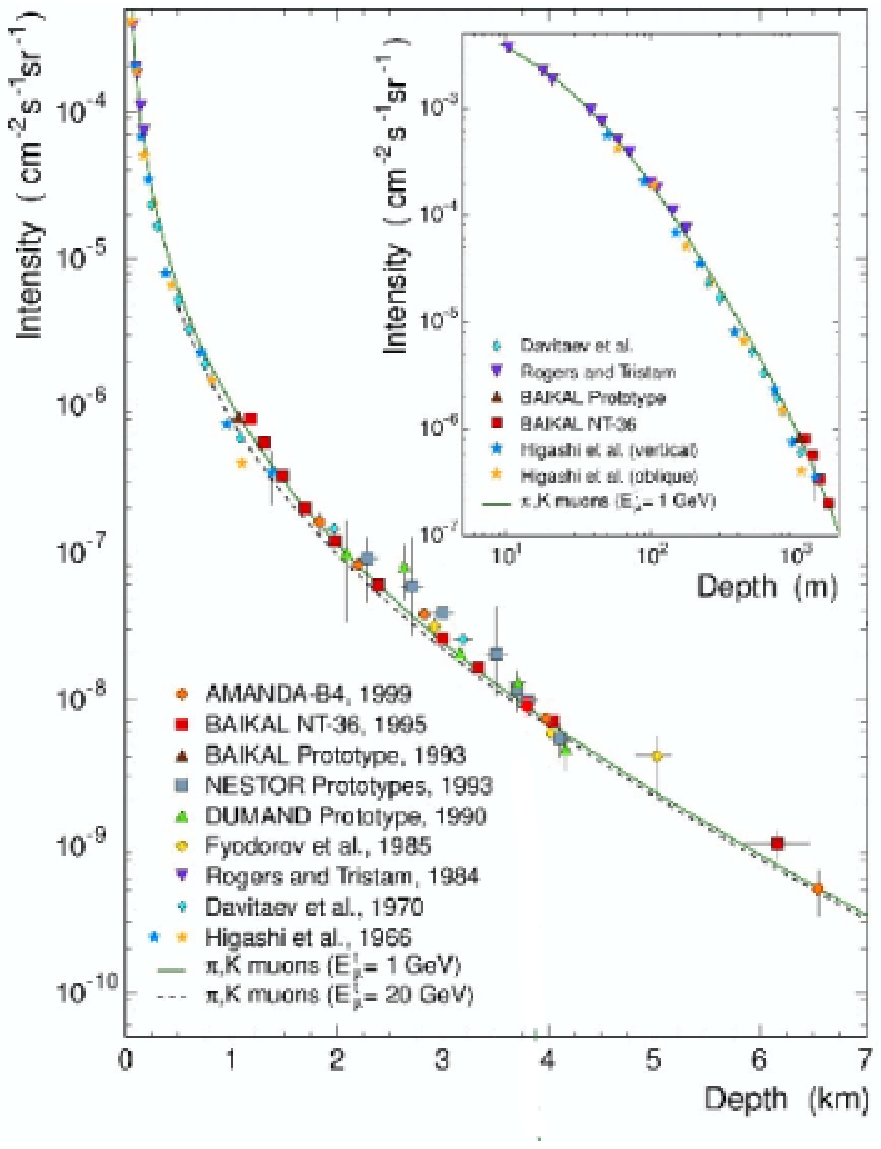,height=0.8\linewidth,width=\linewidth}
\caption{Vertical atmospheric muon intensity of various experiments. NESTOR measurements pointed at by arrow.
Insert plot, ditto in double-log scale.
} 
\label{belias_fig4}
\end{minipage}
\end{figure*}
The star deployed at $3800m$ had a diameter of $12m$ and $12$ OMs, $6$ up-/down-looking OM pairs at the end of each arm. The 
baseline rate of a PMT in the deep sea was measured to be of the order of $50kHz$ per PMT and constant with time and independent of trigger criteria. 
A typical PMT pulse height distribution recorded in deep-sea (fig.~\ref{fig2}) is well described by the thermionic PMT noise and single and double photoelectron distributions as measured in calibration runs in the laboratory~\cite{Aggouras:1}.

At periods, high PMT rate bursts were observed above the baseline of $50kHz$ for a prolonged duration of $1-10s$. These are attributed to bioluminuous activity of marine organisms. In requiring a majority trigger, the effect of the bio-activity on the active experimental time has been estimated, see figure~\ref{belias_fig3}, and found to be in the order of $1\%$ at the NESTOR site, in agreement with several previous findings using autonomous free drop measurements~\cite{Sotiriou:1}.

The atmospheric muon flux was found to be in good agreement with MC predictions, based on the atmospheric muon model of Okada~\cite{Okada:1}. Combined with previous NESTOR measurements~\cite{Resvanis:1}, the flux of atmospheric muons arriving at the detector depth per unit solid angle $d\Omega$, time $dt$ and area $dS$ can be parameterized as $dN/(d\Omega dt dS)=I_0 cos^\alpha (\theta)$ and the vertical atmospheric muon intensity at 3800m depth was found to be, $I_0 = 9.0 \times 10^{-9} \pm 0.7 \times 10^{-9}(stat) \pm 0.4 \times 10^{-9}(syst)cm^{-2}s^{-1}sr^{-1}$, with  $\alpha = 4.7 \pm 0.5(stat) \pm 0.2(syst)$. Figure~\ref{belias_fig4} shows the vertical atmospheric muon intensity as measured by NESTOR along measurements of other experiments,~\cite{Aggouras:2,Tsirigotis:1}. 

\begin{figure*}[t]
\begin{minipage}[t]{0.48\linewidth}
\centering\epsfig{file=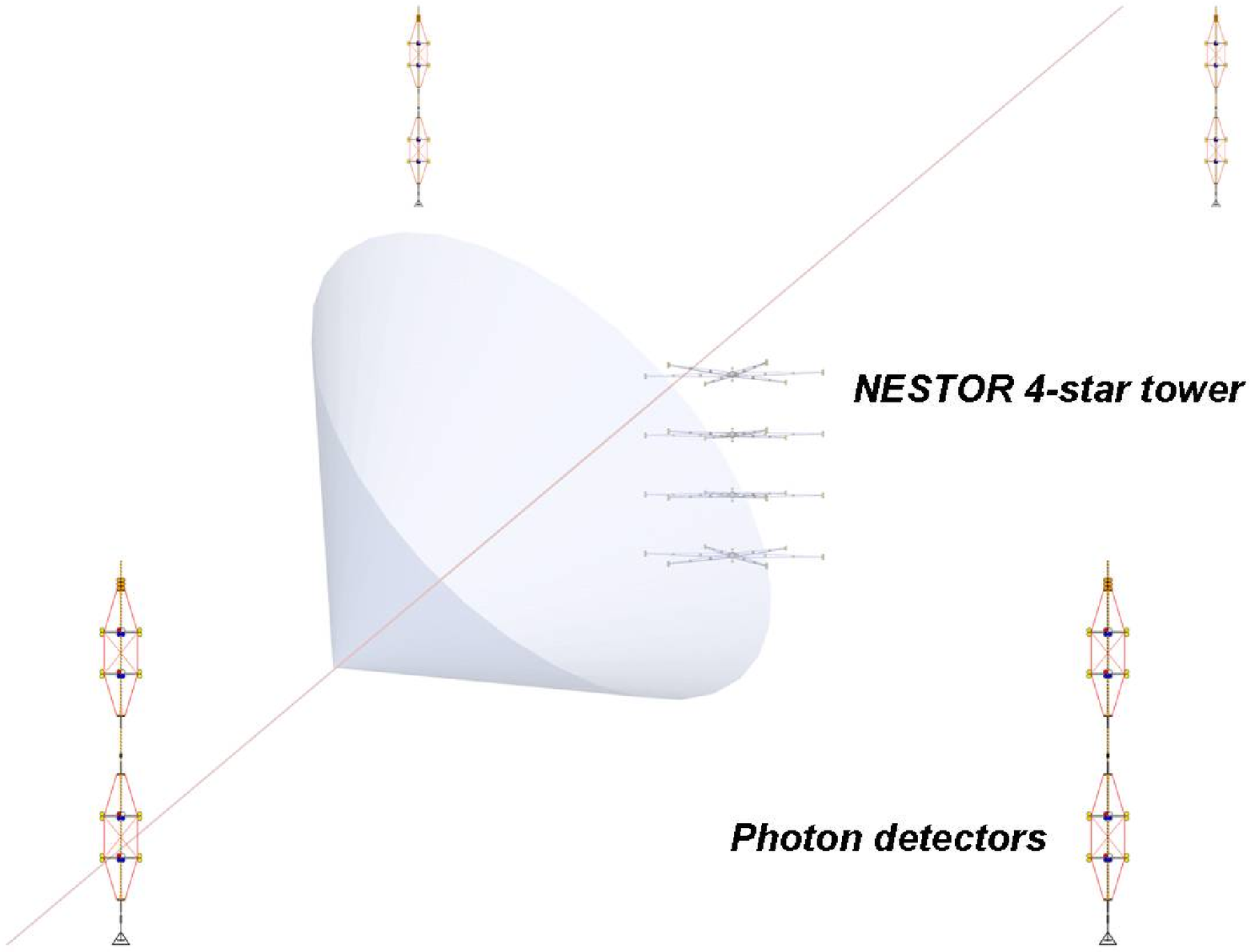,height=0.8\linewidth,width=\linewidth}
\caption{The NuBE array showing 4 strings surrounding the NESTOR tower. 
Also shown is a downward going muon with its Cherenkov cone.}
\label{fig5}
\end{minipage}\hfill
\begin{minipage}[t]{0.48\linewidth}
\centering\epsfig{file=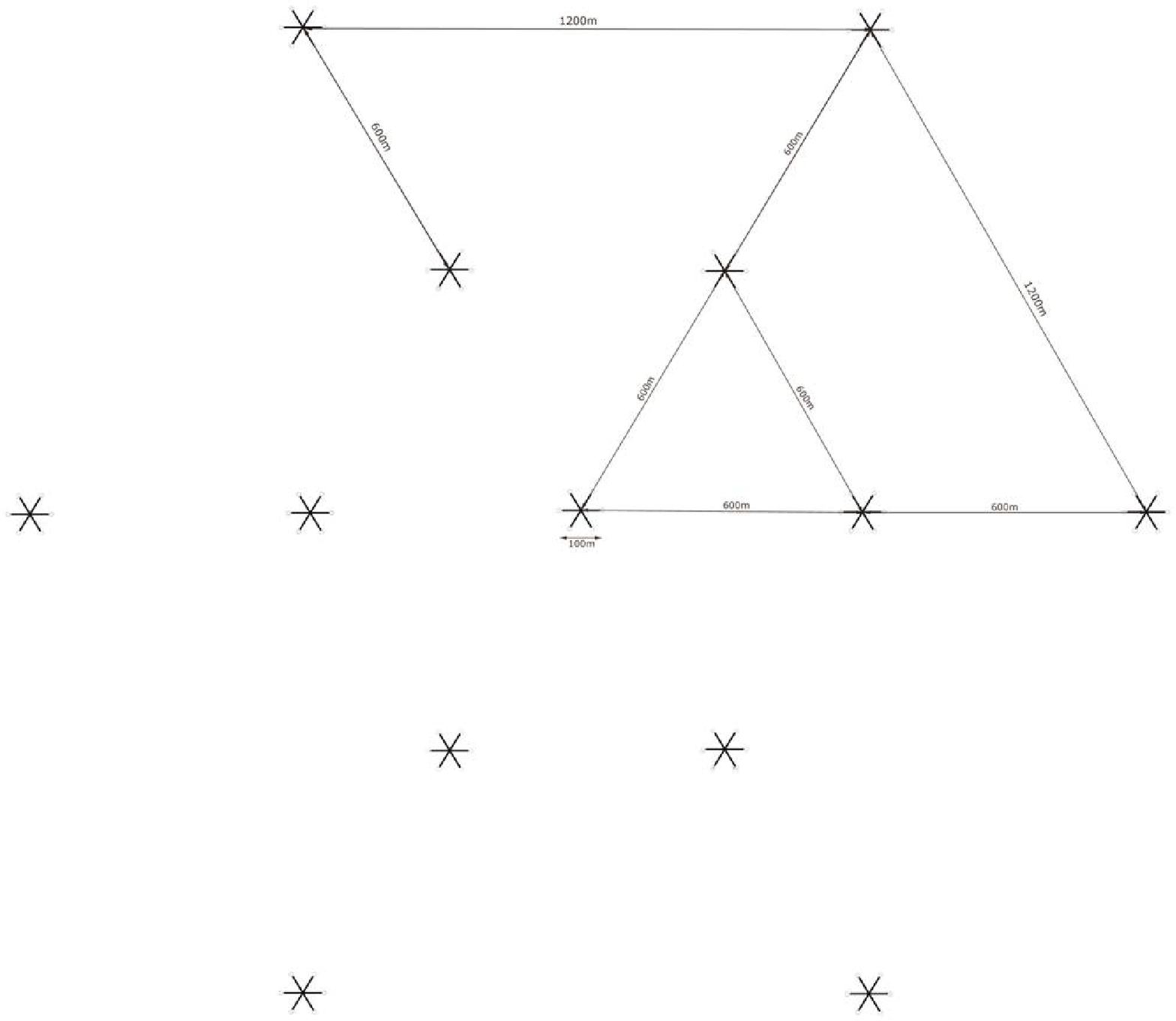,height=0.8\linewidth,width=\linewidth}
\caption{Concentric arrangement of NESTOR towers with variable density detector geometry can cover energy ranges from GeV to PeV of neutrino induced particles.} 
\label{fig6}
\end{minipage}
\end{figure*}

\section{\label{sec:nube} Neutrino Burst Experiment}
Bursts of very-high energy neutrinos (E$>10^{14}eV$) are expected to accompany Gamma-Ray Bursts (GRBs) originating in relativistic fireballs. The relativistic fireball model has been used by Waxman and Bahcall~\cite{Waxman_Bahcall:1} to predict that about $10-100$ neutrino-induced muons per year could be detected in coincidence with GRBs in a $1 km^2$-area detector. The particles produced by those neutrino interactions have distinct signatures; muons producing showers (catastrophic bremsstrahlung) of many km long range; electrons or neutral-current events (cascades) producing short range but very bright showers.

NuBE~\cite{nube:1}, proposes to build a high-energy neutrino detector to detect neutrinos of energy $E>10TeV$ and temporal coincidences (in the order of $100s$) between neutrino events and gamma-ray bursts as observed by satellites. The proposed experiment can be accomplished by deploying a NESTOR tower of four stars surrounded by four autonomous strings, as shown in figure \ref{fig5}. A NuBE string will consist of two vertically separated photon-detector nodes, with each node made of two clusters of Optical Modules. Eight Optical Modules will form a cluster, arranged such as to provide omnidirectional coverage.
 NuBE strings at distances of $300m$ from the NESTOR tower will yield an effective are of $> 1km^2$ omnidirectional in sensitivity for neutrino energies $E_\nu > 10TeV$. For GRB durations of about $10-100s$, an absolute timing accuracy of a few seconds per year is achievable, with relative timing accuracy of $0.3\mu s$ in the strings and about $5ns$ in the tower. 
While the primary identification with GRBs lies in time correlations, the angular resolution capability of FWHM $\approx14^o$, see figure 6 and 9 in~\cite{Aggouras:2}, provides further verification of the correlation with GRBs or other astrophysical sources. 

\section{\label{sec:km3net} The KM3NeT}
A deep-sea $km^3$-scale observatory for high energy neutrino astronomy and associated platform for deep-sea science study, 
called KM3NeT~\cite{KM3NeT:1}, is in progress in Europe with 37 participating institutions. A $km^3$-scale detector in the Northern hemisphere will be complementary to the IceCube detector~\cite{IceCube:1} at the South Pole. 

Figure~\ref{fig6} depicts a birds eye view of a possible $km^3$ size detector~\cite{Resvanis:4}, made of 13 NESTOR towers of $100m$ diameter, each made of a stack of 12 stars.

The central tower would be surrounded by other towers on radii in the order of $600m$ from each other. 
Variable density detector arrays can cover different energy ranges from GeV to PeV of neutrino induced particles.
The expertise of designing, deploying and operating the NESTOR stars and strings as proposed for the NuBE aids in the development of the KM3NeT project.

\vspace*{-0.2cm}
\section{\label{sec:conclsions} Conclusions}
The NESTOR collaboration has operated neutrino detector modules at a depth of about $4000m$ in the Mediterranean Sea. The good agreement with the published measurements of the vertical muon intensity in deep-sea demonstrate the precise understanding of the detector response, calibration, simulation and efficiencies. The Neutrino Burst Experiment is planned to measure time correlations of $\nu's$ and $\gamma$ rays of GRBs and to contribute to the development of a $km^3$ sized neutrino telescope in the Mediterranean Sea.

\vspace*{-0.2cm}
\enlargethispage{1cm}

%%%%%%%%%%%%%%%%%%%%%%%%%%%%%%%%%%%%%% reset.txt counters %%%%%%%%%%%%%%
%%
%%%%%%% do not change below here  %%%%%%%%%%%%%%%%%%%%%%%%%%%%%
%%
\setcounter{section}{0}
\setcounter{subsection}{0}
\setcounter{figure}{0}
\setcounter{table}{0}
\setcounter{equation}{0}
\newpage

\begin{frontmatter}


\title{Ultra High Energy Neutrino Signature in Top-Down Scenario}

% use optional labels between square brackets to link authors explicitly 
%to addresses:
% If more than one author, keep a comma between the author tags

\author[address1]{R. Aloisio},

\address[address1]{INFN - Laboratori Nazionali del Gran Sasso, I-67010
  Assergi (AQ), Italy }

\begin{abstract}
Neutrinos are the best candidates to test the extreme Universe and ideas
beyond the Standard Model of particle Physics. Once produced, neutrinos do 
not suffer any kind of attenuation by intervening radiation fields like the 
Cosmic Microwave Background and are not affected by magnetic fields. In this 
sense neutrinos are useful messengers from the far and young Universe. In the
present paper we will discuss a particular class of sources of Ultra High 
Energy Cosmic Rays introduced to explain the possible excess of events with 
energy larger than the Graisen-Zatsepin-Kuzmin cut-off. These sources, 
collectively called top-down, share a common feature: UHE particles are
produced in the decay or annihilation of superheavy, exotic, particles. 
As we will review in the present paper, the largest fraction of Ultra 
High Energy particles produced in the top-down scenario are neutrinos.
The study of these radiation offers us a unique opportunity to test the 
exotic mechanisms of the top-down scenario.
\end{abstract}

% \begin{keyword}

%\PACS 
% \end{keyword}

\end{frontmatter}

\section{\label{sec_in} Introduction}

The Ultra High Energy (UHE) neutrino detection is one the most important 
step forward in the Cosmic Ray (CR) Physics. The discovery of neutrinos with 
energy larger than $10^{17}$ eV will start the neutrino astronomy, enabling 
the observation of the most distant and powerful sources in the Universe. 
The existence of neutrinos with such high energy is intimately related to 
the observation of Ultra High Energy Cosmic Rays (UHECR). 

Soon after the discovery of the Cosmic Microwave Background (CMB) radiation 
it was shown that the flux of UHECR, with an energy larger than $10^{18}$ eV, 
should be characterized by a sharp steepening at energies $\sim 10^{20}$~eV, 
due to the absorption processes on the CMB radiation. This effect 
is the well known Graisen-Zatsepin-Kuzmin (GZK) cut-off \cite{GZK}. 
After a few decades of observations the detection of the GZK steepening 
is still one of the major open problems in UHECR physics and the experimental 
data are not conclusive. The 11 Akeno Grand Air Shower Array (AGASA) events 
with energy larger than $10^{20}$ eV \cite{expCR} contradict the expected 
suppression of the UHECR spectrum. On the other hand the HiRes data seem to 
be consistent with the GZK cut-off picture \cite{expCR}. If the UHECR 
primaries are protons and if they propagate rectilinearly, as the claimed 
correlation with BL-Lacs at energy $4-8\times 10^{19}$ eV implies \cite{TT}, 
than their sources must be seen in the direction of the highest energies 
events with energies up to $2-3 \times 10^{20}$ eV detected by HiRes, 
Fly's Eye and AGASA \cite{expCR}. At these energies the proton attenuation 
length is only about $20-30$ Mpc and no counterparts in any frequency band 
was observed in the direction of these UHECR events. This is a strong 
indication that CR particles with energies larger than $10^{20}$ eV may 
have a different origin from those with lower energies. 

Models of origin of UHECR fall into two categories, top-down and 
bottom-up. In the bottom-up scenario UHECR originates from cosmic 
accelerators. In these accelerators particles of relatively low energy are 
brought to UHE through multiple interactions at the source. The most 
promising accelerators known in nature are based on the diffusive shock 
acceleration mechanism, in which the particle acceleration is realized through 
multiple interactions with a shock front. This mechanism works fairly well in 
Super Nova Remnants (SNR) that are believed to be the responsible for the 
acceleration of Galactic Cosmic Rays of energy $E<10^{18}$ eV (for a review 
see \cite{Hillas}). At the highest energies different bottom-up scenario have 
been proposed, among them, the most promising, are those in which acceleration 
is realized through the interaction with a relativistic shock front in Active 
Galactic Nuclei (AGN) and Gamma Ray Bursts (GRB). In the framework of 
bottom-up 
models the observed UHECR flux should show the predicted GZK steepening and 
UHE neutrinos are produced by the interaction of UHECR with different 
backgrounds, at the source and during their journey to us. These are the 
so-called cosmogenic neutrinos (first proposed in \cite{BereZats}) that we 
will not discuss here.

The presence of an excess of events as claimed by AGASA inspired the 
introduction of several exotic models for the production of UHECR. These 
models, collectively called top-down, explain the excess of AGASA and give 
also a clear explanation for the lacking of any counterpart of the highest 
energy events. Many different ideas have been proposed among top-down models:
strongly interacting neutrinos \cite{nu} and new light hadrons \cite{gluino} 
as unabsorbed signal carriers, $Z$-bursts \cite{Z}, Lorentz-invariance 
violation \cite{Lorentz}, Topological Defects (TD) (see \cite{TD} for a 
review), and Superheavy Dark Matter (SHDM) (see \cite{SHDM} for a review).

In the present paper we will concentrate our attention on the two last models 
that show common features: UHE particles are produced in the decay of 
superheavy particles, that we shall call collectively $X$ particles,
with a typical mass of the order of the Grand Unified energy scale 
$M_{GUT}\simeq 10^{24}$ eV. In the case of TD the $X$ particle once
produced, by the internal dynamics of the defect or through the interaction
of different defects, immediately decays. While in the case of SHDM the $X$
particle itself is long-lived contributing to the Dark Matter of the universe. 

From the point of view of elementary particle physics the $X$ particle 
decay process proceed in a way similar to $e^+e^-$ annihilation into hadrons:
two or more off-mass-shell quarks and gluons are produced and they initiate 
a QCD cascade. Finally the partons are hadronized at the confinement radius. 
Most of the hadrons in the final state are pions and thus the typical
prediction of all these models is the dominance of neutrinos and photons at the
highest energies $E \ge 5\times 10^{19}$~eV. It is important to stress here
that these models predict neutrino fluxes most likely within reach of the 
first generation neutrino telescopes such as AMANDA, and certainly detectable 
by future kilometer-scale neutrino observatories \cite{GHT}.

\section{\label{sec_1} Hadrons spectrum in X decay}

The first step to determine the neutrino flux produced in the 
decay of $X$ particles is the determination of the hadron 
spectrum. Moreover, this evaluation is particularly important because 
it represents a direct signature of the production mechanism that, in 
principle, can be detected experimentally. As discussed in the introduction, 
the mass of the decaying particle, $M_X$, that represents the total CMS 
energy $\sqrt{s}$, is in the range $10^{13}$ -- $10^{16}$~GeV. 

The existing QCD Monte Carlo (MC) codes become numerically unstable at much 
smaller energies, e.g., at $\sqrt{s} \sim 10^7$~GeV and the computing time 
increases rapidly going to larger energies. In this section we will briefly 
review the main results obtained, in the computation of the top-down spectrum 
of UHE particles, using two different computational techniques: one based on a 
new MC scheme \cite{ABK,BereKach} and the other based on the 
Dokshitzer-Gribov-Lipatov-Altarelli-Parisi (DGLAP) evolution equations 
\cite{ABK,previous}. In both cases SUSY is included in the computation.

Monte Carlo simulations are the most physical approach for high
energy calculations which allow to incorporate many important physical
features as the presence of SUSY partons in the cascade and
coherent branching. The perturbative part of a QCD Monte Carlo simulation 
is quite standard with the inclusion of SUSY. For the non-perturbative 
hadronization part an original phenomenological approach is used in 
Ref.~\cite{ABK}. The fragmentation of a parton $i$ into an hadron $h$ is 
expressed through perturbative fragmentation function of partons 
$D_i^j(x,M_X)$, that represents the probability of fragmentation of a parton 
$i$ into a parton $j$ with momentum fraction $x=2p/M_X$, convoluted with the 
hadronization functions $f_j^h(x,Q_0)$ at scale $Q_0$, that is understood 
as the fragmentation function of the parton $i$ into the hadron $h$ at the 
hadronization scale $Q_0\simeq 1.4$ GeV \cite{ABK}. To obtain the 
fragmentation functions of hadrons one has:

\begin{equation}
D_i^h(x,M_X)=
\sum_{j=q,g}\int_x^1\frac{dz}{z}D_i^j(\frac{x}{z},M_X)f_j^h(z,Q_0)
\label{hfunc}
\end{equation}

\noindent where the hadronization functions do not depend on the scale
$M_X$. This important property of hadronization functions allows the  
determination of $f_i^h(x,Q_0)$ from the available LEP data, $D_i^h(x,M_X)$ 
at the scale $M_X=M_Z$. 

The fragmentation functions $D_i^h(x,M_X)$ at a high scale $M_X$ can
be calculated also evolving them from a low scale, e.g. $M_X=M_Z$, where they 
are known experimentally or with great accuracy using the MC scheme. This 
evolution is described by the Dokshitzer-Gribov-Lipatov-Altarelli-Parisi 
(DGLAP) equation \cite{DGLAP} which can be written as

\begin{equation}
\partial_t D_i^h=\sum_j\frac{\alpha_s(t)}{2\pi}P_{ij}(z)\otimes
D_j^h(x/z,t)\,,
\label{DGLAP-eq}
\end{equation}

\noindent where $t=\ln(s/s_0)$ is the scale, $\otimes$ denotes the convolution
$f\otimes g=\int_z^1 dx/x f(x)g(x/z)$, and $P_{ij}$ is the splitting function 
which describes the emission of parton $j$ by parton $i$. Apart from the 
experimentally rather well determined quark fragmentation function 
$D_q^h(x,M_Z)$, also the gluon fragmentation function $D_g^h(x,M_Z)$ 
is needed for the evolution of Eq.~(\ref{DGLAP-eq}). The gluon FF can be taken 
either from MC simulations or from fits to experimental data, in particular
to the longitudinal polarized $e^+e^-$ annihilation cross-section and
three-jet events. The first application of the DGLAP method for the 
calculation of hadron spectra from decaying superheavy particles has been 
made in Refs.~\cite{previous}. The most detailed calculations have been 
performed by Barbot and Drees \cite{previous}, where more than 30 different 
particles were allowed to be cascading and the mass spectrum of the SUSY 
particles was taken into account. The results obtained with the two different 
techniques discussed above agree fairly well \cite{ABK}. The accuracy in the 
hadron spectrum calculations has reached such a level that one can consider 
the spectral shape as a signature of the model. The predicted hadron spectrum 
is approximately $\propto dE/E^{1.9}$ in the region of $x$ relevant for UHECR 
observations.

\section{\label{sec_2} Spectra of Neutrinos, Photons and Nucleons}

The spectra of neutrinos and photons produced by the decay of
superheavy particles are of practical interest in high energy
astrophysics and can be computed from the decay of charged pions
\cite{ABK}. The FFs for charged pions and protons+antiprotons 
can be determined, following \cite{ABK}, from the FFs of hadrons 
$D_h$ simply introducing the ratios $\varepsilon_N(x)$ and 
$\varepsilon_{\pi}(x)$ as: $D_N(x) = \varepsilon_N(x) D_h(x)$ and 
$ D_{\pi}(x) = \varepsilon_{\pi}(x) D_h(x)$. The spectra of pions and 
nucleons at large $M_X$ have approximately the same shape as the hadron 
spectra, and one can use in this case $\varepsilon_{\pi}=0.73\pm 0.03$ and 
$\varepsilon_{N}=0.12\pm 0.02$ \cite{ABK}.   

An interesting feature of the up-dated calculations performed in 
\cite{ABK} and by Barbot and Drees in \cite{previous} is the ratio of 
photons to nucleons, $\gamma/N$. At $x\sim 1\times 10^{-3}$ this ratio 
is characterized by a value of 2 -- 3 only \cite{ABK}. This result has an 
important impact for SHDM and topological defect models because the fraction 
of nucleons in the primary radiation increases. However, in both models 
photons dominate (i.e. their fraction becomes $\ge 50\%$) at 
$E\ge (7-8)\times 10^{19}$~eV. 

Let us now concentrate our attention on UHECR from superheavy dark matter
(SHDM) \cite{BKV97} and topological defects (TD) \cite{HiSch}.
The comparison of the UHECR spectrum obtained with the AGASA data, will 
provide us with the correct neutrino flux normalization.

Production of SHDM particles naturally occurs in a time-varying 
gravitational field of the expanding universe at the post-inflationary 
stage. The relic density of these particles is mainly determined
(at fixed reheating temperature and inflaton mass) by their
mass $M_X$. The range of practical interest is $(3 - 10)\times
10^{13}$ GeV, at larger masses the SHDM is a subdominant component of
the DM. SHDM is accumulated in the Galactic halo with the overdensity 
$\delta= \frac{\bar{\rho}_X^{\rm halo}}{\rho_X^{\rm extr}}=
\frac{\bar{\rho}_{\rm DM}^{\rm halo}}{\Omega_{\rm CDM}\rho_{\rm cr}}$,
where $ \bar{\rho}_{\rm DM}^{\rm halo}\approx 0.3$ GeV/cm$^3$, 
$\rho_{\rm cr}=1.88\times 10^{-29}h^2$ g/cm$^3$ and 
$\Omega_{\rm CDM}h^2=0.135$ \cite{WMAP}. With these numbers, 
$\delta \approx 2.1\times 10^5$. Because of this large local overdensity,
UHECRs from SHDM have no GZK cutoff.

Clumpiness of SHDM in the halo can provide the observed small-angle
clustering. The ratio $r_X=\Omega_X (t_0/\tau_X)$ of relic
abundance $\Omega_X$ and lifetime $\tau_X$ of the $X$ particle is
fixed by the observed UHECR flux as $r_X\sim 10^{-11}$. 
In the most interesting case of gravitational production of $X$
particles, their present abundance is determined by their
mass $M_X$ and the reheating temperature $T_R$.
Choosing a specific particle physics model one can fix also the life-time 
of the $X$ particle. There exist many models in which SH particles can
be quasi-stable with lifetime $\tau_X \gg 10^{10}$~yr.
The measurement of the UHECR flux, and thereby of $r_X$, selects from
the three-dimensional parameter space $(M_X, T_R, \tau_X)$ a
two-dimensional subspace compatible with the SHDM hypothesis.

In Figure \ref{fig} (left panel) we have performed a fit to the AGASA data 
using the photon flux from the SHDM model and the proton flux from uniformly 
distributed astrophysical sources. For the latter we have used the 
non-evolutionary model of \cite{dip}. The photon flux is normalized to 
provide the best fit to the AGASA data at $E\geq 4\times 10^{19}$~eV. 

\begin{figure*}[t]
\begin{minipage}[t]{0.48\linewidth}
\centering\epsfig{file=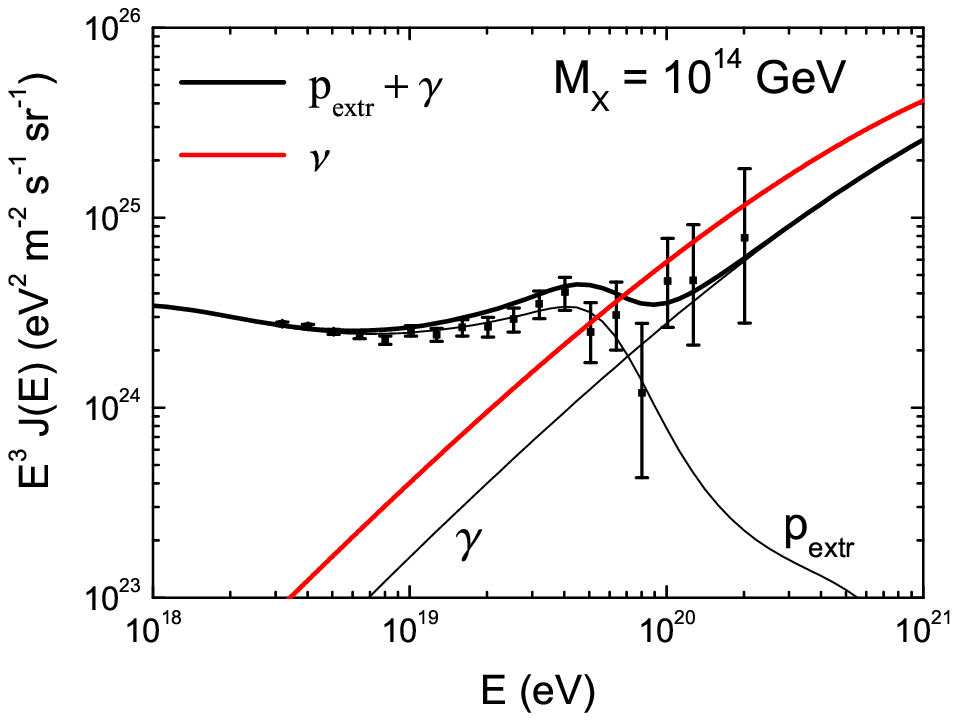,height=\linewidth,width=\linewidth}
\end{minipage}\hfill
\begin{minipage}[t]{0.48\linewidth}
\centering\epsfig{file=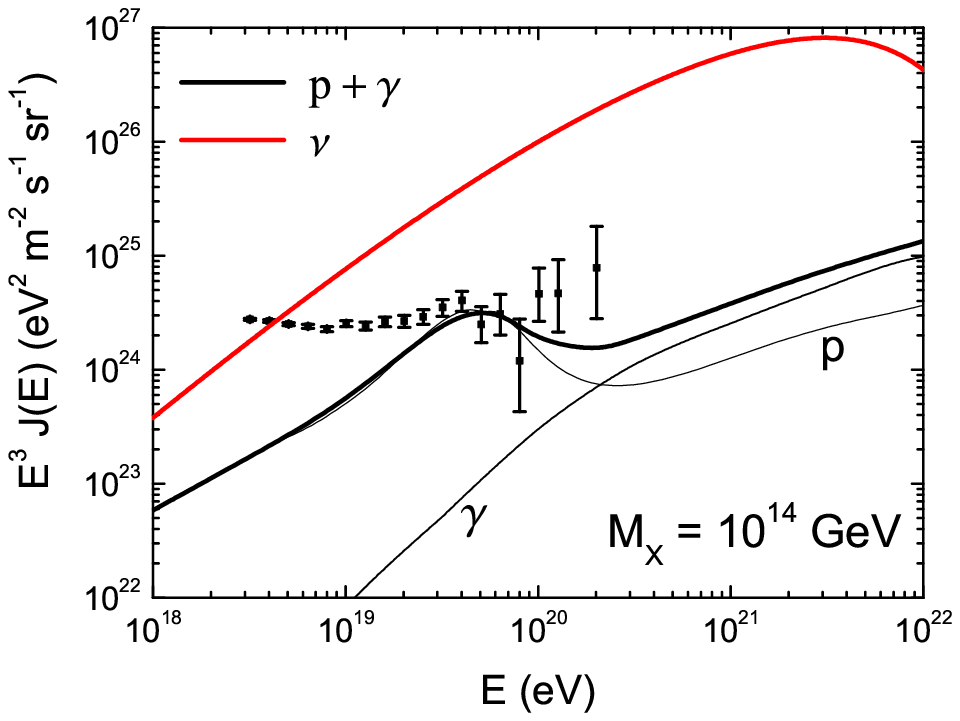,height=\linewidth,width=\linewidth}
\end{minipage}
\caption{[Right Panel] Comparison of SHDM prediction with the AGASA data. 
The calculated spectrum of SHDM photons is shown by the label $\gamma$ and 
by the label $p_{extr}$ the spectrum of extragalactic protons (uniformly 
distributed astrophysical sources). The sum of these two spectra is shown 
by the thick black curve. The red thick line is the SHDM neutrino flux.
[Left Panel] Diffuse spectra from necklaces. The red thick curve shows 
neutrino flux, the black thick curve is the sum of protons and photons fluxes 
produced by necklaces (labeled thin black lines).}
\label{fig}
\end{figure*}

One can see  from the fit of Figure \ref{fig} (left panel), that the SHDM
model can explain only the excess of AGASA events at 
$E \ge 1\times 10^{20}$~eV: depending on the SHDM spectrum
normalization and the details of the calculations for the extragalactic
protons, the flux from SHDM decays becomes dominant only  
above $(6-8)\times 10^{19}$~eV.

Topological Defects (for a review see \cite{Xpart} and reference therein)
can naturally produce UHE particles. 
The following TD have been discussed as potential sources of UHE
particles: superconducting strings, ordinary strings,
monopolonium (bound monopole-antimonopole pair), monopolonia
(monopole-antimonopole pairs connected by a string), networks of 
monopoles connec\-ted by strings, vortons and necklaces (see Ref.~\cite{Xpart}
for a review and references).
Monopolonia and vortons are clustering in the Galactic halo and their 
observational signatures for UHECR are identical to SHDM. However the friction 
of monopolonia in cosmic plasma results in monopolonium lifetime much 
shorter than the age of the universe. 
Of all other TD which are not clustering in the Galactic halo, the most 
favorable for UHECR are {\em necklaces}. 

Necklaces are hybrid TD produced in the symmetry breaking pattern
$G \rightarrow H \times U(1) \rightarrow H \times Z_2$. At the first
symmetry breaking monopoles are produced, at the second one each
(anti-) monopole get attached to two strings. This system resembles
ordinary cosmic strings with monopoles playing the role of
beads. Necklaces exist as the long strings and loops.
The symmetry breaking scales of the two phase transitions, $\eta_m$
and $\eta_s$, are the main parameters of the necklaces. They determine
the monopole mass, $m \sim 4\pi \eta_m/e$, and the mass of the string per
unit length $\mu \sim 2\pi \eta_s^2$. The evolution of necklaces is
governed by the ratio $r\sim m/\mu d$, where $d$ is the average
separation of a monopole and antimonopole along the string. As it is
argued in Ref.~\cite{neckl}, necklaces evolve towards configuration with
$r\gg 1$. Monopoles and antimonopoles trapped in the necklaces inevitably 
annihilate in the end, producing heavy Higgs and gauge bosons ($X$ particles) 
and then hadrons. The rate of $X$ particles production in the universe can be
estimated as \cite{neckl} $\dot{n}_X \sim \frac{r^2 \mu}{t^3 M_X}$,
where $t$ is the cosmological time.

The photons and electrons from pion decays initiate e-m cascades and
the cascade energy density can be calculated as
$\omega_{\rm cas}=\frac{1}{2}f_{\pi}r^2\mu \int_0^{t_0} \frac{dt}{t^3}
\frac{1}{(1+z)^4}=\frac{3}{4}f_{\pi}r^2\frac{\mu}{t_0^2}$,
where $z$ is the redshift and $f_{\pi}\sim 1$ is the fraction of the 
total energy release transferred to the cascade. 
The parameters of the necklace model for UHECR are restricted by the 
EGRET observations~\cite{EGRET} of the diffuse gamma-ray flux.  
This flux is produced by UHE electrons and photons from
necklaces due to e-m cascades initiated in collisions with CMB photons. In the
range of the EGRET observations, $10^2 - 10^5$~MeV, the predicted spectrum
is $\propto E^{-\alpha}$ with $\alpha=2$ \cite{cascade}. The EGRET
observations determined the spectral index as $\alpha=2.10\pm 0.03$
and the energy density of radiation as $\omega_{\rm obs} \approx
4\times 10^{-6}$~eV/cm$^3$. The cascade limit consists in the 
bound $\omega_{\rm cas}\leq \omega_{\rm obs}$.
According to the recent calculations, the Galactic contribution of gamma 
rays to the EGRET observations is larger than estimated earlier, and the 
extragalactic gamma-ray spectrum is not described by a power-law with 
$\alpha=2.1$. In this case, the limit on the cascade radiation with $\alpha=2$ 
is more restrictive and is given by 
$\omega_{\rm cas} \leq 2\times 10^{-6} {\rm eV/cm}^3$; we shall use this limit 
in further estimates. Using $\omega_{\rm cas}$ with $f_{\pi}= 1$ and 
$t_0=13.7$~Gyr \cite{WMAP} we obtain from the limit on the cascade radiation 
$r^2\mu \leq 8.9\times 10^{27}$~GeV$^2$.

The important and unique feature of necklaces is their small separation $D$, 
which ensures an high density. The distance $D$ is given by 
$D \sim r^{-1/2}t_0$ \cite{neckl}; since $r^2\mu$ is limited by e-m cascade 
radiation we can obtain a lower limit on the separation between necklaces as
$D \sim \left (\frac{3f_{\pi}\mu}{4t_0^2\omega_{\rm cas}} \right )^{1/4}t_0
> 10(\mu/10^6~{\rm GeV}^2)^{1/4}~{\rm kpc}$,
this small distance is a unique property of necklaces allowing the
unabsorbed arrival of particles with the highest energies.
The fluxes of UHECR from necklaces are shown in Figure \ref{fig} (right panel).
We used in the calculations $r^2\mu = 4.7\times 10^{27}$~GeV$^2$ which 
corresponds to $\omega_{\rm cas}= 1.1\times 10^{-6}$~ eV/cm$^3$, i.e. 
twice less than allowed by the bound on $\omega_{\rm cas}$. The mass of the $X$
particles produced by monopole-antimonopole annihilations is taken as  
$M_X= 1\times 10^{14}$~GeV. From  Figure \ref{fig} (right panel) one can see 
that the necklaces model for UHECR can explain only the highest energy part 
of the spectrum, with the AGASA excess somewhat above the prediction. Thus 
UHE particles from necklaces can serve only as an additional component in the 
observed UHECR flux. This result has a particular impact on the possible UHE 
neutrino detection. In fact, the necklaces model is only under constrained by 
the available UHECR data, in this context only a clear UHE neutrino 
observation with a typical spectrum as in figure \ref{fig} (right panel) 
can confirm (or falsify) the model.
  
\section*{Acknowledgments}
%% Keep the small font tag for the acknowledegments
{\small 
I am grateful to V. Berezinsky and M. Kachelrie{\ss} with whom the 
present work was developed. 
}

%%%%%%%%%%%%%%%%%%%%%%%%%%%%%%%%%%%%%% reset.txt counters %%%%%%%%%%%%%%
%%
%%%%%%% do not change below here  %%%%%%%%%%%%%%%%%%%%%%%%%%%%%
%%
\setcounter{section}{0}
\setcounter{subsection}{0}
\setcounter{figure}{0}
\setcounter{table}{0}
\setcounter{equation}{0}
\newpage

%%%%%%%%%%%%%%%%%%%%%%%%%%%%%%%%%%%%%%%%%%%%%%%%%%% Title, authors and addresses
\begin{frontmatter}

\title{Radio detection of UHE  Neutrinos off the Moon}

\author[KVI]{O. Scholten},
\author[KVI]{J. Bacelar},
\author[ASTRON]{R. Braun},
%\email{rbraun@astron.nl}
\author[ASTRON,RUG]{A.~G. de Bruyn},
\author[ASTRON,RU]{H. Falcke},
\author[ASTRON,UvA]{B. Stappers},
\author[ASTRON,UvA]{R.~G. Strom}
%\email{strom@astron.nl}

\address[KVI]{Kernfysisch Versneller Instituut, University of Groningen,\\
9747 AA, Groningen, The Netherlands}
\address[ASTRON]{ASTRON, 7990 AA Dwingeloo, The Netherlands}
\address[RUG]{Kapteyn Institute, University of Groningen, 9747
AA, Groningen, The Netherlands}
\address[RU]{Department of Astrophysics,
IMAPP, Radboud University, \\6500 GL Nijmegen, The Netherlands}
\address[UvA]{Astronomical Institute `A. Pannekoek',
University of Amsterdam, 1098 SJ, The Netherlands}
\begin{center}{\small{(scholten@kvi.nl)}}\end{center}

\begin{abstract}
  When high-energy cosmic rays impinge on a dense dielectric medium, radio waves
are produced through the Askaryan effect. At wavelengths comparable to the
length of the shower produced by an Ultra-High Energy cosmic ray or neutrino,
radio signals are an extremely efficient way to detect these particles. This
approach offers, for the first time, the realistic possibility of measuring UHE
neutrino fluxes below the Waxman-Bahcall limit.
\end{abstract}

% \begin{keyword}

%\PACS
% \end{keyword}

\end{frontmatter}

%%%%%%%%%%%%%%%%%%%%%%%%%%%%%%%%%%%%%%%%%%%%%%%%%%%%%% MAIN TEXT
\section{\label{sec:intro} Introduction}

Askaryan predicted as early as 1962~\cite{Ask62} that particle showers in dense
media produce coherent pulses of microwave \v{C}erenkov radiation. Recently
this prediction was confirmed in experiments at accelerators~\cite{Sal01} and
extensive calculations have been performed on the development of showers in
dense media to yield quantitative predictions for this effect~\cite{Zas92}. The
pulses emitted when UHE particles strike the lunar regolith are detectable at
Earth with radio telescopes, an idea first proposed by Dagkesamanskii and
Zheleznyk~\cite{Dag89}. Several experiments have since been
performed~\cite{Han96,Gor04} to find evidence for UHE neutrinos. All of these
experiments have looked for this coherent radiation near the frequency where
the intensity of the emitted radio waves is expected to reach its maximum.
Since the typical lateral size of a shower is of the order of 10~cm the peak
frequency is of the order of 3~GHz.

We have proposed~\cite{Sch06} a different strategy to look for the radio waves
at considerably lower frequencies where the wavelength of the radiation is
comparable in magnitude to typical longitudinal size of showers. The lower
intensity of the emitted radiation, which implies a loss in detection
efficiency, is compensated by the increase in detection efficiency due to the
near isotropic emission of coherent radiation. The net effect is an increased
sensitivity by several orders of magnitude, for the detection of UHE cosmic
rays and neutrinos~\cite{Sch06}.

\section{Model for Radio Emission}

There exist two rather different mechanisms for radio emission from showers
triggered by UHE cosmic rays or neutrinos. One is the emission of radio waves
from a shower in the terrestial atmosphere. Here the primary mechanism is the
synchrotron acceleration of the electrons and positrons in the shower due to
the geomagnetic field, called geosynchrotron radiation, which has recently been
confirmed with new digital radio techniques~\cite{Fal05}. The second mechanism,
known as the Askaryan effect~\cite{Ask62}, applies to showers in dense media,
such as ice, salt, and lunar regolith, where the front end of the shower has a
surplus of electrons. Since this cloud of negative charge is moving with a
velocity which exceeds the velocity of light in the medium,
\v{C}erenkov radiation is emitted. For a wavelength in the radio-frequency
range, coherence builds up and the intensity of the emitted radiation reaches a
maximum.

The intensity of radio emission from a hadronic shower, with energy $E_s$, in
the lunar regolith, in a bandwidth $\Delta\nu$ at a frequency $\nu$ and an
angle $\theta$, can be parameterized as~\cite{Sch06}
\beq
 F(\theta,\nu,E_s)
 = {3.86 \times 10^4\; e^{-Z^2}\Delta\nu \over 100\mbox{ MHz}}
  \Big( {\sin{\theta}\over \sin{\theta_c}}
  {E_s \over 10^{20} \mbox{ eV} }
% \nonumber \\ &&\hspace*{-1.5em}  \times
  {d_{moon} \, \nu \over d \, \nu_0 (1+(\nu/\nu_0)^{1.44})} \Big)^2
  \mbox{Jy} \;,
 \label{shower-l}
\eeq
where $Z
 = (\cos{\theta} -1/n)
 \Big({n\over \sqrt{n^2-1}}\Big)\Big({180\over \pi \Delta_c}\Big)
$, $\nu_0=2.5$~GHz, $d$ is the distance to the observer, and $d_{moon}=3.844
\times 10^8$~m is the average Earth-Moon distance. The angle at which the
intensity of the radiation reaches a maximum, the
\v{C}erenkov angle, is related to the index of refraction ($n$) of the medium,
$\cos{\theta_c}=1/n$.

The spreading of the radiated intensity around the \v{C}erenkov angle,
$\Delta_c$, is, on the basis of general physical arguments, inversely
proportional to the shower length and the frequency of the emitted radiation.
Based on the results given in Ref.~\cite{Alv01} it can be parameterized as
\beq
\Delta_c = 4.32^\circ
 \Big({1\over \nu\,[\mbox{GHz}] }\Big)
 \Big({L(10^{20}\mbox{eV})\over L(E_s) }\Big) \;,
\label{del_c_had'}
\eeq
where $L(E_s)$ is the shower length which depends on the energy. In
\figref{ang-spread} the spreading angle is compared at two frequencies with
different phenomenological descriptions and with different analytical models.
At the higher frequency all descriptions give the same result while at the
lower frequency the prediction of \ref{shower-l} is consistent with the
analytical models.

In our simulations we have taken into account the attenuation of radio waves in
the Lunar regolith. As a mean value for the attenuation length for the radiated
power we have taken $\lambda_r= (9/\nu$[GHz])~m~\cite{Olh75}.

\begin{figure*}[t]
\begin{minipage}[t]{0.48\linewidth}
\centering \includegraphics[width=\linewidth,bb=50 147 515 650,clip]%
{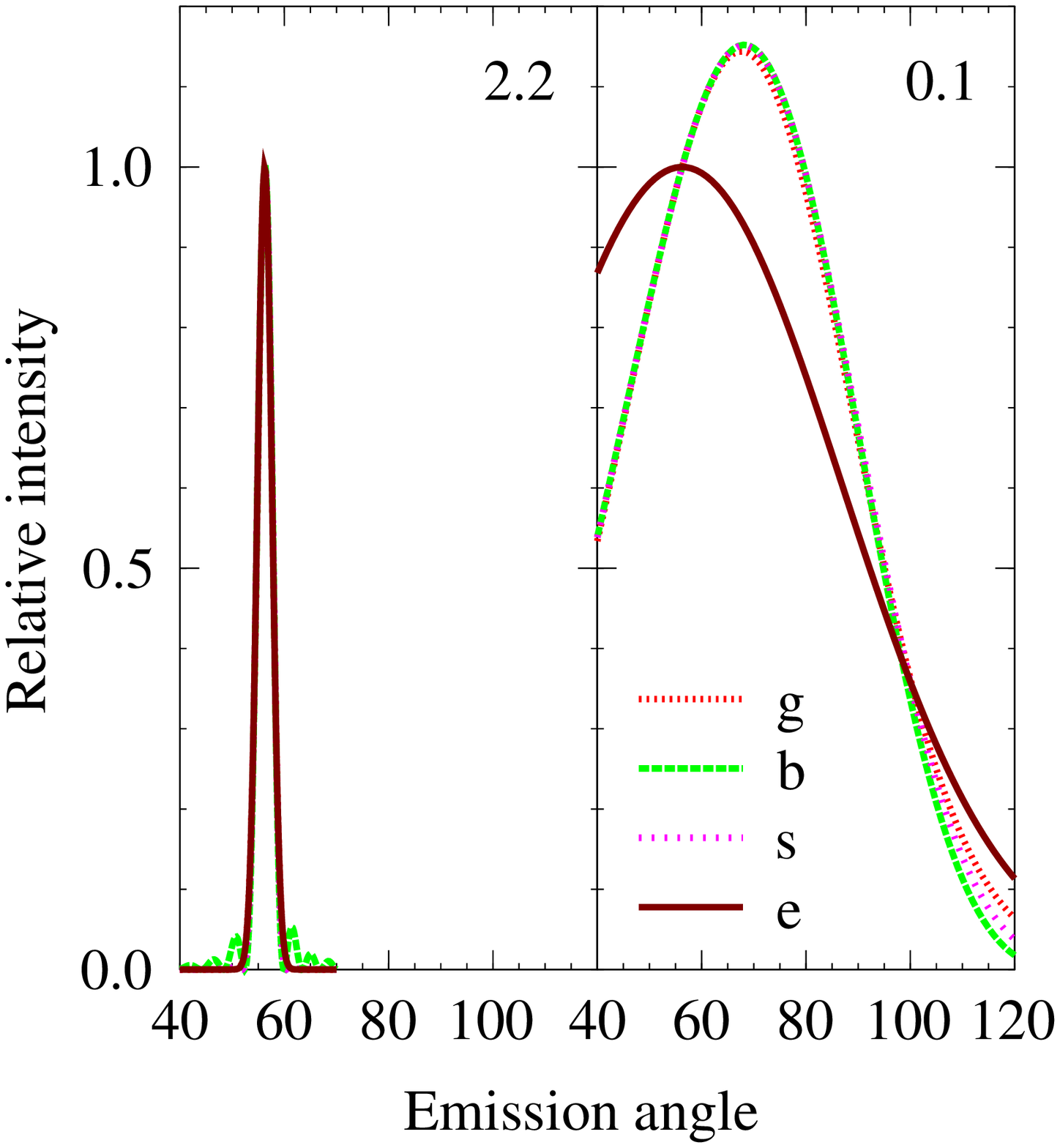}
\caption[fig9]{The angular spread around the \v{C}erenkov angle.
The left (right) hand displays the results for 2.2 GHz
(100MHz) respectively. }
  \figlab{ang-spread}
\end{minipage}
\hfill
\begin{minipage}[t]{0.48\linewidth}
\centering \includegraphics[width=\linewidth,bb=36 137 515 672,clip]%
{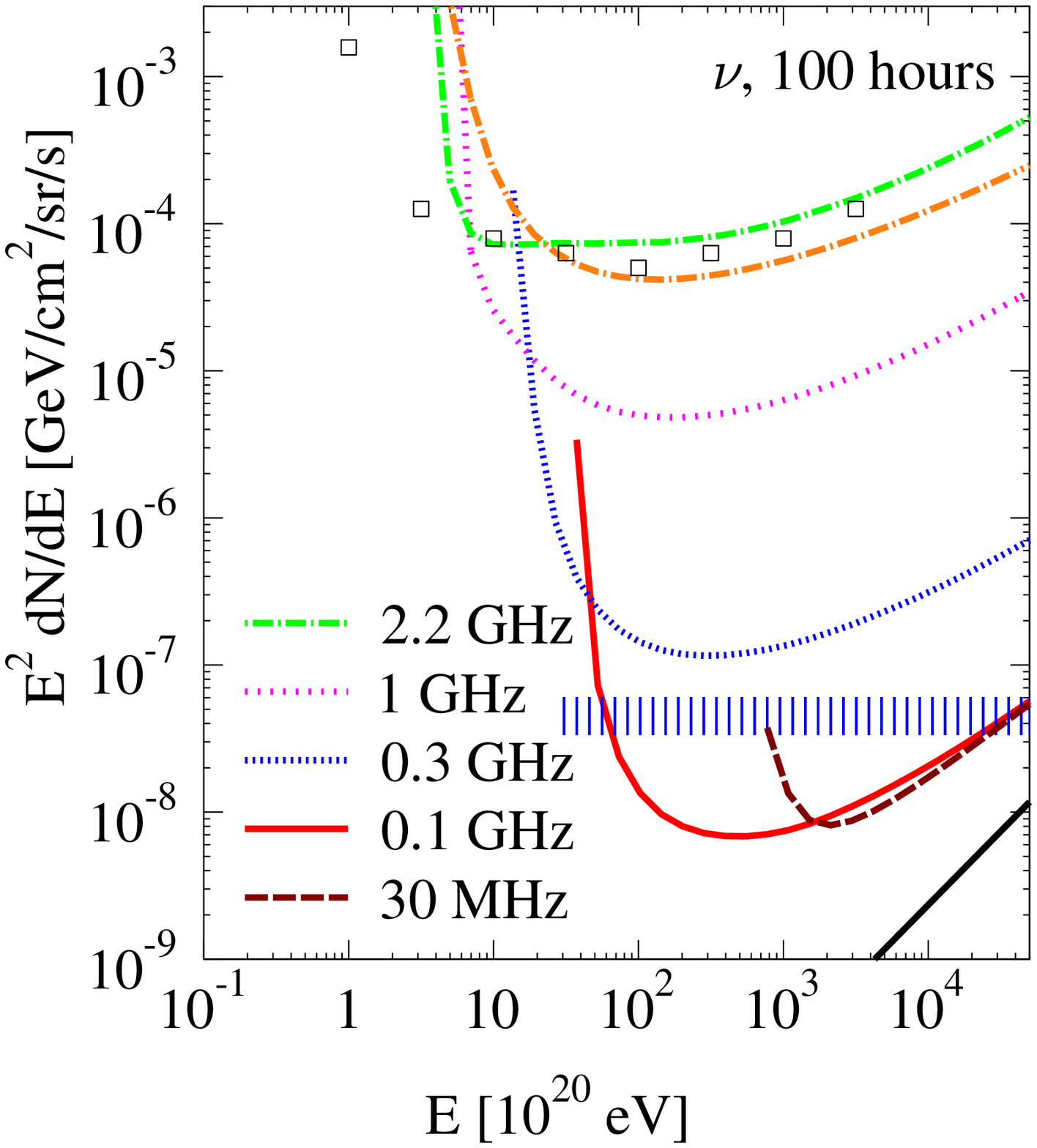}
\caption[fig4]{Flux limits (assuming a null observation) for UHE neutrinos
as can be determined in a 100 hour observation.}
  \figlab{Flux-n}
\end{minipage}
\end{figure*}

In \figref{Flux-n} we compare the detection limits for UHE neutrinos at
different frequencies with the results obtained from the GLUE
experiment~\cite{Gor04}. One sees that with decreasing frequency one loses
sensitivity at lower-energies. This follows directly from
\ref{shower-l} since with decreasing frequency the maximum signal strength
decreases. If the energy is more than a factor 4 above this threshold value the
detection limit improves rapidly with decreasing radio-frequency until one
reaches a frequency of 100~MHz where one obtains the optimum sensitivity.
Decreasing the frequency even lower provides no gain since the detection limit
has already reached the optimum, given by the heavy black line.

Our result at 2.2~GHz lie close to that of the GLUE experiment. The
dashed-dotted curve in \figref{Flux-n} shows the results of a calculation where
we have reproduced the simulation for the GLUE experiment, i.e.\ including (in
a somewhat simplified manner) the effects of averaging over lunar-surface
slopes of 10$^\circ$. This result lies close to the published limits (the open
squares in \figref{Flux-n}).

\section{WSRT \& LOFAR predictions\seclab{RealObs}}

The Westerbork Synthesis Radio Telescope (WSRT) consists of fourteen 25~m
parabolic dishes located on an east-west baseline extending over 2.7~km. The
low frequency band which concerns us here covers 115-170~MHz~\cite{WSRT-l}.
Each WSRT element has two receivers with orthogonal dipoles enabling
measurement of all four Stokes parameters. In tied-array mode the system noise
at low frequencies is $F_{noise}=$600~Jy. To observe radio bursts of short
duration, the new pulsar backend (PuMa II) is used. It can provide
dual-polarization baseband sampling of eight 20~MHz bands, enabling a maximum
time resolution of approximately the inverse of the bandwidth. In the
configuration which we propose to use, four frequency bands will observe the
same part of the moon with the remaining four a different section. In total,
coverage of about 50\% of the lunar disk can be achieved.

An even more powerful telescope to study radio flashes from the moon will be
the LOFAR array~\cite{LOFAR}. With a collecting area of about 0.05~km$^2$ in
the core (which can cover the full moon with an array of beams), LOFAR will
have a sensitivity about 25 times better than that of the WSRT.  LOFAR will
operate in the frequency band from 115-240~MHz where it will have a sensitivity
of about $F_{noise}=$20~Jy.

\begin{figure*}[t]
\begin{minipage}[t]{0.48\linewidth}
\centering
 \includegraphics[width=\linewidth,bb=27 137 515 672,clip]{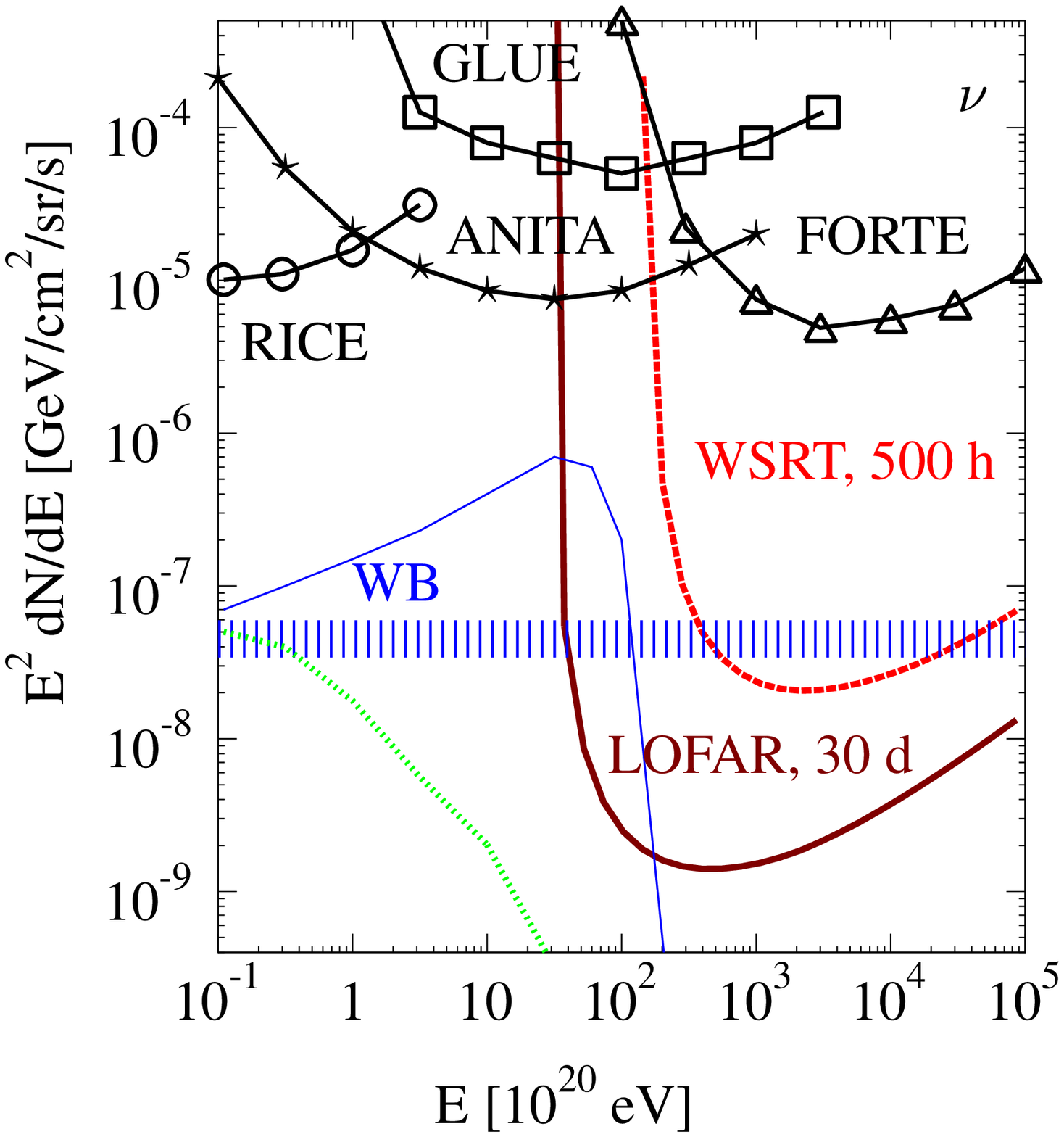}
\caption[fig6]{Flux limits on UHE neutrinos as can be determined
with WSRT and LOFAR observations (see text) are compared with various models.}
  \figlab{LOFAR}
\end{minipage}
\hfill
\begin{minipage}[t]{0.48\linewidth}
\centering
 \includegraphics[width=\linewidth,bb=27 137 515 672,clip]{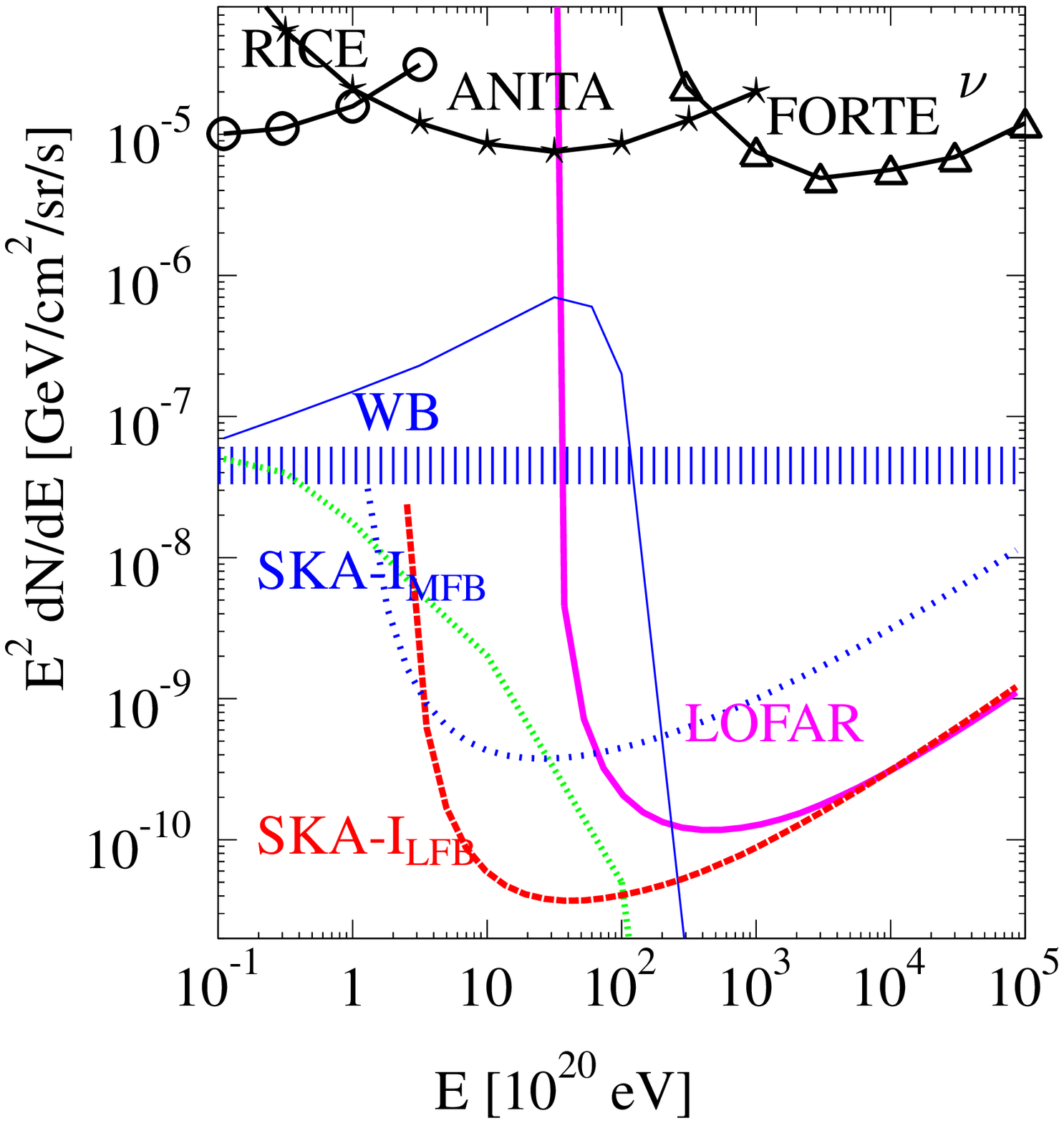}
\caption[fig6]{Flux limits on UHE cosmic rays as can be determined
in a 1 year observation with the LOFAR antenna system compared to that with SKA
in 2 frequency bands.}
  \figlab{SKA}
\end{minipage}
\end{figure*}

The probability for finding during the complete observation time a signal
simultaneously in four frequency bands with a power of 20$\times F_{noise}$ per
frequency band, where $F_{noise}$ is the noise level per polarization
direction, as a random fluctuation of the background noise is less than 0.001
(equal to $3
\sigma$ significance). For this reason we have assumed in the calculations a
detection threshold of 25$\times F_{noise}$ for both the WRST and the LOFAR
telescopes.

A simulation for LOFAR, taking $\nu=120$~MHz, bandwidth of $\Delta\nu=20$~MHz,
and an observation time of 30 days is shown in \figref{LOFAR} for neutrinos.
The results are compared with the limits that can be obtained from a presently
proposed observation for 500 hours at the WSRT observatory assuming a 50\% Moon
coverage. The predicted spectrum of GZK neutrinos is taken from
Ref.~\cite{Eng01}. The prediction of top-down (TD) neutrinos is based on the
decay of a topological defect with a mass of $10^{24}$~eV. The Waxman-Bahcall
(WB) limit~\cite{Bah01}, based on theoretical arguments, is an upper limit on
the neutrino flux which is consistent with the data on the fluxes of UHE cosmic
rays. The limits are also compared with those from the  RICE~\cite{Kra03},
GLUE~\cite{Gor04}, ANITA~\cite{Bar06}, and FORTE~\cite{Leh04} experiments which
are calculated as model independent limits similar to our limits. The limit
from the RICE~\cite{Kra03} experiment has been calculated assuming different
power-law spectra for the neutrinos. In general such a limit lies below the
model-independent limit~\cite{Leh04}.

\section{Outlook}

Observations at the WSRT are in progress. 200 hours observation time has been
assigned to the experiment with a promise for an additional 300 hours. In the
first observations we have shown that the subtraction of RFI is well under
control. So-far we have not seen evidence for a signal from the moon and we are
in the process of setting improved limits.

For the longer term future observations are planned with the  Lunar Orbiting
Radio Detector (LORD) and with the Square Kilometer Array (SKA) telescope. The
LORD project is a collaboration between Russia (LPI, Lavochkin Association,
MSU, JINR) and Sweden (ISP)~\cite{LORD} which plans to launch a satellite
(around 2009) in an orbit around the Moon. For the detection of the radio pulse
the satellite will carry an antenna system which is sensitive to a wide band,
100-1000 MHz. The SKA project~\cite{SKA} is a collaboration between
institutions in 18 countries. The SKA is an aperture synthesis radio telescope
where the location of the site will be decided in 2007. The system will operate
in a very wide frequency band, from 100 MHz to 25 GHz. The detection
sensitivity of SKA for UHE particles is even greater than that of LORD. To show
the possibilities with these new antennas, the predicted sensitivities are
given for observations in the low, 100-300 MHZ, (LFB) and the medium, 300-500
MHz (MFB) frequency bands with SKA. The exciting possibility of a real
measurement of the GZK-neutrino flux with several hundreds of count will become
available.

\section*{Acknowledgments}
%% Keep the small font tag for the acknowledegments
{\small This work was performed as part of the research programs of the
Stichting voor Fundamenteel Onderzoek der Materie (FOM) and of ASTRON, both
with financial support from the Nederlandse Organisatie voor Wetenschappelijk
Onderzoek (NWO). We gratefully acknowledge discussions with
J.~Alvarez-Mu\~{n}iz on different aspects of shower development in dense media.
}

%%%%%%%%%%%%%%%%%%%%%%%%%%%%%%%%%%%%%% reset.txt counters %%%%%%%%%%%%%%
%%
%%%%%%% do not change below here  %%%%%%%%%%%%%%%%%%%%%%%%%%%%%
\setcounter{section}{0}
\setcounter{subsection}{0}
\setcounter{figure}{0}
\setcounter{table}{0}
\setcounter{equation}{0}
\newpage

\begin{frontmatter}

\title{Probing the variation of relic neutrino masses with extremely high-energy cosmic neutrinos}

\author[address1]{Lily~Schrempp},

\address[address1]{Deutsches Elektronen-Synchrotron DESY, Notkestra\ss e  85, 22607 Hamburg, Germany}
\begin{center}{\small{(lily.schrempp@desy.de)}}\end{center}

%\address[address2]{Department of Something Else, Univesity2}

\begin{abstract}

We analyze the prospects for testing the cosmic neutrino background and its interpretation as source of Neutrino Dark Energy with the neutrino telescope LOFAR.  

\end{abstract}

% \begin{keyword}

%\PACS 
% \end{keyword}

\end{frontmatter}

%%%%%%%%%%%%%%%%%%%%%%%%%%%%%%%%%%%%%%%%%%%%%%%%%%%%%% MAIN TEXT

\vspace{-6.mm}
\section{\label{sec:introd} Introduction}

Certainly, one of the most challenging questions in modern Cosmology and Particle physics is, what is the nature of Dark Energy? Strong observational evidence hints at the existence of this smooth, exotic energy component which drives the apparent accelerated expansion of the universe. Recently, Fardon, Nelson and Weiner~\cite{Fardon:2003eh} have shown that Big Bang relic neutrinos -- the analog of the cosmic microwave photons -- are promoted to a natural Dark Energy candidate if they interact through a new non-Standard Model force. Due to this scalar force, the homogeneously distributed relic neutrinos can form a negative pressure fluid and thus exhibit just the right properties to act as Dark Energy. As a further consequence of this new interaction, the neutrino mass becomes a function of neutrino energy density which decreases as the universe expands. Thus, intriguingly, the neutrino mass is not a constant but is promoted to a dynamical quantity. After discussing the details of this so-called Mass Varying Neutrino (MaVaN) scenario, I will consider an astrophysical possibility of testing it with extremely high energy-cosmic neutrinos ($\E$). A more detailed discussion of the results may be found in Ref.~\cite{Ringwald:2006ks}.   
\vspace{-2.5mm} 
\section{\label{sec:MaVaNs} Mass Varying Neutrinos}

I will concentrate on a concrete realization of the non-Standard Model neutrino interaction as preferred in the literature~\cite{Fardon:2003eh,Fardon:2005wc} which implements the seesaw mechanism~\cite{Gell-Mann}. Generically, varying mass particle scenarios exhibit a so-called dark sector which only consists of Standard Model singlets. In the considered case of MaVaNs it contains a light scalar field, the acceleron ${\mathcal{A}}$, which has an associated fundamental potential $V_0({A})$. The acceleron interacts with a second field of the dark sector, a right-handed neutrino $N$, through a Yukawa coupling, $\kappa{\mathcal{A}}NN$, and thus generates its mass $M_N({\mathcal{A}})=\kappa {\mathcal{A}}$. The dark force mediated by the acceleron is transmitted to the active neutrino sector via the seesaw mechanism which in addition provides a natural explanation for the smallness of the left-handed neutrino mass $m_\nu$. At scales well below $100$ GeV the Lagrangian contains a Majorana mass term~\cite{Fardon:2003eh,Ringwald:2006ks},
\begin{equation}
{\mathcal{L}}\supset \frac{m_D^2}{M_N({\mathcal{A}})}\nu^2+h.c.+V_0({\mathcal{A}}),\,\,\mbox{where}\,\,m_\nu(\mathcal{A})= \frac{m_D^2}{M_N({\mathcal{A}})},
\end{equation}
for the active left-handed neutrino, where the Dirac type mass $m_D$ originates from electroweak symmetry breaking and $M_N\gg m_D$ has been assumed. Consequently, the neutrino mass $m_\nu(\mathcal{A})$ is light and, since it is generated by the value of the acceleron, neutrinos interact through a new force mediated by ${\mathcal{A}}$.

The coupling leads to a complex interplay between the acceleron and the neutrinos which links their dynamics. Since the neutrino energy density is a function of the neutrino mass $m_\nu({\mathcal{A}})$, it becomes an indirect function of the value of the acceleron, $\rho_\nu(m_\nu({\mathcal{A}}),z)$. As a direct consequence, it stabilizes the acceleron by contributing to its effective potential,
\begin{eqnarray}
V_{\rm eff}({\mathcal{A}},z)&=&\rho_\nu(m_\nu({\mathcal{A}}),z)+V_0({\mathcal{A}}),\,\,\mbox{where}\\
\rho_{\nu}(m_\nu({\mathcal{A}}),z)&=&\frac{T_{\nu}(z)^4}{\pi^2} \int\limits_0^{\infty}\frac{dy\, y^2 \sqrt{y^2+\left(\frac{m_{\nu}({\mathcal{A}})}{T_{\nu}(z)}\right)^2}}{e^y+1},
\end{eqnarray}
with $z$ denoting the redshift and $T_{\nu_0}(z)=T_{\nu_0}(1+z)$ the neutrino temperature with $T_{\nu_0}\sim 1.69\times 10^{-4}$ eV. Since cosmic expansion causes a dilution of the neutrino energy density, also $V_{\rm eff}$ evolves with time. Assuming the curvature scale of $V_{\rm eff}$ and thus the mass of $\mathcal{A}$ to be much larger than the Hubble scale, $\partial V^2_{\rm eff}/\partial{\mathcal{A}}^2= m^2_{\mathcal{A}}\gg H^2$, the adiabatic solution to the equations of motions apply~\cite{Fardon:2003eh}. As a result, the acceleron instantaneously tracks the minimum of its effective potential, the total energy density of the coupled system, and thus varies on cosmological time scales. Consequently, the neutrino mass $m_\nu(\mathcal{A})$ which is generated from the value of the acceleron, is not a constant but promoted to a dynamical quantity. As long as $\partial m_\nu/\partial {A}$ does not vanish, its time variation is determined by 
\begin{equation}
\label{minimum}
\frac{\partial V_{\rm eff}(\mathcal{A})}{\partial\mathcal{A}}=\frac{\partial m_\nu(\mathcal{A})}{\partial\mathcal{A}}\left(\left.\frac{\partial \rho_\nu(m_\nu,z)}{\partial  m_\nu}\right|_{m_\nu=m_\nu(\mathcal{A})}+\left.\frac{\partial V_0(m_\nu)}{\partial m_\nu}\right|_{m_\nu=m_\nu(\mathcal{A})}\right)=0.
\end{equation} 
Fig.~\ref{Mass}a shows the evolution of both the effective potential $V_{\rm eff}(\mathcal{A},z)$ as well as its minimum due to changes in $\rho_\nu(z)$. 

In the following I will address a possible shortcoming of MaVaN models and how it can be avoided. Since MaVaNs attract each other through the force mediated by the acceleron, they are possibly subject to a phenomenon similar to gravitational instabilities of cold dark matter (CDM)~\cite{Afshordi:2005ym}. As long as the neutrinos are still relativistic, their random motions can prevent a collapse. However, when they turn non-relativistic the system can become unstable leading to the possible formation of so-called `neutrino nuggets'~\cite{Afshordi:2005ym}. If the neutrinos really clump, they would redshift like CDM with $w\sim 0\neq -1$ and thus cease to act as Dark Energy. However, firstly, certain constraints on the scalar potential $V_0$ and on the function $m_\nu({A})$, can lead to stable MaVaN models even in the highly non-relativistic regime~\cite{Takahashi:2006jt}. Secondly, neutrino oscillation experiments allow one neutrino to be still relativistic today, which could be responsible for cosmic acceleration until the present time~\cite{Afshordi:2005ym,Fardon:2005wc}. In Ref.~\cite{Fardon:2005wc} the latter case emerges naturally after a straightforward super-symmetrization of the standard MaVaN scenario. However, the modified model relies on a slightly different acceleration mechanism known from Hybrid inflation~\cite{Linde:1993cn} to be considered in the following. A scalar field keeps a second scalar field in a metastable minimum due to its large value and the energy density stored in the false minimum can drive cosmic acceleration. In the MaVaN hybrid model the first scalar field can be identified with the acceleron, which is driven to larger values due to the stabilizing effect of the fermionic neutrino background. It keeps the scalar partner ${N}$ of the lightest neutrino, naturally present in a super-symmetric theory, in a false minimum until it reaches a critical value ${\mathcal{A}}_{\rm crit}$. The combined scalar potential $V_0({\mathcal{A}},{N})$ appears as dark energy with $w\sim -1$ and can therefore drive accelerated expansion. This hybrid model provides a microscopic origin for a quadratic scalar potential $V_0({\mathcal{A}})\propto {\mathcal{A}}^2$ and thus according to eq.~\ref{minimum} fixes the neutrino mass evolution. After generalizing eq.~\ref{minimum} to include three neutrino species, the evolution of the neutrino masses in the low redshift regime is found to be well approximated by a simple power law~\cite{Ringwald:2006ks},

\begin{figure*}[t]
\begin{minipage}[t]{0.48\linewidth}
\centering\epsfig{file=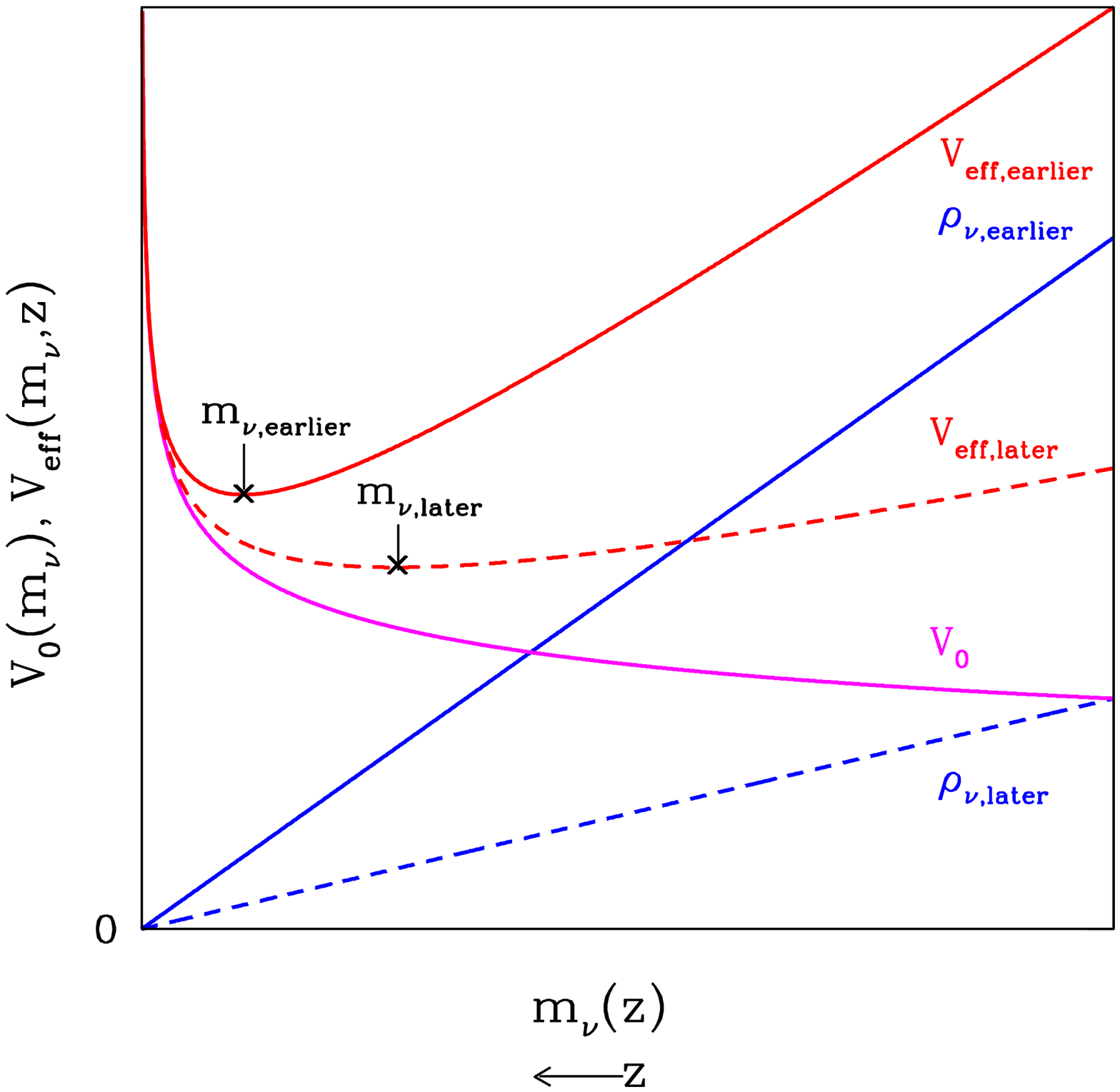,height=4cm,width=5cm}
\end{minipage}\hfill
\begin{minipage}[t]{0.48\linewidth}
\centering\epsfig{file=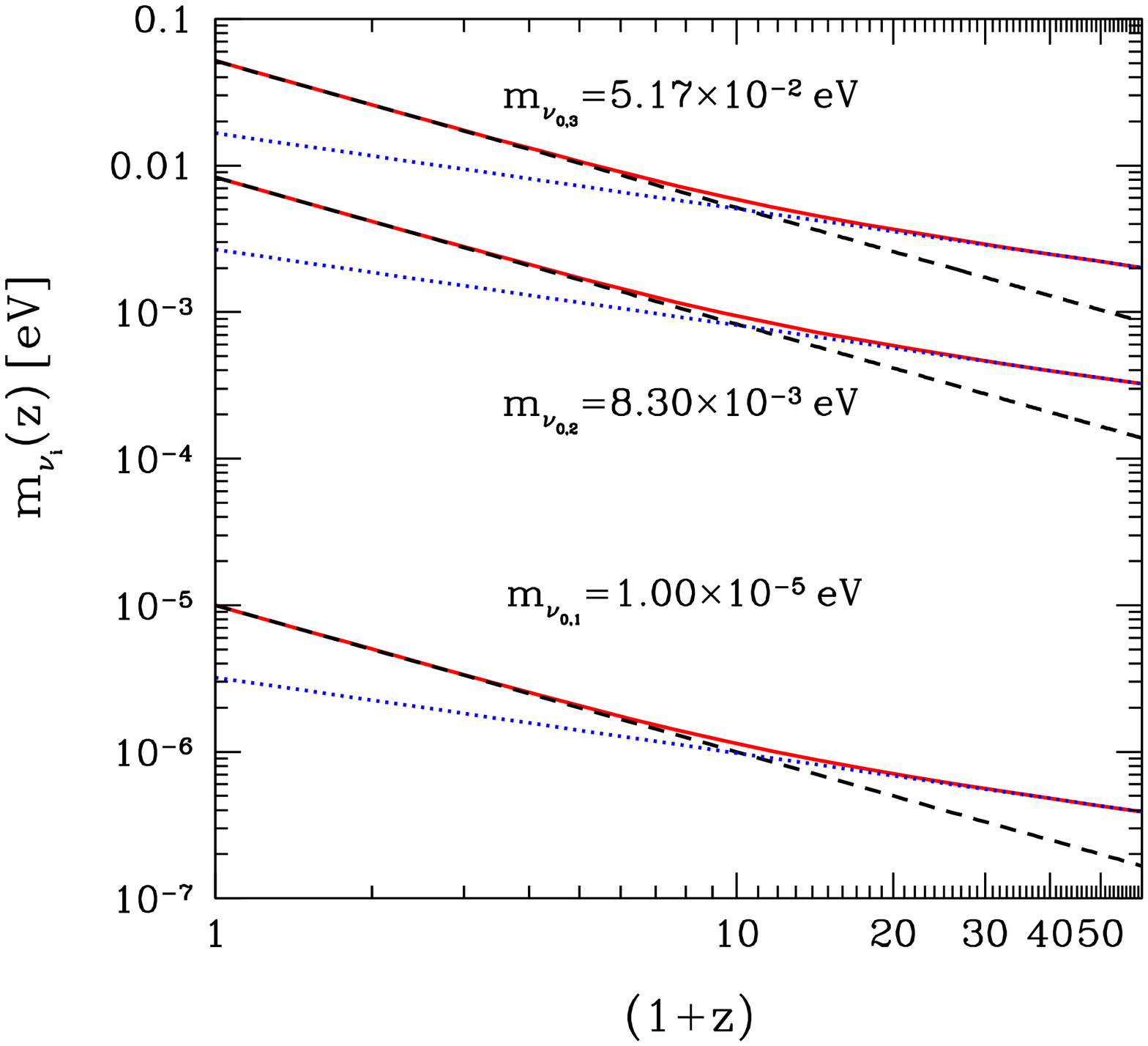,height=4cm,width=5cm}
\end{minipage}
\caption{a) Evolution of the effective potential $V_{\rm eff}(m_\nu,z)$ and the neutrino mass $m_\nu(z)$ due to changes in the neutrino energy density $\rho_\nu(z)$; b) Exact $m_{\nu_i}(z)$ (solid), approximated by $m_{\nu_i}(z)\propto (1+z)^{-1}$ and $m_{\nu_i}(z)\propto (1+z)^{-1/2}$ (dashed and dotted) in the low and high redshift regime, respectively, assuming $m_{\nu_{0,1}}=10^{-5}\,{\rm eV}$ and a normal neutrino mass hierarchy such that neutrino oscillation experiments fix $m_{\nu_{0,2}}=8.3\times 10^{-3}\,{\rm eV}$ and $m_{\nu_{0,3}}=5.17\times 10^{-2}\,{\rm eV}$ today.}\label{Mass}
\end{figure*}

\begin{equation}
\label{mvary}
m_{\nu_i}(z)\simeq m_{\nu_i,0}(1+z)^{-1},\,\,\mbox{where}\,\,m_{\nu_i,0}=m_{\nu_i}(0)\,\,\mbox{and}\,\,i=1,2,3,
\end{equation}
as shown in fig.~\ref{Mass}b, where the lightest neutrino is assumed to be still relativistic today, $m_{\nu_1}=10^{-5}\,{\rm eV}\ll T_{\nu_0}$. Finally, we are in a position to do MaVaN phenomenology. In the following I will discuss a possible astrophysical test~\cite{Ringwald:2006ks} for the neutrino mass variation involving extremely high-energy cosmic neutrinos.
  
\begin{figure*}[t]
\centering\epsfig{file=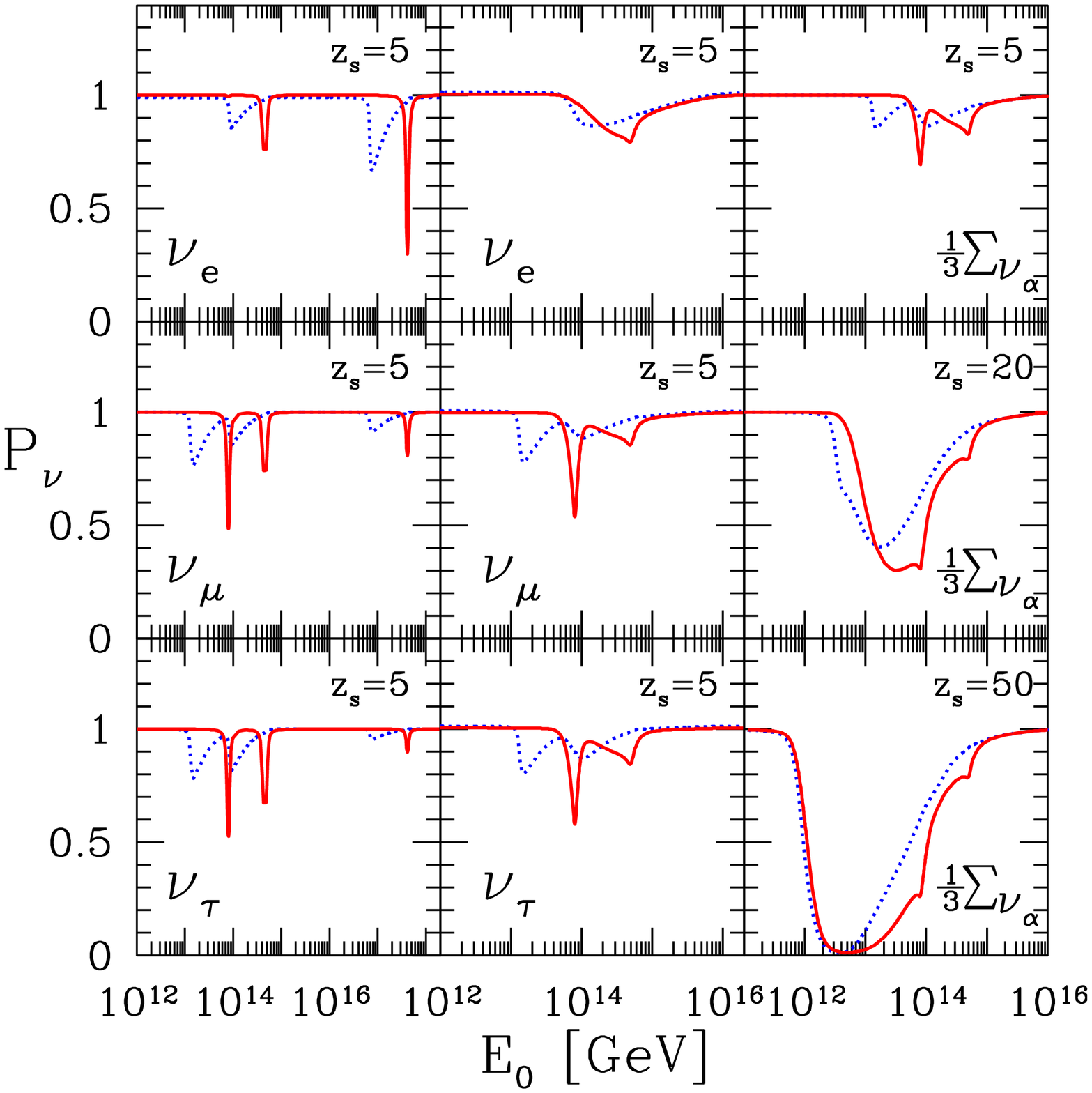,height=5.6cm,width=8.5cm}
\caption{a), b) (first two columns): Flavor survival probability
  $P_\nu=P_{\nu_{\alpha}}$, $\alpha=e,\mu,\tau$, without and with
  inclusion of thermal background effects, respectively, integrated
  back to $z_s=5$; c)(third column) Normalized sum $P_\nu=\frac{1}{3}\sum P_{\nu_{\alpha}}$ including thermal effects for $z_{\rm s}=5$, $z_{\rm s}=20$ and $z_{\rm s}=50$ from top to bottom. All panels assume a normal neutrino mass hierarchy with $m_{\nu_{0,1}}=10^{-5}$ eV and varying (solid) and constant (dotted) neutrino masses as a function of their energy $E_0$ at earth.}\label{Dips}
\end{figure*}
\vspace{-2.5mm} 
\section{\label{sec:EHECnu} Extremely high-energy cosmic neutrinos} 

We are living in exciting times for extremely high-energy cosmic neutrinos~\cite{Ringwald:2005wa,Ringwald:2006ks}. Existing and planned observatories cover an energy range of $10^7\,{\rm GeV}<E_0<10^{17}$ GeV and promise appreciable event samples (cf.~\cite{Ringwald:2005wa,Ringwald:2006ks} and references therein). Thus it seems timely to consider the diagnostic potential of $\E$'s for astrophysical processes. A particular example will be discussed in the following. If the energy of an $\E$ coincides with the resonance energy, $E_{i}^{\rm res}=\frac{M^2_Z}{2m_{\nu_{i}}}=4.2\times 10^{12}\,\,\left(\frac{\rm eV}{m_{\nu_i}}\right) {\rm GeV},\,i=1,2,3,$ of the process $\nu\bar{\nu}\rightarrow Z$, 
%stated in the rest system of the relic neutrinos,  
an $\E$ can annihilate with a relic anti-neutrino and vice versa into a $Z$ Boson~\cite{Weiler:1982qy} with mass $M_Z$. This exceptional loss of transparency of the cosmic neutrino background ($\C$) with respect to cosmic neutrinos is expected to lead to seizable absorption dips in the diffuse $\E$ fluxes to be detected at earth (cf. fig.~\ref{Dips}). Independent of the nature of neutrino masses, their resolution would constitute the most direct evidence of the $\C$ so far. 
In addition, since the annihilation process is sensitive to the neutrino mass and thus to its possible time variation, it could serve as a test for Neutrino Dark Energy (MaVaNs). 

In fig.~\ref{Dips} the survival probability of an $\E$ both for varying and constant neutrino masses is plotted as a function of its energy $E_0$ as measured at earth. It encodes the physical information on possible annihilation processes (for details see~\cite{Ringwald:2006ks}) and takes values between $0$ and $1$. In order to disentangle the different influences on the $\E$ survival probability, let us first assume the relic neutrinos to be at rest (cf. fig.~\ref{Dips}a). In this case the absorption features for constant neutrino masses are only subject to the effect of cosmic redshift which causes an $\E$ emitted at its source at $z_{\rm s}$ with energy $E$ to arrive at earth with the red-shifted energy $E_0=E/(1+z_{\rm s})<E$. As a result, the survival probability is reduced in the interval $E_{0,i}^{\rm res}/(1+z_{\rm s})<E_0<E_{0,i}^{\rm res}$ and the absorption minima are thus redshift distorted. In case of a possible mass variation, however, the survival probability exhibits sharp spikes at the resonance energies $E_{0,i}^{\rm res}$ (cf. fig.~\ref{Dips}a). This can be understood by recalling the mass behavior in the low redshift regime stated in eq.~\ref{mvary}. The neutrino masses $m_{\nu_i}(z)$ introduce a $z$ dependence into the resonance energies, $E_{i}^{\rm res}(z)=M^2_Z (1+z)/(2m_{\nu_{0,i}})=E_{0,i}^{\rm res}(1+z)$ which is compensated by the effect of cosmic redshift. Accordingly, all annihilations occurring at $0<z<z_{\rm s}$ contribute to an absorption dip at $E_{i}^{\rm res}(z)/(1+z)=E_{0,i}^{\rm res}$. As a second step let us take into account that the relic neutrinos are moving targets with a Fermi-Dirac momentum distribution. In fig.~\ref{Dips}b the corresponding Fermi-smearing both for constant and varying neutrino masses results in a thermal broadening (and thus merging) of the dips produced by the mass eigenstates (cf. fig.~\ref{Dips}a). 

The plot of the flavor summed survival probability in fig.~\ref{Dips}c shows that the dip depth increases with $z_s$ independent of the nature of neutrino masses. However, generically, the MaVaN features are clearly shifted to higher energies and the minima deeper than for neutrinos with constant mass.   
%\vspace{-5mm} 
\section{\label{sec:outlook} Outlook} 
\begin{figure*}[t]
\begin{minipage}[t]{0.48\linewidth}
\centering\epsfig{file=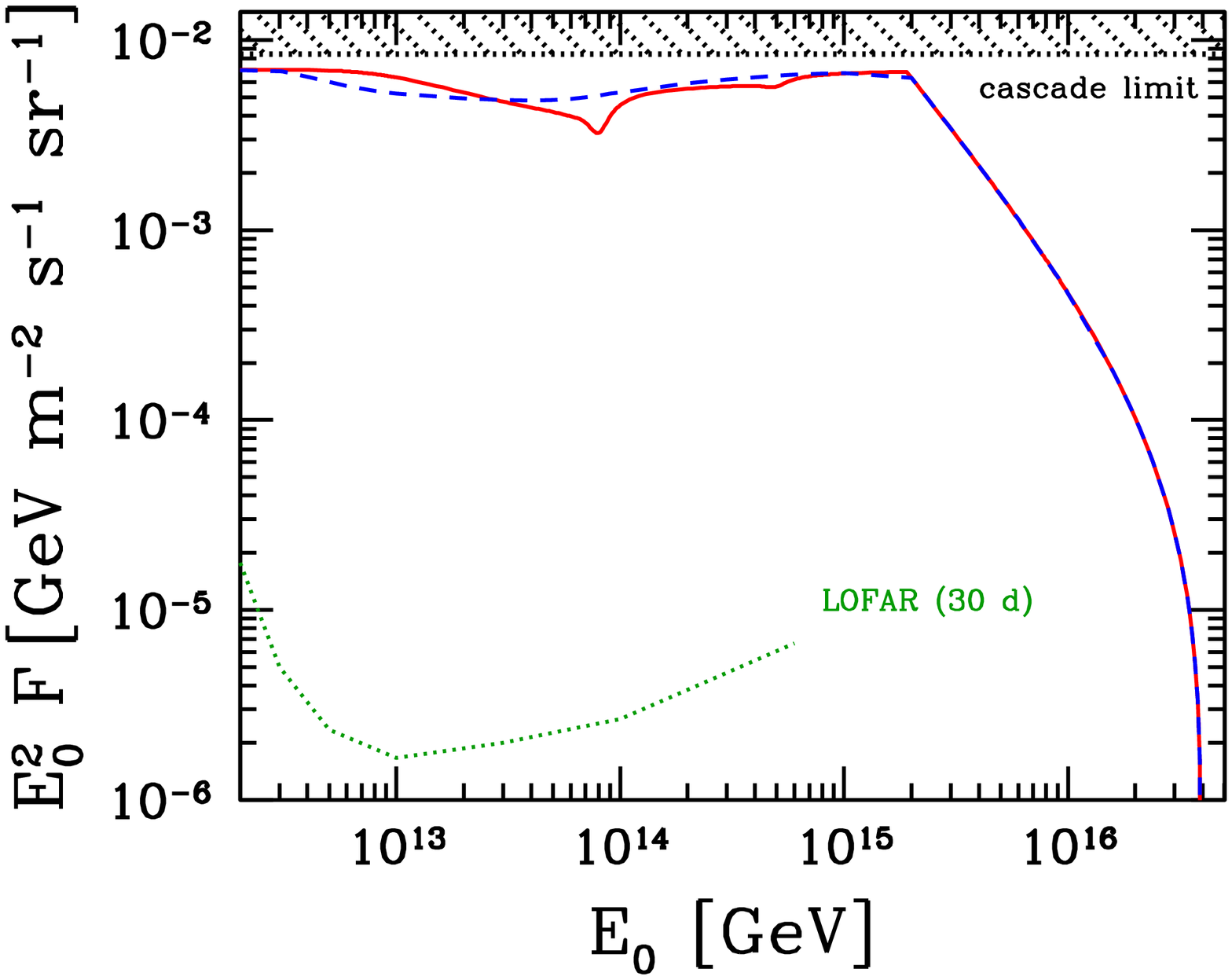,height=4.2cm,width=5cm}
\end{minipage}\hfill
\begin{minipage}[t]{0.48\linewidth}
\centering\epsfig{file=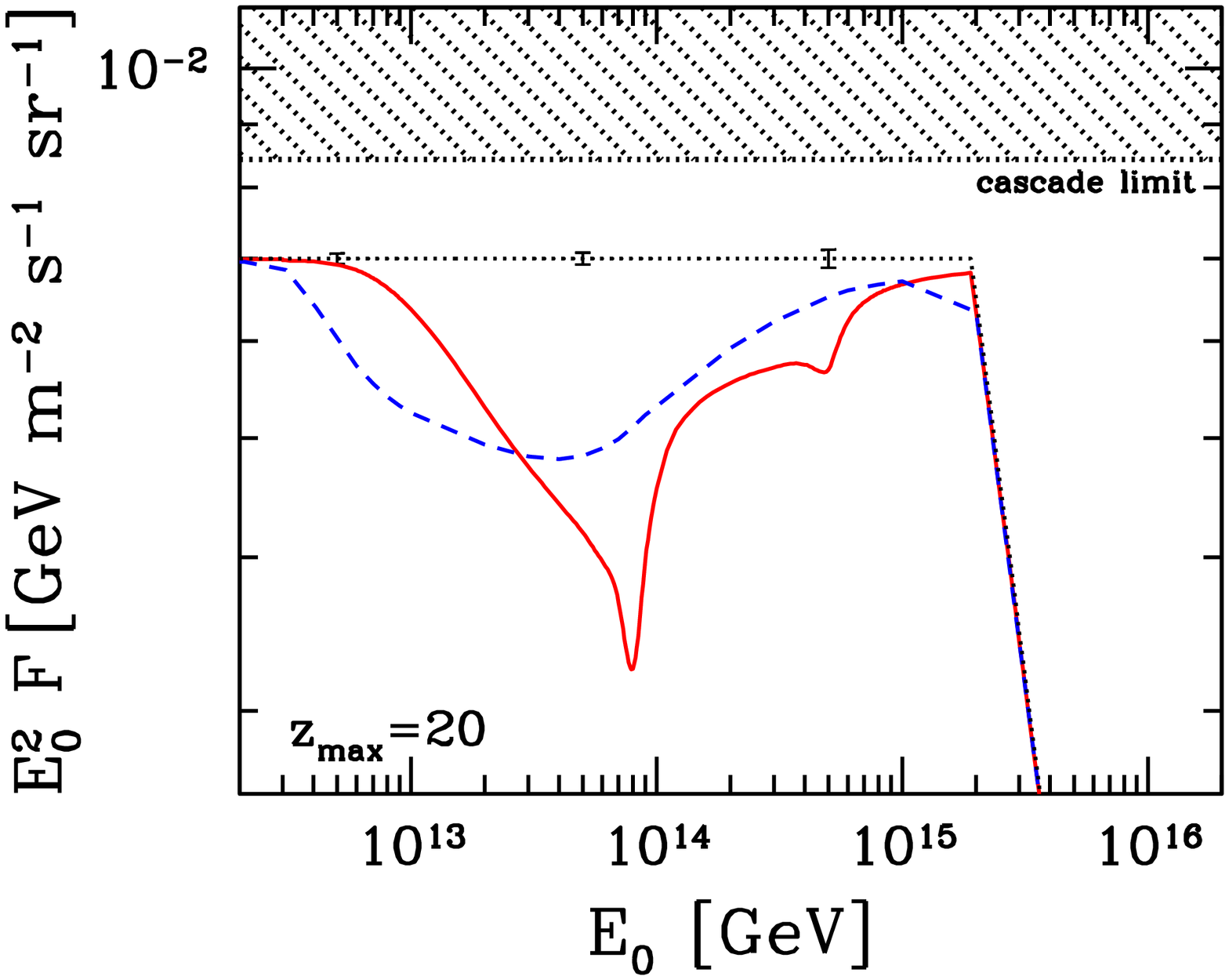,height=4.2cm,width=5cm}
\end{minipage}
\caption{$E^2_0 F$ with $F=\sum F_{\nu_{\alpha}}+\sum F_{\bar{\nu}_{\alpha}}$ for a flux saturating today's observational bound for varying (solid) and constant (dashed) neutrino masses and for $z_{\rm s}=20$, assuming a normal neutrino mass hierarchy with $m_{\nu_{0,1}}=10^{-5}$ eV. a) Together with projected sensitivity of LOFAR expressed in terms of the diffuse neutrino flux per flavor, corresponding to one event per energy decade and indicated duration; b) Together with error bars indicating the corresponding statistical accuracy.}\label{Astro}
\end{figure*}
The fig.~\ref{Astro} shows the expected $\E$ flux from an astrophysical source at a redshift $z_{\rm s}=20$ both for constant was well as for varying neutrino masses normalized to saturate today's observational bound. In fig.~\ref{Astro}a it is plotted together with the projected sensitivity of the neutrino telescope LOFAR~\cite{Scholten:2005pp} which corresponds to maximally $3500$ events detected per energy decade and indicated duration. This translates into very small statistical error bars as included in fig,~\ref{Astro}b, a blow up of fig.~\ref{Astro}a. Accordingly, independent of the nature of neutrino masses, the detection of absorption dips with LOFAR and thus the most direct evidence for the existence of the $\C$ so far can be expected within the next decade. Furthermore, if LOFAR achieves a decent energy resolution, the variation of neutrino masses and thus the interpretation of the $\C$ as source of Dark Energy could be tested.  
\enlargethispage{1cm}
\vspace{-3.5mm}
\frenchspacing
%\bibliography{schrempp}

%\addcontentsline{toc}{section}{Bibliographie}

%\bibliographystyle{phaipmc}

%%%%%%%%%%%%%%%%%%%%%%%%%%%%%%%%%%%%%% reset.txt counters %%%%%%%%%%%%%%
%%
%%%%%%% do not change below here  %%%%%%%%%%%%%%%%%%%%%%%%%%%%%
\setcounter{section}{0}
\setcounter{subsection}{0}
\setcounter{figure}{0}
\setcounter{table}{0}
\setcounter{equation}{0}
\newpage

%%%%%%%%%%%%%%%%%%%%%%%%%%%%%%%%%%%%%%%%%%%%%%%%%%% Title, authors and addresses
\begin{frontmatter}


\title{Black Holes at Neutrino Telescopes: \\
Bounds on TeV-Gravity }

\author[address1]{Marek Kowalski}
\address[address1]{Humboldt University, Berlin, Germany}

\begin{abstract}
An exciting prediction of  higher dimensional, TeV-scale 
quantum-gravity models 
is the copious production of microscopic black holes in the interactions of high energy particles. Such black holes could be produced at the LHC, as well as in the interaction of high energy cosmic neutrinos.
In this short review we discuss the current bounds obtained from the 
non-observation of neutrino-induced black holes in neutrino detectors 
 as well as  future perspectives for their detection.

\end{abstract}

% \begin{keyword}

%\PACS 
% \end{keyword}

\end{frontmatter}

%%%%%%%%%%%%%%%%%%%%%%%%%%%%%%%%%%%%%%%%%%%%%%%%%%%%%% MAIN TEXT
\section{\label{sec:intro} Introduction}
The possibility that black holes will be produced in the 
collision of two light particles at center-of-mass (cm) energies above the Planck scale is an exciting and realistic perspective  in the context of theories with
$\delta = D-4\geq 1$ large compact~\cite{Arkani-Hamed:1998rs} 
or warped~\cite{Randall:1999ee} extra dimensions 
and a low fundamental Planck scale 
$M_D\,\gwig$ 1 TeV characterizing quantum 
gravity.
In these theories one might expect the copious production of black holes 
in high energy collisions
at cm energies above 
$M_D$~\cite{Argyres:1998qn}.
Correspondingly, the Large Hadron Collider (LHC) 
expected  to have a first pilot run in late 2007, may turn into a factory of black holes
at which their production and evaporation may be studied in 
detail~\cite{Giddings:2001bu,Dimopoulos:2001hw} (see \cite{Landsberg06} for a recent review and a more complete list of references). 

However, even before the LHC starts operating, important measurements can still be made with high-energy neutrino telescopes. In fact, among the strongest constraints today are obtained from a (zero-)measurement with neutrino telescopes! 
Black hole production and subsequent decay in the scattering of ultra-high energy cosmic neutrinos can be detected by large scale neutrino-detectors \cite{kowalski02,jami}, such as IceCube \cite{ic} or RICE \cite{rice}.
Alternatively, black holes produced in the scattering on nucleons in the 
atmosphere may initiate quasi-horizontal air showers 
far above the Standard Model rate~\cite{Feng:2002ib,Emparan:2001kf,Ringwald:2001vk,Anchordoqui:2001cg}.

 The current bounds and the perspective for the future are topic of this short 
review. In addition we discuss some experimental signatures of black hole production which will allow to discriminate against standard model background.

\section{\label{sec:eqs} Black hole creation and decay}
The original cross-section calculations 
for black hole production follow simple, 
semi-classical reasoning.
It is assumed that 
at trans-Planckian parton-parton cm energies squared, $\hat s\gg M_D^2$,  
and at impact parameters smaller than the Schwarzschild radius $r_S$ of a $(4+\delta )$-dimensional black 
hole with mass $M_{\rm bh}=\sqrt{\hat s}$, a black hole will form.
The cross section for black hole production is then:
\begin{equation}
\label{sig_bh_geom}
\hat\sigma^{\rm bh} (\hat s ) \approx \pi\,r_S^2 
\left( M_{\rm bh}=\sqrt{\hat s} \right)
\end{equation}
with the  Schwarzschild radius,
\[
r_S =\frac{1}{M_D}
\left[
\frac{M_{\rm bh}}{M_D}
\left(
\frac{2^\delta \pi^{\frac{\delta -3}{2}}\,\Gamma\left( \frac{3+\delta}{2}\right)}
{2+\delta}
\right)
\right]^{\frac{1}{1+\delta}}
\,.
\]
Frequently one introduces the ratio $x=M_{\rm bh}/M_D$ and demands that $x\gg1$ to ensure that  the
semi-classical cross-section estimation is justified.

The contribution of black hole production to the neutrino-nucleon  cross section
 is then
obtained as the convolution 
\begin{equation}
\label{sig_nuN_bh_pdf}
\sigma_{\nu N}^{\rm bh} (\hat s) =
\sum_i \int^1_0%_{M_{\rm bh}^{{\rm min}\,2}/s}^1
{\rm d}x\,f_i (x,\mu )\,\hat \sigma_{\nu i}^{\rm bh} (x \hat s)
\, .
\end{equation}

The sum 
extends over all partons in the nucleon, with parton distribution functions 
$f_i(x,\mu )$ and 
factorization scale $\mu$. The results are not very sensitive on the parton distribution function or the exact value of the factorization scale~\cite{Ringwald:2001vk}. The cross-sections are displayed in Fig.\ \ref{kowalski_fig1} for several different extra dimensions and $M_{\rm bh}/M_D>5$. 

There are many ongoing efforts to make the somewhat simplistic 
geometric cross-section calculation more accurate. These include the incorporation of gray body \cite{graybody}, form factor \cite{formfactor} or impact parameter effects \cite{impactpar}. Their discussion is beyond the scope of this review, however, with a few exceptions they do not alter the cross-sections by more than a factor of a few (see \cite{Landsberg06} for further reading.) 

\begin{figure*}[t]
\centering\epsfig{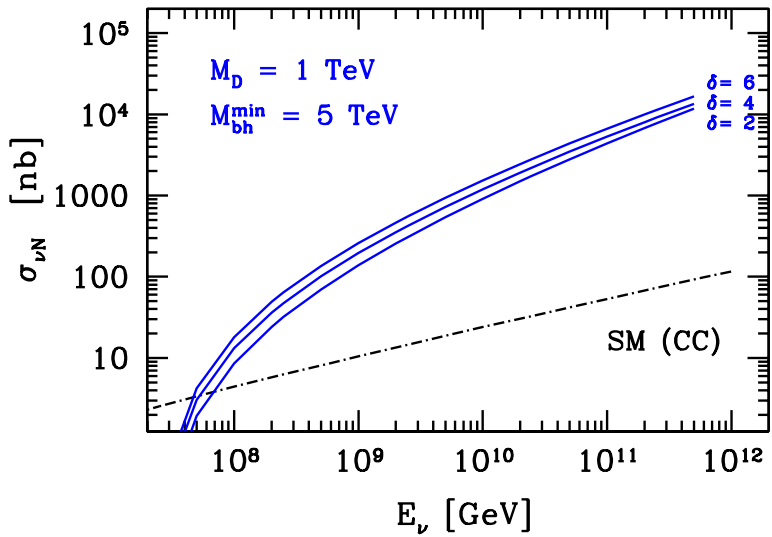}
\caption{ Cross section $\sigma_{\nu N}^{\rm bh}$, Eq.~(\ref{sig_nuN_bh_pdf}), for black hole production in 
neutrino-nucleon scattering, 
for $M_D = 1$ TeV, $M_{\rm bh}^{\rm min} = 5$ TeV, and $\delta = 2,4,6$ extra dimensions (solid lines, from bottom to top).   
Also shown is the Standard Model (SM) charged current (CC) neutrino-nucleon 
cross section (dashed-dotted line). Plot taken from \cite{Ringwald:2001vk}}
\label{kowalski_fig1}
\end{figure*}

Once a TeV-mass black hole is created, it will decay primarily through Hawking 
radiation on extremely short time-scales $\sim10^{-27}$~s, thereby producing a large number of $N\sim{\mathcal O}(20)$ hard quanta \cite{kowalski02,Giddings:2001bu,Dimopoulos:2001hw}. Since the temperature associated with the black hole is so large, $T\sim M_{\rm bh}/N$, the full spectrum of standard model particles will  be generated in a rather flavor democratic way.

\section{The Neutrino Beam}
Clearly, an essential ingredient to probe the TeV-gravity scenarios 
is a neutrino beam with high enough neutrino energies  
($\sim10^9$~GeV). So far no such 
flux has been detected. However, there are a number of model predictions for 
astrophysical neutrino productions.
More or less guaranteed, and therefore comprising a reasonable lower bound, are 
the so-called 
cosmogenic neutrinos which are produced when ultra-high energy cosmic protons
inelastically scatter off the cosmic microwave background radiation
in processes such as $p\gamma\to \Delta\to n\pi^+$, where the produced pion 
subsequently decays.
Estimates of these fluxes  can be found in 
Refs.~\cite{Yoshida:1993pt,Protheroe:1996ft,Engel:2001hd,kkss}, some 
of which 
are shown in Fig.~\ref{kowalski_fig2}.  While these model predictions might be viewed as  
a lower bound, a theoretical upper bound \cite{Mannheim:2001wp} is also 
included in  Fig.~\ref{kowalski_fig2}.

\begin{figure*}[t]
\centering\epsfig{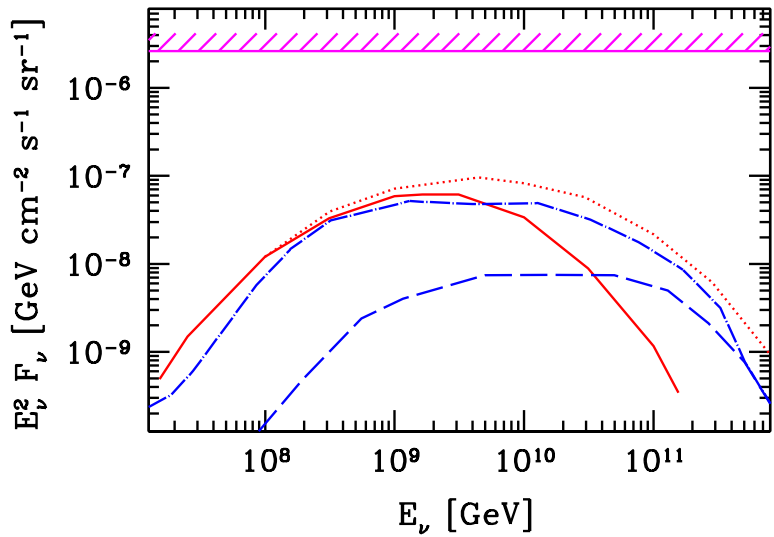}
\caption{Predictions of the cosmogenic neutrino flux (the sum over all flavors). Long-dashed (long-dashed--dotted) line: Flux from 
Ref.~\cite{Yoshida:1993pt} for cosmological evolution parameters 
$m=2$, $z_{\rm max}=2$ ($m=4$, $z_{\rm max}=4$).
Solid (dotted) line: Flux from Ref.~\cite{Protheroe:1996ft}, 
assuming a maximum energy of $E_{\rm max}=3\cdot 10^{20(21)}$ eV for the 
ultra-high energy cosmic rays.
Shaded: Theoretical upper limit of the neutrino flux from ``hidden'' %hadronic 
astrophysical sources that are
non-transparent to ultra-high energy nucleons~\cite{Mannheim:2001wp}. The theoretical upper limit is by now surpassed by experimentally upper bounds from AMANDA \cite{amanda_diffuse}. Plot taken from \cite{Ringwald:2001vk}}
\label{kowalski_fig2}
\end{figure*}

\section{Current Constraints}

\begin{figure*}[t]
\centering\epsfig{file=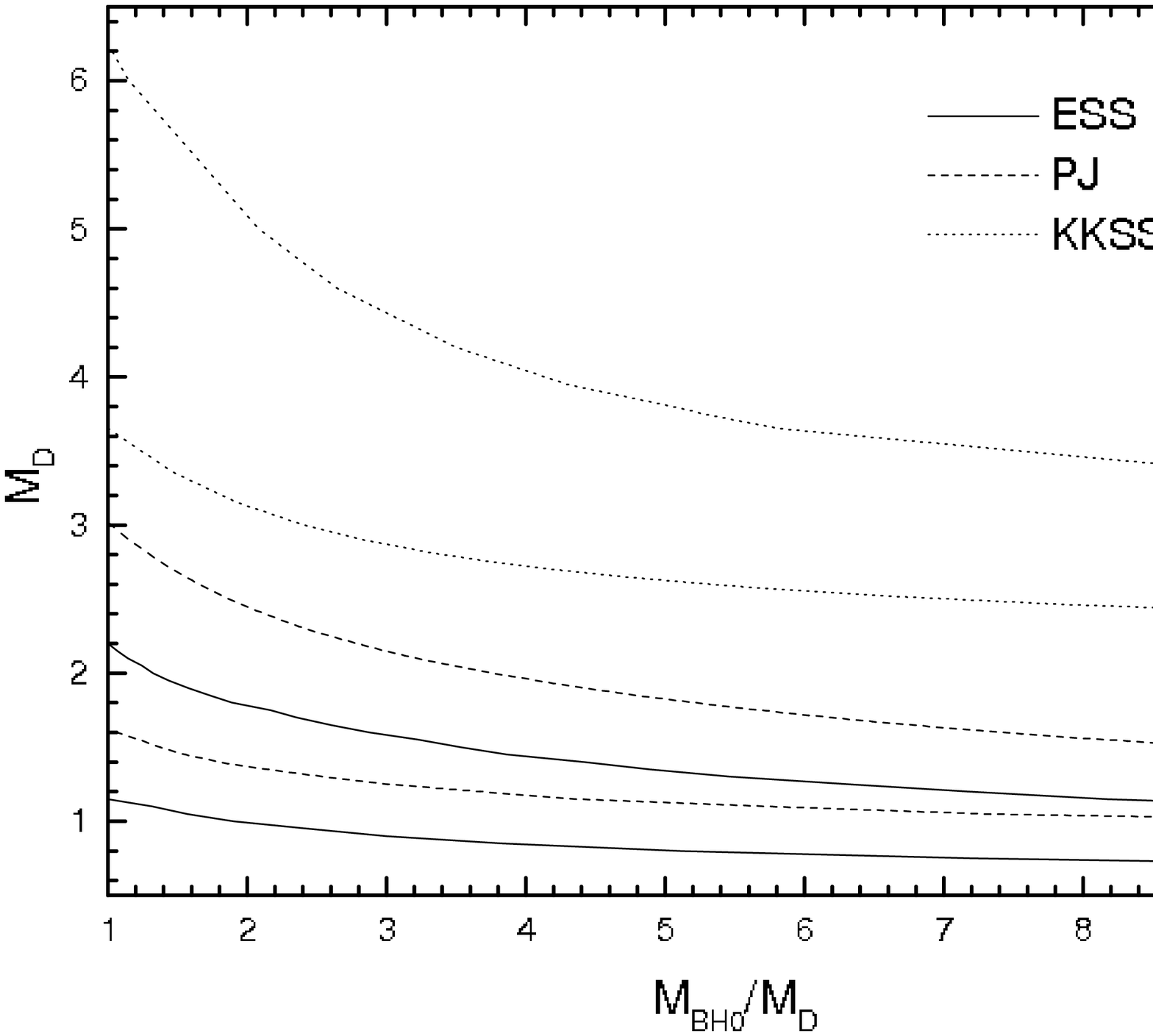,width=0.7\linewidth}
\caption{ Lower bounds on $M_{D}$ as a function of the ratio of the minimum
black hole formation threshold, $M_{\rm bh}^{min}$, to $M_{D}$. The upper curve for each flux
model is the lower bound when the geometrical cross-section of Eq. \ref{sig_bh_geom} 
is used for the black hole
cross section. The lower curve corresponds to the lower bound once
impact parameter effects are included. ESS, KKSS  and PJ refer to Refs. \protect\cite{Engel:2001hd},\cite{Protheroe:1996ft} and \protect\cite{kkss}, respectively. Plot taken from \cite{Hussain:2005dm}.}
\label{kowalski_fig3}
\end{figure*}
Several dedicated neutrino telescopes and air shower detectors 
look for high-energy neutrinos with energies above 10~PeV. 
Concerning black hole production it was shown \cite{Feng:2002ib,Ringwald:2001vk,Anchordoqui:2001cg} that already now sensible 
constraints can be obtained from the non-observation of horizontal showers
by the Fly's Eye collaboration~\cite{Baltrusaitis:1985mx} and the 
Akeno Giant Air Shower Array (AGASA) collaboration~\cite{Chiba:1992nf}.  
These constraints turn out to be 
competitive with other currently available constraints on TeV-scale gravity 
which are mainly based on interactions associated with
Kaluza-Klein gravitons, 
according to which a fundamental Planck scale as low as $M_D = {\mathcal O}(1)$ TeV is still allowed for $\delta\geq 5$ \cite{Anchordoqui:2001cg}. Constraints from one year operation of the Pierre-Auger-Observatory should allow 
to constrain  $M_D$ up to $\sim 2$~TeV \cite{Feng:2002ib,Ringwald:2001vk,Anchordoqui:2001cg}  assuming the geometrical cross-section to hold down to $M_{\rm bh}^{min} = M_D$ and a cosmogenic neutrino flux according to the more conservative model of \cite{Protheroe:1996ft}.

Dedicated neutrino telescopes such as IceCube \cite{ic} or RICE \cite{rice} have other means to 
search for ultra high-energy neutrino interactions. 
At the energies relevant for black hole production the Earth is opaque to 
neutrinos, hence one only expects neutrinos from above the horizon.  
Due to the high energy, the background of atmospheric muons becomes 
insignificant and one can search for neutrino-induced through-going muons as 
well as for neutrino-induces showers in or near the detector. Details of the 
signature are discussed in the next section. It was estimated that with a $\rm km^3$ sized neutrino detector constraints of the order of a  $M_D$ $\sim 2$~TeV can 
be obtained, both from contained events as well as through-going muon events \cite{kowalski02,jami}. Recently the RICE collaboration has reported the results 
from a  search for very high energy neutrinos using a detector of 
20 radio antennas distributed within an instrumented volume of just 
$\sim 0.01~{\rm km}^3$ \cite{rice}. However, due to the km-long attenuation length of the  radio signal in  cold polar ice, the effective volume is significantly 
larger than the instrumented volume. Already at 1~EeV energies, the volume for hadronic cascades exceeds one cubic kilometer. The large volume, combined with an analyzed 
live-time of nearly two years and the fact that no neutrino-induced event candidate was found \cite{rice} was used to put constrain on the fundamental 
Planck mass $M_D$ \cite{Hussain:2005dm}. These constraints are shown in Fig.\ \ref{kowalski_fig3} for three different cosmogenic neutrino flux models and two different calculations of the black hole production cross-section. Assuming the geometrical cross-section calculation, $M_{\rm bh}/M_D>1$  and a flux from \cite{Protheroe:1996ft} they obtain a bound on $M_D$ of 
3~TeV. The bound weakens to about 1.5~TeV, once the cross-section calculation 
includes effects due to the impact parameter \cite{impactpar}. Note that these results have only a small dependence on the number of extra dimensions. If the model predictions for cosmogenic neutrino are to be believed, these bounds on the 
fundamental mass scale are currently the most restrictive available for $\delta \geq 5$.
\section{The Future}
Several future projects are currently planned which will have orders of 
magnitude larger sensitivities for neutrinos of  ultra-high energies. 
Radio detectors such as ANITA \cite{anita} or SALSA \cite{salsa}, or a hybrid extension of IceCube with radio and acoustic sensors \cite{Besson:2005re} will have effective volumes of ${\mathcal O}(100~{\rm km}^3$), and will therefore be capable to detect dozens events due to standard model interactions of cosmogenic neutrinos.

Individual black 
hole events will consist of a large hadronic component, a somewhat smaller 
electro-magnetic component, as well as muons and taus produced in the event.
 The sensitivity to probe non-standard model interactions will depend on 
observable signatures to discriminate against the standard model background. 
While it is a challenge for sparse high energy neutrino detectors to 
identify the relative contributions, some promising ideas exist nevertheless. The LPM effect will lead to a significant stochasticly variable elongation of electro-magnetic showers above $\sim 10^8$~GeV energies, while it has less of an influence for hadronic cascades. If simultaneously intensity and longitudinal profile can be measured, one might discriminate the two contributions. The contribution of taus can be identified through its decay---a second shower displaced relative to the primary vertex \cite{Cardoso:2004zi}. The possibility to identify ``soft'' muons from a black hole event was studied in \cite{Anchordoqui:2006fn}. In a large hybrid detector which combines optical with radio and acoustic sensitivity, all the above signatures could be used for identifying black hole events.
One can also test the statistical properties of an observed event distribution. Clearly, a sudden rise in the energy spectrum will be a first indication of additional non-standard cross-sections. An essentially 
model independent signature of additional interaction channels would be an observed 
 suppression of the angular distribution of high energy neutrino events towards the horizon \cite{Anchordoqui:2005pn}.

Summarizing it can be said, that while 
current bounds from neutrino telescopes on $M_D$ are already probing an important mass-regime, detectors currently under construction have the sensitivity to improve the bounds significantly, or better yet, discover the first micro black holes.

\section*{Acknowledgments}

{\small 
I would like to thank the organizers of the EPNT workshop for the
invitation to this very interesting meeting. 
In addition I would like to thank A. Ringwald and H. Tu for a very stimulating collaborators on the subject of microscopic black holes. Financial support was obtained through the Deutsche Forschungsgemeinschaft (DFG).

}

%%%%%%%%%%%%%%%%%%%%%%%%%%%%%%%%%%%%%% reset.txt counters %%%%%%%%%%%%%%
%%
%%%%%%% do not change below here  %%%%%%%%%%%%%%%%%%%%%%%%%%%%%
\setcounter{section}{0}
\setcounter{subsection}{0}
\setcounter{figure}{0}
\setcounter{table}{0}
\setcounter{equation}{0}
\newpage

%%%%%%%%%%%%%%%%%%%%%%%%%%%%%%%%%%%%%%%%%%%%%%%%%%% Title, authors and addresses
\begin{frontmatter}


\title{Non-standard Neutrino Oscillations at Icecube}

% use optional labels between square brackets to link authors explicitly to addresses:
% If more than one author, keep a comma between the author tags

\author[address1]{M.C. Gonzalez-Garcia},
\address[address1]{
  Instituci\'o Catalana de Recerca i Estudis Avan\c{c}ats (ICREA), \\
  Departament d'Estructura i Constituents de la Mat\`eria,
  Universitat de Barcelona,\\
  Diagonal 647, E-08028 Barcelona, Spain\\
{\rm and:}  C.N. Yang Institute for Theoretical Physics\\
  Stony Brook University, 
  Stony Brook, NY 11794-3840, USA}
\begin{abstract}
In this talk I review the potential of Icecube for revealing 
physics beyond the standard model in the oscillation of
atmospheric neutrinos~\cite{ouricecube}.
\end{abstract}

% \begin{keyword}

%\PACS 
% \end{keyword}

\end{frontmatter}

\section{Introduction}
With its high statistics data~\cite{skatmlast}
Super--Kamiokande (SK) established beyond doubt that the observed
deficit in the $\mu$-like atmospheric events is due to oscillations, a
result supported by the K2K and MINOS long-baseline (LBL)
experiments~\cite{k2kprl,MINOSdat}.

Mass oscillations are not the only possible
mechanism for atmospheric (ATM) $\nu_\mu \to \nu_\tau$ flavour
transitions. These can be also generated by a variety
of nonstandard physics characterized by the presence of
an unconventional $\nu$ interaction 
that mixes neutrino flavours~\cite{npreview}.  Examples include
violations of the equivalence principle (VEP)~\cite{VEP,qVEP},
non-standard neutrino interactions with matter~\cite{NSI}, neutrino
couplings to space-time torsion fields~\cite{torsion}, violations of
Lorentz invariance (VLI)~\cite{VLI1} and of CPT
symmetry~\cite{VLICPT1,VLICPT2}. 
In contrast to the $E$ energy dependence of the conventional
oscillation length, new physics can produce neutrino oscillations with
wavelengths that are constant or decrease with energy. 
~\cite{yasuda1,flanagan}. At present these scenarios cannot explain the
data\cite{oldatmfitnp} and a combined
analysis of the ATM and LBL data can be performed to 
constraint them even as  subdominant
oscillation effects~\cite{ouratmnp}.

IceCube, with energy
reach in the $0.1 \sim 10^4$\,TeV range for ATM neutrinos, is
the ideal experiment to search for new physics. For most of this
energy interval standard $\Delta m^2$ oscillations are suppressed and
therefore the observation of an angular distortion of the ATM
neutrino flux or its energy dependence provide a clear signature for
the presence of new physics mixing neutrino flavours~\cite{ouricecube}.

\section{Propagation in Matter of High Energy  Oscillating Neutrinos}
\label{sec:formaprop}
We concentrate on $\nu_\mu$--$\nu_\tau$ flavour mixing
mechanisms for which the propagation of 
$\nu$'s ($+$) and
$\bar\nu$'s ($-$) is governed by the following
Hamiltonian~\cite{VLICPT2}:
\begin{eqnarray} \label{eq:hamil}
    {\rm H}_\pm \equiv
    \dfrac{\Dmq}{4 E}
    \mathbf{U}_\theta
    \begin{pmatrix}
	-1 & ~0 \\
	\hphantom{-}0 & ~1
    \end{pmatrix}
    \mathbf{U}_\theta^\dagger
    + \sum_n
    \sigma_n^\pm \dfrac{\Dlt_n\, E^n}{2}
    \mathbf{U}_{\xi_n,\pm\eta_n}
    \begin{pmatrix}
	-1 & ~0 \\
	\hphantom{-}0 & ~1
    \end{pmatrix}
    \mathbf{U}_{\xi_n,\pm\eta_n}^\dagger \;, 
\end{eqnarray}
$\Dmq$ is the mass--squared difference between the two neutrino
mass eigenstates, $\sigma_n^\pm$ accounts for a possible relative
sign of the new physics (NP) 
effects between $\nu$'s and $\bar\nu$'s and
$\Dlt_n$ parametrizes the size of the NP terms. 
By $\eta_n$ we denote the possible non-vanishing relative phases. 
\begin{eqnarray}
\label{eq:rotat}
    \mathbf{U}_\theta =
    \begin{pmatrix}
	\hphantom{-}\cos\theta & ~\sin\theta \\
	-\sin\theta & ~\cos\theta
    \end{pmatrix}\,,
    \qquad
    \mathbf{U}_{\xi_n,\pm\eta_n} =
    \begin{pmatrix}
	\hphantom{-}\cos\xi_n\hphantom{e^{-i\eta_n}} 
	& ~\sin\xi_n e^{\pm i\eta_n} 
	\\
	-\sin\xi_n e^{\mp i\eta_n} 
	& ~\cos\xi_n\hphantom{e^{-i\eta_n}}
    \end{pmatrix}\,;
\end{eqnarray}

If NP strength is constant along the neutrino trajectory the oscillation
probabilities take the form ~\cite{VLICPT2}:
\begin{eqnarray}
&& \!\!\!\!\!\! \!\!\!\!\!\!  \!\!  
P_{\nu_\mu \to \nu_\mu} = 1 - P_{\nu_\mu \to \nu_\tau} =
    1 - \sin^2 2\Theta \, \sin^2 \left( 
    \frac{\Dmq L}{4E} \, \mathcal{R} \right)
\nonumber    \\
&&    
\!\!\!\!\!\!\!\!\!\!\!\!\!\!      
\sin^2 2\Theta = \frac{1}{\mathcal{R}^2} \left(
    \sin^2 2\theta + R_n^2 \sin^2 2\xi_n
    + 2 R_n \sin 2\theta \sin 2\xi_n c\eta_n \right) \,,
\nonumber    \\
&&    
\!\!\!\!\!\!\!\!\!\!\!\!\!\!      
\mathcal{R} =\sqrt{1 + R_n^2 + 2 R_n \left( \cos 2\theta \cos 2\xi_n
      + \sin 2\theta \sin 2\xi_n c\eta_n \right)}\;, 
\;\;\;\;
    R_n = \sigma_n^+ \frac{\Dlt_n E^n}{2} \, \frac{4E}{\Dmq} \,,
\nonumber 
\end{eqnarray}
where, for simplicity, we have assumed  scenarios 
with one NP source characterized by a unique $\Delta\delta_n$.
$c\eta_n=\cos\eta_n$ 

Eq.~\ref{eq:hamil} describes, for example, flavour mixing due to
new tensor-like interactions for which $n=1$ leading to a
contribution to the oscillation wavelength inversely proportional to 
the neutrino energy. This is the case for 
$\nu_\mu$'s and $\nu_\tau$'s of different masses in the
presence of violation of the equivalence principle  due to non-
universal coupling of the neutrinos, $\gamma_1\neq \gamma_2$
to the gravitational potential
$\phi$~\cite{VEP},  
so  $\Dlt_1 = 2 |\phi|(\gamma_1- \gamma_2) \equiv 2 |\phi| \Delta\gamma $.
$\nu_1$ and $\nu_2$ are related to $\nu_\mu$ and $\nu_\tau$ by a rotation
$\xi_1=\xi_{ vep}$.

For constant potential $\phi$, this mechanism is phenomenologically
equivalent to the breakdown of Lorentz invariance resulting from different
asymptotic values of the velocity of the neutrinos, $c_1\neq c_2$,
$ \Dlt_1 = (c_1- c_2)\equiv\delta c/c$, 
with $\nu_1$ and $\nu_2$ being related to $\nu_\mu$ and $\nu_\tau$ by
a rotation $\xi_1=\xi_{vli}$~\cite{VLI1}.  

For vector-like interactions, $n=0$,  the oscillation wavelength 
is energy-independent. This
may arise, for instance, from a non-universal coupling of the
neutrinos, $k_1\neq k_2$ so $\Dlt_0= Q (k_1- k_2)$
($\nu_1$ and $\nu_2$ is related to the
$\nu_\mu$ and $\nu_\tau$ by a rotation $\xi_0=\xi_Q$), to a space-time
torsion field $Q$~\cite{torsion}.
Violation of CPT resulting from Lorentz-violating effects such 
as the operator,  
$\bar{\nu}_L^\alpha b_\mu^{\alpha\beta} \gamma_\mu
\nu_L^\beta$,  also leads to an
energy independent contribution to the oscillation
wavelength~\cite{VLICPT1,VLICPT2}  which is a function of 
the eigenvalues of the Lorentz violating CPT-odd
operator, $b_i$, $
\Dlt_0 = b_1-b_2 $,
and  the rotation angle, $\xi_0=\xi_{\not\text{CPT}} $, between the
corresponding eigenstates $\nu_i$ and the flavour states
$\nu_\alpha$.

%At present the strongest limits on NP neutrino oscillations
%arise from the non-observation of departure from the $\Delta m^2$
%oscillation behaviour in ATM neutrinos at SK and the
%confirmation of $\nu_\mu$ oscillations with the same 
%oscillation parameters from LBL experiments~\cite{ouratmnp}.

For most of the neutrino energies considered here, 
$\Delta m^2$ oscillations are suppressed and the NP effect is
directly observed. Thus the results will be independent
of the phase $\eta_n$ and we can chose the NP parameters in the range
$\Dlt_n \geq 0$,   $0 \leq \xi_n \leq \pi/4 $.

The Hamiltonian of Eq.~(\ref{eq:hamil}) describes the coherent
evolution of the $\nu_\mu$--$\nu_\tau$ ensemble for any neutrino
energy.  High-energy neutrinos propagating in the Earth can also interact
inelastically with the Earth matter either by charged current (CC)
and neutral current (NC) and as a
consequence the neutrino flux is attenuated.  This attenuation is
qualitatively and quantitatively different for $\nu_\tau$'s and
$\nu_\mu$'s. $\nu_\mu$'s are absorbed by CC  interactions while 
$\nu_\tau$'s are regenerated because they produce a $\tau$ that
decays into another tau neutrino
before losing energy ~\cite{hs}. As a consequence, for each $\nu_\tau$
lost in CC interactions, another $\nu_\tau$ appears (degraded in
energy) from the $\tau$ decay and the Earth never becomes opaque to
$\nu_\tau's$.  Furthermore, a 
secondary flux of $\bar\nu_\mu$'s is also generated in the leptonic
decay $\tau\rightarrow \mu\bar\nu_\mu\nu_\tau$~\cite{kolb}.

Attenuation and regeneration effects of incoherent neutrino fluxes can
be consistently described by a set of coupled partial
integro-differential cascade equations (see for example~\cite{reno}
and references therein) or by a Monte Carlo simulation of the neutrino
propagation in matter~\cite{hs,kolb,crotty}. For astrophysical $\nu$'s, 
because of the long distance traveled  
from the source, the oscillations average out and at the Earth the neutrinos 
can be treated as an incoherent superposition of
mass eigenstates.

For ATM neutrinos this is not the case because oscillation,
attenuation, and regeneration effects occur simultaneously when the
neutrino beam travels across the Earth's matter. 
For conventional neutrino oscillations this
fact can be ignored because the neutrino energies covered by current
experiments are low enough for attenuation and regeneration effects to
be negligible. But for non-standard scenario oscillations,
future experiments probe high-energy neutrinos for which the
attenuation and regeneration effects have to be accounted for
simultaneously. 

In order to do so it is convenient to use the density matrix 
formalism to describe neutrino flavour oscillations. 
The evolution of the neutrino ensemble is determined
by the Liouville equation for the density matrix 
$\rho(t)=\nu(t)\otimes \nu(t)^\dagger$
\begin{equation}
\frac{d{\rho}}{dt}=-i[{\rm H}, {\rho}] \, ,
\end{equation}
where ${\rm H}$ is given by Eq.~(\ref{eq:hamil}). The survival probability
is  given by 
$P_{\mu\mu}(t)={\rm Tr}[\Pi_{\nu_\mu}\,\rho(t)]$, where  
$\Pi_{\nu_\mu}=\nu_\mu\otimes \nu_\mu$ is the $\nu_\mu$ state projector, 
and with initial condition  $\rho(0)=\Pi_{\nu_\mu}$. An equivalent 
equation can be written for the $\bar\nu$ density matrix.

In this formalism attenuation effects due to CC and 
NC interactions 
can be introduced by relaxing the condition ${\rm Tr}(\rho)=1$. 
In this case 
\begin{equation}
\frac{d{\rho(E,t)}}{dt}=-i[{\rm H}(E), \rho(E,t)]
-\sum_\alpha \frac{1}{2\lambda^\alpha_{\rm int}(E,t)}
\left\{\Pi_\alpha,\rho(E,t)\right\} \, ,
\end{equation}
where $[\lambda^\alpha_{int}(E,t)]^{-1}\equiv
[\lambda^\alpha_{\rm CC}(E,t)]^{-1}+
[\lambda_{\rm NC}(E,t)]^{-1}$, $[\lambda^\alpha_{\rm CC}(E,t)]^{-1}
=n_T(x)\, \sigma^{\alpha}_{\rm CC}(E)$, and 
$[\lambda_{\rm NC}(E,t)]^{-1}=n_T(x)\, \sigma_{\rm NC}(E)$
($n_T(x)$ is the number density of nucleons at the point $x=ct$). 

$\nu_\tau$ regeneration and neutrino energy degradation can be accounted
for by coupling these equations to the  shower equations for the $\tau$ flux,
$F_\tau(E_\tau,t)$ (we denote by $F$ the differential fluxes 
$d\phi/(dE \, d\cos\theta)$).
For convenience we define the {\sl neutrino flux density matrix} 
$F_\nu(E,x)=F_{\nu_\mu}(E,x_0)\rho(E,x=c\,t)$   
where $F_{\nu_\mu}(E,x_0)$ is the initial neutrino flux:
\begin{eqnarray}
\frac{d{F_\nu}(E_\nu,x)}{dx}
&=&-i[{\rm H}, F_\nu(E_\nu,x)]
-\sum_\alpha \frac{1}{2\lambda^\alpha_{\rm int}(E_\nu,x)}
\left\{\Pi_\alpha,F_\nu(E_\nu,x)\right\}\nonumber \\
& &+ 
\int_{E_\nu}^\infty  \frac{1}{\lambda_{\rm NC}(E'_\nu,x)}
F_\nu(E'_\nu,x) 
\frac{d N_{\rm NC}(E'_\nu,E_\nu)}{d E_\nu} dE'_\nu  \nonumber \\
& & +
\int_{E_\nu}^\infty   \frac{1}{\lambda^\tau_{\rm dec}(E_\tau,x)}
 F_\tau(E_\tau,x) 
 \frac{d N_{\rm dec} (E_\tau,E_\nu)}{d E_\nu} dE_\tau\, \Pi_\tau \nonumber \\
& & + 
{\rm Br_{\mu}}\,
\int_{E_\nu}^\infty   
\frac{1}{\lambda^\tau_{\rm dec}(E_\tau,x)}\, 
\bar{F}_\tau(\bar E_\tau,x) 
 \frac{d \bar{N}_{\rm dec} 
(\bar{E}_\tau,E_\nu)}{d E_\nu} d\bar{E}_\tau\, \Pi_\tau \, , 
\label{eq:nshower}  \\
\frac{d F_\tau(E_\tau,t)}{d\,x}&=&-\frac{1}{\lambda^\tau_{dec}(E_\tau,x)}
F_\tau(E_\tau,x) \nonumber \\
&&
+ 
\int_{E_\tau}^\infty 
\frac{1}{\lambda^\tau_{\rm CC}(E_\nu,t)} {\rm Tr}[\Pi_\tau\, {F_\nu}(E_\nu,t)] 
 \frac{d N_{\rm CC}(E_\nu,E_\tau)}{d E_\tau} d E_\nu\, .  
\label{eq:tshower}
\end{eqnarray}
$\lambda^\tau_{\rm dec}(E_\tau,x)=\gamma_\tau\, c\, \tau_\tau$. 
$\tau_\tau$ is the $\tau$ lifetime and  
$\gamma_\tau=E_\tau/m_\tau$ is its gamma factor. 
$\frac{d N_{\rm NC}(E'_\nu,E_\nu)}{d E_\nu}\equiv
\frac{1}{\sigma_{\rm NC}(E'_\nu)}  
\frac{d\sigma_{\rm NC}(E'_\nu,E_\nu)}{dE_\nu}$ and 
$\frac{d N_{\rm CC}(E_\nu,E_\tau)}{d E_\tau}\equiv
\frac{1}{\sigma^{\tau}_{\rm CC}(E_\nu)} 
\frac{d\sigma^\tau_{\rm CC}(E_\nu,E_\tau)}{dE_\tau}$ can
be easily computed. The $\tau$ decay distributions 
$\frac{d N_{\rm dec} (E_\tau,E_\nu)}{d E_\nu}$
and $\frac{d \bar N_{\rm dec} (\bar E_\tau,E_\nu)}
{d E_\nu}$  can be found in Refs.~\cite{reno,gaisserbook}.
 
The third term in Eq.~(\ref{eq:nshower}) represents the neutrino
regeneration by NC interactions and the fourth term represents the
contribution from $\nu_\tau$ regeneration, $\nu_\tau
\rightarrow\tau^-\rightarrow\nu_\tau$, describing the energy
degradation in the process.  The secondary $\nu_\mu$ flux from
$\bar\nu_\tau$ regeneration, $\bar\nu_\tau \rightarrow\tau^+
\rightarrow\bar\nu_\tau\, \mu^+\, \nu_\mu$, is described by the last
term where we denote by over-bar the energies and fluxes of the
$\tau^+$.  $\rm Br_\mu=0.18$ is the branching ratio for this decay.
In Eq.~(\ref{eq:tshower}) the first term gives the loss of taus due to
decay and the last term gives the $\tau$ generation due to CC
$\nu_\tau$ interactions. In writing these equations we have neglected
the tau energy loss, which is only relevant at much higher energies.

An equivalent set of equations can be written for the $\bar\nu$
flux density matrix and for the $\tau^+$ flux. Both sets 
are coupled due to the secondary $\nu$ flux term.

We solve this set of ten coupled evolution equations that describe
propagation through the Earth numerically using the matter density
profile of the Preliminary Reference Earth Model and
obtain the neutrino fluxes in the vicinity of the detector
$\frac{d\phi_{\nu_\alpha}(E,\theta)}
{dE\, d\cos\theta}={\rm Tr}[F_\nu(E,L=2R\cos\theta)\, \Pi_\alpha]\,$ .

In Fig.~\ref{fig:neucasc}  we illustrate the 
interplay between the different terms in Eqs.~(\ref{eq:nshower}) 
and~(\ref{eq:tshower}). The figure covers the example of VLI-induced 
oscillations with $\delta c/c=10^{-27}$ and 
maximal $\xi_{vli}$ mixing. 
The upper panels show the final $\nu_\mu$ and $\nu_\tau$ fluxes
for vertically upgoing neutrinos after traveling the full length of the
Earth for the initial conditions
$d{\Phi(\nu_\mu)_0}/{dE_\nu}=d{\Phi(\bar\nu_\mu)_0}/{dE_\nu}\propto
E^{-1}$ and 
$d{\Phi(\nu_\tau)_0}/{dE_\nu}=d{\Phi(\bar\nu_\tau)_0}/{dE_\nu}=0$.

The figure illustrates that the attenuation in the Earth suppresses
the neutrino fluxes at higher energies. The effect of the attenuation
in the absence of oscillations is given by the dotted thin line 
in the left panel. Even in the presence of oscillations 
this effect can be well described by an overall exponential suppression 
~\cite{gaisserbook,ls} both for $\nu_\mu$'s and the oscillated $\nu_\tau$'s.
In other words, we closely reproduce the curve for 
``oscillation + attenuation" simply by multiplying the initial flux by 
the oscillation probability and an exponential damping factor: 
\begin{equation}
\frac{d\phi_{\nu_\alpha}(L=2R \cos\theta)}{dE d\cos\theta}=
\frac{d\phi_{\nu_\mu,0}}{dE d\cos\theta}\,
P_{\mu\alpha}(E,L=2R \cos\theta) \,
\exp[-X(\theta)(\sigma_{\rm NC}(E)+\sigma_{\rm CC}^\alpha(E))] \, ,
\label{eq:fluxapp}
\end{equation}
where $X(\theta)$ is the column density of the Earth.

The main effect of energy degradation by NC interactions (the third
term in Eq.~(\ref{eq:nshower})) that is not accounted for in the
approximation of Eq.(\ref{eq:fluxapp}) is the increase of the flux in
the oscillation minima (the flux does not vanish in the minimum)
because higher energy neutrinos end up with lower energy as a
consequence of the NC interactions.  The difference between the
dash-dotted line and the dashed line is due to the interplay between
the $\nu_\tau$ regeneration effect (fourth term in
Eq.~(\ref{eq:nshower})) and the flavour oscillations.  As a
consequence of the first effect, we see in the right upper panel that the
$\nu_\tau$ flux is enhanced because of the regeneration of higher
energy $\nu_\tau$'s,
$\nu_\tau(E)\rightarrow\tau^-\rightarrow\nu_\tau(E'<E)$, that
originated from the oscillation of higher energies $\nu_\mu$'s.  In
turn this excess of $\nu_\tau$'s produces an excess of $\nu_\mu$'s
after oscillation which is seen as the difference between the dashed
curve and the dash-dotted curve in the left upper panel.  Finally the
secondary effect of $\bar\nu_\tau$ regeneration (last term in
Eq.~(\ref{eq:nshower})), $\bar\nu_\tau(E)\rightarrow\tau^+\,\rightarrow\mu^+\,\bar\nu_\tau\,\nu_\mu
(E'<E)$, results into the larger $\nu_\mu$ flux  (seen 
in the left upper panel as the difference between the dashed and the thick 
full lines). This, in turn, gives an enhancement in the 
 $\nu_\tau$ flux after oscillations as seen in the right upper panel.
\begin{figure}[t]
\begin{center}
\includegraphics[width=3in]{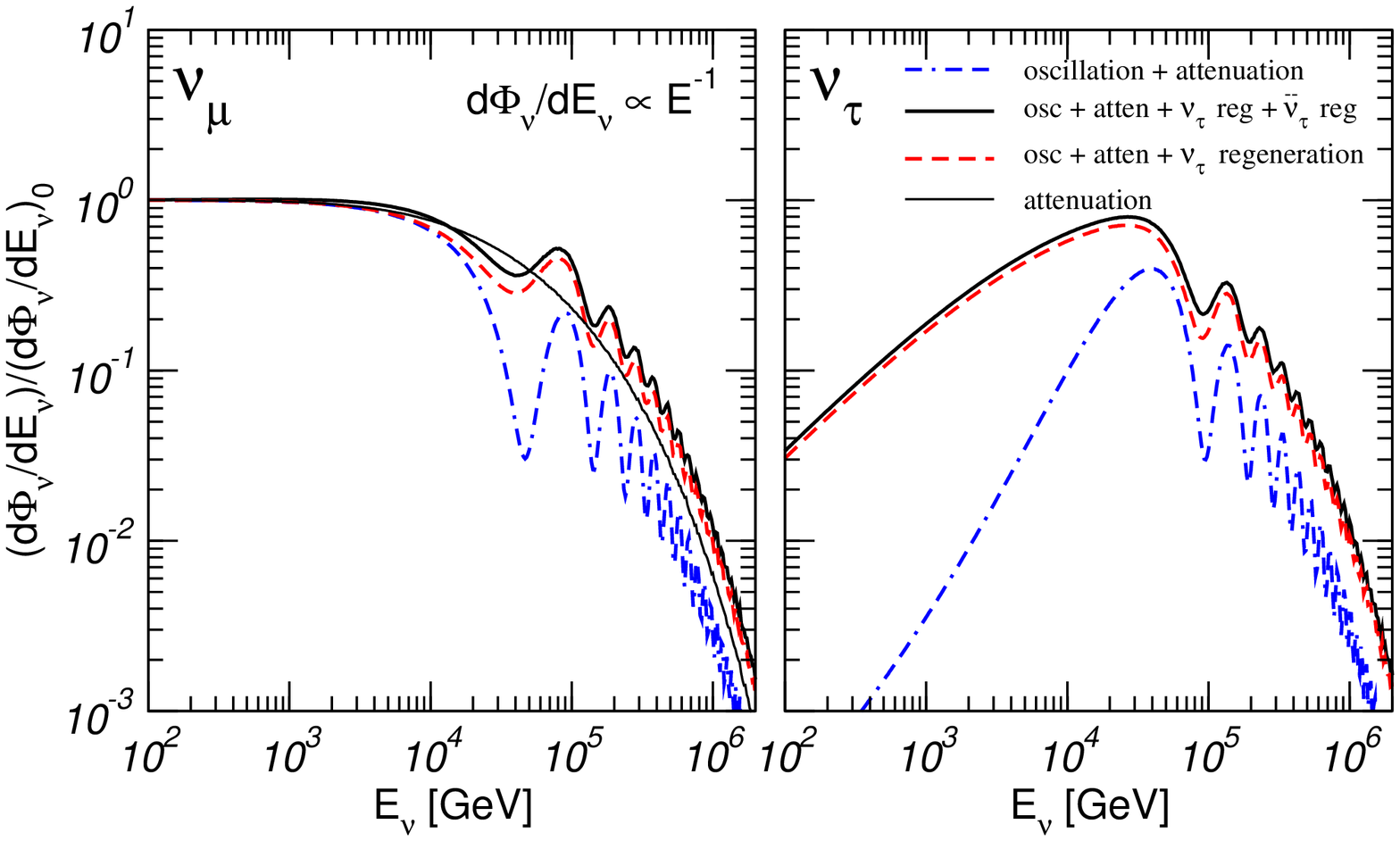}\\
\includegraphics[width=3in]{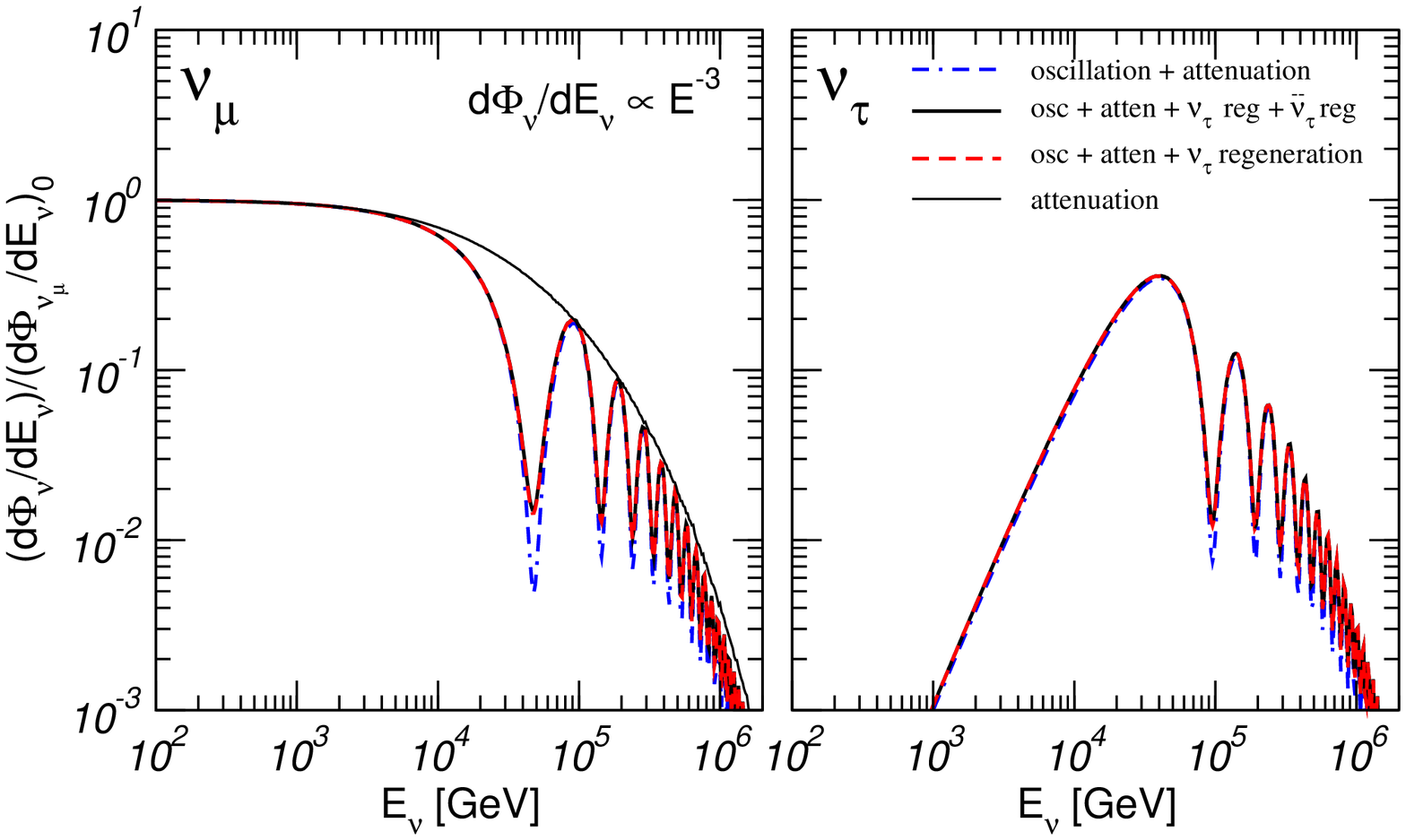}
\end{center}
\caption{\label{fig:neucasc} 
Vertically upgoing neutrinos after traveling the full length of the
Earth taking into account the effects due to VLI oscillations, 
attenuation in the Earth, $\nu_\tau$ regeneration and secondary
$\bar\nu_\tau$ regeneration (see text for details).} 
\end{figure}

The lower panels show the final 
fluxes for an atmospheric-like energy spectrum
$d{\Phi(\nu_\mu)_0}/{dE_\nu}=d{\Phi(\bar\nu_\mu)_0}/{dE_\nu}\propto
E^{-3}$ and
$d{\Phi(\nu_\tau)_0}/{dE_\nu}=d{\Phi(\bar\nu_\tau)_0}/{dE_\nu}=0$.
In this case  regeneration effects result in the degradation of the neutrino
energy and the more steeply falling the neutrino energy
spectrum, the smaller the contribution to the total
flux. As a result the final fluxes can be
relatively well described by the approximation in
Eq.(\ref{eq:fluxapp}).

\section{Example of Physics Reach: VLI-induced Oscillations}
\label{sec:results}
The expected number 
of $\nu_\mu$ induced events at IceCube can be obtained by 
a semianalytical calculation as:
\begin{eqnarray}
N^{\nu_\mu}_{\rm ev}
&=& T \int^{1}_{-1} d\cos\theta\,  
\int_0^\infty dl'_{min}\, \int^\infty_{l'_{min}} dl\,
\int_{m_\mu}^\infty dE_\mu^{\rm fin}\,
\int_{E_\mu^{\rm fin}}^\infty dE_\mu^0\, 
\int_{E_\mu^0}^\infty dE_\nu \\ \nonumber
&&\frac{d\phi_{\nu_\mu}}{dE_\nu d\cos\theta}(E_\nu,\cos\theta)
\frac{d\sigma^\mu_{CC}}{dE_\mu^0}(E_\nu,E_\mu^0)\, n_T\, 
F(E^0_\mu,E_\mu^{\rm fin},l)\, A^0_{eff}\, \, .
\label{eq:numuevents}
\end{eqnarray}
$\frac{d\phi_{\nu_\mu}}{dE_\nu d\cos\theta}$ is the differential muon
neutrino neutrino flux in the vicinity of the detector after evolution
in the Earth matter obtained as described in the previous section. We
use the neutrino fluxes from Honda~\cite{honda:04} 
extrapolated to match at higher energies the fluxes from
Volkova~\cite{volkova}.  At high energy prompt neutrinos from charm
decay are important and it is evaluated for   
two different models of charm production: the recombination quark parton
model (RQPM) developed by Bugaev {\sl et al}~\cite{rqpm} and the model
of Thunman {\sl et al} (TIG)~\cite{tig} that predicts a smaller rate.
$\frac{d\sigma^\mu_{CC}}{dE_\mu^0}(E_\nu,E_\mu^0)$ is the differential
CC interaction cross section producing a muon of energy $E_\mu^0$. 
$T$ is the exposure time of the detector. Equivalently,
muon events arise from $\bar\nu_\mu$ 
interactions that are evaluated by an equation similar to
Eq.(\ref{eq:numuevents}).

After production with energy $E_\mu^0$, the muon ranges out in the
rock and in the ice surrounding the detector and looses energy.  We
denote by $F(E^0_\mu,E_\mu^{\rm fin},l)$ the function that describes
the energy spectrum of the muons arriving at the detector.  
We compute the function $F(E^0_\mu,E_\mu^{\rm fin},l)$ by propagating
the muons to the detector taking into account energy losses due to
ionization, bremsstrahlung, $e^+e^-$ pair production and nuclear
interactions according to Ref.~\cite{ls}. 

The details of the detector are encoded in the effective area
$A^0_{eff}$ for which we make a phenomenological parametrization to
simulate the response of the IceCube detector after events that are
not neutrinos have been rejected (
referred to as ``level 2" cuts in Ref.~\cite{IceCube}).  The explicit
form of $A^0_{eff}$ cn be found in Ref.\cite{ouricecube}.

Together with $\nu_\mu$-induced muon events, oscillations also
generate $\mu$ events from the CC interactions of the 
$\nu_\tau$ flux  which reaches the detector producing a
$\tau$ that subsequently decays as 
$\tau\rightarrow \mu \bar \nu_\mu \nu_\tau$ and produces a $\mu$ in the 
detector: 
\begin{eqnarray}
\!\!\!\!\!\!\!\!\!\!\!\!\!\!\!\!
N^{\nu_\tau}_{\rm ev}
&=& T \int^{1}_{-1} d\cos\theta\,  
\int_0^\infty dl'_{min}\, \int^\infty_{l'_{min}} dl\,
\int_{m_\mu}^\infty dE_\mu^{\rm fin}\,
\int_{E_\mu^{\rm fin}}^\infty dE_\mu^0\, 
\int_{E_\mu^0}^\infty dE_\tau 
\int_{E_\tau}^\infty dE_\nu \\ \nonumber
&&
\!\!\!\!\!\!\!\!\frac{d\phi_{\nu_\tau}}{dE_\nu d\cos\theta}(E_\nu,\cos\theta)
\frac{d\sigma^\mu_{CC}}{dE_\tau}(E_\nu,E_\tau)\, n_T\, 
\frac{dN_{dec}}{dE_\mu^0}(E_\tau,E_\mu^0)
F(E^0_\mu,E_\mu^{\rm fin},l)\, A^0_{eff}\,  \; ,
\label{eq:nutauevents}
\end{eqnarray}
where $\frac{d N_{\rm dec} (E_\tau,E_\mu^0)}{d E_\mu^0}$ 
can be found in Ref.~\cite{gaisserbook}. Equivalently we compute
the number of  $\bar\nu_\tau$-induced muon 
events.

Neutrino oscillations introduced by NP effects result in
an energy dependent distortion of the zenith angle distribution 
of ATM muon events. We quantify this effect in IceCube 
by evaluating the expected angular and $E_\mu^{fin}$ distributions in 
the detector using Eqs.~(\ref{eq:numuevents}) and 
~(\ref{eq:nutauevents}) in conjunction with the fluxes obtained after 
evolution in the Earth for different sets of NP oscillation parameters.
\begin{figure}[t]
\begin{center}
\includegraphics[width=3in]{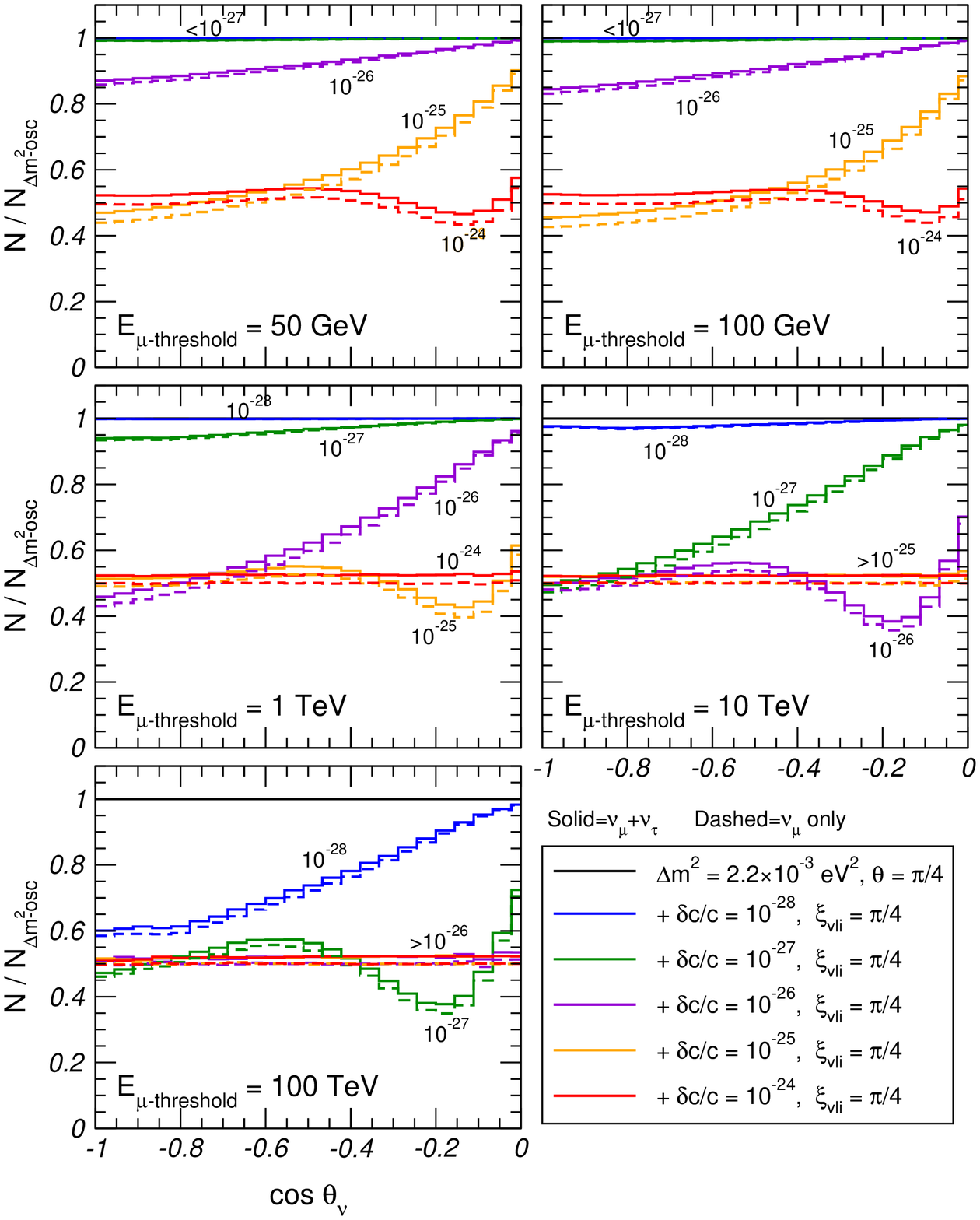}
\includegraphics[width=3in]{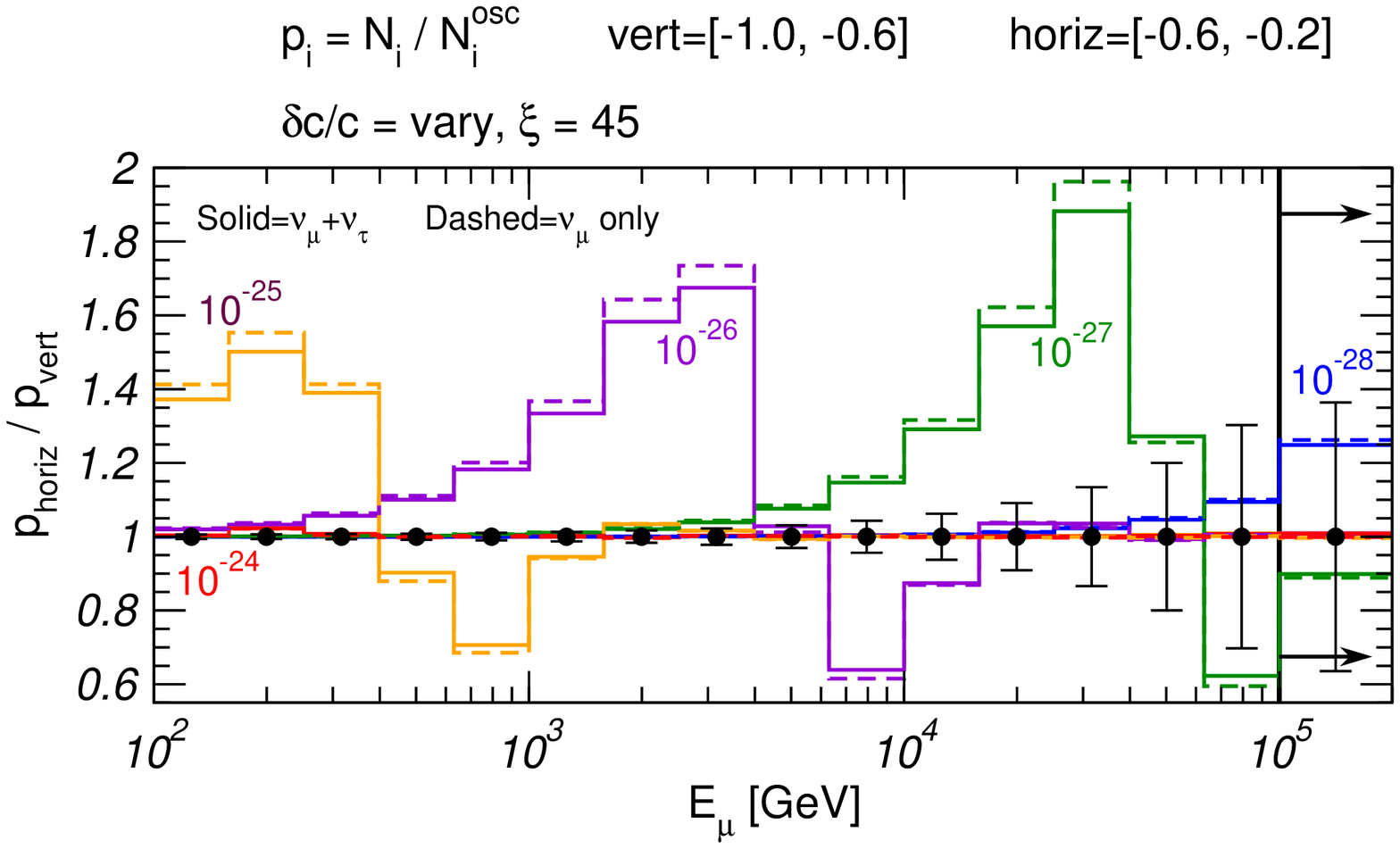}
\end{center}
\caption{\label{fig:zenith0} 
{\bf Upper panels}:
Zenith angle distributions for muon induced events for different
values of the VLI parameter $\delta c/c$ and maximal mixing
$\xi_{vli}=\pi/4$ for different threshold energy 
$E_\mu^{fin}>E_{\rm threshold}$ normalized
to the expectations for pure $\Delta m^2$ oscillations . 
{\bf Lower panels}:
The predicted horizontal-to-vertical
double ratio in Eq.(\ref{eq:dblratio}) for different values of $\delta
c/c$. The data points in the figure show the expected statistical
error corresponding to the observation of no NP effects in 10 years of
IceCube.}
\end{figure}

For illustration we concentrate on oscillations resulting from VLI 
that lead to  an oscillation wavelength inversely proportional
to the neutrino energy. The results can be directly applied to 
oscillations due to VEP.   
We show in Fig.~\ref{fig:zenith0} the 
zenith angle distributions for muon induced events for different
values of the VLI parameter $\delta c/c$ and maximal mixing
$\xi_{vli}=\pi/4$ for different threshold energy 
$E_\mu^{fin}>E_{\rm threshold}$ normalized
to the expectations for pure $\Delta m^2$ oscillations. 
The full lines include both the $\nu_\mu$-induced events 
(Eq.(\ref{eq:numuevents})) and $\nu_\tau$-induced events
(Eq.(\ref{eq:nutauevents})) while the last ones are not included
in the dashed curves. 
We see that for a given value of $\delta c/c$ there is a range
of energy for which the angular distortion is maximal. Above
that energy, the oscillations average out and result in
a constant suppression of the number of events. 
Inclusion of the $\nu_\tau$-induced events leads to an overall 
increase of the event rate but slightly reduces the  angular distortion.

In order to quantify the energy-dependent angular distortion we define 
the vertical-to-horizontal double ratio 
\begin{equation}
R_{h/v}(E_\mu^{fin,i})\equiv
\frac{P_{\rm hor}}{P_{\rm ver}}(E_\mu^{fin,i})
=\frac
{\frac
{\displaystyle 
N^{vli}_\mu(E_\mu^{fin,i}, -0.6<\cos\theta<-0.2)}
{\displaystyle 
N^{no-vli}_\mu(E_\mu^{fin,i}, -0.6<\cos\theta<-0.2)}}
{\frac
{\displaystyle 
N^{vli}_\mu(E_\mu^{fin,i}, -1<\cos\theta<-0.6)}
{\displaystyle 
N^{no-vli}_\mu(E_\mu^{fin,i}, -1<\cos\theta<-0.6)}} \; ,
\label{eq:dblratio}
\end{equation} 
where by $E_\mu^{fin,i}$ we denote integration in an energy bin 
of width $0.2\,\log_{10}(E_\mu^{fin,i})$ using that 
IceCube measures energy to 20\% in $\log_{10} E$ for muons. 

In what follows we will use the double ratio in
Eq.~(\ref{eq:dblratio}) as the observable to determine the sensitivity
of IceCube to NP-induced oscillations. We have chosen a double ratio
to eliminate uncertainties associated with the overall normalization
of the ATM fluxes at high energies.  It is worth noticing that
using this observable relies on the fact that the zenith angular
dependence of the effective area is well understood.  

\begin{figure}[t]
\begin{center}
\includegraphics[width=3.in]{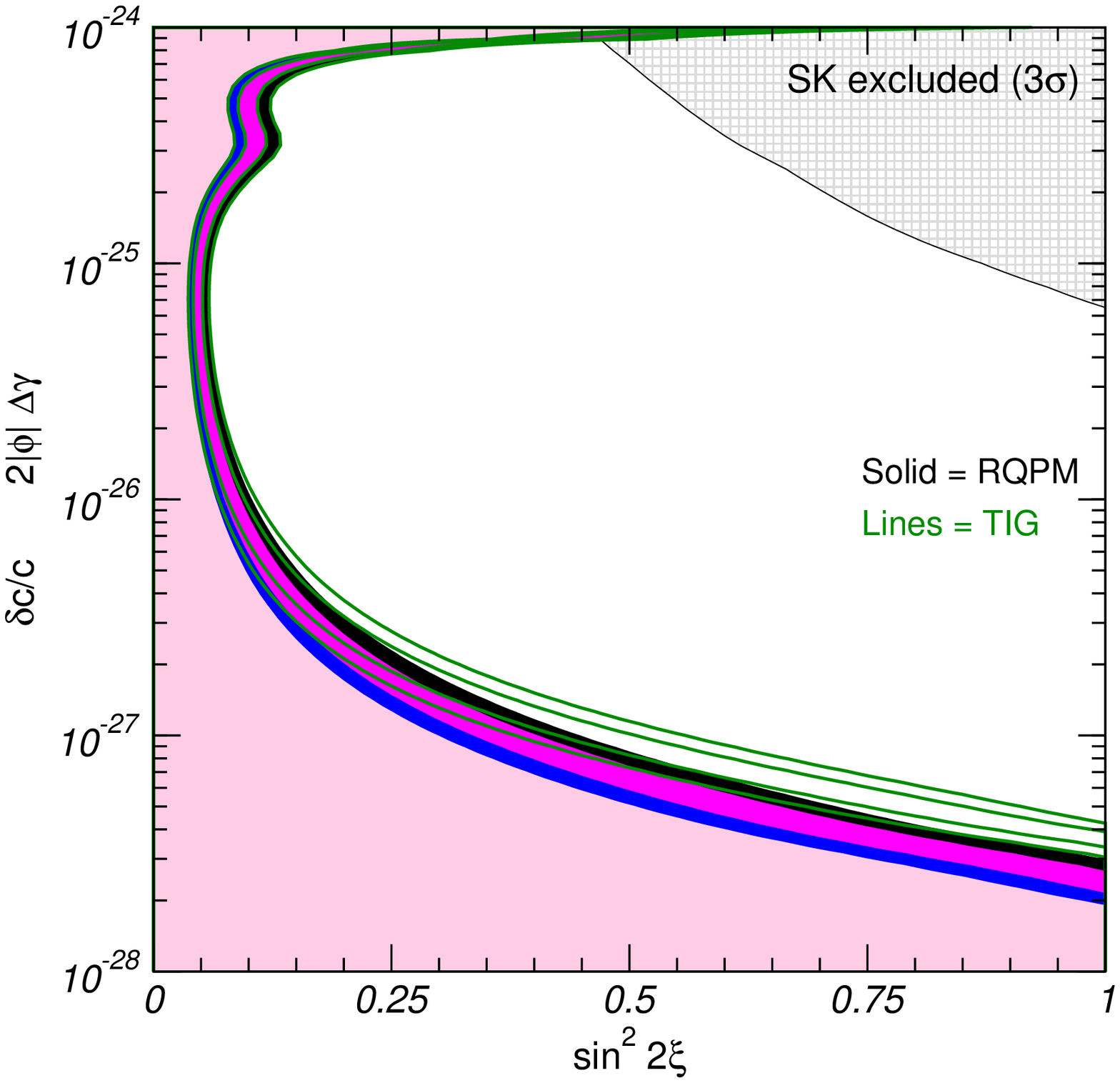}
\end{center}
\caption{\label{fig:chisq} 
Sensitivity limits
in the $\delta c/c, \xi_{\rm vli}$ at 90, 95, 99 and 3 $\sigma$ CL.
The hatched area in the upper right corner is the present $3\sigma$
bound from the analysis of SK data in Ref.~\cite{ouratmnp}.}
\end{figure}
In Fig.~\ref{fig:zenith0} we plot the expected value of this ratio
for different values of $\delta c/c$. As mentioned above, IceCube
measures energy to 20\% in $\log_{10} E$ for muons. Accordingly, we
have divided the data in 16 $E_\mu^{fin}$ bins: 15 bins between
$10^2$ and $10^5$ GeV and one containing all events above $10^{5}$
GeV.  In the figure the full lines include both the $\nu_\mu$-induced
events (Eq.(\ref{eq:numuevents})) and $\nu_\tau$-induced events
(Eq.(\ref{eq:nutauevents})) while the last ones are not included in
the dashed curves. As described above, the net result of including the
$\nu_\tau$-induced events is a slight decrease of the maximum expected
value of the double ratio. The data points in the figure show the
expected statistical error corresponding to the observation of no NP
effects in 10 years of IceCube.
In order to estimate the expected sensitivity we assume that no NP effect 
is observed and define a simple $\chi^2$ function including only 
the statistical errors. 

We show in Fig.~\ref{fig:chisq} 
the sensitivity limits
in the $[\delta c/c, \xi_{\rm vli}]$-plane at 90, 95, 99 and 3 $\sigma$ CL
obtained from the condition 
$\chi^2(\delta c/c, \xi_{\rm vli})<\chi^2_{max}({\rm CL,2dof})$.
We show in the figure the results obtained
using the RQPM model and the 
TIG model. The difference is about 
50\%  in the strongest bound on $\delta c/c$. The figure illustrates the improvement on the present bounds
by more than two orders of magnitude even within the context of this very 
conservative analysis. The loss of sensitivity at large 
$\delta c/c$ is a consequence of the use of a double ratio as
an observable. Such an observable is insensitive to NP effects 
if $\delta c/c$ is large enough for the oscillations
to be always averaged leading only to an overall suppression.

When data becomes available a more realistic analysis is likely to lead to
a further improvement of the sensitivity. 
\vskip 0.3cm

\section*{Acknowledgments}

%% Keep the small font tag for the acknowledegments
{\small 
This work was supported in part by the National Science Foundation
grant PHY0354776 and in part by Spanish Grant No
FPA-2004-00996.
}

%%%%%%%%%%%%%%%%%%%%%%%%%%%%%%%%%%%%%% reset.txt counters %%%%%%%%%%%%%%
%%
%%%%%%% do not change below here  %%%%%%%%%%%%%%%%%%%%%%%%%%%%%
\setcounter{section}{0}
\setcounter{subsection}{0}
\setcounter{figure}{0}
\setcounter{table}{0}
\setcounter{equation}{0}
\newpage

%%%%%%%%%%%%%%%%%%%%%%%%%%%%%%%%%%%%%%%%%%%%%%%%%%% Title, authors and addresses
\begin{frontmatter}


\title{ Testing Lorentz Invariance using Atmospheric Neutrinos and
  AMANDA-II }

\author[address1]{J. L. Kelley}
\author[address2]{for the IceCube Collaboration}

\address[address1]{Department of Physics, University of Wisconsin, Madison,
  WI 53706, U.S.A.}
\address[address2]{http://icecube.wisc.edu}

% \linenumbers

\begin{abstract}
  Several phenomenological models of physics beyond the Standard Model
  predict flavor mixing in the neutrino sector in addition to
  conventional mass-induced oscillations.  In particular, violation of
  Lorentz invariance (VLI) results in neutrino oscillation effects
  parametrized by the maximal attainable velocity difference $\delta c/c$.
  We report on a study of the sensitivity of the AMANDA-II detector to such
  effects using distortions in the spectrum of high-energy
  atmospheric neutrinos.  For maximal mixing and six years of simulated
  data, the preliminary sensitivity of AMANDA-II to VLI of this type is
  $\delta c/c < 2.1 \times 10^{-27}$ at the 90\% confidence level.
\end{abstract}

\end{frontmatter}

% \linenumbers

%%%%%%%%%%%%%%%%%%%%%%%%%%%%%%%%%%%%%%%%%%%%%%%%%%%%%% MAIN TEXT
\section{\label{sec:intro} Introduction}

  Flavor oscillations in the neutrino sector provide an interesting method
  to test phenomenological models of physics beyond the Standard Model.
  While mass-induced oscillations of atmospheric neutrinos are on firm
  experimental footing \cite{superk04,soudan03,macro02},
  subdominant effects may yet be present.  In particular, violation of
  Lorentz invariance (VLI) can result in oscillations at high energies and
  can distort the atmospheric neutrino spectrum.

  The AMANDA-II detector, a subdetector of the IceCube experiment, is an
  array of 677 optical modules buried in the ice at the geographic South Pole
  which detects the \v{C}erenkov radiation from charged particles produced in
  neutrino interactions with matter \cite{andres01}.  In particular, muons
  produced in charged-current $\nu_{\mu}$ and $\bar{\nu}_{\mu}$ interactions
  deposit light in the detector with a track-like topology, allowing us
  to use directional reconstruction to reject the large background of
  down-going atmospheric muon events.  After suitable quality selection
  criteria are applied, AMANDA-II accumulates atmospheric neutrino
  candidates above 50 GeV at a rate of $\approx 4$ per day
  \cite{ackermann05}. While conventional oscillations are suppressed at
  these energies, VLI 
  effects can be detected or constrained by their influence on the zenith
  angle distribution and energy-correlated observables.
  
\section{\label{sec:instructions} Phenomenology }

Various new physics scenarios can result in neutrino flavor mixing beyond
conventional oscillations.  We focus here on
oscillations induced by differing maximally attainable velocities (MAVs) in
the neutrino sector.  MAV eigenstates can be distinct from flavor
eigenstates, resulting in oscillations characterized by the MAV difference
$\delta c / c = (c_{1} - c_{2}) / c$.  

Conventional and VLI oscillations can be combined in a two-family scenario,
with the following survival muon 
neutrino survival probability as a function of energy $E$ and baseline $L$
(in energy units) \cite{coleman99,glashow04,ggm04}:

\begin{equation}
  P_{\nu_{\mu} \rightarrow \nu_{\mu}} = 1\ -\ \sin^2 2\Theta\ \sin^2 \left(
  \frac{\Delta m^2L}{4E}\ \mathcal{R}\right)\ ,
\end{equation}

\noindent where

\begin{equation}
  \sin^2 2\Theta = \frac{1}{\mathcal{R}^2}(\sin^2 2\theta + R^2 \sin^2 2\xi
  + 2R\sin 2\theta \sin 2\xi \cos \eta)\ ,
\end{equation}
\begin{equation}
  \mathcal{R} = \sqrt{1 + R^2 + 2R(\cos 2\theta \cos 2\xi + \sin 2\theta
    \sin 2\xi \cos \eta)}\ ,
\end{equation}
\noindent and
\begin{equation}
  R = \frac{\delta c}{c}\frac{E}{2}\frac{4E}{\Delta m^2}\ .
\end{equation}

Standard oscillations are characterized by the mass-squared
difference $\Delta m^2$ and mixing angle $\theta$, while VLI
oscillation parameters include the velocity difference $\delta c/c$, the mixing
angle $\xi$, and the phase $\eta$.  If we take both conventional and VLI
mixing to be maximal ($\theta = \xi = \pi/4$) and set $\cos \eta = 1$, this
reduces to the following:

\begin{equation}
\label{max_prob}
  P_{\nu_{\mu} \rightarrow \nu_{\mu}} \mathrm{(maximal)} = 1\ -\ \sin^2 \left(
  \frac{\Delta m^2L}{4E} + \frac{\delta c}{c}\frac{LE}{2}\right)\ .
\end{equation}

Note the different energy dependence of the two effects.  For atmospheric
neutrinos, the zenith angle functions as a
surrogate for the baseline $L$, allowing path lengths up to the diameter of
the Earth.  Figure \ref{fig_survival} shows the survival probability as a
function of neutrino energy and zenith angle for the maximal case, as in
equation~(\ref{max_prob}).

\begin{figure*}[hbt]
\centering\epsfig{file=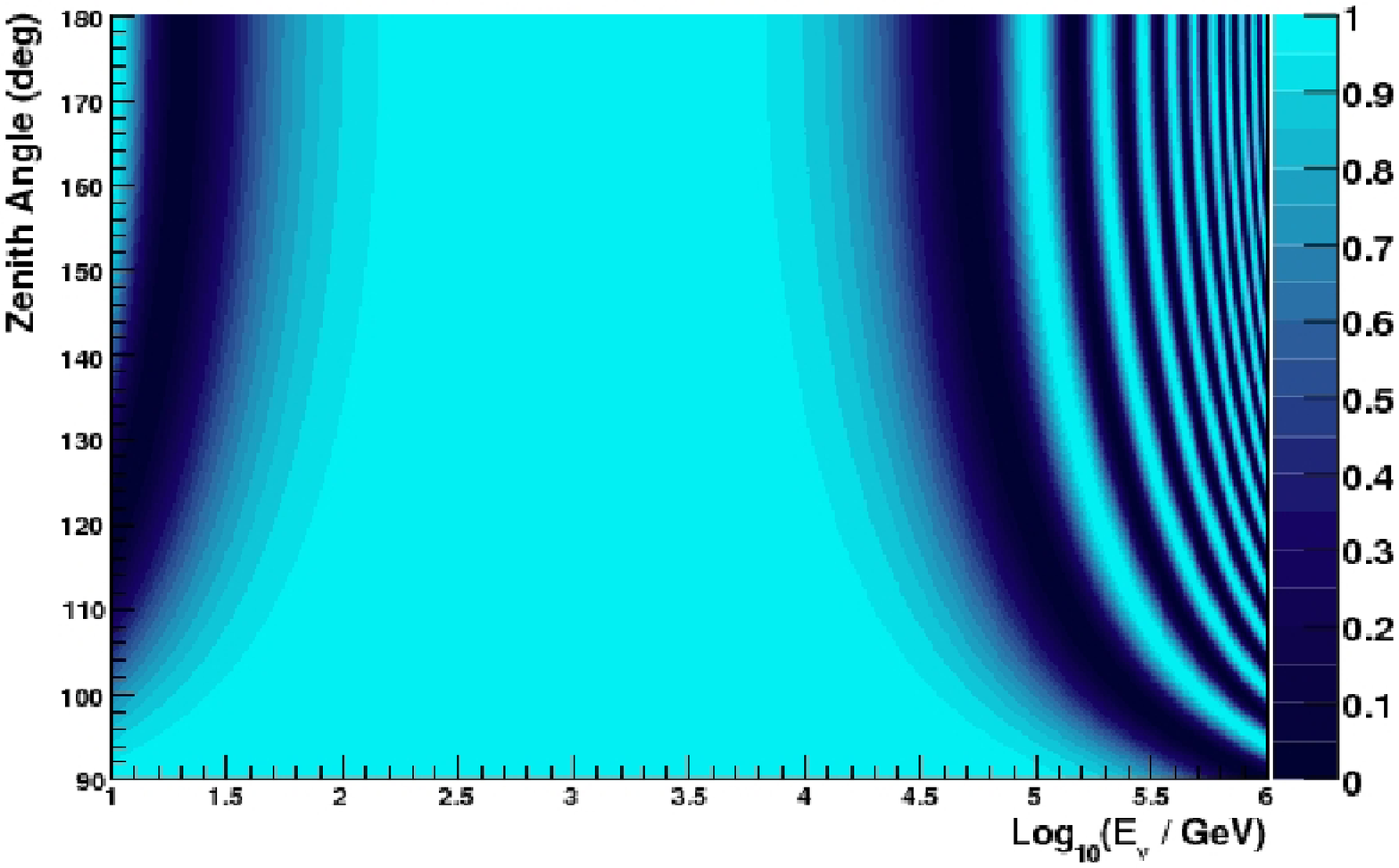,height=8cm}
\caption{Atmospheric $\nu_{\mu}$ survival probability as function of neutrino energy and
zenith angle.  Conventional oscillations are present at low energies, while
high-energy oscillations are due to VLI (maximal mixing, $\delta c /
c = 10^{-27}$).}
\label{fig_survival}
\end{figure*}

\section{\label{sec:plots} Analysis Methodology }

First, to obtain a clean sample of atmospheric neutrinos, we must separate these
from the large background of atmospheric muons.  Selecting events with a
reconstructed zenith angle below the horizon allows rejection of many such
events, but we must generally apply further quality criteria to eliminate
mis-reconstructed muons.  For this study, we have used the selection
criteria from the 2000-03 AMANDA-II point source search \cite{ackermann05}
and examine only zenith angles $>$ 100\textdegree.

Next, our goal is to measure or constrain the energy-dependent angular
distortions caused by VLI effects.  While AMANDA-II has an angular
resolution of a few degrees \cite{ptsrc05}, reconstruction of the neutrino
energy is more difficult and fundamentally limited by the stochastic losses
of the muon.  Instead, we use a well-simulated energy-correlated
observable, the number of triggered optical modules ($N_{ch}$).

Now, to determine values of the parameters ${\theta_i}$ of our hypothesis (in the simplest
one-dimensional case, just $\delta
c/c$) that are allowed or excluded at some confidence level, we follow the
likelihood prescription described by Feldman and Cousins \cite{feldman98}:\\

\begin{itemize}
\item{For each point in the parameter space ${\theta_i}$, we sample many times
from the parent Monte Carlo distributions of the observable(s) (MC
``experiments'').}\\

\item{For each MC experiment, we calculate the log likelihood ratio 

\begin{equation}
\Delta \mathcal{L} = -2 \ln \mathsf{L}_{i} + 2 \ln \mathsf{L}_{i,
    best}\ , 
\end{equation}

\noindent where $\mathsf{L}_{i}$ is the Poisson probability that the MC experiment is
derived from the parent distribution at ${\theta_i}$ (other likelihood
formulations are possible).}\\

\item{For each point ${\theta_i}$, we find the value $\Delta \mathcal{L}_{crit}$
at which, say, 90\% of MC experiments have a lower $\Delta \mathcal{L}$.}\\

\item{Finally, we compare the $\Delta \mathcal{L}$ of the data (or in our case, a simulated data set
generated under the null hypothesis) with the critical surface $\Delta
\mathcal{L}_{crit}$, and regions of the parameter space at which $\Delta
\mathcal{L} > \Delta \mathcal{L}_{crit}$ are excluded at that confidence
level.  For a one-dimensional parameter space, this can likely be interpreted an
upper limit, and one can calculate a median sensitivity by iterating over a
number of simulated data sets.}\\

\end{itemize}

\noindent As noted in \cite{feldman98}, the likelihood formulation has a
number of desirable features compared to a standard $\chi^2$ approach, the
most significant being proper coverage.

\section{\label{sec:sensitivity} Sensitivity of AMANDA-II }

We have performed a Monte Carlo study using six years of simulated
AMANDA-II data: an integrated exposure of 1200 days, approximately 5100
events below the horizon under the null hypothesis (conventional
oscillations only).  For this initial study, we have tested only the
$N_{ch}$ distribution across a one-dimensional parameter space, varying the
VLI strength $\delta c/c$.  To anticipate the impact of the inclusion of
systematic errors in the future, we have left free the normalization of the
atmospheric neutrino flux ({\em i.e.} treating it as a nuisance parameter).
We have not included the zenith angle distribution in this analysis, as we
have not yet accounted for systematic uncertainties in the shape of the
spectrum.  The curves of $\Delta \mathcal{L}_{crit}$ for the 90\%, 95\%,
and 99\% confidence levels are shown in Figure \ref{fig_llh}, along with
the likelihood ratio for a single simulated data set.

\begin{figure*}[t]
\centering\epsfig{file=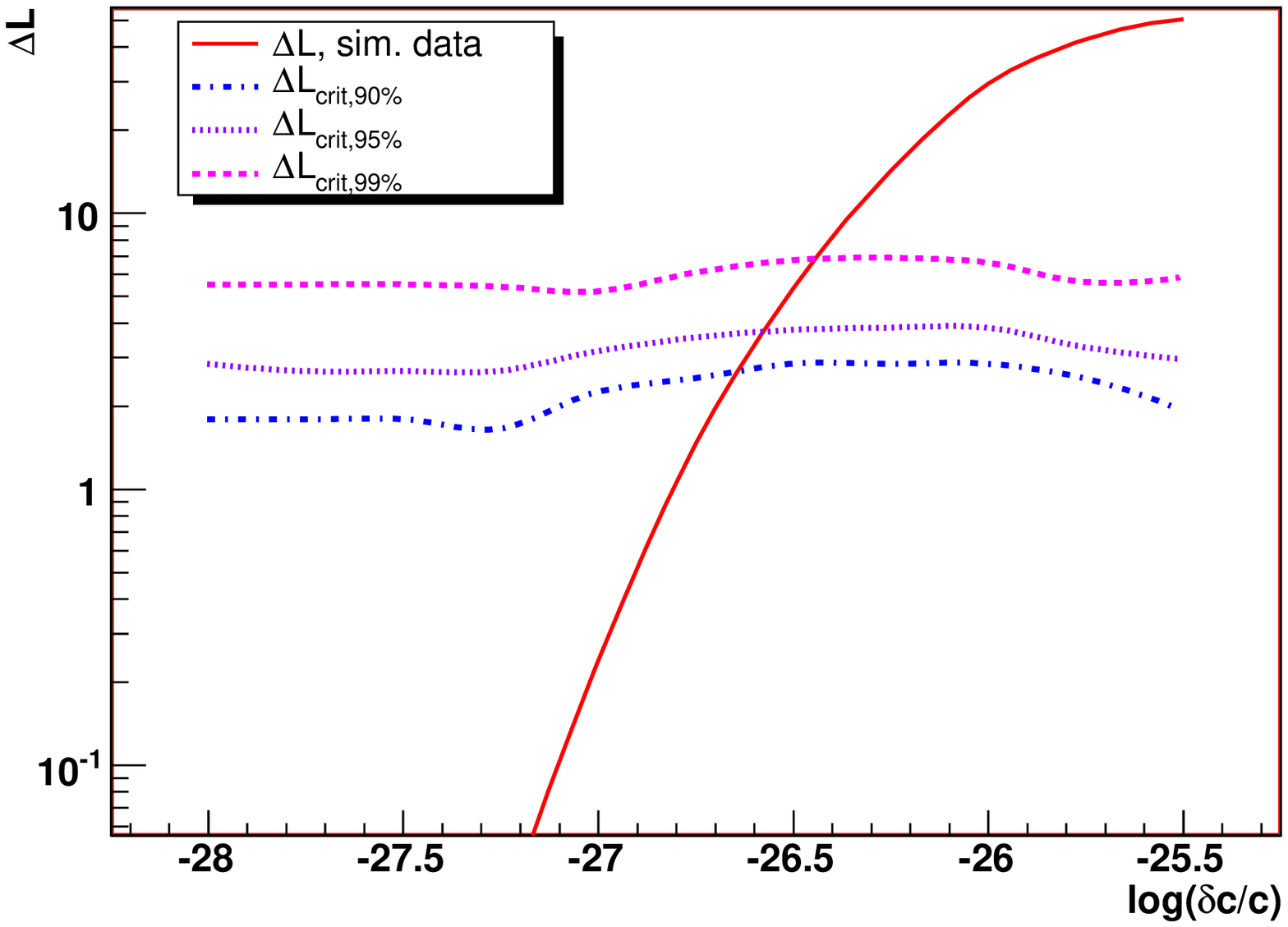,height=7cm}
\caption{Likelihood ratio for VLI effects using the shape of the $N_{ch}$
  distribution, for values of the parameter $\delta c/c$.  The critical
  curves for various confidence levels are shown, along with $\Delta \mathcal{L}$
  for a simulated six-year data set.  Values of $\delta c/c$ to the right of the
  point of intersection with the critical curve are excluded.}
\label{fig_llh}
\end{figure*}

Assuming maximal mixing ($\sin 2\xi = 1$) and phase $\cos \eta = 1$, we
find a median sensitivity 
of $\delta c / c < 2.1 \times 10^{-27}$ at the 90\% confidence level.
Existing experimental limits include the MACRO result of $\delta c/c < 2.5
\times 10^{-26}$ 
\cite{macro05} and the limit by Gonz\'{a}lez-Garc\'{i}a and Maltoni using the
Super-Kamiokande + K2K data, $\delta c/c < 2.0 \times 10^{-27}$ \cite{ggm04}.  

\section{\label{sec:conclusions} Conclusions and Outlook }
\enlargethispage{0.5cm}
Using its large sample of atmospheric neutrinos, AMANDA-II is
capable of detecting or constraining high-energy new physics effects in
the neutrino sector.  The Monte Carlo study presented here indicates a
sensitivity to VLI effects competitive with existing limits, and a number of
improvements (such as testing multiple observables) and
optimizations (event selection criteria and the binning of the
observables) are forthcoming.  We anticipate applying this analysis in the
near future to the AMANDA-II data collected during 2000-2005.  Furthermore,
the same methodology can also be applied to constrain other  
physics beyond the Standard Model, such as violations of the equivalence
principle \cite{gasperini89} or quantum decoherence resulting from interactions
of neutrinos with the background space-time foam
\cite{ellis84,morgan04,anchordoqui05}.

The next-generation IceCube detector, with an instrumented volume of 1
$\mathrm{km}^3$, will allow unprecedented sensitivity to these same
effects.  In 10
years of operation, IceCube will collect a sample of over 700 thousand
atmospheric neutrinos and will be sensitive at the 90\%
confidence level to VLI effects at the level of $\delta c/c < 2.0 \times 10^{-28}$
\cite{halzen05}. This high-statistics
sample will also provide an 
opportunity to test other phenomenological models of physics beyond the
Standard Model.

\section*{Acknowledgments}
%% Keep the small font tag for the acknowledegments
{\small 
  The author wishes to thank A. Olivas for presenting this work at the
  conference. 
}

%%%%%%%%%%%%%%%%%%%%%%%%%%%%%%%%%%%%%% reset.txt counters %%%%%%%%%%%%%%
%%
%%%%%%% do not change below here  %%%%%%%%%%%%%%%%%%%%%%%%%%%%%
\setcounter{section}{0}
\setcounter{subsection}{0}
\setcounter{figure}{0}
\setcounter{table}{0}
\setcounter{equation}{0}
\newpage

%%%%%%%%%%%%%%%%%%%%%%%%%%%%%%%%%%%%%%%%%%%%%%%%%%% Title, authors and addresses
\begin{frontmatter}


\title{Emergent Relativity: Neutrinos as Probe of the Underlying Theory}

% use optional labels between square brackets to link authors explicitly to addresses:
% If more than one author, keep a comma between the author tags

\author[address1]{F.R. Klinkhamer}

\address[address1]{Institute for Theoretical Physics,
             University of Karlsruhe (TH), 76128 Karlsruhe, Germany}

\begin{abstract}
Neutrinos allow for a test of the hypothesis that
the fermions of the Standard Model have Fermi-point splitting,
analogous to the fermionic quasi-particles of
certain condensed-matter systems. If present,
the corresponding Lorentz-violating terms in the Hamiltonian
may provide a new source
of \DS{T} and \DS{CP} violation in the leptonic sector, which is
not directly related to mass.
\end{abstract}

% \begin{keyword}

%\PACS
% \end{keyword}

\end{frontmatter}

%%%%%%%%%%%%%%%%%%%%%%%%%%%%%%%%%%%%%%%%%%%%%%%%%%%%%% MAIN TEXT

\section{\label{FRK-sec:intro} Introduction}

The basic idea of this talk is to suggest
neutrinos as  a probe of radically new physics.
Of course, this is a long-shot $\ldots$ but worth trying.

One example of such new physics would be related to the
concept of \emph{emergent symmetries}
\cite{FroggattNielsen1991,Bjorken2001,Laughlin2003,Volovik2003}.
Lorentz invariance, for example, would  not be a fundamental
symmetry but  an emergent phenomenon at low energies.

In order to be specific, we start from an
analogy with quantum phase transitions in fermionic atomic gases or
superconductors and consider the hypothesis
\cite{KV-JETPL2004,KV-IJMPA2005,KV-JETPL2005} that the fermions
of the Standard Model have tiny Lorentz-violating effects due to
Fermi-point splitting (abbreviated FPS and explained below).

If Fermi-point splitting would indeed occur for the quarks and
leptons of the \SM, then
neutrinos may provide a \emph{unique} window to the underlying theory
\cite{K-JETPL2004,K-IJMPA2006,K-PRD2005,K-PRD2006}.
Specifically, there could be new effects in neutrino oscillations,
possibly showing significant \DS{T} and \DS{CP} violation
(and perhaps even \DS{CPT} violation).
The aim of this talk is to sketch some of the potential FPS effects
but we refer, in particular, to the contribution of
M.C. Gonz\'{a}lez-Garc\'{i}a
in these Proceedings for a more general discussion of non-standard
neutrino oscillations.

The outline of this write-up is as follows.
In Sec.~\ref{FRK-sec:FPS-in-cond-mat}, some background
on condensed matter physics is  given and, in Sec.~\ref{FRK-sec:FPS-hypothesis},
a possible application to elementary particle physics is discussed.
In Sec.~\ref{FRK-sec:Simple-neutrino-model},
a simple but explicit neutrino model with both Fermi-point splittings
and mass differences is introduced.
In Sec.~\ref{FRK-sec:Neutrino-oscillations},
some interesting results on neutrino oscillations from this model are reviewed.
In Sec.~\ref{FRK-sec:Outlook}, concluding remarks are presented.

\section{\label{FRK-sec:FPS-in-cond-mat} Fermi-point splitting
          in atomic and condensed-matter systems}

Ultracold quantum gases of
fermionic atoms (e.g., ${}^6\text{Li}$ at nano-Kelvin temperatures)
are extremely interesting systems, especially as they can have
\emph{tunable} interactions by way of magnetic-field Feshbach
resonances. In the so-called BEC--BCS crossover region
of these systems, a BCS--type condensate has recently been
observed for $s$--wave pairing \cite{Regal-etal2004}.
As usual, BEC stands for Bose--Einstein condensate and
BCS for the superconductivity triumvirate
Bardeen, Cooper, and Schrieffer.

For the BEC--BCS crossover region in systems with
$p$--wave pairing, there is the
prediction \cite{KV-JETPL2004,KV-IJMPA2005} that a
\emph{quantum phase transition}
between a vacuum state with fully-gapped fermionic spectrum and a
vacuum state with topologically protected Fermi points (gap nodes)
occurs. Here, we only give a simple illustration of this new type of
quantum phase transition and refer the reader
to Ref.~\cite{Volovik2006} for an extensive review.

The \BN~Hamiltonian for fermionic quasi-particles in the
axial state of $p$--wave pairing is given by
\beq
H_\text{BN}= \left(
\begin{array}{cc}
|{\bf p}|^2/ (2m)  -q &\;\; c_\perp\,{\bf p}\cdot (\widehat{\bf e}_1+ i\,
\widehat{\bf e}_2)
\\[0mm]
c_\perp\,{\bf p}\cdot (\widehat{\bf e}_1- i\, \widehat{\bf e}_2) &\;\;
-|{\bf p}|^2/ (2m) +q
\end{array} \right),
\label{FRK-BogoliubovNambuH}
\eeq
with
$m$ the mass of the fermionic atom (considered is the direction
of atomic spin, which experiences the Feshbach resonance),
$(\widehat{\bf e}_1,\, \widehat{\bf e}_2,\,\widehat{\bf l}\,)$ an
orthonormal triad, $\widehat{\bf l}$
the direction of the orbital momentum of the pair, $c_\perp$ the
maximum transverse speed, and
$q$ a parameter controlled by the magnetic
field near the Feshbach resonance.

The energy spectrum of this Hamiltonian is readily calculated:
\beq
E_\text{BN}^2 ({\bf p}) = \big(|{\bf p}|^2/(2m)-q \big)^{\!2} +\,
          c_\perp^2\,\big|{\bf p}\times \widehat{\bf l}\:\big|^2 .
\label{FRK-BogoliubovNambuE}
\eeq
Clearly, there are two regimes. For parameter $q<0$, on the one hand,
there is a BEC regime  with mass gap, $E \ne 0$.
For parameter $q>0$, on the other hand, there is a BCS
regime with two Fermi points  in momentum space,
\beq
{\bf b}_1=+p_F \: \widehat{\bf l}\,,\quad
{\bf b}_2=-p_F \: \widehat{\bf l}\,,\quad
p_F\equiv \sqrt{2 m q}\,,
\eeq
at which the energy function vanishes,
$E({\bf p})=0$  for ${\bf p}={\bf b}_a$ with $a=1,2$.

There is then a quantum phase transition at $q=0$,
with a mass gap for $q<0$ and a space-like splitting of Fermi points
($\Delta{\bf b}\equiv{\bf b}_1-{\bf b}_2 \ne 0$)
for $q>0$; see Fig.~\ref{FRK-FIG-FPS}.

This example also clarifies the
concept of emergent relativity mentioned in the Introduction.
Consider momenta close to one of the two Fermi points,
for example, ${\bf p}={\bf b}_1+{\bf k}$ with $|{\bf k}| \ll p_F$.
Then, the energy \ref{FRK-BogoliubovNambuE} becomes
\beq
E_\text{BN}^2 \sim
(p_F/m)^2\:k_\parallel^2 + c_\perp^2\:k_\perp^2 \sim \, \widetilde{c}^2 \,
\big(\, \widetilde{k}_\parallel^2 + \widetilde{k}_\perp^2\, \big)\,,
\label{FRK-mass-shell}
\eeq
after the following rescalings:
\beq
k_\parallel \equiv {\bf k} \cdot \,\widehat{\bf l}
            \equiv (\,\widetilde{c} \,m/p_F) \;\widetilde{k}_\parallel \,,\quad
k_\perp \equiv |\,{\bf k} \times \widehat{\bf l}\,|
            \equiv (\,\widetilde{c}/c_\perp)\; \widetilde{k}_\perp \,,
\eeq
which would be appropriate for a local observer
made of the \emph{same} quasi-particles \cite{Volovik2003,Volovik2006}.
In terms of the rescaled momentum $\widetilde{\bf k}$, relation
\ref{FRK-mass-shell} corresponds precisely to the mass-shell condition
of a massless relativistic particle.

\begin{figure*}[t]
\begin{center}
\includegraphics[width=0.65\textwidth]{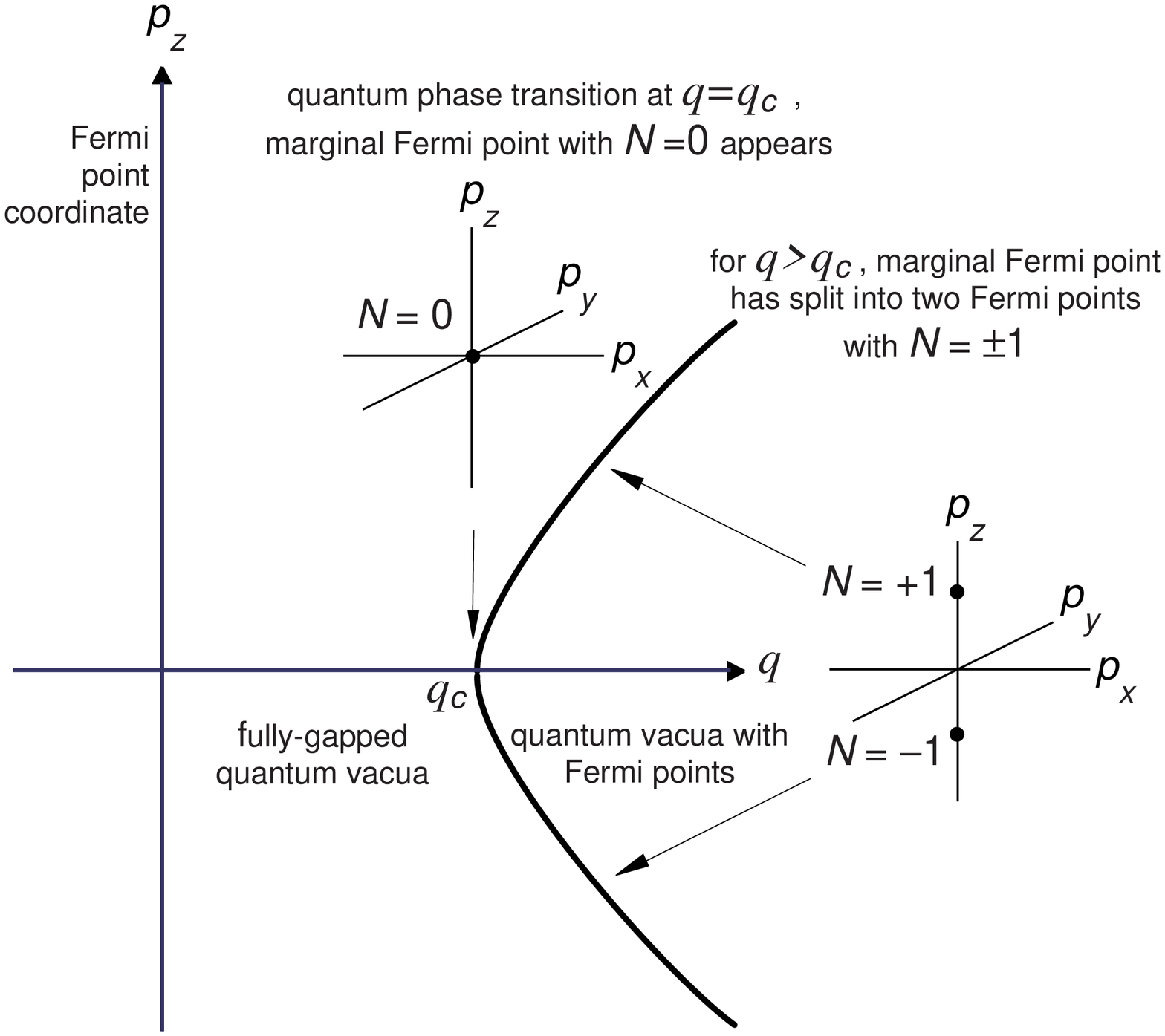}  %%[width=8cm]
\caption{
Quantum phase transition at $q=q_c$ between a quantum vacuum with mass gap and
one with topologically-protected Fermi points (gap nodes).
At $q=q_c$, there appears a marginal Fermi point with topological charge $N=0$
(inset at the top).
For $q > q_c$, the marginal Fermi point has split into two Fermi points
characterized by nonzero topological invariants $N = \pm 1$
(inset on the right). A system
described by Hamiltonian \ref{FRK-BogoliubovNambuH},
for $\widehat{\bf l}=(0,0,1)$, has critical parameter $q_c=0$.}
\label{FRK-FIG-FPS}
\end{center}
\end{figure*}

\section{\label{FRK-sec:FPS-hypothesis} FPS hypothesis
         for elementary particle physics}

Based on the analogy with certain condensed-matter systems
discussed in Sec.~\ref{FRK-sec:FPS-in-cond-mat}, the following
hypothesis has been put forward \cite{KV-JETPL2004,KV-IJMPA2005}:
perhaps the chiral fermions of the Standard Model also have
Fermi-point splitting (FPS).
Specializing to time-like splittings ($\Delta b_0 \ne 0$)
and vanishing Yukawa coupling constants
(i.e., vanishing fermion masses),
their dispersion relations would be given by:
\beq
\big( E_{a,f}({\bf p}) \big)^2  =
\Big(\,c\, |{\bf p}| + b_{0a}^{(f)}\,\Big)^2 \,,
\label{FRK-SMdispLaw-timelike}
\eeq
where $a$ labels the 16 types of massless left-handed Weyl fermions
(including a left-handed antineutrino) and $f$ the $N_\text{fam}$
fermion families (henceforth, we take $N_\text{fam}=3$).
The maximum velocity of the fermions is assumed to be universal
and equal to the velocity of light \emph{in vacuo}, $c$.
Note that we still speak about Fermi-\emph{point} splitting
even though the energy \ref{FRK-SMdispLaw-timelike} for
$b_{0a}^{(f)}<0$ gives rise to a Fermi \emph{surface}.

One possible FPS pattern is given by the following \emph{factorized}
Ansatz \cite{KV-IJMPA2005}:
\beq
b_{0a}^{(f)} =  Y_a \; \widetilde{b}_{0}^{(f)}\,,
\label{FRK-SMb0pattern}
\eeq
where $Y_a$ are the known hypercharge values of the fermions
and $\widetilde{b}_{0}^{(f)}$ three unknown energy scales.
Independent of the particular FPS
pattern, the dispersion relations of massless left-handed neutrinos
and right-handed antineutrino would be
\beq
\big( E_{\nu_L,f}({\bf p}) \big)^2  =
\Big(\,c\, |{\bf p}| +    b_0^{(f)}\,\Big)^2  \,,\quad
\big( E_{\bar\nu_R,f}({\bf p})\big)^2  =
\Big(\,c\, |{\bf p}| + s\,b_0^{(f)}\,\Big)^2 \,,
\label{FRK-DispLaw-nu}
\eeq
where a value $s = 1$ respects \DS{CPT} and $s=-1$ violates it.

More generally, one may consider for large momentum $|{\bf p}|$:
\beq
 E({\bf p})       \sim
c\, |{\bf p}| \pm b_0 +  m^2 c^4/(2\, c\,|{\bf p}|) +
\text{O} \big(\,1/|{\bf p}|^{2} \,\big)\,.
\label{FRK-DispLaw-b0m}
\eeq
The conclusion is then that the search for possible
FPS effects prefers neutrinos with the highest possible momentum.

At this point, two questions on energy scales arise.
First, what is known experimentally?
The answer is: not very much, apart from the following
upper bounds:
\beq
|b_0^{(e)}|  \lesssim 1\;\mathrm{keV}\,,\quad
\sum_{f=1}^{3}\; m_f \lesssim 100\;\mathrm{eV}\,,
\eeq
from low-energy neutrino physics \cite{DiGrezia-etal2005}
and cosmology, respectively.

Second, what can be said theoretically about the expected
energy scale of FPS?
The answer is: little to be honest, but perhaps
the following speculation may be of some value. For definiteness,
start from a particular emergent-physics scenario
with two energy scales \cite{KV-JETPL2005}:
\begin{itemize}
\item
$E_\text{LV}$ of the fundamental Lorentz-violating fermionic theory;
\item
$E_\text{comp}$ as the compositeness scale of the Standard Model gauge bosons.
\end{itemize}
Taking the LEP values of the gauge coupling constants,
the renormalization-group equations for $N_\text{fam}=3$ give
these numerical values:
\beq
E_\text{comp} \sim 10^{13}\;\mathrm{GeV} \,, \quad
E_\text{LV}\sim 10^{42}\;\mathrm{GeV} \,.
\eeq
The speculation, now, is that perhaps ultrahigh-energy Lorentz
violation \emph{re-enters} at an ultralow energy scale:
\beq
|b_0|\stackrel{?}{\sim} E_\text{comp}^2/E_\text{LV} \sim 10^{-7}\;\mathrm{eV}\,.
\eeq
If correct, this motivates the search for FPS effects at the sub--eV level.

\section{\label{FRK-sec:Simple-neutrino-model} Simple FPS neutrino model}

A general neutrino model with \emph{both} Fermi-point splittings (FPS)
and mass differences (MD) has many mixing angles and complex
Dirac phases to consider (not to mention possible Majorana phases).
In order to get an idea of
potentially new effects, consider a relatively simple FPS--MD
neutrino model \cite{K-PRD2005,K-PRD2006} having
\begin{itemize}
\item
  a standard neutrino mass sector with ``optimistic'' values
  for $\theta_{13}$ and $\delta$;
\item
  %a FPS sector with large mixing angles, equidistant splittings,
  %and  a nonzero Dirac phase $\omega$.
  a FPS sector with large mixing angles, energy splittings,
  and Dirac phase $\omega$.
 \end{itemize}
Specifically, the mass sector has the following
mass-square-difference ratio, mixing angles, and Dirac phase:
\begin{subequations}\label{FRK-FPS-MD-model}
\beq
R_m \equiv
\fracnew{\Delta m_{21}^2}{\Delta m_{32}^2}
\equiv \fracnew{m_2^2-m_1^2}{m_3^2-m_2^2} =\frac{1}{30}\,,\:\:
\theta_{21} = \theta_{32} = \frac{\pi}{4}\,,\:\:
\sin^2 2 \theta_{13}=\frac{1}{20}\,,\:\: \delta=\frac{\pi}{2}\,,
\label{FRK-FPS-MD-model-masssector}
\eeq
and the FPS sector has energy-difference ratio, mixing angles, and Dirac phase:
\beq
R \equiv \fracnew{\Delta b_0^{(21)}}{\Delta b_0^{(32)}}
         \equiv \fracnew{b_0^{(2)}- b_0^{(1)}}{b_0^{(3)}- b_0^{(2)}}=1\,,\;\;
\chi_{21} = \chi_{32} = \chi_{13} = \omega = \frac{\pi}{4}\,.
\label{FRK-FPS-MD-model-FPSsector}
\eeq
\end{subequations}

For later use, we also define two additional models.
The first additional model is a \emph{pure} FPS
model \cite{K-IJMPA2006} with trimaximal couplings
($\chi_{21}$ $=$ $\chi_{32}$ $=$  $\pi/4$ and
$\chi_{13}$ $=$ $\arctan\sqrt{1/2}\,$),
complex phase $\omega$, and FPS ratio $R$.
At sufficiently high energies,
the model for $R=1$ and  $\omega =\pi/4$
is close to the FPS--MD model mentioned above.

The second additional model is a \emph{pure} MD model with
a mass-square-difference ratio $R_m  =1/30$ and the following more or less
realistic values for the mixing angles and Dirac phase:
$\sin^2 2 \theta_{23}=1$, $\sin^2 2
\theta_{12}=0.8$, $\sin^2 2 \theta_{13}=0.2$, and $\delta=0$.

In the rest of this contribution, these three models will be referred to as
the FPS--MD model, the FPS model, and the MD model, respectively.

\begin{figure*}[t]
\begin{center}
\includegraphics[width=\textwidth]{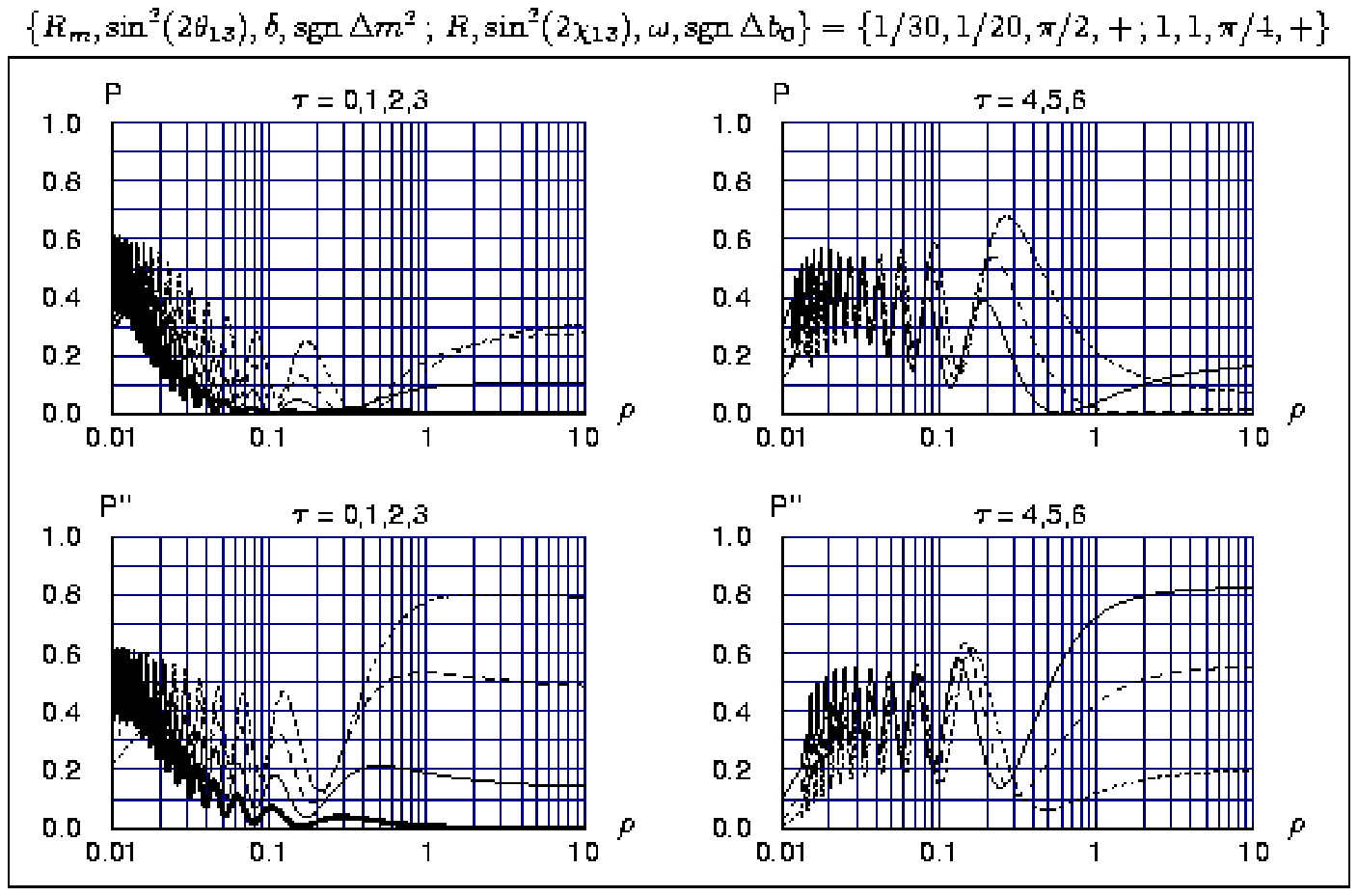}  %%[width=12cm]
\end{center}
\caption{Vacuum probabilities from the FPS--MD model
(\ref{FRK-FPS-MD-model}ab)
as a function of the dimensionless parameters $\rho$ and $\tau$,
defined by Eqs.~(\ref{FRK-rho-tau}ab).
Top panels: $P \equiv P(\nu_\mu\rightarrow \nu_e)$.
Bottom panels: $P^{\prime\prime}
\equiv P(\nu_e\rightarrow \nu_\mu)$.
If \DS{CPT} invariance holds, also $P=P(\overline{\nu}_e\rightarrow
\overline{\nu}_\mu)$ and $P^{\prime\prime}=P(\overline{\nu}_\mu\rightarrow
\overline{\nu}_e)$.
Shown are constant--$\tau$ slices, where the heavy-solid curves
in the two left panels correspond to $\tau=0$ (pure mass-difference model)
and the other thin-solid, long-dashed, and short-dashed curves for positive $\tau$
correspond to $\tau = 1,2,0 \pmod 3$, respectively.}
\label{FRK-FIG-nu-oscill}
\end{figure*}

\section{\label{FRK-sec:Neutrino-oscillations}
         FPS effects in neutrino oscillations}

Consider now a high-energy ($E_\nu \sim c\, |{\bf p}|$)
neutrino beam traveling over a distance $L$.
Neutrino oscillations from the FPS--MD model of
Sec.~\ref{FRK-sec:Simple-neutrino-model}
are then determined by two dimensionless parameters:
\begin{subequations}\label{FRK-rho-tau}
\beq
\rho \equiv
\fracnew{2\,E_\nu\,\hbar c}{L\,|\Delta m^2_{31}|\,c^4} \approx
  0.98 \; \Bigg( \fracnew{E_\nu}{20\;\mathrm{GeV}} \Bigg)
 \Bigg( \fracnew{3000\;\mathrm{km}}{L} \Bigg)\,
 \Bigg( \fracnew{2.7\times 10^{-3}\;\mathrm{eV}^2/c^4}{|\Delta m^2_{31}|}\Bigg),
\label{FRK-rho}
\eeq
\beq
\tau \equiv L\,|\Delta b_0^{(31)}|/(\hbar c) \approx
3.0 \; \Bigg(\fracnew{L}{3000\;\mathrm{km}} \Bigg)
  \Bigg( \fracnew{|\Delta b_0^{(31)}|}{2.0\times 10^{-13}\;\mathrm{eV}} \Bigg).
\label{FRK-tau}
\eeq
\end{subequations}
for numerical values of $L$ and $E_\nu$  appropriate to a neutrino factory
\cite{Geer1997}.
Possible new effects in neutrino oscillations from FPS may occur as
\begin{itemize}
\item
energy dependence of the vacuum mixing angle $\Theta_{13}$
\cite{K-PRD2005};
\item
novel source of \DS{T}, \DS{CP}, and perhaps \DS{CPT} violation
\cite{K-PRD2006};
\item
modified flavor ratios for high-energy cosmic neutrinos
\cite{K-JETPL2004,K-IJMPA2006}.
\end{itemize}
In this contribution, we discuss only the last two effects.

Figure \ref{FRK-FIG-nu-oscill} shows that,
provided the FPS parameter $\Delta b_0^{(31)}$ is large enough
for given baseline $L$, the probabilities of
time-reversed processes can be different by several tens of percents:
$P(\nu_\mu\rightarrow \nu_e)$ $\approx$ $20\,\%$ versus
$P(\nu_e\rightarrow \nu_\mu)$ $\approx$ $80\,\%$
at $\rho \sim 1$ and $\tau \sim 3$, for example.
For the record, standard mass-difference neutrino oscillations
($\tau=0$) give more or less equal probabilities at $\rho \sim 1$:
$P(\nu_\mu\rightarrow \nu_e)$ $\approx$
$P(\nu_e\rightarrow \nu_\mu)$ $\approx$ $0$.
In short, there could be strong \DS{T}--violating
(and \DS{CP}--violating) effects at the high-energy end of
the neutrino spectrum from FPS or other emergent-physics dynamics.

Next, turn to the pure FPS model and also, for comparison, to the
pure MD model, both defined in Sec.~\ref{FRK-sec:Simple-neutrino-model}.
Pion and neutron sources then give the averaged  event ratios
shown in Table~\ref{FRK-table1}, with the clearest difference
between the two models for the case of a neutron
source. In principle, these results may be
relevant to high-energy cosmic neutrinos
but it remains to be seen whether or not
present experiments (e.g., AMANDA and IceCube)
can access this type of information.

\begin{table*}[t]
\caption{\label{FRK-table1}  Averaged  event ratios
$(N_e:N_\mu:N_\tau)$ from pion and neutron sources
for pure Fermi-point-splitting (FPS) and mass-difference (MD)
neutrino models as defined in Sec.~\ref{FRK-sec:Simple-neutrino-model}.
The MD event ratios are taken from Ref.~\cite{Hooper-etal2005}.}
\renewcommand{\tabcolsep}{0.75pc}  % enlarge column spacing
\renewcommand{\arraystretch}{1.1}  % enlarge line spacing
\centering\begin{tabular}{|c|c|c|}
\hline
& $\pi:\text{initial ratios}=(1:2:0)$  & $n: \text{initial ratios}=(1:0:0)$\\
\hline
$\text{FPS} \;(\omega)$ & $( 6 : 7+\cos 2\omega : 5-\cos 2\omega )$
& $( 1 : 1 : 1 )$ \\
$\text{FPS} \;(\pi/4)$  & $(0.33 : 0.39 : 0.28)$  &  $(0.33 : 0.33 : 0.33)$ \\
$\text{MD}$ & $(0.36 : 0.33 : 0.31)$ & $(0.56 : 0.26 : 0.18)$\\
\hline
\end{tabular}
\end{table*}

\vspace{.5\baselineskip}%%FRK
\section{\label{FRK-sec:Outlook} Outlook}

From a phenomenological perspective,
the Fermi-point-splitting (FPS) hypothesis suggests the
following three research directions:
\begin{itemize}
\item
the possible energy dependence of the vacuum mixing angle $\Theta_{13}$ from
FPS, which can be tested by neutrino experiments at a superbeam
or neutrino factory;
\item
the possibility of a new source of leptonic \DS{CP} violation, which
impacts on the physics of the early universe (e.g., the creation
of baryon and lepton number);
\item
the possible modification of the propagation of high-energy cosmic
neutrinos by FPS effects, which may be of relevance to
present and future neutrino telescopes.
\end{itemize}
From a more theoretical perspective, the outstanding issues are:
\begin{itemize}
\item
the precise nature of the conjectured re-entrance mechanism
of Lorentz violation at ultralow energy from Lorentz violation at ultrahigh
energy (condensed-matter physics can perhaps provide some guidance);
\item
the explanation of the large hierarchies of basic scales
(e.g., for mass or FPS).
\end{itemize}
But apart from these theoretical ideas, experiment may, of course,
suggest entirely different directions  $\ldots$

%%\newpage
\vspace*{-0.5\baselineskip}%%FRK
\section*{Acknowledgments}
%% Keep the small font tag for the acknowledegments
{\small
It is a pleasure to thank the EPNT06 organizers and the
Department of Theoretical Physics of Uppsala University for
hospitality and G.E. Volovik for a most stimulating collaboration.
}

\vspace*{-0.5\baselineskip}%%FRK

%%%%%%%%%%%%%%%%%%%%%%%%%%%%%%%%%%%%%% reset.txt counters %%%%%%%%%%%%%%
%%
%%%%%%% do not change below here  %%%%%%%%%%%%%%%%%%%%%%%%%%%%%
\setcounter{section}{0}
\setcounter{subsection}{0}
\setcounter{figure}{0}
\setcounter{table}{0}
\setcounter{equation}{0}
\newpage

%%%%%%%%%%%%%%%%%%%%%%%%%%%%%%%%%%%%%%%%%%%%%%%%%%% Title, authors and addresses
\begin{frontmatter}


\title{Probing TeV gravity at neutrino telescopes}

% use optional labels between square brackets to link authors explicitly to addresses:
% If more than one author, keep a comma between the author tags

\author[address1]{J.I. Illana}, %\ead{illana@ugr.es},
\author[address1]{M. Masip},    %\ead{masip@ugr.es},
\author[address2]{D. Meloni}    %\ead{meloni@roma1.infn.it}

\address[address1]{CAFPE and Depto.~de F{\'\i}sica Te\'orica y del Cosmos, U. de Granada, 18071 Granada, Spain}
\address[address2]{INFN and Dipto.~di Fisica, Universit\`a degli Studi di Roma ``La Sapienza", 00185 Rome, Italy}

\begin{abstract}
Models with extra dimensions and the fundamental scale at the TeV could imply signals in large neutrino telescopes due to gravitational scattering of cosmogenic neutrinos in the detection volume. Apart from the production of microscopic black holes, extensively studied in the literature, we present 
gravity-mediated interactions at larger distances, that can be calculated in the eikonal approximation. In these elastic processes the neutrino loses a small fraction of energy to a hadronic shower and keeps going. The event rate of these events is higher than that of black hole formation and the signal is distinct: no charged leptons and possibly multiple-bang events.
\end{abstract}

%\begin{keyword}
%\end{keyword}

\end{frontmatter}

%%%%%%%%%%%%%%%%%%%%%%%%%%%%%%%%%%%%%%%%%%%%%%%%%%%%%% MAIN TEXT

\section{Motivation: cosmogenic neutrinos and TeV gravity}

Cosmogenic neutrinos, produced in the scattering of protons off cosmic microwave background photons, have access to TeV physics in interactions with terrestrial nucleons at center of mass energies $\sqrt{s}=\sqrt{2m_NE_\nu}\gsim10$~TeV. If the fundamental scale of gravity is $M_D\sim1$~TeV \cite{ADD}, which may happen in $D>4$ space-time dimensions, these $\nu N$ interactions are transplanckian, $\sqrt{s}>M_D$.

The only consistent theory known so far in such a regime, string theory, tells us that the interactions are soft in the ultraviolet. The scattering amplitudes vanish except in the forward region, an effect that can be understood as the destructive interference of string excitations \cite{SR}. The forward amplitudes are dominated by the zero mode of the string, corresponding to the exchange of a gauge particle of spin 1, ${A}\sim gs/t$, for open strings, or a graviton of spin 2, ${A}\sim (1/M^2_D) s^2/t$, for closed strings. Therefore, one expects that gravity dominates in transplanckian collisions.

It must be noticed that present bounds on $M_D$ from colliders (LEP, Tevatron) \cite{colliders} or astrophysics and cosmology (supernovae cooling) \cite{SN} come from processes at energies below $M_D$ and are indirect, since they actually constrain the energy emitted to Kaluza-Klein gravitons of mass $M\propto R^{-1}$ in the $n=D-4$ compact extra dimensions, which is a function of the compactification radius $R$. Those bounds rely on the assumption of all extra dimensions being large (the effective and fundamental Planck scales then relate through $M_P^2\propto R^n M_D^{2+n}$) that can be evaded in more sophisticated compactification models \cite{Giudice:2004mg}. In contrast, transplanckian collisions probe $M_D$ directly and independently of compactification details.

\section{Gravitational interactions}

We have shown that gravitational interactions are the only relevant in the transplanckian regime of energies. In impact parameter space, one must keep in mind two critical values: the Planck length $\lambda_D\sim M_D^{-1}$ and the Schwarzschild radius $R_S(s)\sim(\sqrt{s}/M_D)^{1/(n+1)}M_D^{-1}$. There are two types of interactions.

Short-distance interactions, with impact parameter $b\lsim R_S$, in which the colliding particles (a neutrino and a parton inside the nucleon) collapse into a black hole (BH) correspond to the exchange of strongly coupled gravitons of high momentum (non-linear gravity). The collapse involves strongly coupled gravity and is not calculable perturbatively. Most analyses are based on a geometric cross section \cite{BH} $\hat\sigma_{\rm BH}\simeq\pi R_S^2(\hat s)$ for the partonic process, with $\hat s=xs$. If $\sqrt{\hat s}\gg M_D$, namely $R_S\gg M_D^{-1}$, one expects that this estimate will not be off by any large factors \cite{Giddings:2004xy}. However, most of the BHs produced in the scattering of an ultrahigh energy neutrino off a parton are light, with masses just above $M_D$, since the $\nu N$ cross section is dominated by the low $x$ region. In this regime the amount of gravitational radiation emitted during the collapse or the topology of the singularity are important effects that add uncertainty to the geometric estimate.

Long-distance interactions, with $b\gg R_S$, have to do with the exchange of weakly coupled gravitons of low momentum (linearized gravity) \cite{eik}. In transplanckian collisions quantum gravity acts inside the event horizon ($R_S>\lambda_D$). Therefore, these elastic interactions are due to classical gravity. They are characterized by a small deflection angle $\theta^*$ in the center of mass (CM) frame, 
\begin{eqnarray}
\theta^*\sim\frac{\sqrt{\hat s}}{M_D^{n+2}b^{n+1}}
\sim\left(\frac{R_S}{b}\right)^{n+1}\ll 1\Rightarrow
y=q^2/\hat s=\frac{1}{2}(1-\cos\theta^\star)\ll 1\ .
\end{eqnarray}
The elastic collision of a neutrino and a parton that exchange $D$-dimensional gravitons is then described by the {\em eikonal} amplitude resumming an infinite set of ladder and cross-ladder diagrams in the limit in which the momentum $q$ carried by each graviton is smaller than the CM energy or, in terms of the fraction of energy lost by the incoming neutrino, $y=(E_\nu-E'_\nu)/E_\nu\ll1$.
In this limit the amplitude is independent of the spin of the colliding particles.
Essentially, ${A}_{\rm eik}$ is the exponentiation of the Born
amplitude in impact parameter space \cite{riccardo}:
\begin{eqnarray}
{A}_{\rm eik}(\hat s,t)=\frac{2 \hat s}{i}
\int {\rm d}^2b\; e^{i\mathbf{q}\cdot\mathbf{b}}\;
\left(e^{i\chi (\hat s,b)}-1\right)
\equiv 4\pi\hat sb_c^2F_n(b_cq)\ ,
\end{eqnarray}
where $\chi (\hat s,b)$ is the eikonal phase,
\begin{eqnarray}
\chi(\hat s,b) = \frac{1}{2s}\int\frac{{\rm d}q}{(2\pi)^2}
{\rm e}^{-i{\bf q}\cdot{\bf b}}{A}_{\rm Born}(\hat s,q^2)
\equiv \left(\frac{b_c}{b}\right)^n
\end{eqnarray}
and a new scale $b_c$ appears,
\begin{eqnarray}
b_c(\hat s)=\left[\displaystyle\frac{(4\pi)^{\frac{n}{2}-1}}{2}\Gamma\left(
\displaystyle\frac{n}{2}\right)\frac{\hat s}{M_D^{n+2}}\right]^\frac{1}{n}\ .
\end{eqnarray}
The total partonic cross sections can be obtained from the amplitudes above using the optical theorem. They are $\hat\sigma_{\rm eik}\propto b_c^2\sim \hat s^\frac{2}{n}$, growing faster with energy than the BH cross sections $\hat\sigma_{\rm BH}\propto R_S^2\sim \hat s^\frac{1}{n+1}$.

Therefore, in transplanckian collisions one may consider two types of processes \cite{us}: elastic (long-distance) {\em soft} processes where the neutrino transfers to the partons a small fraction $y<y_{\rm max}$ of its energy and keeps going, and shorter distance ($b<R_S$) {\em hard} processes where the neutrino loses in the collision most of its energy, possibly collapsing into a BH. We take $y_{\rm max}=0.2$, the typical inelasticity of a standard model (SM) interaction, but any value of the order of 0.1 yields similar results. On the other hand, there is a $y_{\rm min}=E_{\rm thres}/E_\nu$ determined by the threshold energy $E_{\rm thres}$ transfered to a parton that produces an observable hadronic cascade.
The corresponding $\nu N$ cross sections are obtained by convolution with the parton distribution functions $f_{\{q,\bar q,g\}}(x,\mu)$ with the appropriate energy scale $\mu$ \cite{riccardo,us}. 

To estimate the relative frequency of both type of processes \cite{us}, consider a  $10^{10}$~GeV neutrino that scatters off a nucleon with $E_{\rm thres}=100$~TeV and $M_D=1$~TeV for $n=2\;(6)$ extra dimensions. The number of eikonal interactions before the neutrino gets destroyed is the ratio of interaction lengths $L_{\rm BH}/L_{\rm eik}=\sigma_{\rm eik}/\sigma_{\rm BH}=12.5\;(1.64)$. For a SM interaction $L_{\rm SM}=440$~km while $L_{\rm BH}=17$~km (4 km) in ice. The total energy lost by the neutrino in these eikonal interactions and the energy lost to graviton radiation are relatively small: $E^{\rm loss}_{\rm eik}=5.9\times10^7$~GeV ($1.2\times10^8$~GeV) and  $E_{\rm loss}^{\rm rad}=9.2\times10^7$~GeV ($1.2\times10^8$~GeV) in 1 km of ice.

\begin{figure*}[t]
\includegraphics[width=0.32\linewidth]{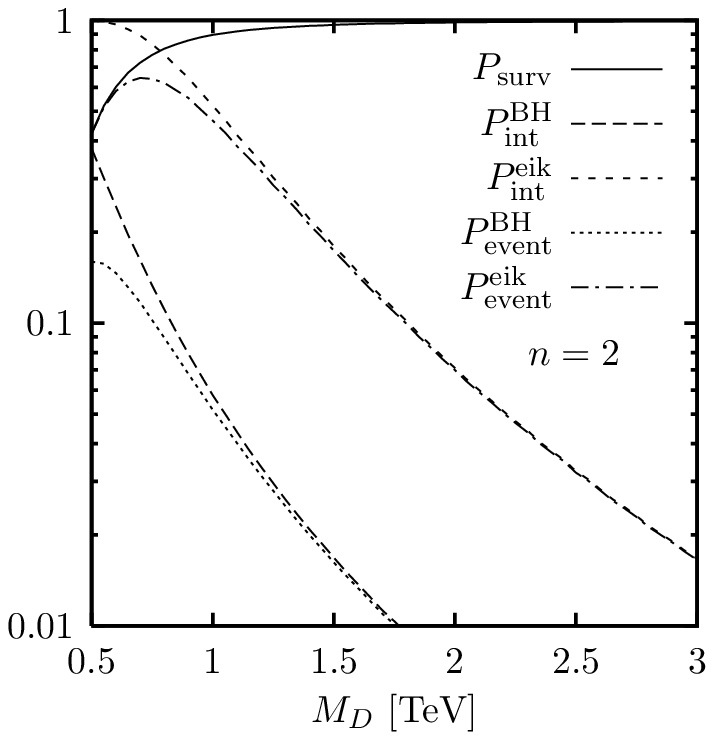}
\hfill
\includegraphics[width=0.32\linewidth]{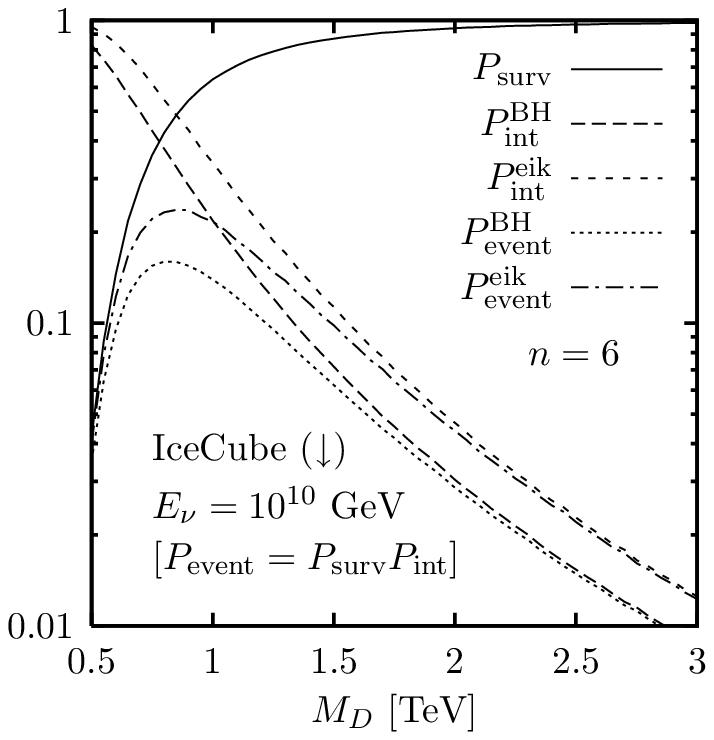}
\hfill
\includegraphics[width=0.32\linewidth]{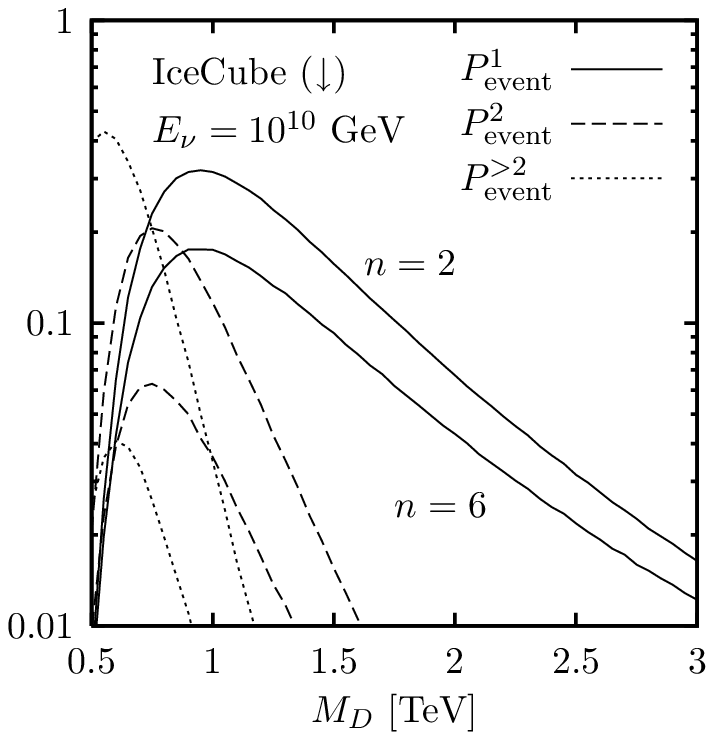}
\caption{
Probabilities defined in the text for a $10^{10}$~GeV neutrino reaching IceCube from $\theta_z=0$ as a function of $M_D$ for $n=2$ and $n=6$.}
\label{illana_fig1}
\end{figure*}

\section{Signals at neutrino telescopes}

The flux of cosmogenic neutrinos, yet unobserved, depends on the production rate of primary nucleons of energy around and above the GZK cutoff. It is correlated with proton and photon fluxes that must be consistent, respectively, with the number of ultrahigh energy events at AGASA and HiRes \cite{Anchordoqui:2002hs} and with the diffuse $\gamma$-ray background measured by EGRET \cite{Sreekumar:1997un}. We base our analysis on the two neutrino fluxes described in \cite{semikoz}. The first one saturates the observations by EGRET, whereas for the second one the correlated flux of $\gamma$-rays contribute only a 20\% to the data, with the nucleon flux normalized in both cases to AGASA/HiRes. The {\it higher} flux predicts 820 downward neutrinos of each flavor with energy between $10^8$~GeV and $10^{11}$~GeV per year and km$^2$, versus 370 for the {\it lower} one. The spectrum has a peak at neutrino energies between $10^9$~GeV and $10^{10}$~GeV.

When a neutrino hits a nucleon it will start a hadronic shower. The total number of hadronic events in a neutrino telescope of cross sectional area $A$ in a time $T$ is
\begin{eqnarray}
N_{\rm events}=2\pi AT\int {\rm d}E_\nu\sum_{\nu_i,\bar \nu_i} \frac{{\rm d}\phi_{\nu_i}}{{\rm d}E_\nu}\int {\rm d}\cos\theta_z P_{\rm surv}P_{\rm int}\ ,
\end{eqnarray}
where $P_{\rm surv}$ is the probability that the neutrino survives to reach the detector and $P_{\rm int}$ the probability that it interacts inside the detector:

\begin{eqnarray}
P_{\rm surv}(E_\nu,\theta_z)&=&{\rm e}^{-x(\theta_z)N_A(\sigma_{\rm SM}+\sigma_{\rm BH})}\ , \quad
P_{\rm int}(E_\nu)\approx 1-{\rm e}^{-L\rho_{\rm ice} N_A\sigma^{\nu N}_{\rm int}}\ ,
\end{eqnarray}
with $x$ the column density of material, $\theta_z$ the zenith angle and $L$ the longitudinal detector size. When $L$ is larger than the interaction length $L_0$, there may be multiple-bang events. Neglecting the energy lost by the neutrino in each interaction, the probability of $N$ bangs and the average, and most probable, number of bangs are, respectively
\begin{eqnarray}
P_N(L)&=&{\rm e}^{-L/L_0}\frac{(L/L_0)^N}{N!}\ ,\quad 
\langle N\rangle=\sum_{N=1}^\infty NP_N=L/L_0\ .
\end{eqnarray}

The different probabilities for a typical cosmogenic neutrino of $10^{10}$~GeV reaching vertically IceCube are shown in Fig.~\ref{illana_fig1} for illustration. Double-bang events could also be produced by SM interactions (the decay of a tau created in a first interaction) or in the BH evaporation. For the double-bang tau event to be contained inside a detector like IceCube (1 km of length with 125 m between strings), the energy of the tau lepton must be between $2.5\times 10^{6}$ GeV and $10^{7}$ GeV. In this case, the probability is only $6.8\times10^{-5}$.

The energy distribution of the hadronic cascades and the total number of black hole and eikonal events at AMANDA (0.03 km$^2$ and a length of 700 m) and IceCube (1 km$^3$) for the neutrino fluxes introduced above are given in Figs.~\ref{illana_fig2} and \ref{illana_fig3}, respectively. In the SM we expect 1.32 (0.50) contained events per year in IceCube for the higher (lower) flux. Of those, 0.38 (0.14) would come from a neutral current and 0.94 (0.36) from a charged current.

\begin{figure*}[t]
\begin{minipage}[t]{0.48\linewidth}
\includegraphics[width=\linewidth]{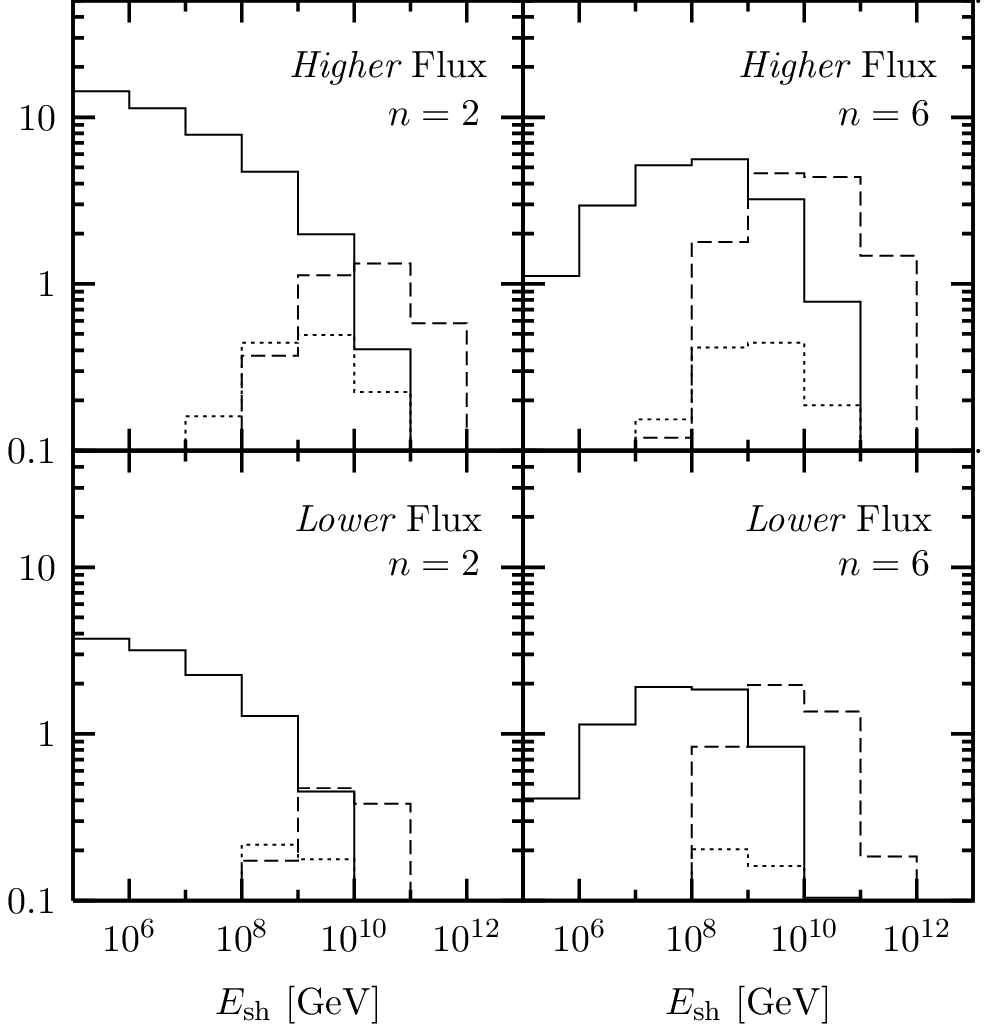}
\caption{
Energy distribution (events per bin) of the eikonal (solid), BH (dashed)
and SM (dotted) events in IceCube per year for the {\it higher} and
the {\it lower}
cosmogenic fluxes, $M_D=2$~TeV and $n=2,\ 6$.}
\label{illana_fig2}
\end{minipage}\hfill
\begin{minipage}[t]{0.48\linewidth}
\includegraphics[width=\linewidth]{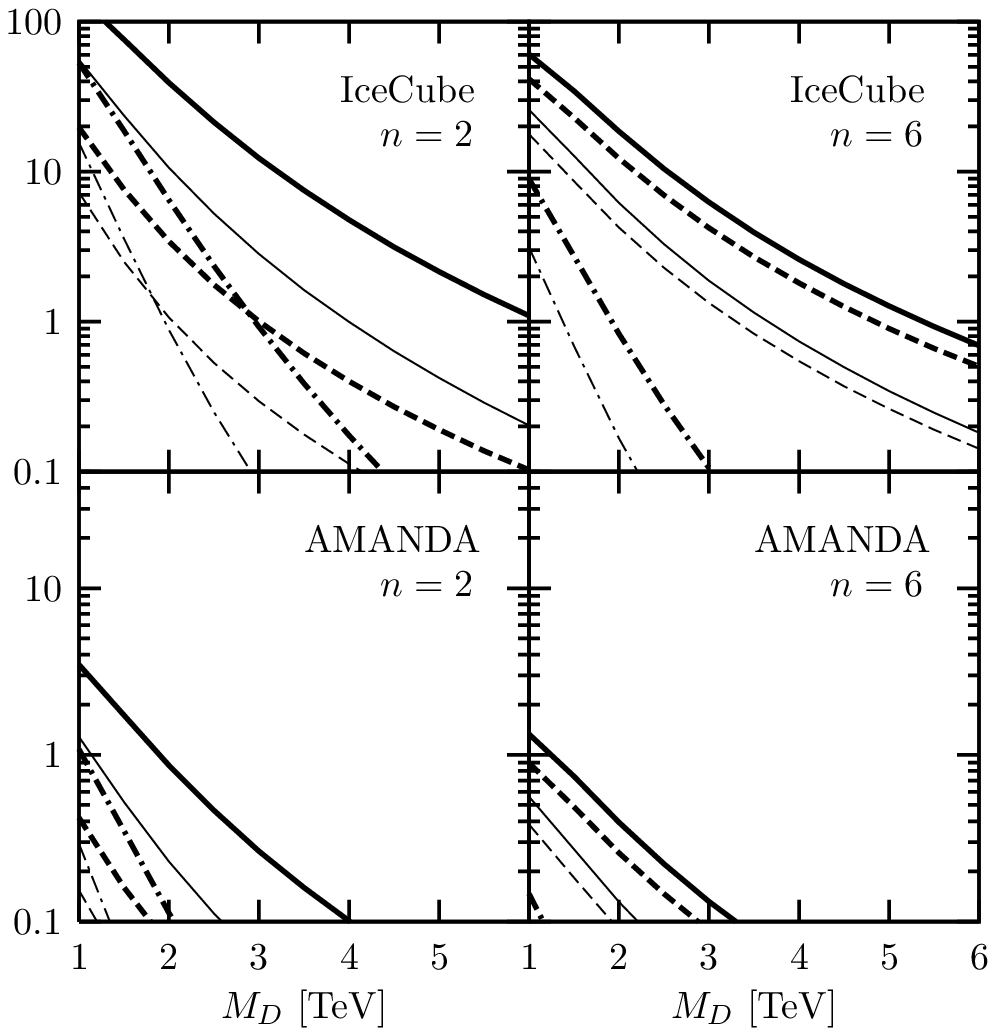}
\caption{
Contained events per year in IceCube and AMANDA for the
{\it higher} (thick) and the {\it lower} (thin) cosmogenic
fluxes and $n=2,\ 6$.
We show eikonal (solid), multi-bang (dashed-dotted) and BH (dashed)
events.}
\label{illana_fig3}
\end{minipage}
\end{figure*}
\vspace*{0.5cm}

\section{Conclusions}

Cosmogenic neutrinos directly probe TeV gravity in transplanckian collisions with nucleons at Earth if there are $D>4$ space-time dimensions. Two types of interactions take place. Hard processes, when the impact parameter is smaller than the Schwarzschild radius of the neutrino-parton system, produce mostly light black holes with theoretically uncertain cross sections. At larger distances, soft elastic processes occur in which the neutrinos lose a small fraction of energy to a hadronic shower, in a well known regime described by the eikonal approximation.

The latter turn out to be dominant and produce a clear signal in large neutrino telescopes: contained hadronic showers without charged leptons. Furthermore, this signal cannot be confused with ordinary SM events due to an unexpectedly high neutrino flux because in the SM 24\% of the events are accompanied by muons and multiple-bang events are very suppressed, in contrast to the elastic gravitational events. The values of the fundamental scale of gravity that IceCube could reach are comparable to those to be explored at the LHC.

%\section*{Acknowledgments}
%% Keep the small font tag for the acknowledegments
%{\small 
%This work has been supported by MEC of Spain (FPA2003-09298-C02-01) and Junta %de Andaluc\'\i a (FQM-101).
%}

%%%%%%%%%%%%%%%%%%%%%%%%%%%%%%%%%%%%%% reset.txt counters %%%%%%%%%%%%%%
%%
%%%%%%% do not change below here  %%%%%%%%%%%%%%%%%%%%%%%%%%%%%
\setcounter{section}{0}
\setcounter{subsection}{0}
\setcounter{figure}{0}
\setcounter{table}{0}
\setcounter{equation}{0}
\newpage

%%%%%%%%%%%%%%%%%%%%%%%%%%%%%%%%%%%%%%%%%% Title, authors and addresses
\begin{frontmatter}

% use the thanksref command within \title, \author or \address for
% footnotes; 
% use the corauthref command within \author for corresponding author
% footnotes; 

\title{Inferring Neutrino Cross Sections Above $10^{19}$~eV}

% use optional labels between square brackets to link authors
% explicitly to addresses: 
% If more than one author, keep a comma between the author tags

\author[address1]{Sergio Palomares-Ruiz}

\address[address1]{Institute for Particle Physics Phenomenology,
  University of Durham, Durham DH1 3LE, UK}

\begin{abstract}
Extremely high energy neutrinos propagating in the atmosphere or in
the Earth can originate horizontal or up-going air-showers,
respectively. We calculate the acceptances (event rate/flux) for
detecting both types of events by fluorescence detectors, both
space-based as with the EUSO and OWL proposals, and ground-based, as
with Auger, HiRes and Telescope Array.  We depict them as a function
of the neutrino-nucleon cross section, $\sig$, and show that from the
ratio of these two classes of events, the inference of $\sig$ above
$10^{19}$~eV appears feasible, assuming that a neutrino flux exists at
these energies. Our semi-analytic calculation includes realistic
energy-losses for tau leptons and Earth-curvature effects. We also
consider constraints on shower development and identification and the
effects of a cloud layer. 

\end{abstract}

% \begin{keyword}

%\PACS 
% \end{keyword}

\end{frontmatter}

%%%%%%%%%%%%%%%%%%%%%%%%%%%%%%%%%%%%%%%%%%%%%%%%%%%%%% MAIN TEXT
\section{\label{sec:intro} Introduction}
Above the Greisen-Zatsepin-Kuzmin (GZK) energy of $\egzk\sim 5\times
10^{19}$~eV~\cite{GZK:66} ultra-high energy neutrinos  are probably
the only propagating primaries. Moreover, in contrast to cosmic-rays,
they point back to their astrophysical sources carrying
information not accessible with other primaries. The detection of
ultra-high energy neutrinos also allows studies of the fundamental
properties of neutrinos themselves, as for instance the
neutrino-nucleon cross section at energies beyond the reach of our
terrestrial accelerators.

In this talk, based on Ref.~\cite{PRIW}, we study the potential for
cosmic-ray experiments designed to track ultra-high energy air-showers
by monitoring their fluorescence yield, to detect horizontal
air-showers (HAS) and up-going air-showers (UAS) induced by a cosmic 
neutrino flux and show the ability of these experiments to infer the
neutrino-nucleon cross section, $\sig$, at energies above $10^{19}$~eV,
from the ratio of their UAS and HAS events. Such energies are orders
of magnitude beyond the energies accessible to man-made terrestrial
accelerators. From the point of view of QCD, such a cross section
measurement would be an interesting microscope into the world of
small-x parton evolution. Deviations from QCD-motivated
extrapolations~\cite{GQRS} could reduce the cross section due to
saturation effects~\cite{saturation} or enhance it by the existence of 
new physics thresholds~\cite{newphysics}.

In Ref.~\cite{KW} it was shown that by comparing the HAS and UAS event
rates the neutrino-nucleon cross section may be inferred. The
calculation of Ref.~\cite{KW} gave an approximate result for the 
dependence of the UAS event rates on the neutrino-nucleon
cross section. In this talk, following the results of Ref.~\cite{PRIW},
we improve upon Ref.~\cite{KW} in several ways, as we show below. On
the other hand, the prospects of inferring the neutrino-nucleon
cross section at neutrino telescopes such as IceCube or at the Auger
observatory have been studied in Ref.~\cite{otherdetectors}.

\section{\label{sec:airshower} Air-shower rates and constraints on
  shower-development}

Ultra-high energy neutrinos are expected to arise from the decay of
pions and subsequently muons produced in astrophysical
sources~\cite{pimuchain} (for the case of production from neutron 
decays see, eg, Ref.~\cite{antinubeam}). After propagating for many
oscillation lengths and due to the maximal mixing between $\numu$ and
$\nutau$ inferred from terrestrial oscillation experiments, all
flavors are populated. Thus, a detector optimized for $\nue$ or
$\numu$ or $\nutau$ can expect a measurable flux from cosmic neutrinos.

The weak nature of the neutrino-nucleon cross section means that
HAS begin low in the atmosphere, where the target is most dense, and
thus that the event rate for neutrino-induced HAS is proportional to
the cross section. Following Ref~\cite{PRIW}, for the case of HAS
event rates we will only consider $\nu_e$ charged current interactions.

For a neutrino-induced UAS, the dependence on the neutrino
cross section is more complicated. The Earth itself is opaque for
neutrinos with energies exceeding about a PeV of energy. However,
``Earth-skimming'' neutrinos, those with a short enough chord length
through the Earth, will penetrate and exit, or penetrate and
interact. In particular, there is much interest in the Earth-skimming
process $\nutau\rarr\tau$ in the shallow Earth, followed by $\tau$
decay in the atmosphere to produce an observable shower. In
Ref.~\cite{KW} it was shown that the rate for the Earth-skimming
process $\nutau\rarr\tau$ is {\sl inversely} proportional to
$\sig$. The inverse dependence of UAS rate on $\sig$ is broken by
the $\tau\rarr$~{\sl shower} process in the atmosphere. As the
cross section decreases, the allowed chord length in the Earth
increases, and the tau emerges with a larger angle from the Earth's
tangent plane. This in turn provides a smaller path-length in air in
which the tau may decay and the resulting shower may evolve. This
effect somewhat mitigates the inverse dependence of the UAS on
$\sig$. 

The main aim of the study in Ref.~\cite{PRIW} was to provide a
detailed and improved extension of the idea introduced in
Ref.~\cite{KW}. Hence, here we include the energy dependences of the
tau energy-losses in the Earth, and of the tau lifetime in the
atmosphere. For the energy-losses, we distinguish between tau
propagation in rock and in water. In the case of the UAS, the
path length of the pre-decayed tau may be so long that the Earth's
curvature represents a non-negligible correction, that we include. We
also consider the partial loss of visibility due to cloud layers. On
the issue of shower development, we incorporate the dependence of
atmospheric density on altitude and add some conditions for the
showers to be observable. Shower detection will require
that within the field of view, the length of the shower track
projected on a plane tangent to the Earth's surface exceeds some
minimum length, $\lmin$. In addition, a minimum column density,
$\dmin$, beyond the point of shower initiation is required for the
shower to develop in brightness. On the other hand, after a maximum
column density, $\dmax$, the shower particles are below threshold for
further excitation of the $N_2$ molecules which provide the observable
fluorescence signal. Therefore, visible showers end at
$\dmax$. Finally, the fluorescent emission per unit length of the
shower will decline exponentially with the air density at 
altitude. We will take $\zthin= 24$~km as the altitude beyond which
the signal becomes imperceptible. Regarding the choice of $\dmin$ and
$\dmax$, they are inferred from the observed longitudinal development
profiles of ultrahigh-energy cosmic ray showers. On the other hand, we
assign a relatively small value to $\lmin$ to maximize the observable
event rate. For a summary of the different values adopted to obtain
the results, full details on the analytic description of the effects
of these parameters on the event rates and comparison with prior work,
we defer the reader to Ref.~\cite{PRIW}.

\section{\label{sec:results} Results}
In this section, we present the results of our semi-analytical
approach and take the product of area and solid angle $\sim 10^6\,{\rm
  km}^2\;{\rm sr}$, i.\ e.\ that of the EUSO design report~\cite{EUSO}.

%%%%%%%%%%%%%%%%%%%%%%%%%%%%%%%%%%%%%%%%%%%%%%%%%%%%%%%
\begin{figure*}[t]
\includegraphics[width=1.\textwidth]{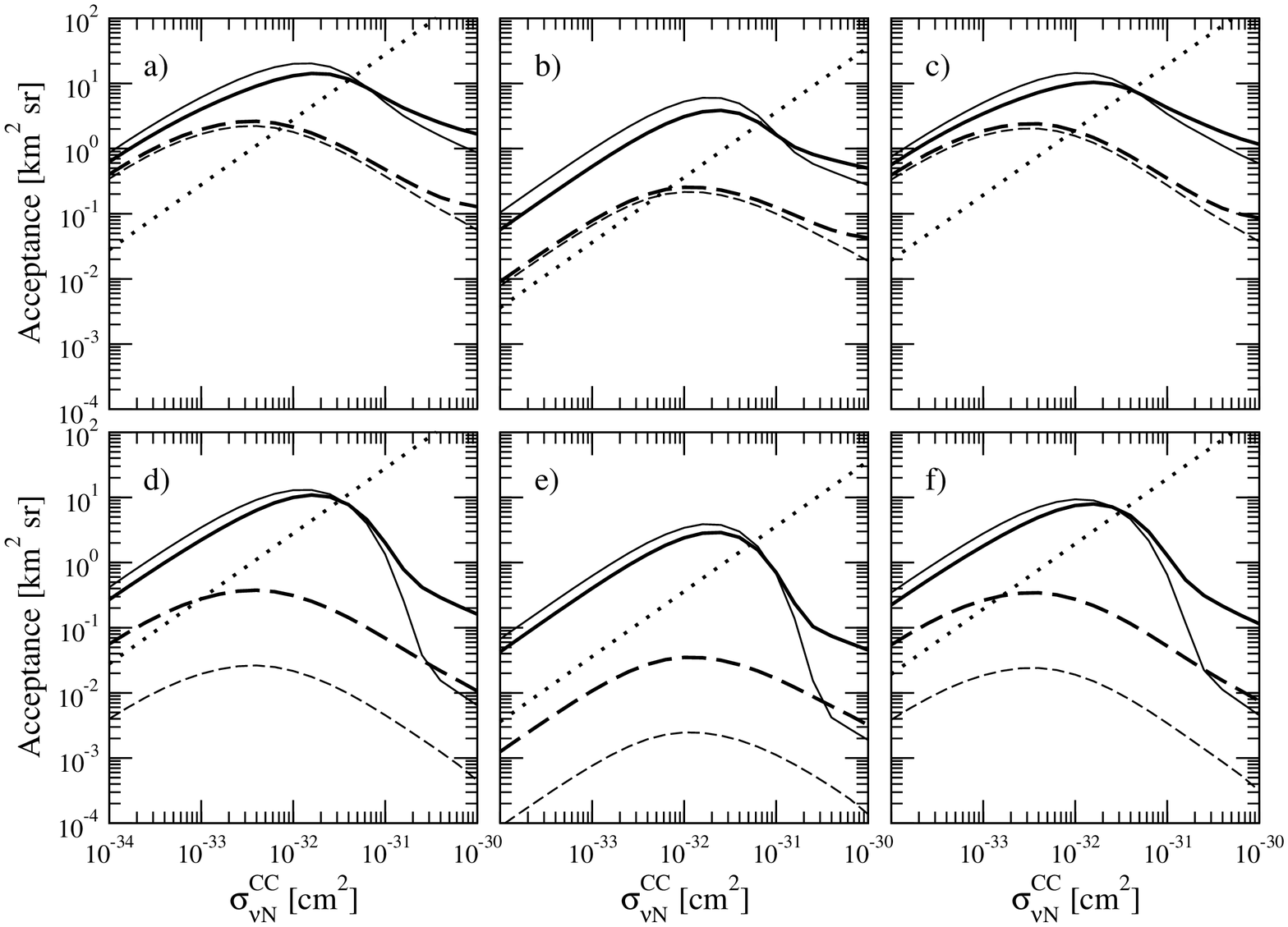}%
\caption{\label{combiEPNT} 
  Acceptances with $l_{\rm{min}}$ fixed at 5~km, $\dmin$ at $400 {\rm
  g}/{\rm cm}^2$, and $\dmax$ at $1200 {\rm g}/{\rm cm}^2$. The curves
  correspond to HAS (dotted line) and UAS over ocean with
  $E_\nu=10^{21}$~eV (thick solid line), ocean with $E_\nu=10^{20}$~eV
  (thin solid line), land with $E_\nu=10^{21}$~eV (thick dashed line),
  and land with $E_\nu=10^{20}$~eV (thin dashed line). Panels are for
(a) space-based (or ground-based) detectors in the absence of clouds
  with $\Esh=10^{19}$~eV;
(b) ground-based detectors in the presence of a cloud layer at
  $z_{\rm{cloud}} = 2$ km with $\Esh=10^{19}$~eV;
(c) spaced-based detectors in the presence of a cloud layer at
  $z_{\rm{cloud}} = 2$ km with $\Esh=10^{19}$~eV;
(d) space-based (or ground-based) detectors in the absence of clouds
  with $\Esh=5 \times 10^{19}$~eV;
(e) ground-based detectors in the presence of a cloud layer at
  $z_{\rm{cloud}} = 2$ km with $\Esh=5 \times 10^{19}$~eV;
(f) spaced-based detectors in the presence of a cloud layer at
  $z_{\rm{cloud}} = 2$ km with $\Esh=5 \times 10^{19}$~eV.
} 
\end{figure*}
%%%%%%%%%%%%%%%%%%%%%%%%%%%%%%%%%%%%%%%%%%%%%%%%%%%%%%%

In Fig.~\ref{combiEPNT} are plotted UAS (solid and dashed) and HAS
(dotted) acceptances in units of (${\rm km}^2$-sr), versus fixed
values of $\sig$. Within the approximations followed in
Ref.~\cite{PRIW}, for the ideal case of a cloudless sky (panels a and
d) there is no difference between the acceptances for ground-based and
space-based detectors. However, there are significant up-down
differences when the sky is covered by clouds (panels b, c, e and
f). In this latter case, we model the cloud layer as infinitely thin
with altitude $\zc$, but with an infinite optical depth so that
showers are completely hidden on the far side of the cloud layer. 

The HAS acceptances depend on neutrino energy only via $\sig (E_\nu)$,
and rise linearly with $\sig$. Plotted against fixed $\sig$, then, the
straight-line HAS curves (dotted) are universal curves valid for any
$E_\nu$ exceeding the trigger threshold $\Esh$. The UAS acceptances
have a complicated dependence on $E_\nu$; it arises from the energy
dependences of $\nu$ propagation in the Earth, tau propagation in the
Earth, and path-length of the tau in the atmosphere before it decays,
the latter also affecting the visible shower characteristics. We can
clearly see that the UAS acceptance (and so also the rate) is
typically an order of magnitude larger when neutrinos traverse a layer
of ocean water, compared to a trajectory where they only cross
rock. Thus, the UAS event rate is enhanced over the ocean relative to
over land. The value of this enhancement depends on the shower
threshold-energy $\Esh$ of the detector (upper versus lower panels)
and on the neutrino-nucleon cross section in a non-trivial
way. On the other hand, quantitatively, the ground-based acceptances
are quite reduced by the low-lying clouds, whereas the space-based
acceptances are not, as one would expect. The suppression of the
ground-based acceptance is most severe for small cross sections, for
which the tau leptons emerge more vertically and disappear into the
clouds before their eventual shower occurs and develops. Ground-based
UAS acceptances are reduced by up to an order of magnitude over water,
and even more over land. Ground-based HAS acceptances are reduced by
an order of magnitude. For space-based detectors, the UAS acceptance
is reduced little by clouds at 2~km. Larger neutrino cross sections
lead to more tangential tau-showers which may hide below a low-lying
cloud layer. We see that UAS reductions are a factor of 2 for the
larger cross sections shown, and less for the smaller values of
cross section.

We obtain benchmark event rates by multiplying our calculated
acceptances with a benchmark integrated flux of one neutrino per
$({\rm km}^2\,{\rm sr}\,{\rm yr})$. The result is a signal exceeding
an event per year for an acceptance exceeding a (${\rm km}^2$-sr).
Thus we see that this benchmark flux gives a HAS rate exceeding 1/yr
if $\sig$ exceeds $10^{-32}\,{\rm cm}^2$; and an UAS rate exceeding
1/yr over water for the whole cross section range with
$\Esh=10^{19}$~eV, and over land if $\sig \lsim 10^{-31}\,{\rm
  cm}^2$. When $\Esh$ is raised, however, the UAS signal over land is 
seriously compromised, while  UAS rates over the ocean are little
changed, HAS rates are unchanged, as long as $\Eth$ exceeds $E_\nu$.

We call attention to the fact that for UAS over both ocean and land,
there is a maximum in the UAS acceptance at cross section values $\sig
\sim (1-2) \times 10^{-32}~\rm{cm}^2$ and $\sig \sim (0.3-0.5) \times
10^{-32}~\rm{cm}^2$, respectively. For cross sections similar or
smaller than those at the maximum, the acceptance for UAS is larger 
than that for HAS; conversely, for cross sections above those at the
maximum, HAS events will dominate UAS events. The cross section value
at the maximum lies just below the extrapolation of the Standard Model
cross section, which for the two initial neutrino energies considered,
$10^{20}$~eV and $10^{21}$~eV, is $0.54 \times 10^{-31} \rm{cm}^2$ and
$1.2 \times 10^{-31} \rm{cm}^2$, respectively. If this extrapolation
is valid, then one would expect comparable acceptances (and event
rates) for UAS over water and for HAS. If the true cross section
exceeds the extrapolation, then HAS events will dominate UAS events;
if the true cross section is suppressed compared to the extrapolation,
then UAS events will dominate HAS events. Importantly, the very
different dependences on the cross section of the HAS and UAS
acceptances offers a practical method to measure $\sig$. One has
simply to exploit the ratio of UAS-to-HAS event rates. Furthermore,
the shape of the UAS acceptance with respect to $\sig$ establishes the
``no-lose theorem''~\cite{PRIW,KW}, which states that although a large
cross section is desirable to enhance the HAS rate, a smaller
cross section still provides a robust event sample due to the
contribution of UAS.  The latter is especially true over ocean.

\section{Conclusions}
In this talk we have presented a mostly analytic calculation of the
acceptances of space-based and ground-based fluorescence detectors of
air-showers at ultra-high energies. Included in the calculation are the
dependences of the acceptances on initial neutrino energy,
trigger-threshold for the shower energy, composition of Earth (surface 
rock or ocean water), and several shower parameters (the minimum and
maximum column densities for shower visibility, and the tangent length
of the shower). Also included in the calculation are suppression of the
acceptances by cloud layers and by the Earth's curvature. And most
importantly, also included are the dependences on the unknown
neutrino-nucleon cross section. The dependence is trivial and linear
for HAS, but nontrivial and nonlinear for UAS.

The merits of the analytic construction are two-fold: it offers an
intuitive understanding of each ingredient entering the calculation;
and it allows one to easily re-compute when different parameters are
varied. While a Monte Carlo approach may be simpler to implement, it
sacrifices some insight and efficiency.

The differing dependences of HAS and UAS on $\sig$ enable two very
positive conclusions: (1) the ``no-lose theorem'' is valid, i.\ e.\ 
that acceptances are robust for the combined HAS plus UAS signal
regardless of the cross section value; (2) and an inference of $\sig$
above $10^{19}$~eV is possible if HAS and UAS are both measured.

\section*{Acknowledgments}
%% Keep the small font tag for the acknowledegments
{\small 
I would like to thank T.~Weiler for enlightening discussions and for a
fruitful collaboration.
}

%%%%%%%%%%%%%%%%%%%%%%%%%%%%%%%%%%%%%% reset.txt counters %%%%%%%%%%%%%%
%%
%%%%%%% do not change below here  %%%%%%%%%%%%%%%%%%%%%%%%%%%%%
\setcounter{section}{0}
\setcounter{subsection}{0}
\setcounter{figure}{0}
\setcounter{table}{0}
\setcounter{equation}{0}
\newpage

\begin{frontmatter}

\title{Exotic Neutrino Interactions in Cosmic Rays}

\author[address1]{Markus~Ahlers}

\address[address1]{Deutsches Elektronen-Synchrotron DESY, Notkestra\ss e  85, 22607 Hamburg, Germany}
\begin{center}{\small{(markus.ahlers@desy.de)}}\end{center}

\begin{abstract}
The spectrum of extra-galactic cosmic rays (CRs) is expected to follow the Greisen-Zatsepin-Kuzmin (GZK) cutoff at about $5\times10^{10}$~GeV which results from energy losses of charged nuclei in the cosmic microwave background. So far the confrontation of this feature with CR data is inconclusive. In the absence of close-by sources a power-law continuation of the spectrum might signal the contribution of new physics. We have investigated the statistical significance of a model where exotic interactions of cosmogenic neutrinos are the origin of super-GZK events. A strong neutrino-nucleon interaction is favored by CR data, even if we account for a systematic shift in energy calibration.
\end{abstract}

\end{frontmatter}
\section{Introduction}
The appearance of extremely high energetic (EHE) cosmic rays (CRs) is a mystery. Simple geometric arguments show that the energy of nuclei originating in cosmic accelerators should be limited corresponding to the size and magnetic field of the accelerating environment (Hillas criterion)~\cite{Hillas:1985is}. Additional constraints arise from energy losses in the source that has to be balanced by the acceleration rate~\cite{Protheroe:2004rt}. In particular, the energy loss length of synchrotron radiation decreases with the third power of the magnetic field (at constant gyro-radius) and limits the efficiency of small candidate acceleration sites. The very few sources that seem to be capable of accelerating protons up to $10^{12}$~GeV include radio galaxy lobes and relativistically moving sources like jets of active galactic nuclei or gamma ray bursts.

Not only the acceleration mechanism of these particles seems to be problematic, but also their propagation in the interstellar medium. Resonant pion photo-production in the cosmic microwave background (CMB) limits the range of EHE particles to a few $10$~Mpc. It was first emphasized by Greisen~\cite{Greisen:1966jv}, Zatsepin, and Kuzmin~\cite{Zatsepin:1966jv} (GZK) that this should be signaled by a cutoff in the CR spectrum at about $5\times10^{10}$~GeV if extra-galactic sources dominate at these energies. So far, the confrontation of this effective cutoff with CR data is inconclusive. In particular, the observations of two large exposure experiments AGASA~\cite{Takeda:2002at} and HiRes~\cite{Bird:1994wp} seem to imply conflicting results.

The ``excess'' of super-GZK events reported by the AGASA collaboration has led to speculations about a different origin of EHE CRs. Berezinsky and Zatsepin proposed that cosmogenic neutrinos~\cite{Beresinsky:1969qj} produced in the decay of the GZK pions could explain these events assuming a rapid rise of the neutrino-nucleon interaction. We have followed this idea in Ref.~\cite{Ahlers:2005zy} and investigated the statistical significance of scenarios with strongly interacting neutrinos combining CR data from AGASA and HiRes and neutrino limits from horizontal events at AGASA and contained events at RICE.

The AGASA excess of super-GZK events could also be the result of a relative systematic error in energy calibration of $\pm30\%$~\cite{DeMarco:2003ig}. This interpretation is fueled by a recent re-analysis of the AGASA data~\cite{SHINOZAKI}. The preliminary results indicate that the energy of highly inclined showers has previously been over-estimated. If verified, this would result in a re-calibration of the spectrum according to a relative energy shift of $-10\%$ to $-15\%$. 

Remarkably, this energy shift agrees with predictions from a matching procedure of CR spectra assuming an early onset in extra-galactic proton dominance at about $10^9$~GeV~\cite{Aloisio:2006wv}. The calibration by the ``dip'' in these proton spectra resulting from electron-positron pair production on the CMB predicts a $-10\%$ and $20\%$ energy shift of the AGASA and HiRes data, respectively. We have incorporated this re-calibration into our analysis and show also the effect of the preliminary data from the Pierre Auger Observatory (PAO)~\cite{Sommers:2005vs} on our results.

\begin{figure*}[t]
\begin{minipage}[c]{\linewidth}
\centering\includegraphics[width=\linewidth]{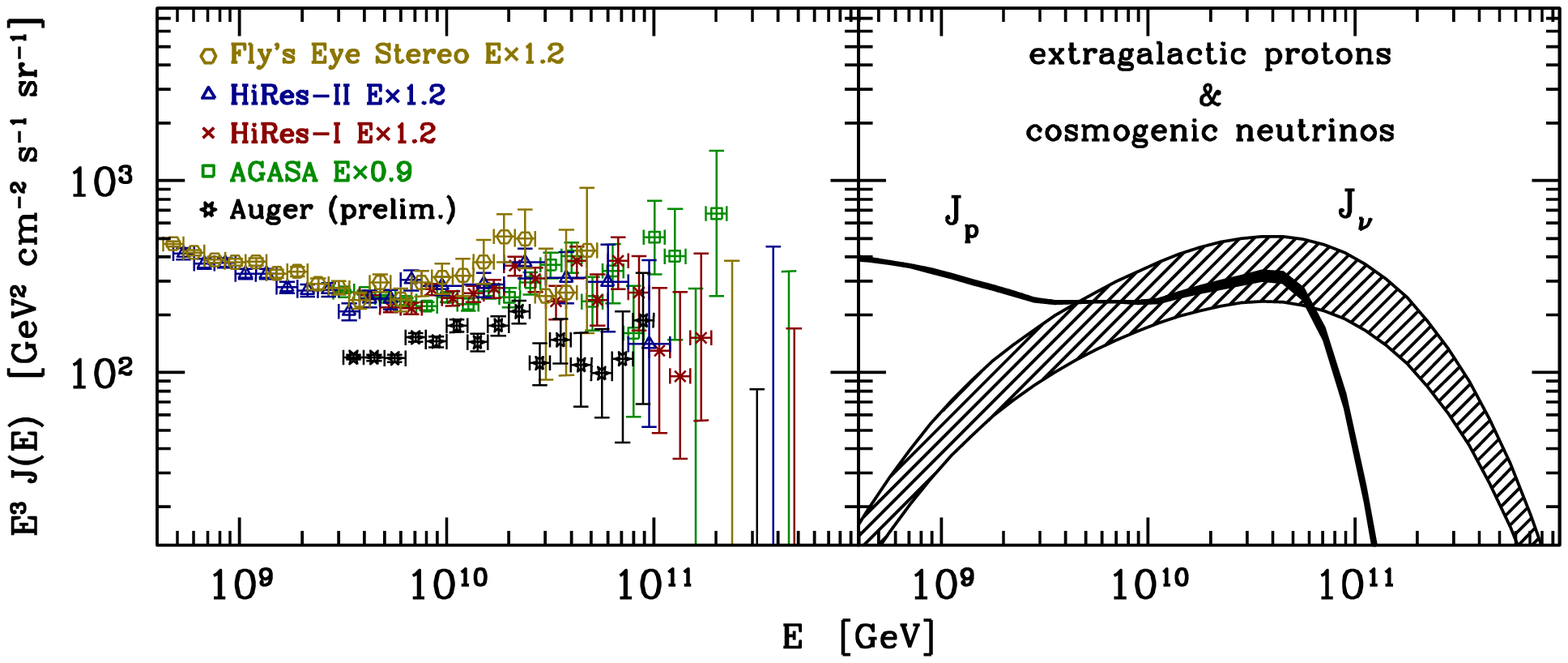}
\caption[]{{\rm Left panel:} The cosmic ray spectra from AGASA, Fly's Eye Stereo, Hires-I/II (with energy shift given in Ref.~\cite{Aloisio:2006wv}), and PAO. {\rm Right panel:} The flux of extra-galactic protons and cosmogenic neutrinos from CMB interactions corresponding to the 99\% CL of the fit.}
\label{ahlers_fig1}
\end{minipage}
\end{figure*}

\section{Cosmic Sources of Protons and Neutrinos}

We assume that extra-galactic protons dominate the CR spectrum above $5\times10^8$~GeV. A convenient parameterization of a spatially homogeneous and isotropic source luminosity is given by
\begin{gather}\label{fluxp}
  \mathcal{L}_{\rm CR}({z,E}) \propto{(1+z)^n}\,{E^{-\gamma}\,e^{-\frac{E}{E_\mathrm{max}}}},\qquad
 z_{\rm min}<z<z_{\rm max},
\end{gather}
which accounts for an evolution with redshift $z$ and an exclusion of nearby ( $z_\mathrm{min}$) and early ($z_\mathrm{max}$) sources. In our analysis we have keep these parameters fixed at $z_\mathrm{min} = 0.012$, corresponding to $r_\mathrm{min} \approx 50\, \mathrm{Mpc}$, and $z_\mathrm{max} = 2.0$.  The power-law injection is effectively limited to energies below $E_\mathrm{max}$, which we also fix at $10^{12}$~GeV.

The flux of protons originating at distant sources is subject to energy redshift and collisions with the interstellar photon background during propagation. The dominating interactions of UHE protons are $e^+e^-$ pair production and meson photo-production in the CMB (see e.g.~\cite{Engel:2001}). These effects are taken into account by means of propagation functions, which have been provided by the authors of Ref.~\cite{Fodor:2000yi}. The pions, which are resonantly produced at the GZK cut-off, decay into electron and muon neutrinos via the reaction chain $\pi^+\rightarrow \mu^+\nu_\mu \rightarrow\nu_\mu\bar \nu_\mu \nu_e\, e^+$ and the conjugate process. This constitutes a flux of UHE cosmogenic neutrinos which can be taken as a ``guaranteed source'' assuming extra-galactic protons as the highest energy CRs.

Depending on the environment of the proton accelerators neutrinos might also emerge in the decay of pions produced by  photo-production processes during acceleration~\cite{Waxman:1998yy}. The flux of extra-galactic neutrinos have been estimated in Ref.~\cite{Ahlers:2005sn} assuming a low cross-over to extra-galactic protons at about $5\times10^8$~GeV. In contrast to a transition at the ``ankle'' this flux of neutrinos from the source dominate the total neutrino flux below $10^9$~GeV and exhausts current experimental neutrino bounds at energies of about $10^7$~GeV. The total neutrino flux beyond the GZK cutoff is however dominated by cosmogenic neutrinos and we will only use this contribution for our statistical evaluation. 

\section{Exotic Neutrino Interactions}

The flux of cosmogenic neutrinos associated with a low cross-over to extra-galactic protons is comparable to the flux of protons at the GZK cut-off as can be seen from Fig.~\ref{ahlers_fig1}. A contribution of these neutrinos as super-GZK events requires a strong deviation from the feeble neutrino-nucleon interactions predicted by the Standard Model. In order to avoid large contributions in neutrino experiments the transition to a anomalously large cross section has to be very rapid. 
 
The realization of such a behavior has been proposed in scenarios beyond the (perturbative) Standard Model (SM), {\it e.g.} arising through compositeness, through electroweak sphalerons, through string excitations in theories with a low string and unification scale, through Kaluza-Klein modes from compactified extra dimensions, or through black hole and $p$-brane production, respectively. For details of these models we refer to some recent reviews, Refs.~\cite{Fodor:2004tr}. Skepticism about these scenarios has been raised in Ref.~\cite{Burdman:1997yg} based on limitations from s-wave unitarity or the naturalness of the couplings. However, these considerations do not necessarily apply for non-probative aspects like instant-ons.

In the following we will use a flexible parameterization of a strong neutrino-nucleon inelastic cross section ($\sigma_{\nu N}^{\rm new}$) focusing on three characteristic parameters: (i) the energy scale $E_{\rm th}$ of the new underlying physics, (ii) the amplification $\mathcal{A}$ compared to the SM predictions, and (iii) the width $\mathcal{B}$ of the transition between weak and strong interaction. A mathematical convenient parameterization is given by
\begin{equation}\label{cs}
  \log_{10}\left(\frac{\sigma_{\nu N}^{\rm new}}{\mathcal{A}\,\sigma_{\nu N}^{\rm SM}}\right)= \frac{1}{2}\left[1+\tanh\left(\log_\mathcal{B}\frac{E_\nu}{E_{\rm th}}\right)\right]\,\,.
\end{equation}
In general, experiments distinguish different CR primaries by their characteristic shower development in matter. For the sake of simplicity we will assume that the characteristics of the showers induced by strongly interacting neutrinos are indistinguishable from those induced by protons.  In particular, we assume for both primaries (i) a complete conversion of the incident energy into the shower, and (ii) equal detection efficiencies at the highest energies.

\begin{figure*}[t]
\begin{minipage}[c]{\linewidth}
\centering\includegraphics[width=\linewidth]{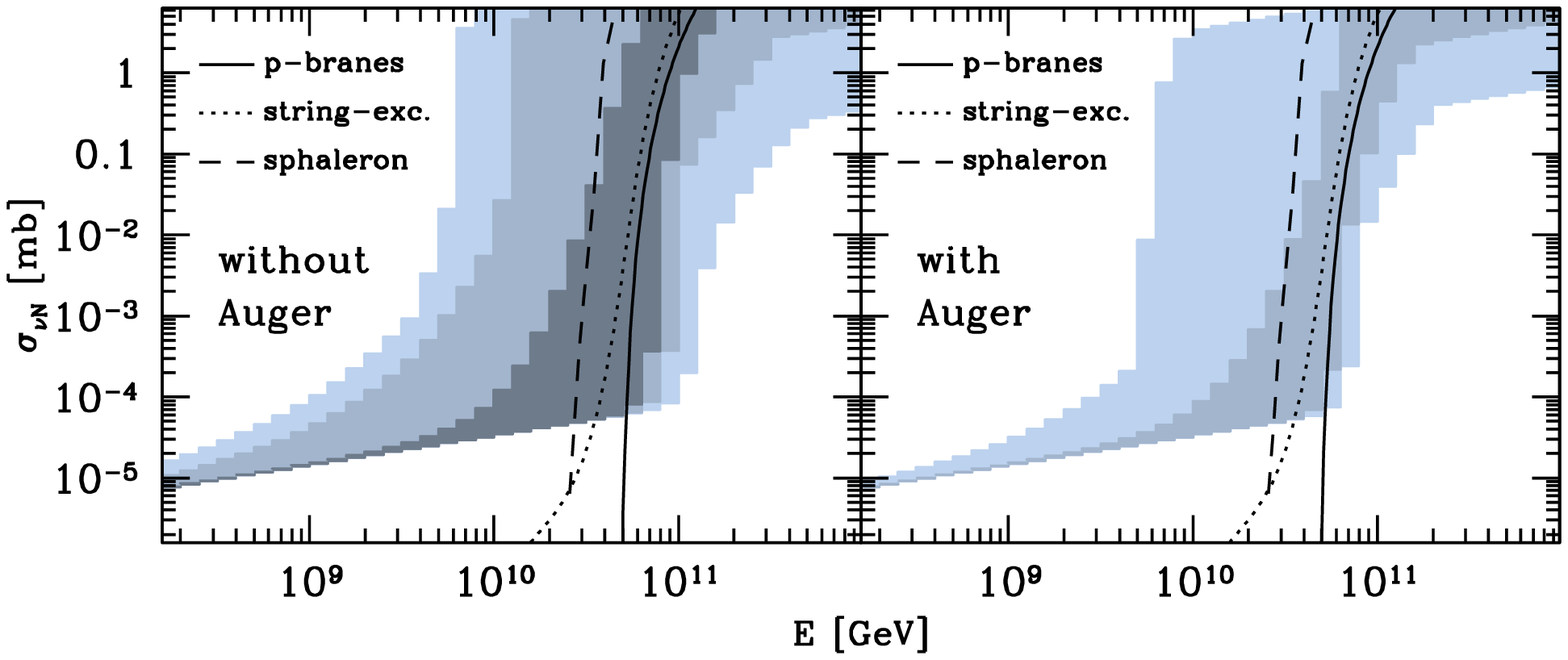}
\caption[]{The range of the cross section within the 95\%, 90\% and 68\% CL (outer to inner bands) obtained by the re-calibrated AGASA, Fly's Eye Stereo, and HiRes-I/II data, with (right panel) and without (left panel) the preliminary Auger data. Also shown are theoretical predictions of the neutrino-nucleon cross section enhanced by electroweak sphalerons, p-branes, and string excitations (see Ref.~\cite{Ahlers:2005zy} for details).}
\label{ahlers_fig2}
\end{minipage}
\end{figure*}

\section{Quantitative Analysis}

Following the procedure of Ref.~\cite{Ahlers:2005zy} we have evaluated the CR spectra of AGASA, Fly's Eye Stereo, HiRes-I/II, and PAO (preliminary data) assuming extra-galactic proton dominance above and $5\times10^8$~GeV and a contribution of strongly interacting neutrinos as super-GZK events. In contrast to Ref.~\cite{Ahlers:2005zy} we have re-calibrated the energy by a multiplicative factor $0.9$ and $1.2$ for the AGASA and HiRes data, respectively, following Ref.~\cite{Aloisio:2006wv}. For consistency with neutrino experiments we have used the search results of horizontal events at AGASA and contained events at RICE~\cite{Kravchenko:2003tc}. Figure~\ref{ahlers_fig2} shows the range of the cross section corresponding to the 68\%, 90\%, and 95\% confidence level (CL) from a goodness-of-fit test assuming cosmogenic neutrino fluxes from a low crossover scenario. The details of the statistical analysis and the approximations involved can be found in Ref.~\cite{Ahlers:2005zy}.

Compared to our previous results~\cite{Ahlers:2005zy} the re-calibrated spectrum is in a much better statistical agreement with the combined flux of protons and neutrinos from exotic interactions. This is mainly due to the fact that the calibration method ``optimizes'' the CR data to extra-galactic proton spectra below the GZK cutoff. However, within this particular model the re-calibrated data still favors an additional contribution from strongly interacting cosmogenic neutrinos. At the 95\% CL the corresponding neutrino-nucleon cross section should exhibit a steep increase by an amplification factor of $\mathcal{A}>10^3$. For $\mathcal{A}<10^5$ the transition should be very rapid ($\mathcal{B}<10$) at about $5\times10^{11}$~GeV. The allowed parameter space further shrinks if we also include the preliminary PAO data.

As an illustration of the result, we have considered three models of a rapidly increasing neutrino-nucleon cross section based on electroweak sphalerons~\cite{Han:2003ru}, $p$-branes~\cite{Anchordoqui:2002it} and string excitations~\cite{Burgett:2004ac}. The details can be found in Ref.~\cite{Ahlers:2005zy}. A separate fit of the source luminosity (Eq.~(\ref{fluxp})) with these cross section using the re-calibrated data gives a statistical acceptance at the 81\%, 65\%, and 60\% CL (without PAO), respectively. If we also include the PAO data in the fit, these values change to 98\%, 90\%, and 88\%, respectively, which is in reasonable good agreement with the results of Fig.~\ref{ahlers_fig2}.

\section{Conclusions}

We have shown that in the absence of close-by sources exotic interactions of cosmogenic neutrinos might extend the cosmic ray spectrum beyond the GZK cut-off. Such a behavior is predicted in various extensions of the Standard Model, but might also be due to non-perturbative aspects like electro-weak sphalerons. In extension to our analysis of Ref.~\cite{Ahlers:2005zy} we have presented here the results of a fit to the AGASA and HiRes data, shifted according to the ``dip''-calibration from extra-galactic proton spectra~\cite{Aloisio:2006wv}. With these modifications a contribution of neutrinos is still favored by the data. The inclusion of preliminary data of the Pierre Auger Observatory weakens the results of the goodness-of-fit test.

\frenchspacing
%\\bibliography{refs}
%\bibliographystyle{h-physrev3_mcite}

\setcounter{section}{0}
\setcounter{subsection}{0}
\setcounter{figure}{0}
\setcounter{table}{0}
\setcounter{equation}{0}
\newpage

\end{document}